\documentclass[10pt,aps,prb,twocolumn,showpacs,eqsecnum,superscriptaddress,floatfix]{revtex4-1}

\usepackage{graphicx}
\usepackage{epstopdf}
\usepackage{color}\usepackage{enumerate}
\usepackage{amsmath}\usepackage{amssymb}\usepackage{stmaryrd}\usepackage{wasysym}
\usepackage{multirow}
\usepackage{float}
\usepackage{longtable,booktabs}
\usepackage{diagrams}
\usepackage{hyperref}\usepackage{url}
\usepackage{subfigure}%
\usepackage{dsfont}

\newcommand{\hexeq}{$\varhexagon$-eq.}

\begin{document}

\title{Theory of Twist Liquids: Gauging an Anyonic Symmetry}

\author{Jeffrey C. Y. Teo}\email{jteo@virginia.edu}
\affiliation{Department of Physics, University of Virginia, Virginia 22904, USA}
\affiliation{Department of Physics and Institute for Condensed Matter Theory, University of Illinois at Urbana-Champaign, 1110 West Green Street, Urbana,  Illinois 61801-3080, USA}
\author{Taylor L. Hughes}
\affiliation{Department of Physics and Institute for Condensed Matter Theory, University of Illinois at Urbana-Champaign, 1110 West Green Street, Urbana,  Illinois 61801-3080, USA}
\author{Eduardo Fradkin}
\affiliation{Department of Physics and Institute for Condensed Matter Theory, University of Illinois at Urbana-Champaign, 1110 West Green Street, Urbana,  Illinois 61801-3080, USA}
\date{\today}


\begin{abstract} Topological phases in $(2+1)$-dimensions are frequently equipped with global symmetries, like conjugation, bilayer or electric-magnetic duality, that relabel anyons without affecting the topological structures. Twist defects are static point-like objects that permute the labels of orbiting anyons. {\em Gauging} these symmetries by quantizing defects into dynamical excitations leads to a wide class of more exotic topological phases referred as {\em twist liquids}, which are generically non-Abelian. We formulate a general gauging framework, characterize the anyon structure of twist liquids and provide solvable lattice models that capture the gauging phase transitions. We explicitly demonstrate the gauging of the $\mathbb{Z}_2$-symmetric toric code, $SO(2N)_1$ and $SU(3)_1$ state as well as the $S_3$-symmetric $SO(8)_1$ state and a non-Abelian chiral state we call the ``4-Potts" state. \end{abstract}

\maketitle

\section{Introduction}
\label{sec:introduction}

Topologically ordered phases in two dimensions are long range entangled states of quantum matter that support fractional quasiparticle excitations with anyonic statistics (for detailed explanations of these concepts see Refs.[\onlinecite{Wilczekbook,Fradkinbook,Wenbook}]). Unlike ordinary phases of matter that are characterized by an order parameter and broken symmetries, topological phases can be featureless, exhibit no apparent local order parameters,
and are not associated with spontaneous symmetry breaking. 
Recently the concept of a topological phase has been broadened 
to include symmetry protected/enriched topological phases (SPT/SET)~\cite{ChenGuLiuWen12, LuVishwanathE8, MesarosRan12, EssinHermele13, WangPotterSenthil13, BiRasmussenSlagleXu14, Kapustin14}, the most notable example being the fermionic topological insulator phases protected by time-reversal symmetry.\cite{HasanKane10, QiZhangreview11, KaneMele2D1, Molenkamp07, MooreBalents07, Roy07, FuKaneMele3D, QiHughesZhang08, Hasan08}

 Many topologically ordered phases can also manifest a new type of discrete symmetry known as an {\em anyonic symmetry} (AS).\cite{Kitaev06,EtingofNikshychOstrik10,Bombin,YouWen,BarkeshliJianQi,TeoRoyXiao13long,khan2014,BarkeshliBondersonChengWang14} For a given topological order there is a corresponding set of anyonic quasiparticles (QPs), and the AS group acts on the set of QPs to permute the anyon labels. This is similar to, for example, the permuting of a discrete set of ground states in a conventional symmetry broken phase by the discrete symmetry operations.
Until recently, these types of anyonic symmetries had rarely manifested any physically measurable consequences.  However, the appearance of the idea of twist defects in topological phases \cite{Kitaev06, EtingofNikshychOstrik10, barkeshli2010, Bombin, Bombin11, KitaevKong12, kong2012A, YouWen, YouJianWen, PetrovaMelladoTchernyshyov14, BarkeshliQi, BarkeshliQi13, BarkeshliJianQi, MesarosKimRan13, TeoRoyXiao13long, teo2013braiding, khan2014, BarkeshliBondersonChengWang14} has yielded some tantalizing opportunities through which these symmetries may be explored. This has attracted attention because the twist defects, despite being essentially point-like semiclassical objects, can often behave as non-Abelian anyons even when the underlying topological order with AS is Abelian. Thus, the remarkable possibility of generating non-Abelian objects from much simpler Abelian theories has drawn an intense focus on this subject. 

To gain an intuitive understanding of twist defects it is useful to compare the twist defects corresponding to a discrete AS group with conventional magnetic flux excitations of a discrete gauge theory\cite{BaisDrielPropitius92,Bais-2007,Propitius-1995,PropitiusBais96,Preskilllecturenotes,Freedman-2004,Mochon04}. In both cases, when the trajectory of an anyon QP in the theory encircles a flux/defect, there is an associated Aharonov-Bohm effect. For gauge fluxes, this effect yields an action of the gauge group on the QP, which depends on the charge of the quasiparticle and the type of flux. For twist defects, there will instead be an action of the AS group which 
permutes the anyon type of the encircling anyon QP. The major difference is that, unlike gauge fluxes in a discrete gauge theory, twist defects in a topologically ordered phase are not excitations of the underlying quantum Hamiltonian. Instead, they are extrinsic objects that depend on the topological textures of some non-dynamical background fields. When the spatial dependence of the background fields is arranged in a topologically non-trivial defect structure, then exotic bound states can be trapped at the defect location, e.g. non-Abelian anyons. 

As of now there is a large collection of systems in which twist defects have been predicted to exist.  One simple example is a heterostructure formed from the edge of a 2D topological insulator (TI), an s-wave superconductor (SC), and a ferromagnet (FM).\cite{FuKane08,FuKaneJosephsoncurrent09} The interface between the SC and FM on the TI edge can trap protected Majorana zero modes.\cite{Majorana37, Wilczek09, HasanKane10, QiZhangreview11, Beenakker11, Alicea12} Thus, the topological insulator mass gap, the superconducting pairing, and the ferromagnetic order parameter  provide the relevant background fields, and the SC-FM interface on the edge provides the topological texture. More exotic fractional Majorana modes (also known as parafermions) have also been proposed to exist in heterostructures of more complicated fractional topological phases.\cite{ClarkeAliceaKirill,LindnerBergRefaelStern,MChen,mongg2,Vaezi} These defect bound states are  non-Abelian objects that can robustly store quantum information non-locally in space, and could be powerful enough for universal topological quantum computation.\cite{Kitaev97, OgburnPreskill99, Preskilllecturenotes, FreedmanKitaevLarsenWang01, ChetanSimonSternFreedmanDasSarma, Wangbook, SternLindner13} These heterostructure designs amount to an explicit construction of twist defects. 

Another prevalent set of examples of systems with twist defects arises in lattice models and bi-layer systems where the AS group is intertwined with the set of lattice symmetries. In these cases, crystalline defects, which are formed by the slow modulation of the lattice or layer parameters, can act as twist defects. 
Some examples include dislocations in the Kitaev toric code model,\cite{Kitaev97,Kitaev06,Bombin,PetrovaMelladoTchernyshyov14} the Wen plaquette rotor model\cite{YouWen,YouJianWen}, and the tricolor rotor model\cite{BombinMartin06, Bombin11, TeoRoyXiao13long}. Similar twist defects could also manifest as disclination defects in topological nematic states\cite{BarkeshliQi} and bilayer fractional quantum Hall (FQH) states\cite{teo2013braiding,BarkeshliQi13}. Disclinations are topological defects of the underlying lattice supporting the topological phase (or of an emergent nematic order in the topological fluid\cite{Maciejko2013,You2014}), which act as curvature sources in the local geometry that is coupled to the topological phase. These defects can effectively rotate the anyon type of an orbiting quasiparticle according to some anyonic symmetry operation, instead of solely acting to rotate a local coordinate frame.

Until now most of the work on twist defects has focused on their exciting semiclassical properties. However, we are interested in understanding the consequences of their quantum dynamics. Given a parent (topological) phase with a global (anyonic) symmetry we can ask what new phase emerges when defects associated to the non-trivial elements of the  symmetry group become deconfined quantum excitations, i.e. when the symmetry is converted from a global symmetry to a local one. There have already been some descriptions of a quantum dynamical theory of twist defects, but only in the special cases of a bilayer FQH state\cite{BarkeshliWen11}, and an Abelian $\mathbb{Z}_N$ theory\cite{BarkeshliWen10, BarkeshliWen12}. 
In this article we will provide a general framework for quantizing bulk defects in topological phases with anyonic symmetry, and apply this to a wide range of examples. 

Before we summarize the results of the article, let us describe our conceptual approach. Our quantum theory of twist defects can be thought of as a generalization of the deconfined phases of discrete gauge theories. In fact, in the simplest case where we start with a phase without topological order (i.e., the set of anyons is only the vacuum), but with some global symmetry group $G$, then quantum `twist' defects are precisely the same as the conventional magnetic fluxes of a local gauge symmetry group $G$. Once fluxes are dynamical they, along with the gauge symmetry charges, represent anyonic excitations of a new topological phase, known as a quantum double $D(G)$, for the discrete symmetry group $G$.\cite{BaisDrielPropitius92,Bais-2007,Propitius-1995,PropitiusBais96,Preskilllecturenotes,Freedman-2004,Mochon04} The resulting quantum phase arising from converting $G$ from a global symmetry to a local symmetry, has non-trivial topological order as exemplified, for example, by its non-vanishing topological entanglement entropy $S_{\rm topo}=-\log\mathcal{D}$, where the total quantum dimension $\mathcal{D}$ is given by the order of the group $|G|$.\cite{KitaevPreskill06} 

The gauging process itself represents a topological phase transition 
\begin{align}
\begin{diagram}
\stackrel{\mbox{Trivial Boson Condensate}}{\mbox{with Global Symmetry}}&\pile{\rTo^{\mbox{\small Gauging}}\\\lTo_{\mbox{\small Condensation}}}&D(G)
\end{diagram}
\label{DGTgauging}
\end{align} 
through which gauge fluxes become deconfined. The reverse (confinement) transition can be driven by condensing all bosonic gauge charges,~\cite{Bais-2007, BaisSlingerlandCondensation, Kong14} and the resulting confined state is a trivial phase with a global $G$ symmetry. 
While the description of the deconfined phase of discrete gauge theories is well established, the gauging of anyonic symmetries that permute anyon quasiparticles in a parent topological phase is largely unexplored. The phase transition of gauging the anyonic symmetry is in many ways very similar to a quantum version of the melting of classical 2D phases with broken symmetries,\cite{chaikin-1995, Nelson-2002} especially when, as mentioned above, the AS group is intertwined with underlying lattice symmetries. In fact, if we consider a topological phase on a torus, we can heuristically say that an anyonic symmetry is {\em weakly} broken, since the topologically degenerate ground states are in one-to-one correspondence to the anyon types, and are thus {\em not} necessarily invariant under the anyonic symmetry. Hence, the symmetry is ``restored" when twist defects (or gauge fluxes) are deconfined and become  dynamical at the quantum level. This is qualitatively a quantum mechanical analog of the defect proliferation in a thermal Kosterlitz-Thouless disordering transition,\cite{KosterlitzThouless} e.g. between a crystal and a liquid or liquid-crystal. We will therefore refer to the resulting topological phase after gauging an anyonic symmetry as a {\em twist liquid}. Hence, the goal of our article is to give a general construction for twist liquids, and to provide a set of explicit examples to illustrate the construction.

From the nature of the twist defects, it is clear that the promotion of a global AS to a local one will, in general, not result in a twist liquid that is still a discrete gauge theory, even if the initial underlying parent theory with AS is a discrete gauge theory. The easiest way to see this is that the quantum dimensions of twist defects are usually irrational numbers, while the quantum dimensions of anyons in a discrete gauge theory are always integer-valued. Since the twist defects end up as anyonic excitations in the twist liquid, the resulting theory cannot be a discrete gauge theory. Despite this complication, the phase transition still resembles that of the diagram in Eq.~\eqref{DGTgauging} except for the fact that the original phase is also topologically ordered: 
\begin{align}
\begin{diagram}
\stackrel{\mbox{Topological Phase with}}{\mbox{Global Anyonic Symmetry}}&\pile{\rTo^{\mbox{\small Gauging}}\\\lTo_{\mbox{\small Condensation}}}&\mbox{Twist Liquid.}
\end{diagram}
\label{gaugingtransition}
\end{align} 
Additionally, we will show that the transition to the twist liquid phase increases the total quantum dimension by a factor of the order of the anyonic symmetry group \begin{align}
\mathcal{D}_{\mbox{\small twist liquid}}=\mathcal{D}_0\; |G|
\end{align} 
where $\mathcal{D}_0$ is the total quantum dimension of the original topological phase with global AS. This is a generalization of the result one finds when gauging a conventional global symmetry starting from a trivial boson condensate. 

There is also an interesting way to understand the reverse transition from the twist liquid back to the parent state with global AS. Once one has a known twist liquid, it can be reduced, via \emph{anyon condensation},\cite{BaisSlingerlandCondensation, Kong14} back to the original topological phase with a global symmetry. Thus, in some familiar examples, we may already have an educated ansatz for the possible twist liquid phase. However, this approach is not 
{\it a priori} well defined as it requires knowledge of the resulting phase for which one is trying to solve,  e.g. the knowledge of data such as the quasiparticle spins and their quantum dimensions. As we will show, there are further complications since generically there are multiple twist liquid candidates that are related to same parent phase by an anyon condensation process. Hence, a systematic scheme for this gauging transition, including a Hamiltonian model and a general construction of the QPs for the resulting twist liquid phase is therefore 
needed. This is the primary motivation for our work.

As an aside we note that in this article we focus on the structure of anyons in a bulk topological phase. However, in many cases one knows that the anyons of the bulk topological phase are connected to the primary fields of an edge conformal field theory (CFT) by a bulk-boundary correspondence.\cite{FrohlichGabbiani90,MooreRead,ReadRezayi} The gauging of a bulk anyonic symmetry is intimately related to the orbifold of an edge CFT,\cite{bigyellowbook,Ginsparg88,DijkgraafVafaVerlindeVerlinde99,CappelliAppollonio02} where bulk twist defects correspond to edge twist fields. We will mainly focus on the two dimensional bulk -- where the phase transition actually take place -- although all examples considered have CFT orbifold analogues.\cite{chenroyteoinprogress}

\subsection{Summary of results}
\label{sec:summary}


To summarize, in this article, we provide a systematic approach for constructing twist liquids beginning with a general (Abelian) topological phase with global anyonic symmetry. We show that the 
gauging transition 
program suggested by Eq.~\eqref{gaugingtransition} can be modeled by a Levin-Wen string-net lattice model.\cite{LevinWen05} This model can be exactly solved, and has an energy gap at the extreme limits on both sides of the transition, i.e.~when the defect string tension is either absent (representing the deconfined twist liquid phase) or infinite (representing the confined phase with only global symmetry). In our approach, the data we input into the Levin-Wen model (mathematically also known as the {\em Drinfeld center}\cite{Kasselbook,BakalovKirillovlecturenotes}) is the fusion structure of both the 
quasiparticles of the anyon model and the set of twist defects of the underlying topological state with global AS. We will refer to this incipient fusion structure 
as a \emph{defect fusion category}.\cite{EtingofNikshychOstrik10,TeoRoyXiao13long,teo2013braiding,BarkeshliBondersonChengWang14} The input data also includes the fusion rules between defects and quasiparticles, as well as a consistent set of basis transformations known as the $F$-symbols.\cite{Kitaev06, Walkernotes91, Turaevbook, FreedmanLarsenWang00, BakalovKirillovlecturenotes, Wangbook} The output of the model is a twist liquid with dynamical quantum defects/fluxes that carry well-defined exchange and braiding statistics.

To serve as an explicit model to introduce our construction, we will carefully demonstrate the gauging procedure in detail for the simple case of the toric code~\cite{Kitaev97}. This system is equivalent 
to the ultra-deconfined limit of the  $\mathbb{Z}_2$ discrete gauge theory, and has an electric-magnetic anyonic symmetry. Its corresponding twist liquid phase has generally been accepted to be a non-chiral $\mbox{Ising}$ state. We will show that this conclusion can be supported by several alternate, and more conventional, arguments: (i) using the structure of the edge theory when the electric-magnetic symmetry is gauged and appealing to a bulk-boundary correspondence, (ii) arguing via the connection between the toric code and a non-chiral $(p_x+ip_y)\times(p_x-ip_y)$ superconductor~\cite{SchnyderRyuFurusakiLudwig08,Kitaevtable08,QiHughesRaghuZhang09}, and (iii) working backwards by appealing to anyon condensation~\cite{BaisSlingerlandCondensation, Kong14}. We will also illustrate how the resulting twist liquid fits into our more general framework using the Levin-Wen construction~\cite{LevinWen05}. This example serves as a twist liquid prototype, and shows the essential anyon structure that will appear in more general systems. For instance, a non-dynamical twist defect in the toric code is promoted to an Ising anyon excitation~\cite{MooreRead,Kitaev06} in the twist liquid via the gauging procedure, and each of the nine resulting anyons in the twist liquid can be seen as certain flux-charge-quasiparticle composites. 

Although the Levin-Wen model is not chiral, remarkably we show that our gauging construction can also be applied to chiral systems after a doubling procedure that introduces an inert chiral counter-partner with trivial anyonic symmetry action. The total system is then non-chiral and can be represented using the Levin-Wen model. Furthermore since the AS only acts on one of the chiral components we can directly infer the twist liquid structure arising from gauging the AS of the chiral state.
We will demonstrate this by studying the transition between the Abelian, charge-conjugation symmetric $SU(3)_1$ FQH state and the non-Abelian $SU(2)_4$ FQH state, both of which are chiral. 

Another key set of our results focuses on an exceptional case where the anyonic symmetry group of the underlying topological phase is non-Abelian. As shown in Ref.~[\onlinecite{khan2014}], the simplest example of a system with a non-Abelian AS group is the chiral $SO(8)_1$ FQH state. This topological state has been of recent interest as it was proposed to occupy the surface of certain topological paramagnets or bosonic SPTs in three space dimensions.\cite{VishwanathSenthil12,BurnellChenFidkowskiVishwanath13,WangSenthil14,WangPotterSenthil13} Its anyon content has three mutually-semionic fermions, and supports the triality anyonic symmetry group $S_3$, the group of permutations of the three fermions. We will show that gauging a $\mathbb{Z}_2$ or $\mathbb{Z}_3$ subgroup of $S_3$ leads to an $(\mbox{Ising})^2$-like state or an $SU(3)_3$-like state respectively, while gauging the full $S_3$ symmetry gives rise to a new non-Abelian topological state with total quantum dimension $\mathcal{D}=12$ and twelve anyon types.

We also present two more instructive examples: the sixteenfold periodic $SO(N)_1$ theories at level $1$ which have a $\mathbb{Z}_2$ AS when $N$ is even, and a state we call the chiral ``$4$-Potts" state~\cite{Ginsparg88,DijkgraafVafaVerlindeVerlinde99,CappelliAppollonio02} which has an $S_3$ AS. The states in the $SO(N)$ series correspond to Abelian topological phases described by Abelian Chern-Simons theories with $K$-matrices equal to the Cartan matrix of the Lie algebra $so(N)$.\cite{khan2014} Each of these theories has a $\mathbb{Z}_2$ AS which we gauge to generate a family of twist liquids. We illustrate the periodicity of these topological phases, i.e. $SO(N)_1 \equiv SO(N+16)_1,$ and show the interesting connections and phase transitions between the resulting family of twist liquids. On the other hand, the chiral ``$4$-Potts" state is remarkable because it has a non-Abelian triality symmetry, like $SO(8)_1$, but additionally it is already a non-Abelian topological phase \emph{before} gauging the AS. In fact, it is the only non-Abelian parent state that we explicitly consider in our article, and we use it as a test case to compare to our general results for Abelian parent states. We also make some brief comments about the bi-layer toric code system, which are connected to the analysis of the ``$4$-Potts" state. 

In addition to a set of interesting and instructive examples, we also report several (more formal) general results. First, we have classified the quantum anyonic symmetry operations and show that they can form an unusual non-symmorphic structure. We have also illustrated how the twist defect fusion rules and basis transformations (the $F$-symbols) can be almost completely determined by resolving Wilson strings around the twist defects. This efficient computational procedure was presented in two previous works of one of the authors,\cite{TeoRoyXiao13long,teo2013braiding} and will be demonstrated in detail for the toric code example, and selectively for the $SO(8)_1$ example.  Interestingly, we also show that there is additional freedom in the defect category structure of a given AS group that cannot be specified by just knowing how anyons are relabeled by the AS. Moreover, we find there can also be obstructions to defining quantum anyonic symmetries and defect fusion structures. These complications arise from the phase ambiguities and inconsistencies of the quantum versions of the symmetry operators. For example, the quantum representation of the product of two anyonic symmetry operations $\widehat{MN}$ may differ from the product of the two representations $\widehat{M}\widehat{N}$ by a unitary phase
{\it i.e.} a cocycle. This can consequently result in inequivalent, or inconsistent (when an obstruction exists), sets of allowed defect fusion rules. In the case when there are inequivalent sets of defect fusion rules then there are several distinct twist liquids corresponding to the same parent state.

Since we are interested in classifying all the possible twist liquid outcomes from an initial parent state then we must take these complications into account. Remarkably, we find that there is a simple physical interpretation for the multiple twist liquid outcomes arising from a single parent state. To account for the multiple possibilities one must allow the underlying topological parent state, which has a global AS symmetry, to be combined/stacked on top of non-trivial 2D SPTs, or the surface of 3D SPTs, with the same symmetry. The new composite parent state would have the same topological order as the original parent phase since the added layer(s) are only short-range entangled and do not contribute new anyons. However, we show that this can lead to an inequivalent set of basis transformations (differentiated by certain Frobenius-Schur indicators~\cite{FredenhagenRehrenSchroer92,Kitaev06}) or even inconsistency (violation of the pentagon identity) in the defect fusion category. Since the string-net construction is determined by the defect fusion category, these quantum phase ambiguities and inconsistencies give rise to multiple gauging outcomes, as well as possible obstructions to the procedure all together. Similar phenomena were recently proposed for SPT's without topological order.\cite{ChenBurnellVishwanathFidkowski14} By accounting for all of these subtle distinguishing characteristics one can classify the set of possible twist liquid outcomes when given a parent topological phase with an AS group.

This paper is organized as follows. In Section \ref{sec:Zkgaugetheory} we present our gauging procedure by carrying it out explicitly for the toric code. This section is split into several pieces. First, we review the relevant topological phase beginning from a lattice model construction. Then we discuss the resulting defect fusion category.  Finally, we gauge the anyonic symmetry by first using conventional arguments, and then systematically using our construction.  We follow this with a long section which describes the general gauging procedure, i.e.~Section \ref{sec:gauginganyonicsymmetries}. In this section we begin by reviewing the notion of anyonic symmetries and the subtleties involved in promoting these symmetries to quantum symmetry operators (Sec.~\ref{sec:anyonicsymmetries}). Then we show how to construct the pieces of the relevant defect fusion category given a parent topological phase and an anyonic symmetry (Sec.~\ref{sec:twistdefects}). From here we show how to derive the quasiparticle content of the resulting twist liquid (Sec.~\ref{sec:QPstructuretwistliquid}). Then we show that we can use the Levin-Wen string-net construction to model the parent and twist liquid phases and the phase transition between them (Sec.~\ref{sec:stringnet}). Finally, we close this section by discussing the effects of gauging the global symmetry of SPT phases and how this can change the resulting twist liquid when such an SPT is combined with a parent topologically ordered state (Sec.~\ref{sec:gaugingSPT}).

The remaining part of the main body of the article consists of four examples, each of which illustrates a distinct aspect in the gauging procedure. Section~\ref{sec:gauging-em-bilayer-su3} shows the gauging of the chiral  $SU(3)_1$ fractional quantum Hall state. This section demonstrates how the chiral twist liquid, which arises after gauging the charge-conjugation symmetry, can be embedded in a non-chiral string-net model. The sixteen fold $SO(N)_1$ series in Section~\ref{sec:so(N)} shows the relationships between a series of Abelian parent states (for even $N$) and a series of Ising-like twist liquids (for odd $N$). The $SO(8)_1$ state in this series is special since it is the simplest Abelian state with a non-Abelian anyonic symmetry. Section~\ref{sec:so(8)symmetry} demonstrates the gauging of its full $S_3$ triality symmetry. Finally, Section~\ref{sec:4statePotts} considers the $S_3$ symmetric non-Abelian chiral ``4-Potts" phase. It illustrates the most general scenario where the parent state itself is non-Abelian. Building on this we make some comments at the end of this section about gauging the bi-layer symmetry in the bi-layer toric code.

The article is concluded in Section~\ref{sec:conclusion}, where the reader will find an informal summary that highlights the main features while avoiding most technicalities. In fact, we point the reader to that section if they only want a flavor of the contained results. We also include a set of detailed Appendices to which we often refer in the main body of the article. 

During the preparation of this manuscript we learned of 
the 
related 
work of Barkeshli~\emph{et.~al.}\cite{BarkeshliBondersonChengWang14}, and the unpublished work of Fidkowski~\emph{et.~al.}\cite{TarantinoLindnerFidkowskiunpubplished,FidkowskiKitaevunpubplished}. The results we present in this work overlap with parts of these papers, but were determined independently  and offer complementary discussions and points of view.
Generally, the classification and obstructions of global quantum symmetries, as well as the corresponding defect fusion category in Sections~\ref{sec:globalquantumsymmetries} and \ref{sec:twistdefects} are essentially physical descriptions of the mathematical work by Etingof~{\em et.~al.}\cite{EtingofNikshychOstrik10}. The description in terms of the $G$-crossed theory is presented and defined in Ref.\onlinecite{BarkeshliBondersonChengWang14}, and we provide a systematic approach, especially for  general Abelian parent states, to derive the defect objects and their basis transformations. In our work, we note that we have avoided the concept of exchange and braiding of extrinsic defects since they are classical objects. Instead, we show that the spin and braiding properties of the resulting quantum anyons after gauging can be directly determined by a (relative) Drinfeld construction (Sec.\ref{sec:stringnet}). This is nicely controlled by a string-net Hamiltonian that depends only on the fusion properties of defects. These are explicitly demonstrated in the detailed examples studied in this article, all of which illustrate salient and original features of the gauging process. We have tried to note throughout the paper when how our major results overlap and relate with those of Ref. \onlinecite{BarkeshliBondersonChengWang14}.  Unfortunately, we were not able to do the same for  Ref. \onlinecite{TarantinoLindnerFidkowskiunpubplished,FidkowskiKitaevunpubplished} yet, as it is not currently available. The combined results and examples in all three works should establish a comprehensive theory.

\section{Gauging the electric-magnetic symmetry of the Kitaev toric code}
\label{sec:Zkgaugetheory}

\begin{figure}[htbp]
\includegraphics[width=0.3\textwidth]{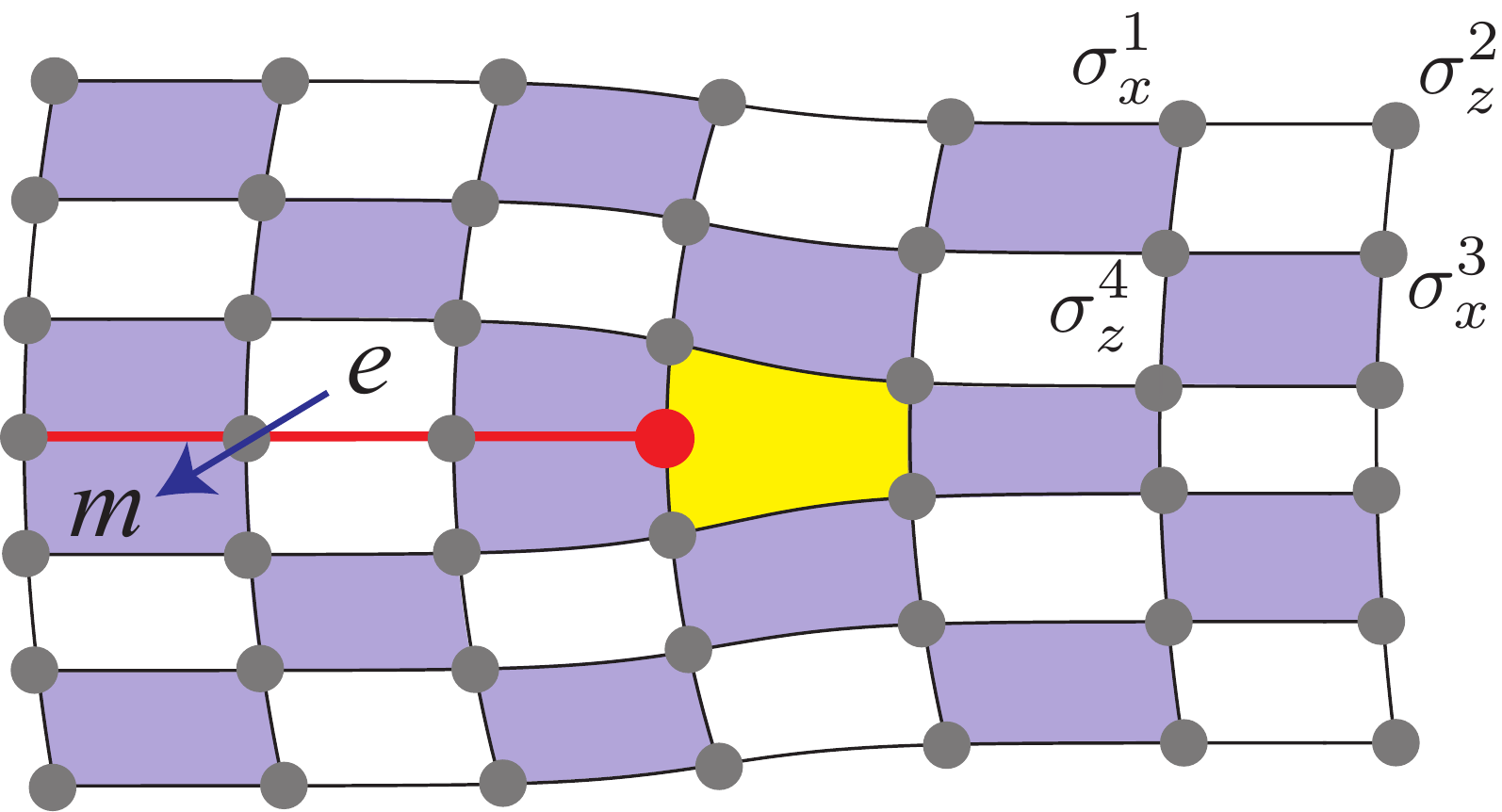}
\caption{Wen plaquette-model (for our purposes this is essentially equivalent to the toric code) with a dislocation (yellow pentagon).}
\label{fig:toriccode}
\end{figure}

The toric code\cite{Kitaev97} is an exactly solvable model that describes the topological phase of a $2+1$-dimensional $\mathbb{Z}_2$ gauge theory\cite{BaisDrielPropitius92,Propitius-1995,PropitiusBais96,Preskilllecturenotes,Freedman-2004,Mochon04} deep in its deconfined phase~\cite{JalabertSachdev91,SenthilMatthew00,MoessnerSondhiFradkin01,ArdonneFendleyFradkin04}. To facilitate the later discussion of twist defects in this system, 
we found it convenient to define the plaquette model proposed by Wen,\cite{Wenplaquettemodel}
a model on a rectangular checkerboard lattice where there is a spin-$1/2$ degree of freedom at each vertex. This model has a topological phase which is equivalent to that found in Kitaev's original model of the toric code. 
The Hamiltonian $H=-\sum_P\hat{P}$ is a sum of plaquette operators of the form $\hat{P}=\sigma_x^1\sigma_z^2\sigma_x^3\sigma_z^4$, i.e. 
a product of spin operators at the surrounding vertices of the plaquette as shown in Fig.~\ref{fig:toriccode}.

Ground states of this model are simultaneous eigenstates of each $\hat{P}$ with $\hat{P}=+1$ for all plaquettes so that the energy is minimized. Quasiparticles are excitations localized at plaquettes with $\hat{P}=-1$. The excitations can only be diagonally transported to an adjacent plaquette with the same color, and this is executed by acting on the quantum state with a spin operator at the connecting vertex. Repeating this motion will leave a string of spin operators along the quasiparticle trajectory that connects plaquettes of the same color. Since the lattice is bi-partite it is easy to see there are two fundamental non-trivial quasiparticles: the charge $e$ and flux $m$ that live on oppositely colored plaquettes. They are bosons but obey mutual semionic statistics, i.e.  dragging one completely around the other will result in a $-1$ braiding phase. The flux and charge can also combine to form a composite quasiparticle $\psi=e\times m$ which, on the lattice, is equivalent to the excitation of two adjacent plaquettes with opposite color. The quasiparticle $\psi$ is a fermion due to the $-1$ twist phase upon a $360^\circ$ rotation of its internal structure.

The reason we chose this lattice model for the $\mathbb{Z}_2$ topological phase is that the bi-colored structure of the checkerboard lattice is intimately related to an electric-magnetic anyonic symmetry of the system. Switching the colors will exchange the labels of the charge and flux $e\leftrightarrow m,$ while keeping the fermion $\psi$ unchanged. This relabeling corresponds to an anyonic symmetry, as the operation does not alter the spin and braiding statistics of the set of quasiparticles. The symmetry is, however, non-local in this model as there is no local operator that can switch between quasiparticles with distinct anyon types. The non-locality can also be seen geometrically by noticing that swapping the plaquette colors in a local patch would violate the checkerboard pattern along the (possibly large) boundary of the patch.

To make a connection with other descriptions in the literature, we note that this topological phase can also be described by an Abelian Chern-Simons theory via the $K$-matrix formalism\cite{WenZee92,Wenedgereview} with the Lagrangian density 
\begin{align}
\mathcal{L}=\frac{1}{4\pi} K_{IJ}\alpha_Id\alpha_J+\alpha_I\mathcal{J}^I
\label{toriccodeCSaction}
\end{align} 
where we have used a 2-component $U(1)$ gauge field $\alpha_I$ with the $2\times2$ $K$-matrix $K=2\sigma_x$. Here $\mathcal{J}^1$ and $\mathcal{J}^2$ correspond to the currents of the charge $e$ and flux $m$ quasiparticles respectively. The action is invariant under the electric-magnetic (duality) symmetry. In this formulation this is equivalent to swapping the two gauge fields, $\alpha_1\leftrightarrow\alpha_2,$ and is represented by the matrix $M=\sigma_x$. The $K$ matrix is unchanged under conjugation by this operation, $MKM^T=K.$

\subsection{Twist defects}
\label{sec:toriccodetwistdefects}

As we discussed, the plaquette model gives rise to an {\em Abelian} topological phase where quasiparticles obey single-channel fusion rules, e.g. 
\begin{equation}
e\times m=\psi, \qquad m\times m=1.
\end{equation}
A quantum state can thus be specified by the anyon types of its quasiparticle excitations and their locations. Any exchange and braiding operations cannot alter the state other than multiplying it with a unitary phase, and therefore these operations mutually commute. This set of quasiparticles represents the parent topological phase.

Remarkably it was shown that this model, and such a $\mathbb{Z}_2$ topological phase in general, can support {\em non-Abelian} objects in the form of twist defects. For this particular lattice realization the twist defects take a simple form, they are realized as dislocations on the rectangular lattice\cite{Kitaev06,Bombin,YouWen,PetrovaMelladoTchernyshyov14} (see Fig.~\ref{fig:toriccode}). We will now review the fusion properties of these twist defects.

\begin{figure}[t]
\includegraphics[width=0.25\textwidth]{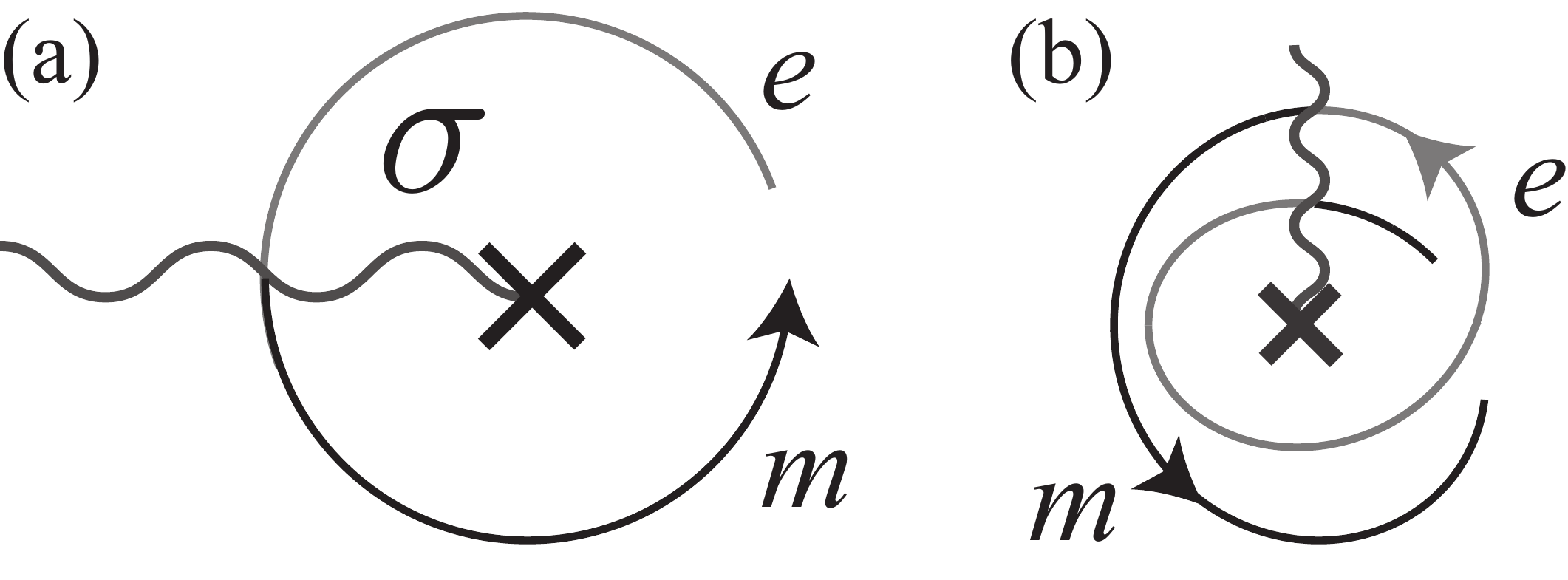}
\caption{(a) A quasiparticle changing type when traveling across a branch cut terminated at a twist defect $\sigma$. (b) The Wilson loop $\Theta$ that distinguishes defect species.}\label{fig:defect1}
\end{figure}

Microscopically, a dislocation of a two-dimensional square lattice is a 
topological singularity of the underlying (square in this case) lattice. Dislocations are characterized by a topological charge known as the Burgers vector. For the square lattice, a dislocation  is a line of atoms (lattice sites) that end at a trivalent vertex adjacent to a pentagon plaquette, i.e. a disclination dipole (see, e.g.~Refs.~[\onlinecite{chaikin-1995,Nelson-2002}]). The line of atoms plays the role of  a ``Dirac string'' (effectively a {\em branch cut}), and its position is definition-dependent (gauge-dependent). The plaquette operator at the pentagon in the Hamiltonian is modified to be $P_{\pentagon}^{\pm}=\pm\sigma_y^0\sigma_x^1\sigma_z^2\sigma_x^3\sigma_z^4$, where the additional $0^{th}$ site is the trivalent vertex, and the sign can be arbitrarily fixed locally at each defect. This operator commutes with all other plaquette operators and the model is still exactly solvable. However, the charge $e$ and flux $m$ quasiparticles can no longer be globally distinguished. This is because the bi-colored checkerboard pattern cannot be globally defined in the presence of a dislocation, and there is a branch cut -- represented by the red line in Fig.~\ref{fig:toriccode} -- originating from the defect where neighboring plaquettes share identical colors. As quasiparticles move diagonally from plaquette to plaquette, they change type across the branch cut according to the electric-magnetic anyon symmetry $e\leftrightarrow m$ (see Figs.~\ref{fig:toriccode} and \ref{fig:defect1}(a)). 

In general, a twist defect in the toric code is a 
topological defect of the lattice that switches the anyon labels of 
quasiparticles being dragged adiabatically around the twist defect according to the electric-magnetic symmetry $e\leftrightarrow m$ (see Fig.~\ref{fig:defect1}(a)). It can be constructed from any defect that violates the checkerboard lattice pattern, such as a dislocation or a disclination with an odd-coordinated vertex.  The sign of the defect plaquette operator $P_{\pentagon}^\pm$ corresponds to two distinct defect species, $\sigma_0$ and $\sigma_1$.\cite{TeoRoyXiao13long,teo2013braiding} The species labels can be more generically distinguished by the Wilson loop operator $\Theta$ (see Fig.~\ref{fig:defect1}(b)) formed by dragging either the $e$ or $m$ quasiparticle around the defect \emph{twice} to form a closed loop. This Wilson loop operator can be absorbed into the ground state, but with a 
remaining eigenvalue phase $i$ or $-i$ depending on the defect species. Defect species can also mutate between $0$ and $1$ by absorbing or emitting an $e$ or $m$ quasiparticle. This is because the additional quasiparticle string emanating from a mutated defect will intersect the double loop operator $\Theta$ and contribute a minus sign. However, adding or subtracting a fermion $\psi$  to/from the defect does not change its species because the $\psi$-string would intersect $\Theta$ twice and give the trivial phase. These results are summarized by the following fusion (or equivalently splitting) rules of the twist defects and quasiparticles
\begin{align}
\sigma_\lambda\times\psi=\sigma_\lambda,\quad\sigma_\lambda\times e=\sigma_\lambda\times m=\sigma_{\lambda+1}
\label{toriccodedefectfusion1}
\end{align} 
where $\lambda=0,1$ (mod 2) denotes the defect species.

We show explicitly in Appendix \ref{app:toricdefect} how to construct the defect fusion category for this phase, but for now we will just summarize the relevant data. 
First we need to give the fusion rules between pairs of defects:
\begin{align}
\sigma_0\times\sigma_0=\sigma_1\times\sigma_1=1+\psi,\quad\sigma_0\times\sigma_1=e+m.\label{toriccodedefectfusion2}
\end{align}
The first equation, for example, states that two defects with identical species can fuse into either the vacuum $1$ or fermion $\psi$ channel. It is worth noting that this matches the fusion of a pair of Ising anyons.~\cite{Kitaev06} 
 
 We also need to list the set of $F$-symbols, which are proven in Ref.~\onlinecite{teo2013braiding}, and are summarized in Table~\ref{tab:Fsymbolstoriccode}. For the definition of the $F$-symbols and a quick review of the relevant technical points regarding topologically ordered phases, refer to Appendix \ref{app:reviewTQFT}. As discussed in Appendix \ref{app:toricdefect}, there are two possible inequivalent, but consistent, choices for the $F$-symbols which differ by signs. The twist liquid of the bare toric code corresponds to one choice, while that of the toric code stacked on a non-trivial $\mathbb{Z}_2$ SPT corresponds to the other choice. We will discuss this more in Section~\ref{sec:gaugingSPT}. 

\begin{table}[htbp]
\begin{tabular}{ll}
\multicolumn{2}{c}{Defect $F$-symbols for the (bare) toric code}\\\hline
$F^{{\bf a}{\bf b}{\bf c}}_{\bf d}$, $F^{{\bf a}{\bf b}\sigma}_{\sigma}$, $F^{\sigma{\bf a}{\bf b}}_{\sigma}$, $F^{{\bf a}\sigma\sigma}_{\bf b}$, $F^{\sigma\sigma{\bf a}}_{\bf b}$ & $1$ \\
$F^{{\bf a}\sigma{\bf b}}_{\sigma}$, $F^{\sigma{\bf a}\sigma}_{\bf b}$ & $(-1)^{a_2b_2}$ \\
$\left[F^{\sigma\sigma\sigma}_{\sigma}\right]_{\bf a}^{\bf b}$ & $\frac{1}{\sqrt{2}}(-1)^{a_2b_2}$
\end{tabular}
\caption{Admissible $F$-symbols for defects in the toric code with quasiparticle decomposition ${\bf a}=e^{a_1}m^{a_2}$, ${\bf b}=e^{b_1}m^{b_2}$. They do not depend on the defect species $\sigma=\sigma_0,\sigma_1$.}\label{tab:Fsymbolstoriccode}
\end{table}

\subsection{Gauging the electric-magnetic anyonic symmetry}
\label{sec:gauging-em}


The toric code has a $\mathbb{Z}_2$ global anyonic symmetry group that acts to exchange the labels of the the charge $e$ and flux $m$ quasiparticles in the effective $\mathbb{Z}_2$ gauge theory. That is, this topological phase has a local $\mathbb{Z}_2$ symmetry and a separate global $\mathbb{Z}_2$ anyonic symmetry. Because the anyonic symmetry exchanges electric-charge with magnetic-flux, we will will also refer to the symmetry as an electric-magnetic or e-m symmetry when convenient.

Now, we are interested in driving the topological system through a quantum phase transition so that a local electric-magnetic symmetry is generated/restored. To understand the type of phase transition we have in mind, perhaps it  is beneficial to draw an analogy with a lattice melting transition where broken continuous translation symmetry is restored by proliferating dislocation defects. If we consider proliferating dislocations in Wen's plaquette model then this would melt the checkerboard lattice and simultaneously smear out the distinction between $e$ and $m,$ thereby making the the electric-magnetic symmetry local. For this particular realization of the deconfined phase of the $\mathbb{Z}_2$ gauge theory the dislocations act as twist defects, which are the fluxes of the anyonic symmetry group. The relabeling operation $e\leftrightarrow m$ of encircling quasiparticles can be regarded as a holonomy representing the action of the $\mathbb{Z}_2$ anyonic symmetry. Initially these twist defects are classical static objects and we seek a quantum dynamical description where they become anyonic excitations of a new topological phase where the electric-magnetic symmetry is {\em gauged}. We refer to this new topological phase as a twist liquid.

In many ways, the twist liquid bears a similarity to a typical $(2+1)$D discrete gauge theory. For example, we will soon show that it also naturally supports gauge charges of the AS group that braid non-trivially around the quantum twist defects, to which we will refer simply as fluxes from now on. However, due to the non-trivial anyons of the original phase, and the possible non-Abelian nature of the twist defects, the twist liquid is generically {\em not} a discrete gauge theory. For instance, fluxes in the twist liquid do not necessarily carry integral quantum dimensions. The rest of this section will be dedicated to proving that the twist liquid derived from the toric code with gauged anyonic e-m symmetry has a non-chiral Ising-type topological order: \begin{align}\begin{diagram}\mbox{Toric Code}&\pile{\rTo^{\mbox{\small Gauging}}\\\lTo_{\mbox{\small Condensation}}}&\mbox{Ising}\times\overline{\mbox{Ising}}.\end{diagram}\label{toriccodetoIsing}\end{align} We will begin our discussion by presenting some complementary indirect evidence that will serve as physical intuition. This will followed by a more systematic approach that we will adapt to more general systems in Section \ref{sec:gauginganyonicsymmetries}.

\subsubsection{Fermionization, fermion parity gauging, and anyon condensation}

The first piece of indirect evidence comes from an analysis of the expected edge theory, and then using the bulk-boundary correspondence to infer the bulk properties from the boundary. For our case, the expectation of a non-chiral Ising theory can be understood by studying the symmetry-enforced edge, i.e. an edge where the electric-magnetic symmetry is explicitly enforced. It is well-known\cite{Wenedgereview,Wenbook} that the boundary of the effective Abelian Chern-Simons bulk theory has a Luttinger liquid description $\mathcal{L}=\frac{1}{4\pi}K_{IJ}\partial_x\phi_I\partial_t\phi_J$, where $I,J=1,2$ index the chiral $U(1)$ bosons $\phi_I$ and $K=2\sigma_x$. The gapless bosons are unstable against two relevant competing condensation mechanisms (mass terms) \begin{align}\mathcal{H}=J_e\cos2\phi_1+J_m\cos2\phi_2\label{IsingtransitionHam}\end{align} where $J_e$ (or $J_m$) is responsible for the condensation of the bosonic charge $e$ (resp.~flux $m$). This leads to an energy gap along the edge when the two competing terms are not balanced, and the two gapped phases, $J_e<J_m$ and $J_e>J_m$, are realized by the ``smooth" or ``rough" edges of  Kitaev's original toric code lattice model\cite{KitaevKong12,bravyiedge}. However, when the electric-magnetic symmetry is strictly enforced, the Hamiltonian is pinned at the self-dual transition point for $J_e=J_m$. At this point Eq.~\eqref{IsingtransitionHam} can be mapped to a transverse field Ising model, or equivalently fermionized into a pair of real fermions.\cite{Tsvelikbook,Fradkinbook} It is well-known that this critical theory is characterized by a non-chiral Ising CFT.\cite{bigyellowbook} If we appeal to the conventional bulk-boundary correspondence this would imply that the bulk theory has the anyon content with non-chiral Ising character as well. 

The second argument is based on a convenient representation of a conventional $s$-wave superconductor. First, one can physically model the anyon content of the toric code by gauging the fermion parity of a conventional two dimensional $s$-wave superconductor (SC).\cite{HanssonOganesyanSondhi04,BondersonNayak13} The anyon content of the toric code is represented by  the superconducting condensate vacuum which corresponds to 1, the fermionic Bogoliubov$-$de Gennes quasiparticle\cite{deGennesbook} which is $\psi$, the deconfined $hc/2e$ quantum flux vortex which is $m$, and an excited Caroli$-$de Gennes$-$Matricon vortex state\cite{CarolideGennesMatricon} which corresponds to $e=m\times\psi.$ The important realization is that an $s$-wave SC is topologically equivalent, and adiabatically connected (if we do not enforce any extra symmetries), to a bilayer $(p_x+ip_y)\times(p_x-ip_y)$ SC as they both have vanishing Chern invariant (or thermal Hall conductivity).\cite{SchnyderRyuFurusakiLudwig08,Kitaevtable08,QiHughesRaghuZhang09} We must remember that, despite the bi-layer representation, only the overall combined fermion parity $(-1)^{F_++F_-}$ is gauged, and the flux vortex $m$ is a bound composite of vortices $\sigma_+\sigma_-$ in both layers. By themselves the $\sigma_{\pm}$ bind local Majorana bound states $\gamma_{\pm},$ but we see that $m=\sigma_+\sigma_-$ is still a conventional boson as the pair of vortex bound Majorana zero modes can couple and be annihilated by interlayer electron tunneling. 

In this representation, a vortex $\sigma_\pm$ on a single layer is the physical realization of a twist defect. We can see this as follows: consider dragging a bilayer vortex $m({\bf x})$ once around the single layer one, say $\sigma_+({\bf y})$. The Majorana zero mode $\gamma_+({\bf x})$ on $m({\bf x})$ in the first layer accumulates a braiding phase of $-1$, while the other zero mode $\gamma_-({\bf x})$ on $m({\bf x})$ is unaltered by the braiding process. As a result the $m({\bf x})$ vortex undergoes a fermion parity pumping process and becomes $e({\bf x})$ after a cycle;\cite{KhanTeoVishveshwaraappearsoon} i.e. the localized vortex bound state constructed from the two local Majorana modes becomes excited: \begin{align}c({\bf x})=\frac{\gamma_-({\bf x})+i\gamma_+({\bf x})}{2}\to\frac{\gamma_-({\bf x})-i\gamma_+({\bf x})}{2}=c^\dagger({\bf x}).\end{align} Thus we see that under this process $m\to e.$ If we started with an excited bilayer vortex $e({\bf x})$ we can see that it would be converted to an $m$ quasi-particle during the same process.

Now when we consider gauging the electric-magnetic symmetry, then single layer vortices $\sigma_\pm$ become deconfined and quantum dynamical. Equivalently, the individual fermion parities $(-1)^{F_\pm}$ of both layers are gauged as a result. A chiral $p_x\pm i p_y$ SC with a dynamical $\mathbb{Z}_2$ gauge field has non-Abelian (chiral) Ising topological order\cite{ReadGreen,Kitaev06}. The non-chiral bilayer system thus belongs to the non-chiral $\mbox{Ising}\times\overline{\mbox{Ising}}$ phase. We notice in passing that if the two layers carry opposite spins, then the combined time reversed $p_x\pm ip_y$ partners corresponds to a non-trivial $\mathbb{Z}_2$ topological SC in class DIII in two dimensions.\cite{SchnyderRyuFurusakiLudwig08,Kitaevtable08,QiHughesRaghuZhang09} The fact that this state has a gapless helical Majorana edge mode at its edge resonates with the previous  discussion of the critical Ising Hamiltonian \eqref{IsingtransitionHam} on the edge, except that the former is protected by time reversal, and the latter is enforced by electric-magnetic symmetry.

The third line of argument comes from reverse engineering our guess that the twist liquid phase is the non-chiral Ising theory. We will see that by condensing\cite{BaisSlingerlandCondensation,Kong14} a non-trivial boson in the twist liquid, one can have a transition back to the toric code with the global e-m symmetry. The non-chiral $\mbox{Ising}\times\overline{\mbox{Ising}}$ phase with quasiparticle structure $\{1,\psi,\sigma\}\times\{\overline{1},\overline{\psi},\overline{\sigma}\}$ has a non-trivial (i.e. not the vacuum) boson $z=\psi\overline{\psi}$. The $z$ quasiparticle plays the role of the $\mathbb{Z}_2$ charge of the electric-magnetic anyonic symmetry, and accumulates a braiding phase of $-1$ around the Ising anyon $\sigma$ (or $\overline{\sigma}$), which, as such, will be identified as the $\mathbb{Z}_2$ flux of the e-m symmetry. 

Upon condensing the fermion pair, the two Ising anyons are confined. The two fermions $\psi$ and $\overline{\psi}$ become identified since they can transform into one another by effectively `absorbing'  the condensate. The super-selection sector $\sigma\overline{\sigma}$ (we will henceforth abbreviate super-selection sector to super-sector) has quantum dimension 2 as it satisfies the fusion rule \begin{align}\sigma\overline{\sigma}\times\sigma\overline{\sigma}=1+\psi\overline{\psi}+\psi+\overline{\psi}.\label{e+mfusion}\end{align} Recall in the $(p_x+ip_y)\times(p_x-ip_y)$ SC, the bilayer vortex $\sigma_+\sigma_-$ has split energy states $e$ and $m$ due to the coupling between the pair of local vortex Majorana zero modes $\gamma_\pm$. Hence, from this intuition we expect the super-sector $\sigma\overline{\sigma}$ must also split into independent dimension 1 sectors, i.e. $\sigma\overline{\sigma}=e+m$, with opposite fermion parity, $e=m\times\psi$. After $\psi\overline{\psi}$ is condensed and identified with the vacuum, the fusion rule \eqref{e+mfusion} becomes \begin{align}(e+m)\times(e+m)=1+1+\psi+\psi.\end{align}  This recovers the fusion rules $e^2=m^2=1$ and $e\times m=\psi$ of the toric code, as well as the braiding statistics \begin{align}R^{ee}_1=R^{mm}_1=R^{(\sigma\overline\sigma)(\sigma\overline\sigma)}_1=1\end{align} where $R^{xx}_1$ is the exchange phase of QP $x$.
From this we can conclude that the condensation of $z=\psi\overline{\psi}$, or equivalently confinement of the Ising anyons $\sigma,\overline{\sigma}$, drives a topological phase transition from the twist liquid to the $\mathbb{Z}_2$ toric code. \begin{align}\begin{diagram}\mbox{Ising}\times\overline{\mbox{Ising}}&\rTo^{\mbox{\small $\psi\overline{\psi}$ Condensation}}&\mbox{Toric code}.\end{diagram}\end{align}

We can also see something else interesting from this construction: we see that the origin of the global electric-magnetic symmetry in the $\mathbb{Z}_2$ toric code arises from the the fact that $e$ and $m$ are the offspring of the same super-sector $\sigma\overline{\sigma}$ in the corresponding twist liquid.  As such, they have identical fusion and statistical properties, which is a requirement of being related by an anyonic symmetry. 

\subsubsection{Drinfeld construction}
\label{Drinfeld-cosntruction}

The above three pieces of evidence are convincing suggestions that the toric code with gauged electric-magnetic anyonic symmetry should be described by the $\mbox{Ising}\times\overline{\mbox{Ising}}$ topological phase. However, there are drawbacks to these arguments. For instance, for the argument based on the edge-theory, the instability of the symmetric edge \eqref{IsingtransitionHam} is irrelevant, in the renormalization group sense, for $K$-matricies $k\sigma_x$ for $k>4$. Hence, without a more careful treatment of edge boundary conditions, it would not allow one to infer the bulk twist liquid after gauging the e-m symmetry of a more general $\mathbb{Z}_k$ gauge theory. Additionally, the arguments concerning the bilayer $p_x\pm ip_y$ SC and fermion pair condensation both rely heavily on the hindsight that the gauging of e-m symmetry would lead to a non-chiral Ising theory; a theory which is already physically well-understood. Thus it is unclear one could find a clean way to generalize such a construction. Hence,  in this section we will discuss a more systematic approach that can be carried out in more general, and less well-known, anyonic symmetric topological phases.

\begin{figure}[ht]
\includegraphics[width=0.30\textwidth]{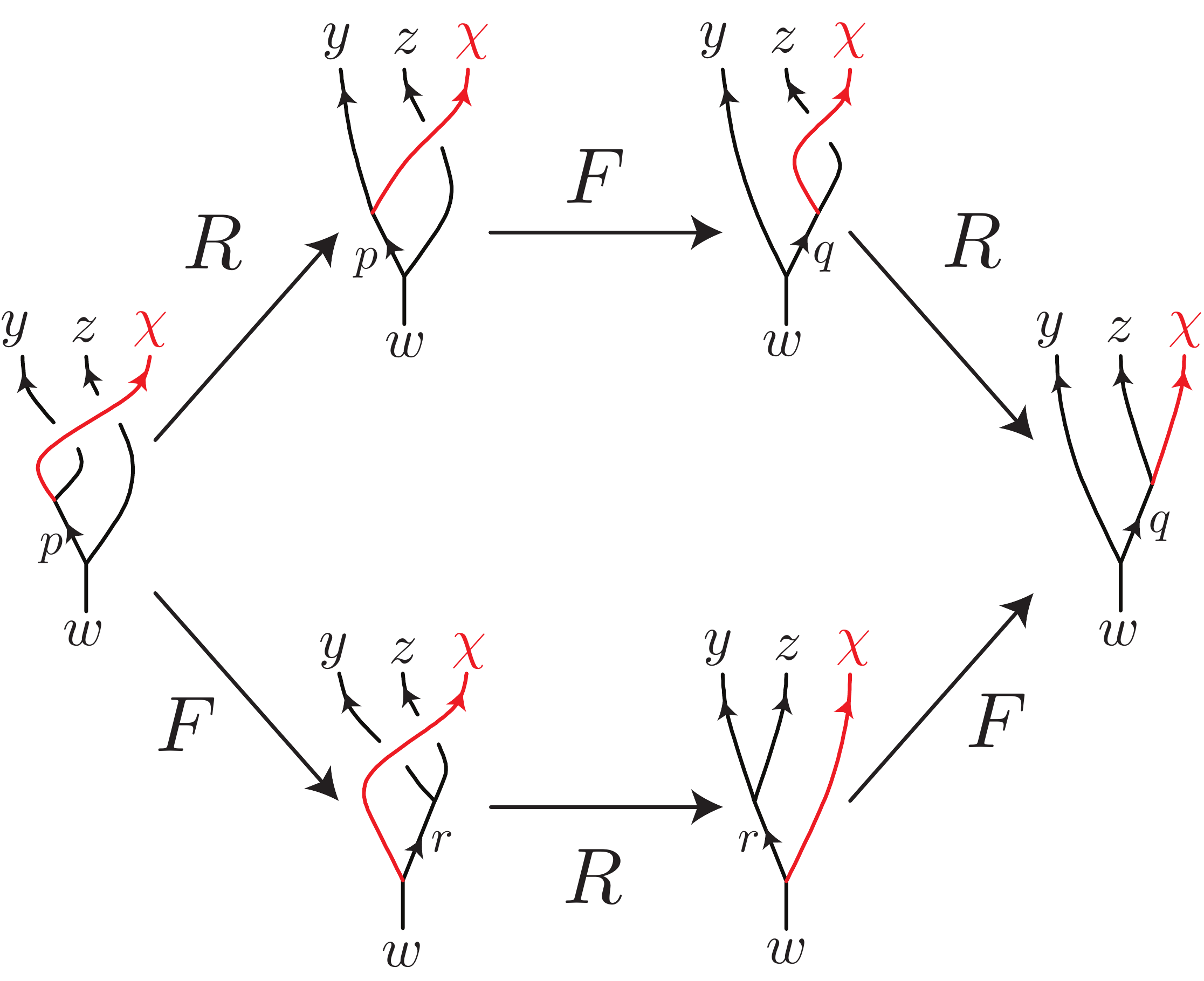}
\caption{Hexagon equation.}\label{fig:hexagon1}
\end{figure}

In Section~\ref{sec:toriccodetwistdefects}, we discussed the fusion structure of twist defects in the toric code. They obey the fusion rules \eqref{toriccodedefectfusion1}, \eqref{toriccodedefectfusion2}, and basis transformations are generated by the $F$-symbols listed in Table~\ref{tab:Fsymbolstoriccode}. These defects become quantum dynamical fluxes in the twist liquid after gauging the electric-magnetic symmetry. Like all anyonic excitations of any topological phase, they satisfy a set of consistent braiding rules. The anti-clockwise exchange of the anyons $x$ and $y$ with a fixed fusion channel $x\times y\to z$ gives an Abelian phase \begin{align}\left|\vcenter{\hbox{\includegraphics[width=0.05\textwidth]{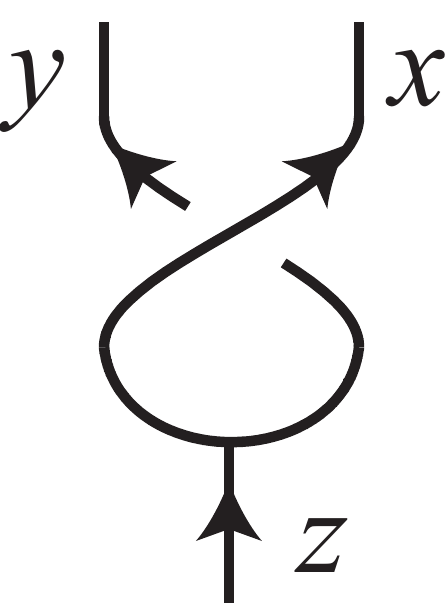}}}\right\rangle=R^{xy}_z\left|\vcenter{\hbox{\includegraphics[width=0.05\textwidth]{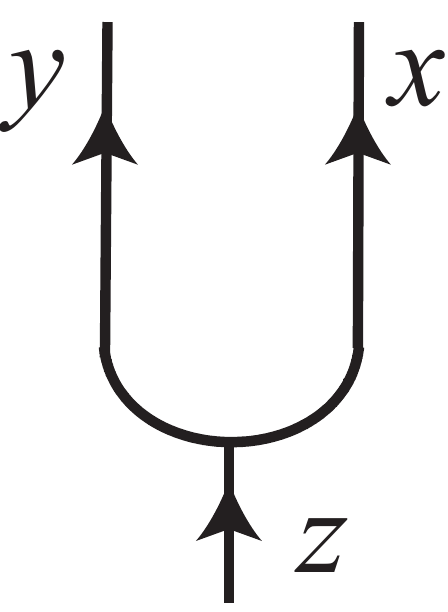}}}\right\rangle .\end{align} These exchange $R$-symbols follow the consistency relations called the {\em hexagon identity}\cite{Kitaev06} (see Eq.~\eqref{hexagoneq} and Fig.~\ref{fig:hexagon1}), which we will abbreviate by the notation \hexeq
\begin{align}R^{\chi y}_p\left[F^{y\chi z}_w\right]_p^qR^{\chi z}_q=\sum_r\left[F^{\chi yz}_w\right]_p^rR^{\chi r}_w\left[F^{yz\chi}_w\right]^q_r .\label{hexagoneq}\end{align}

First we notice that within the full defect fusion category $\mathcal{C}=\{1,e,m,\psi,\sigma_0,\sigma_1\}$, there is a closed sub-category $\mathcal{C}_0=\{1,\psi,\sigma_0\}$ that has the fusion rules $\psi\times\psi=1$, $\psi\times\sigma_0=\sigma_0$ and $\sigma_0\times\sigma_0=1+\psi$, and non-trivial basis transformations (see Table~\ref{tab:Fsymbolstoriccode}) \begin{align}F^{\sigma\psi\sigma}_\psi=F^{\psi\sigma\psi}_\sigma=-1,\quad F^{\sigma\sigma\sigma}_\sigma=\frac{1}{\sqrt{2}}\left(\begin{array}{*{20}c}1&1\\1&-1\end{array}\right)\label{Ising0Fsymbols}\end{align} where the entries for $\left[F^{\sigma\sigma\sigma}_\sigma\right]_x^y$ runs through the intermediate channels $x,y=1,\psi$, and all other admissible $F$-symbols are 1. We are interested in constructing the \emph{Drinfeld center}\cite{Kasselbook,BakalovKirillovlecturenotes,LevinWen05} $Z(\mathcal{C}_0)$ of this sub-category. In Sec.~\ref{sec:stringnet} we will illustrate that using the Drinfeld center $Z(\mathcal{C})$ of the full defect fusion category will produce $Z(\mathcal{C}_0)$ as well as an unwanted redundant copy of toric code (see Eq.~\eqref{Drinfeld=TL}). Generically, instead of taking the Drinfeld center of a defect subcategory we will need to take the full Drinfeld center and then implement constraints to construct a \emph{relative} Drinfeld center. This will be discussed in detail in Sec.~\ref{sec:stringnet}, and explicitly implemented for the examples in Sec.~\ref{sec:gauging-em-bilayer-su3}, \ref{sec:so(N)}, \ref{sec:so(8)symmetry} and \ref{sec:4statePotts}.

To perform this construction we need to find the set of (Drinfeld) anyons, which are the QPs in the twist liquid, and their associated braiding and fusion properties. An anyon $\chi=(x;\mathcal{R}^{x\ast}_\ast)$ in the Drinfield center is described by two data. The quantity $x$ is an object in the \emph{original} fusion category and can, in general, be a linear combination of the labels $1,\psi,\sigma_0$. It dictates the fusion properties of $\chi$. The quantity $\mathcal{R}^{x\ast}_\ast$ is a consistent set of unitary solutions $\{\mathcal{R}^{xy}_z\}$ to the hexagon equations (\hexeq) in Eq.~\eqref{hexagoneq} (see Fig.~\ref{fig:hexagon1}) and characterizes the clockwise exchange between the Drinfeld anyon $\chi$ and a label $y$ in the original category with fixed fusion channel $z.$ The trivial \hexeq~$\mathcal{R}^{x1}_1\mathcal{R}^{x1}_1=\mathcal{R}^{x1}_1$ implies $\mathcal{R}^{x1}_1=1$ for any $x$. To find the twist liquid that results from our defect fusion category we must find all of the irreducible anyon solutions of the \hexeq

First we begin by solving $\mathcal{R}^{x\ast}_\ast$ for $x=1$. Exchanging $1$ with a pair of $\sigma_0$'s give the \hexeq's \begin{align}\mathcal{R}^{1\sigma}_\sigma\mathcal{R}^{1\sigma}_\sigma=\mathcal{R}^{11}_1=\mathcal{R}^{1\psi}_\psi=1.\end{align} There are two sets of solutions \begin{align}1&=\left(1;\mathcal{R}^{1\psi}_\psi=\mathcal{R}^{1\sigma}_\sigma=1\right)\label{Isingsol1}\\ z&=\left(1;\mathcal{R}^{1\psi}_\psi=-\mathcal{R}^{1\sigma}_\sigma=1\right)\label{Isingsolz}\end{align} where $1$ represents the true vacuum of the Drinfeld center, and $z$ is a non-trivial boson that will be identified as the $\mathbb{Z}_2$-charge for the e-m symmetry.

Next we consider the \hexeq's for exchanging $\sigma_0$ with a pair of $\psi$'s \begin{align}-\mathcal{R}^{\sigma\psi}_\sigma\mathcal{R}^{\sigma\psi}_\sigma=\mathcal{R}^{\sigma1}_\sigma=1\end{align}
and with a pair of $\sigma_0$'s \begin{align}\frac{1}{\sqrt{2}}\mathcal{R}^{\sigma\sigma}_1\mathcal{R}^{\sigma\sigma}_1&=\frac{1}{2}\left(\mathcal{R}^{\sigma1}_\sigma+\mathcal{R}^{\sigma\psi}_\sigma\right)\label{TCRsigmasigma}\\\frac{1}{\sqrt{2}}\mathcal{R}^{\sigma\sigma}_1\mathcal{R}^{\sigma\sigma}_\psi&=\frac{1}{2}\left(\mathcal{R}^{\sigma1}_\sigma-\mathcal{R}^{\sigma\psi}_\sigma\right)\\-\frac{1}{\sqrt{2}}\mathcal{R}^{\sigma\sigma}_\psi\mathcal{R}^{\sigma\sigma}_\psi&=\frac{1}{2}\left(\mathcal{R}^{\sigma1}_\sigma+\mathcal{R}^{\sigma\psi}_\sigma\right).\end{align} There are four solutions which we label by\begin{align}\sigma&=\left(\sigma_0;\mathcal{R}^{\sigma\psi}_\sigma=-i,\mathcal{R}^{\sigma\sigma}_1=e^{-\pi i/8},\mathcal{R}^{\sigma\sigma}_\psi=e^{3\pi i/8}\right)\label{TCDrinfeldsigma1}\\\overline{\sigma}&=\left(\sigma_0;\mathcal{R}^{\sigma\psi}_\sigma=i,\mathcal{R}^{\sigma\sigma}_1=e^{\pi i/8},\mathcal{R}^{\sigma\sigma}_\psi=e^{-3\pi i/8}\right)\label{TCDrinfeldsigma2}\\\sigma'&=\left(\sigma_0;\mathcal{R}^{\sigma\psi}_\sigma=-i,\mathcal{R}^{\sigma\sigma}_1=e^{7\pi i/8},\mathcal{R}^{\sigma\sigma}_\psi=e^{-5\pi i/8}\right)\label{TCDrinfeldsigma3}\\\overline{\sigma}'&=\left(\sigma_0;\mathcal{R}^{\sigma\psi}_\sigma=i,\mathcal{R}^{\sigma\sigma}_1=e^{-7\pi i/8},\mathcal{R}^{\sigma\sigma}_\psi=e^{5\pi i/8}\right).\label{TCDrinfeldsigma4}\end{align} These will be identified as the flux-charge composites for the $\mathbb{Z}_2$ e-m symmetry.

The \hexeq's for exchanging $\psi$ with a pair of $\sigma_0$'s \begin{align}\mathcal{R}^{\psi\sigma}_\sigma\mathcal{R}^{\psi\sigma}_\sigma=\mathcal{R}^{\psi\psi}_1=-\mathcal{R}^{\psi1}_\psi=-1\end{align} have two solutions \begin{align}\psi&=\left(\psi;\mathcal{R}^{\psi\psi}_1=-1,\mathcal{R}^{\psi\sigma}_\sigma=-i\right)\label{Isingsolpsi}\\\overline{\psi}&=\left(\psi;\mathcal{R}^{\psi\psi}_1=-1,\mathcal{R}^{\psi\sigma}_\sigma=i\right).\label{Isingsolpsibar}\end{align} 

However, these are not all the irreducible solutions of the \hexeq's. There is an extra indecomposible (i.e. cannot be broken up into contributions from the previously listed anyons) $\mathcal{R}^{x\ast}_\ast$ associated with the super-sector $x=1+\psi$. It has degenerate fusion rules $x\times\sigma=2\sigma,$ and therefore the exchange symbol $\mathcal{R}^{x\sigma}_\sigma$, is a $2\times2$ matrix. The \hexeq's for exchanging $1+\psi$ with a pair of $\psi$'s are \begin{align}\mathcal{R}^{(1+\psi)\psi}_\psi\mathcal{R}^{(1+\psi)\psi}_\psi=\mathcal{R}^{(1+\psi)1}_1=1\\\mathcal{R}^{(1+\psi)\psi}_1\mathcal{R}^{(1+\psi)\psi}_1=\mathcal{R}^{(1+\psi)1}_\psi=1.\end{align} The \hexeq's for exchanging $1+\psi$ with a pair of $\sigma_0$'s are \begin{align}\mathcal{R}^{(1+\psi)\sigma}_\sigma\mathcal{R}^{(1+\psi)\sigma}_\sigma&=\left(\begin{array}{*{20}c}\mathcal{R}^{(1+\psi)1}_1&0\\0&\mathcal{R}^{(1+\psi)\psi}_1\end{array}\right)\label{e+msshexeq1}\\\mathcal{R}^{(1+\psi)\sigma}_\sigma\sigma_z\mathcal{R}^{(1+\psi)\sigma}_\sigma&=\left(\begin{array}{*{20}c}\mathcal{R}^{(1+\psi)\psi}_\psi&0\\0&\mathcal{R}^{(1+\psi)1}_\psi\end{array}\right)\label{e+msshexeq2}\end{align} where \eqref{e+msshexeq1} has overall fusion channel $\sigma\times(\sigma\times(1+\psi))\to1$, and \eqref{e+msshexeq2} has $\sigma\times(\sigma\times(1+\psi))\to\psi$; hence the appearance of the  Pauli matrix $\sigma_z=\mbox{diag}(1,-1)$ comes from the $F$-symbols $F^{\sigma(1+\psi)\sigma}_\psi=(F^{\sigma1\sigma}_\psi,F^{\sigma\psi\sigma}_\psi)=(1,-1)$.

To solve the \hexeq's, we first notice that $\mathcal{R}^{(1+\psi)\psi}_1$ and $\mathcal{R}^{(1+\psi)\psi}_\psi$ must have opposite signs. Otherwise, \eqref{e+msshexeq1} and \eqref{e+msshexeq2} would be identical up to a sign and lead to the contradiction $\mathcal{R}^{(1+\psi)\sigma}_\sigma\sigma_z\mathcal{R}^{(1+\psi)\sigma}_\sigma=\pm\mathcal{R}^{(1+\psi)\sigma}_\sigma\mathcal{R}^{(1+\psi)\sigma}_\sigma$. Next, if \begin{align}-\mathcal{R}^{(1+\psi)\psi}_1=\mathcal{R}^{(1+\psi)\psi}_\psi=1\end{align} then \eqref{e+msshexeq1} and \eqref{e+msshexeq2} require $[\mathcal{R}^{(1+\psi)\sigma}_\sigma,\sigma_z]=0$ and $\mathcal{R}^{(1+\psi)\sigma}_\sigma\mathcal{R}^{(1+\psi)\sigma}_\sigma=\sigma_z$. They have four diagonal solutions \begin{align}\mathcal{R}^{(1+\psi)\sigma}_\sigma=\left(\begin{array}{*{20}c}\pm1&0\\0&\pm i\end{array}\right)\end{align} which are decomposable into $1+\psi$, $1+\overline{\psi}$, $z+\psi$ and $z+\overline{\psi}$ according to \eqref{Isingsol1}, \eqref{Isingsolz}, \eqref{Isingsolpsi} and \eqref{Isingsolpsibar}, and thus lead to no new anyons.

The irreducible solution is given by \begin{align}\mathcal{R}^{(1+\psi)\psi}_1=-\mathcal{R}^{(1+\psi)\psi}_\psi=1\end{align} so that \eqref{e+msshexeq1} and \eqref{e+msshexeq2} require  $\{\mathcal{R}^{(1+\psi)\sigma}_\sigma,\sigma_z\}=0$ and $\mathcal{R}^{(1+\psi)\sigma}_\sigma\mathcal{R}^{(1+\psi)\sigma}_\sigma=\openone$. This has off-diagonal solutions \begin{align}\mathcal{R}^{(1+\psi)\sigma}_\sigma=\left(\begin{array}{*{20}c}0&e^{i\phi}\\e^{-i\phi}&0\end{array}\right).\end{align} The phase $\phi$ is a gauge degree of freedom and can be set to 0 by a $u(2)$ gauge transformation at the degenerate fusion vertex $\sigma\times(1+\psi)=2\sigma$. From this we see that the final anyon in the Drinfeld center is the super-selection sector \begin{align}\mathcal{E}=\left(1+\psi;\mathcal{R}^{(1+\psi)\psi}_1=-\mathcal{R}^{(1+\psi)\psi}_\psi=1,\mathcal{R}^{(1+\psi)\sigma}_\sigma=\sigma_x\right).\end{align}

Collecting the solutions together we have a set of nine anyons that recover the same fusion structure as the $\mbox{Ising}\times\overline{\mbox{Ising}}$ theory by identifying \begin{gather}\overline{\psi}=z\times\psi,\quad\sigma'=z\times\sigma,\quad\overline{\sigma}'=z\times\overline{\sigma}\label{Isingfusionss'}\\\mathcal{E}=z\times\mathcal{E}=\sigma\times\overline{\sigma}.\label{IsingfusionE=ss}\end{gather} These can be verified by combining the $\mathcal{R}$-symbols with appropriate hexagon identities. For instance, \eqref{Isingfusionss'} can be proven by equating $\mathcal{R}^{\overline{\psi}\sigma}_\sigma=\mathcal{R}^{z\sigma}_\sigma\mathcal{R}^{\psi\sigma}_\sigma$ and $\mathcal{R}^{\sigma'\sigma}_\ast=\mathcal{R}^{z\sigma}_\sigma\mathcal{R}^{\sigma\sigma}_\ast$. To see Eq.~\eqref{IsingfusionE=ss},  we note that fusing the supersector $\mathcal{E}$ with the fermion pair $z$ does not alter it because $\mathcal{R}^{\mathcal{E}\psi}_\ast=\mathcal{R}^{z\psi}_\psi\mathcal{R}^{\mathcal{E}\psi}_\ast$. Moreover $\mathcal{R}^{\mathcal{E}\sigma}_\sigma$ and $\mathcal{R}^{z\sigma}_\sigma\mathcal{R}^{\mathcal{E}\sigma}_\sigma$ only differ by a sign and are related by a $U(2)$ gauge transformation.

Now, from our set of anyons we identify the $\sigma$ anyons as the $\mathbb{Z}_2$ fluxes for the electric-magnetic symmetry as they originate from the twist defect. The fermion pair $z=\psi\overline{\psi}$ is the $\mathbb{Z}_2$ charge because of the braiding phase with $\sigma$ \begin{align}R^{z\sigma}_\sigma R^{\sigma z}_\sigma=\mathcal{R}^{z\sigma}_\sigma\mathcal{R}^{\sigma 1}_\sigma=-1.\end{align} (Note that $R^{\sigma z}=\mathcal{R}^{\sigma 1}$ because the second argument of the $R$-symbol only depends on the original fusion category label.) As a result, $\sigma'$ and $\overline{\sigma}'$ are flux-charge composites. 

As a consistency check, we can compute the spins of the nine quasiparticles from the $\mathcal{R}$-symbols. It is clear that the vacuum (1) and the fermion pair ($z$) are bosons as $\theta_1=R^{11}_1=1$ and $\theta_z=R^{zz}_1=\mathcal{R}^{z1}_1=1$. It is also straightforward to see that $\psi$ and $\overline{\psi}$ are fermions from $\theta_\psi=R^{\psi\psi}_1=-1$ and $\theta_{\overline{\psi}}=R^{\overline{\psi}\overline{\psi}}_1=\mathcal{R}^{\overline{\psi}\psi}_1=-1$. For the Ising anyons, we have \begin{align}\theta_\sigma=\frac{R^{\sigma\sigma}_1+R^{\sigma\sigma}_\psi}{d_\sigma}=\frac{e^{-\pi i/8}+e^{3\pi i/8}}{\sqrt{2}}=e^{\pi i/8}.\end{align} Similarly, as $\mathcal{R}^{\overline{\sigma}\sigma}=(\mathcal{R}^{\sigma\sigma})^\ast$, $\mathcal{R}^{\sigma'\sigma}=-\mathcal{R}^{\sigma\sigma}$ and $\mathcal{R}^{\overline{\sigma}'\sigma}=-(\mathcal{R}^{\sigma\sigma})^\ast$, we have $\theta_{\overline{\sigma}}=\theta_\sigma^\ast=e^{-\pi i/8}$, $\theta_{\sigma'}=-\theta_\sigma=e^{-7\pi i/8}$ and $\theta_{\overline{\sigma}'}=-\theta_\sigma^\ast=e^{7\pi i/8}$. Finally the spin of the super-sector ($\mathcal{E}$) is \begin{align}\theta_{\mathcal{E}}&=\frac{1}{d_{\mathcal{E}}}\left(R^{\mathcal{E}\mathcal{E}}_1+R^{\mathcal{E}\mathcal{E}}_z+R^{\mathcal{E}\mathcal{E}}_\psi+R^{\mathcal{E}\mathcal{E}}_{\overline{\psi}}\right)\nonumber\\&=\frac{1}{2}\left(\mathcal{R}^{\mathcal{E}1}_1+\mathcal{R}^{\mathcal{E}\psi}_1+\mathcal{R}^{\mathcal{E}1}_\psi+\mathcal{R}^{\mathcal{E}\psi}_\psi\right)=1\end{align} which confirms that $\mathcal{E}=\sigma\overline{\sigma}$ is a boson.

Interestingly, we notice that the Ising fusion category $\{1,\psi,\sigma_0\}$ with $F$-symbols \eqref{Ising0Fsymbols} contains the $\mathbb{Z}_2$ sub-category $\{1,\psi\}$ with trivial $F$-symbols. We have now exhaustively shown that the Drinfeld center arising from the Ising fusion category gives the non-chiral $\mbox{Ising}\times\overline{\mbox{Ising}}$ phase. It is also known that the Drinfeld center for the $\mathbb{Z}_2$ sub-category describes the $\mathbb{Z}_2$ toric code\cite{LevinWen05}. We can thus induce a phase transition between the two topological states  by adding string tension terms that contribute a local energy to the string type $\sigma_0$. These terms do not commute with the plaquette stabilizers, and in their ``ordered" phase, when the string tensions are strong, they confine all the Ising anyons and condense the bosonic fermion pair $z=\psi\overline{\psi}$ (e.g. there is no longer an exchange phase $\mathcal{R}^{z\sigma}_\sigma=-1$ to distinguish $z$ from the vacuum). From this we see that the system flows to the $\mathbb{Z}_2$ toric code.

Since it will become important later, we note that the gauging result will be altered if a $\mathbb{Z}_2$ SPT~\cite{ChenGuLiuWen12, LuVishwanathE8} is stacked on top of the toric code to form a composite system. The SPT will not change the QP structure of the toric code, but the non-trivial $\mathbb{Z}_2$ topological phase will modify the defect $F$-symbols \eqref{Ising0Fsymbols} by a sign, $F^{\sigma\sigma\sigma}_\sigma\to-F^{\sigma\sigma\sigma}_\sigma$. This sign is identical to the Frobenius-Schur indicator\cite{FredenhagenRehrenSchroer92,Kitaev06} of the Ising anyon, and has the same cohomological classification as the SPT itself (see Sec.~\ref{sec:gaugingSPT} and Ref.[\onlinecite{BarkeshliBondersonChengWang14}]). If one repeats the Drinfeld construction with this composite system then the resulting $\mbox{Ising}\times\overline{\mbox{Ising}}$ phase would contain Ising anyons with spin $h=\frac{3}{16},\frac{5}{16},\frac{11}{16},\frac{13}{16}$ instead of $h=\frac{1}{16},\frac{7}{16},\frac{9}{16},\frac{15}{16}$. (For details, see Sec.~\ref{sec:so(N)}.) We will see much more about this in the next section.

\section{General gauging framework}
\label{sec:gauginganyonicsymmetries}

In this section we present the general procedure to construct new topological phases, which we refer to as as twist liquids, by gauging the anyonic symmetries of an underlying parent topological phase. In the previous sections, we showed how the electric-magnetic symmetry of the toric code can be gauged, and that it results in the $\mbox{Ising}\times\overline{\mbox{Ising}}$ phase, where the e-m symmetry has become a local gauge invariance. This can be generalized to arbitrary topological phases with {\em anyonic symmetries},\cite{Kitaev06,EtingofNikshychOstrik10,Bombin,YouWen,BarkeshliJianQi,TeoRoyXiao13long,khan2014,BarkeshliBondersonChengWang14} i.e. topological phases with symmetry operations which re-label the anyonic excitations, while keeping their fusion and braiding properties unaltered. As we mentioned earlier, these symmetries are non-local (do not confuse the notion of local vs.~global with acting locally vs.~non-locally as we mean here) because the long-range entanglement protects the topological charge of an anyon excitation from being disturbed by any transformations that only act locally within some neighborhood of the excitation. We provide the general characteristics, and classification, of these symmetries in Section~\ref{sec:anyonicsymmetries}.

General anyonic symmetries heuristically resemble broken symmetries in a classical system, at least to some degree~\cite{Kitaev06} For example, they preserve the form of the topological field theory action, but are broken by individual anyon labels. Similar to $(2+1)D$ melting transitions of ordered phases that restore broken symmetries by vortex/defect proliferation,\cite{chaikin-1995,Nelson-2002,KosterlitzThouless} anyonic symmetries can be gauged by promoting defects to quantum dynamical fluxes. One can think of the extrinsic defects,\cite{Kitaev06, EtingofNikshychOstrik10, barkeshli2010, Bombin, Bombin11, KitaevKong12, kong2012A, YouWen, YouJianWen, PetrovaMelladoTchernyshyov14, BarkeshliQi, BarkeshliQi13, BarkeshliJianQi, MesarosKimRan13, TeoRoyXiao13long, teo2013braiding, khan2014, BarkeshliBondersonChengWang14} before gauging, as generalized ``disclinations" that ``rotate" the anyon labels of orbiting quasiparticles (QP) according to some anyonic symmetry group element instead of a point-group symmetry element. Although these defects are semi-classical and static, they have well-defined fusion rules, and form a complete fusion theory -- referred to as a defect fusion category\cite{Kitaev06, Walkernotes91, Turaevbook, FreedmanLarsenWang00, BakalovKirillovlecturenotes, Wangbook,EtingofNikshychOstrik10,BarkeshliBondersonChengWang14} in Section~\ref{sec:twistdefects} -- that extends the fusion structure of the original topological state. We will describe the general structure of the defect fusion category in Section ~\ref{sec:twistdefects}.

Once these defects become dynamical, a different topological phase is generated, i.e.~a twist liquid. Twist liquids are generalizations of $(2+1)D$ discrete gauge theories\cite{BaisDrielPropitius92,Bais-2007,Propitius-1995,PropitiusBais96,Preskilllecturenotes,Freedman-2004,Mochon04}. The quasiparticles in a twist liquid are compositions of the fluxes and charges associated to the gauged anyonic symmetry, as well as (super-sectors of) the quasiparticles of the original topological state. We give a detailed procedure for the accounting of all of the anyons in the twist liquid and their individual quantum dimensions. Furthermore, we arrive at the remarkable result that the total quantum dimension of a twist liquid increases from its initial value in the parent topological state (with only global anyonic symmetry) by a factor of $|G|$, the order of the anyonic symmetry group $G$ (see Eq.~\eqref{TLdimension3}). This implies that the topological entanglement entropy\cite{KitaevPreskill06} in a disk-like spatial partition therefore increases by $\log|G|$ after gauging. We will see that a twist liquid is generically {\em not} a discrete gauge theory, nor does it necessarily contain a discrete gauge theory as a sub-sector. This is in spite of the fact that the underlying parent phase may be a discrete gauge theory, and that we are gauging a discrete symmetry group on top of the underlying topological phase. These claims will be illustrated in Section~\ref{sec:QPstructuretwistliquid}.

We will also show that the phase transition that drives a globally symmetric parent state to a twist liquid (see Eq.~\eqref{gaugingtransition}) can be captured by an exactly solvable string-net model proposed by Levin and Wen\cite{LevinWen05}. The model is constructed using the fusion data provided by the defect fusion category, and the output is a topological state which contains the anyon content of a {\em Drinfeld center} construction.\cite{Kasselbook,BakalovKirillovlecturenotes} We have already provided a simple example of this construction for the toric code in the previous section. The general construction will be explained in Section~\ref{sec:stringnet}. In this model the phase transition is controlled by tuning various string tensions which act to condense charges and confine fluxes/defects. We postulate that the the string-net model, in its deconfined phase, always contains the twist liquid as a sub-sector, and that the exact anyon data of the twist liquid of interest can be determined using the {\em relative Drinfeld construction} by solving a constrained set of \hexeq's. The reason why mention the caveat that it might only be a sub-sector is because the string-net construction always gives rise to time-reversal invariant ``doubled-theories", and in some cases the string-net model may provide the relevant twist liquid and additionally an extra redundant time-reversed partner of the parent state. This is especially true when we deal with anyonic symmetric chiral states. When this issue arises, the relative Drinfeld construction is invoked to systematically remove the redundancy. In Section \ref{sec:gaugingSPT} we show that how families of twist liquids can arise from the same parent state when combined with an SPT phase protected by the same symmetry group as the AS group of the parent phase. An important and instructive list of examples will be demonstrated in Section~\ref{sec:gauging-em-bilayer-su3}, \ref{sec:so(N)}, \ref{sec:so(8)symmetry} and \ref{sec:4statePotts}. In fact, on a first reading, since this section is long and a bit complicated, it may be helpful to skip to the examples section briefly to gain some intuition about the nature of our construction and anyonic symmetries.


\subsection{Anyonic symmetries}
\label{sec:anyonicsymmetries}

Before we introduce the general structure of anyonic symmetry, we will begin by reviewing the effective Chern-Simons description~\cite{WenZee92,Wenedgereview} of Abelian topological phases and their anyon relabeling symmetries. The topological information -- quasiparticle (QP) fusion and braiding statistics -- of an Abelian phase in $(2+1)$D can be characterized by a quantum field theory $Z[\mathcal{J}]=\int[D\alpha({\bf r})]\exp(iS[\mathcal{J}])$ involving an $N$-component set of $U(1)$ gauge fields $\alpha=(\alpha^1,\ldots,\alpha^N)$ with action \begin{align}S[\mathcal{J}]=\frac{1}{4\pi}\int K_{IJ}\alpha^I\wedge d\alpha^J+\alpha^I\mathcal{J}_I\label{CSaction}\end{align} where the $K$-matrix is integer-valued, symmetric, and non-degenerate.

Quasiparticles $\psi^{\bf a}$ are sources for the currents $~\mathcal{J}_1^{a_1},\ldots\mathcal{J}_N^{a_N},$ and are labeled by $N$-dimensional integral vectors ${\bf a}=(a_1,\ldots,a_N)$ on a  lattice $\Gamma^\ast=\mathbb{Z}^N$. At long distances, nearby quasiparticles combine to form single entity and have a fusion structure. For  Abelian theories such as these, the quasiparticle fusion rules coincide with lattice vector addition \begin{align}\psi^{\bf a}\times\psi^{\bf b}=\vcenter{\hbox{\includegraphics[width=0.5in]{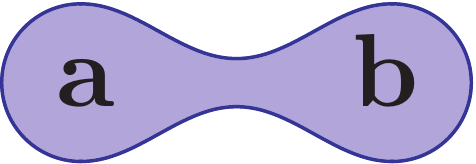}}}=\psi^{{\bf a}+{\bf b}},\end{align} and since this is a topological theory, this is independent of the details of the internal structure of its constituents. The $K$-matrix dictates the braiding statistics of quasiparticles. The braiding phase when dragging $\psi^{\bf a}$ once around $\psi^{\bf b}$ is given by \begin{align}\mathcal{D}S_{{\bf a}{\bf b}}=\vcenter{\hbox{\includegraphics[width=0.7in]{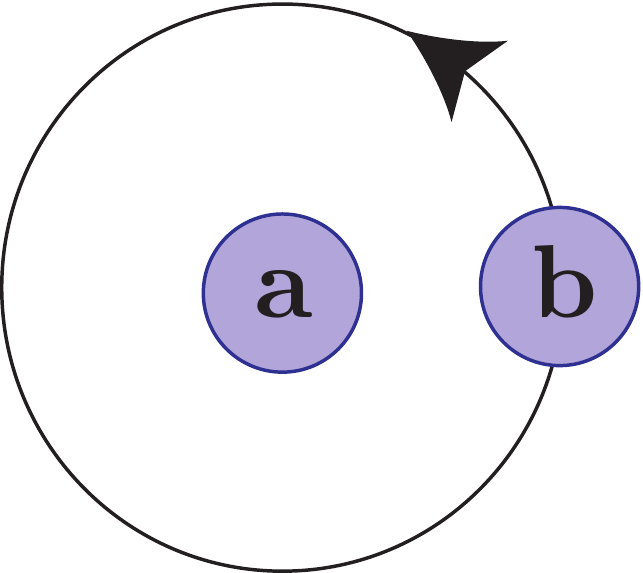}}}=e^{2\pi i{\bf a}^TK^{-1}{\bf b}}\label{AQPbraiding}\end{align} where $\mathcal{D}$ ($=\sqrt{\vert \det (K)\vert}$) is the total quantum dimension.  $\mathcal{D}$ is defined this way so that the $\mathcal{D}^2$-dimensional $S$-matrix in Eq.~\eqref{AQPbraiding} that characterizes anyon braiding is normalized and unitary.
The braiding phase is insensitive to the exact paths of the deconfined quasiparticle pair as long as the linking number of their world-lines is unchanged. 

The exchange statistics of a quasiparticle type is given by the Abelian phase factor \begin{align}\theta_{\bf a}=e^{2\pi ih_{\bf a}}=\vcenter{\hbox{\includegraphics[width=0.5in]{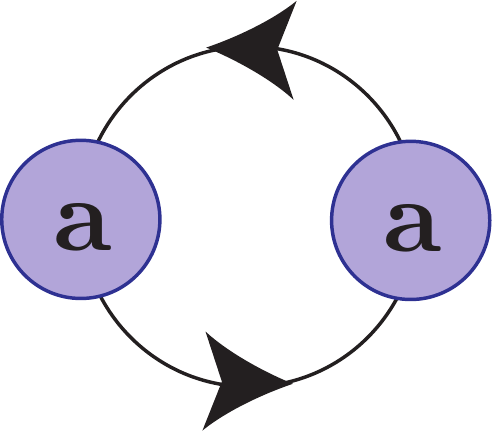}}}=e^{\pi i{\bf a}^TK^{-1}{\bf a}}.\label{AQPexchange}\end{align} From the spin-statistics theorem, the exchange phase $\theta_{\bf a}$ is equivalent to the quasiparticle spin $h_{\bf a}$ (mod 1) defined by the phase gained when a single quasiparticle is rotated by $2\pi$. The collection of {\em local} particles that do not contribute any non-trivial braiding phases in \eqref{AQPbraiding} occupies the sublattice $\Gamma=K\mathbb{Z}^N$ of the full anyon lattice, i.e.~the lattice formed by the image of the $K$-matrix. For simplicity throughout this article, we will assume that all $K$-matrices have even-integer diagonal entries so that all local particles are bosonic with trivial statistics \eqref{AQPexchange}. At zero temperature, the local bosons can condense, and the topological state will contain a finite set of quasiparticle types that are identified up to local bosons, $\psi^{\bf a}\sim\psi^{{\bf a}+K{\bf b}}$. These anyons are labeled by equivalence classes $[{\bf a}]$ in the quotient group $\mathcal{A}=\Gamma^\ast/\Gamma=\mathbb{Z}^N/K\mathbb{Z}^N$, whose order (i.e., number of elements) counts the number of anyon types, and is given by $|\mathcal{A}|=|\det(K)|$. It is also well-known that the number of anyon types is equivalent to the ground state degeneracy on a torus.~\cite{Wentopologicalorder90,WenNiu90}

A Chern-Simons description of a topological phase is not unique. A $K$-matrix would encode the identical fusion and braiding structure even after undergoing a basis transformation: $K\to WKW^T$ where  $W$ is an invertible integer-valued matrix in $GL(N,\mathbb{Z})$. There are special transformations of this type, known as {\em automorphisms}, that leave the $K$-matrix invariant \begin{align}MKM^T=K.\label{automorphism}\end{align} Each such transformation $M$ corresponds to an {\em anyonic symmetry operation}~\cite{khan2014} that permutes the anyons that have the same fusion properties and spin/statistics: \begin{align}(M{\bf a})\times(M{\bf b})=M({\bf a}\times{\bf b}),\quad\theta_{M{\bf a}}=\theta_{\bf a}.\label{ASdef2}\end{align} 

The collection of automorphisms forms a group $\mbox{Aut}(K)$ that classifies the global symmetries of the topological quantum field theory \eqref{CSaction}. Within $\mbox{Aut}(K)$, there lies a sub-collection of trivial basis transformations, called {\em inner automorphisms}, that only rotate quasiparticles up to local particles, $N_0{\bf a}={\bf a}+K{\bf b}$. They act trivially on the anyon labels in $\mathcal{A},$ and form a normal subgroup $\mbox{Inner}(K)$. We are interested in non-trivial anyon relabeling actions, known as {\em outer automorphisms}. These are classified by ``taking out" the trivial inner automorphisms. Mathematically they live in the quotient group \begin{align}\mbox{Outer}(K)=\frac{\mbox{Aut}(K)}{\mbox{Inner}(K)}\label{outer}\end{align} so that the basis transformations $M$ and $N$ correspond to the same anyon relabeling action if they differ by an inner automorphism, i.e.  if $N=MN_0$ where $N_0$ is an inner automorphism, then we say $M$ is equivalent to $N.$

Most anyonic symmetries considered in this article can be described within this simple framework. Additionally, global anyonic symmetries are easy to identify as they are common features of many topological phases. For example, all topological phases have a conjugation symmetry $C:{\bf a}\to\overline{\bf a}=-{\bf a}$, and all bilayer systems support a layer-flipping symmetry that switches anyons living on opposite layers. These both represent global anyonic symmetries that can be represented by this framework. A glossary of the anyonic symmetries studied in detail in this article can be found in Table~\ref{tab:ASexamples}. 
\begin{table}[htbp]
\centering
\begin{tabular}{lll}
Topological & Anyonic & Relabeling\\
Phase & Symmetries & Action\\\hline
All phases & $\mathbb{Z}_2$ conjugation & ${\bf a}\leftrightarrow\overline{\bf a}$\\\noalign{\smallskip}
Bi-layer systems & $\mathbb{Z}_2$ bilayer & ${\bf a}_\uparrow\leftrightarrow{\bf a}_\downarrow$\\
$(K=K_\uparrow\oplus K_\downarrow)$ & symmetry & \\\noalign{\smallskip}
$\mathbb{Z}_k$ gauge theory & $\mathbb{Z}_2$ e-m symmetry & $e^am^b\leftrightarrow m^ae^b$\\
$(K=k\sigma_x)$ & & \\\noalign{\smallskip}
$SO(8)_1$ & Triality & permutation of\\
 & $S_3$-symmetry & fermions $\psi_1,\psi_2,\psi_3$\\\noalign{\smallskip}
Bi-layer toric code & $S_3\times\mathbb{Z}_2$ & $S_3$-permutation of\\
($SO(8)_1^L\times SO(8)_1^R$) & & $\psi_1^{L/R},\psi_2^{L/R},\psi_3^{L/R}$\\
 & & Bi-layer $\psi_i^L\leftrightarrow\psi_i^R$\\
$4$-Potts Phase & $S_3$ & permutation of\\
 & & bosons $\{j_1,j_2,j_3\}$,\\
 & & twist fields $\{\sigma_1,\sigma_2,\sigma_3\}$\\
 & & and $\{\tau_1,\tau_2,\tau_3\}$
\end{tabular}
\caption{Examples of anyonic symmetries in some topological phases and their matrix representations. Details of these topological phases and symmetries can be found in Section~\ref{sec:gauging-em-bilayer-su3}, \ref{sec:so(N)}, \ref{sec:so(8)symmetry} and \ref{sec:4statePotts}.}
\label{tab:ASexamples}
\end{table}


\subsubsection{Global quantum symmetries}
\label{sec:globalquantumsymmetries}

In general, exhausting all anyonic symmetries in a Chern-Simons theory by evaluating $\mbox{Outer}(K)$ can be a combinatorically difficult task. Additionally, Eq.~\eqref{outer} might not even contain all anyonic symmetries of an Abelian topological state. For example, we will illustrate below that there are relabeling symmetries that cannot be represented by an invertible matrix, and we may also want to consider other non-trivial symmetries -- such as those of a symmetry protected topological phase (SPT)\cite{ChenGuLiuWen12, LuVishwanathE8, MesarosRan12, EssinHermele13, WangPotterSenthil13, BiRasmussenSlagleXu14, Kapustin14} -- that do not relabel anyons. Moreover, the concept of symmetry actually extends to non-Abelian phases which do not have a $K$-matrix description. Thus, a more abstract definition of anyonic symmetry is required to address these issues. Although the $K$-matrix treatment would be sufficient in most cases studied in this article, and for most apparent symmetries in ordinary Abelian systems, we here outline the essential ingredients in a more general setting for the sake of completeness, and prospective future directions of study. We should note that one class of systems where the anyonic symmetries have been tabulated is in bosonic Abelian fractional quantum Hall (FQH) states with $K$-matrices given by the Cartan matrices of the ADE series of Lie algebras. In those cases the relevant anyonic symmetry groups are equivalent to the set of spatial symmetries of the Dynkin diagrams corresponding to each Lie algebra\cite{khan2014}.

Let us now give an example of a simple Abelian example that falls outside of the $K$-matrix automorphism paradigm \eqref{automorphism}. One such example is the $U(1)_6$ state with the single-component $K$-matrix $K=12$. This is the same as the bosonic Laughlin $\nu=1/12$ FQH state. Its anyons are labeled by group elements $[m]$ in the cyclic group $\mathbb{Z}_{12}$ with spin/statistics $\theta_m=e^{2\pi im^2/24}$. Apart from the conjugation symmetry that is represented by the automorphism $m\to-m$, there is another hidden relabeling symmetry $[m]\to[5m]$.\cite{khan2014,LuFidkowski14} As 5 is relatively prime with 12, the relabeling action is one-to-one and onto. It preserves the fusion rules because multiplication and addition are associative, and the action is two-fold (i.e. squares to the identity) since $5^2=25\equiv1$ mod 12. The spin/statistics are unaltered since $5^2=25\equiv1$ mod 24. However, unlike charge-conjugation (multiplying by $-1$), transforming by multiplying by $5$ is not an invertible integer, and does not leave the $K$-matrix invariant in \eqref{automorphism}. 

In light of the above example, the automorphism condition \eqref{automorphism}, and the requirement that a symmetry is represented by an invertible integer-valued matrix, are too strong. In a general topological phase $\mathcal{B}$ (a non-degenerate braided fusion category\cite{Kitaev06, Walkernotes91, Turaevbook, FreedmanLarsenWang00, BakalovKirillovlecturenotes, Wangbook} to be precise), a symmetry operation is a bijective relabeling of anyons $M:\mathcal{B}\to\mathcal{B}$ that renames ${\bf a}\to M{\bf a}$ while preserving fusion and spin/statistics \eqref{ASdef2}. Once this is satisfied we can see that the invariance of mutual braidings is a consequence of the {\em ribbon identity} that relates fusion, twisting, and braiding: \begin{align}\theta_{\bf c}=\vcenter{\hbox{\includegraphics[width=0.5in]{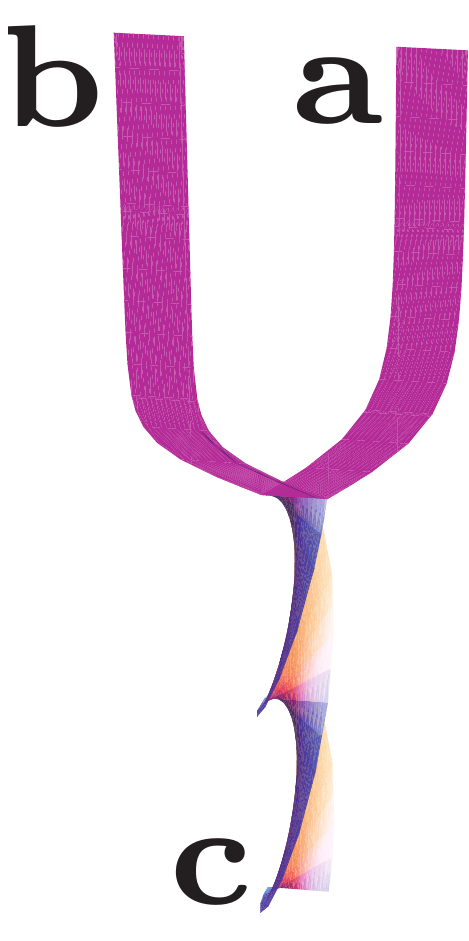}}}=\vcenter{\hbox{\includegraphics[width=0.5in]{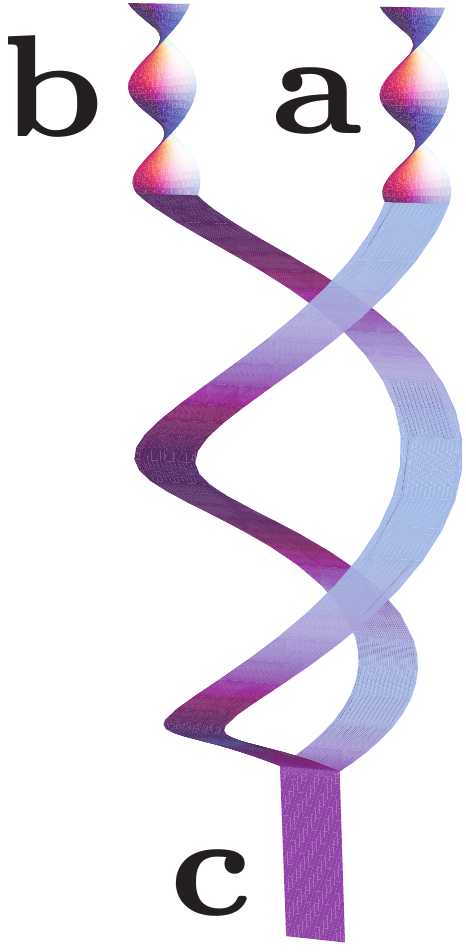}}}=R^{{\bf a}{\bf b}}_{\bf c}R^{{\bf b}{\bf a}}_{\bf c}\theta_{\bf a}\theta_{\bf b}\label{ribbon}\end{align} where $R^{{\bf a}{\bf b}}_{\bf c}R^{{\bf b}{\bf a}}_{\bf c}$ is the gauge independent ($2\pi$) braiding phase between ${\bf a}$ and ${\bf b}$ with a fixed overall fusion channel ${\bf c}$. Heuristically Eq.~\eqref{ribbon} holds because twisting the overall quasiparticle ${\bf c}$ involves twisting its constituents as well as rotating its internal structure. As a result, the braiding $S$-matrix \begin{align}S_{{\bf a}{\bf b}}=\frac{1}{\mathcal{D}}\sum_{\bf c}d_{\bf c}\mbox{Tr}\left(R^{{\bf a}{\bf b}}_{\bf c}R^{{\bf b}{\bf a}}_{\bf c}\right)=\frac{1}{\mathcal{D}}\sum_{\bf c}d_{\bf c}N^{{\bf c}}_{{\bf a}{\bf b}}\frac{\theta_{\bf c}}{\theta_{\bf a}\theta_{\bf b}}\label{braidingS}\end{align} is left unchanged by an anyonic symmetry, i.e. $S_{M{\bf a}M{\bf b}}=S_{{\bf a}{\bf b}}$, where $d_{\bf c}$ is the quantum dimension of the admissible fusion channel ${\bf c},$ and the trace is taken over the fusion degeneracy $N_{{\bf a}{\bf b}}^{\bf c}$. 

Eq.~\eqref{ASdef2}, however, is not the only concern when dealing with anyonic symmetries. In addition, the phase ambiguity of a quantum state gives rise to the subtle possibilities of {\em projectiveness} of, and {\em obstruction} to, defining a global anyonic quantum symmetry operation. In the Heisenberg picture, (local) quasiparticle operators\cite{footnotequantumop} transform according to \begin{align}\psi_{\bf a}\to\psi_{M{\bf a}}=\hat{M}\psi_{\bf a}\hat{M}^\dagger,\label{Mheisenberg}\end{align} or equivalently, in the Schr\"odinger picture (local) quantum states\cite{footnotequantumop} transform by \begin{align}|{\bf a}\rangle\to|M{\bf a}\rangle=\hat{M}|{\bf a}\rangle.\end{align} The unitary representation of symmetries could be {\em projective} in the sense that a trivial operation (or identity auto-equivalence) could give rise to an Abelian phase factor (or natural isomorphism) that alters quantum states: \begin{align}\widehat{(MN)}|{\bf a}\rangle=e^{i\phi_{M,N}({\bf a})}\hat{M}\hat{N}|{\bf a}\rangle.\label{M2cochain}\end{align} An anyonic symmetry is thus not only characterized by its relabeling operation ${\bf a}\to M{\bf a},$ but also a quantum phase $\phi_{M,N}({\bf a})$.

For example, let us take the bosonic Laughlin $\nu=1/2n$ FQH state. The corresponding $U(1)_n$ Chern-Simons theory has $2n$ quasiparticles $1,e,\ldots,e^{2n-1}$ with spins $\theta_{e^m}=e^{2\pi im^2/4n}$. The $\mathbb{Z}_2$ conjugation symmetry flips $\sigma:e^m\to e^{2n-m}$ while preserving fusion and spin. There are two inequivalent ways for the symmetry to act on quantum states, and they are distinguished by the sign, $(-1)^s,$ of the square of the conjugation operator \begin{align}\hat\sigma\hat\sigma|e^m\rangle=e^{i\phi_{\sigma,\sigma}(m)}|e^m\rangle=(-1)^{sm}|e^m\rangle.\label{Laughlinsigmasquare}\end{align} Interestingly, the sign $(-1)^{sm}$ is identical to the braiding phase $\mathcal{D}S_{e^{sn}e^m}$, and therefore the action $\hat\sigma^2$ is can be physically interpreted as (and is equivalent to) a braiding operation around the Abelian anyon $e^{sn}$.
There are consistency relations, which will be shown below, that restrict the Abelian phase factor in \eqref{Laughlinsigmasquare}. We can already see here that the signs are important because they are identified with the braiding phases involving the self-conjugate anyon $e^{sn}$, which itself remains fixed under the symmetry. As a result $\hat\sigma$ and $\hat\sigma^2$ commute, or equivalently in this case, group associativity is preserved, $\hat\sigma(\hat\sigma\hat\sigma)=(\hat\sigma\hat\sigma)\hat\sigma$.

Now we turn to a general topological phase $\mathcal{B}$. Suppose ${\bf c}$ is an Abelian anyon which can be split into a pair of Abelian anyons ${\bf a}\times{\bf b}$. In terms of quantum states, they are related by a {\em splitting state} (defined in Ref.~[\onlinecite{Kitaev06}] and reviewed in Appendix~\ref{app:reviewTQFT}). \begin{align}|{\bf a}\times{\bf b}\rangle=\hat{L}^{\bf ab}_{\bf c}|{\bf c}\rangle,\quad\mbox{for }\hat{L}^{\bf ab}_{\bf c}=\left[\vcenter{\hbox{\includegraphics[height=0.3in]{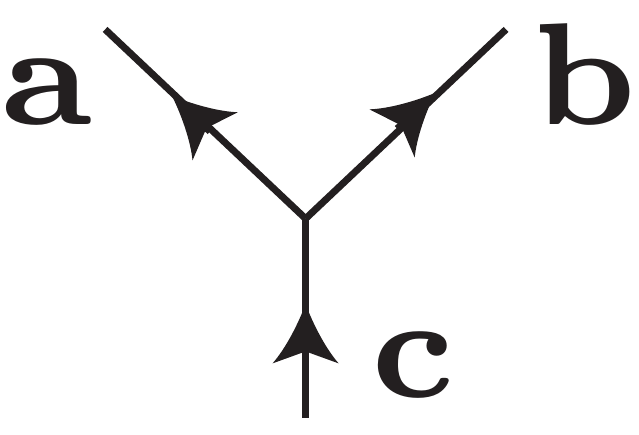}}}\right].\label{splittingstateabc}\end{align} 
Precisely, a splitting state is an equivalence class $[\hat{L}]$ of local operators -- think of string operators -- that connect $|{\bf a}\times{\bf b}\rangle=\hat{L}|{\bf c}\rangle$ for fixed anyons ${\bf a},{\bf b},{\bf c}$.  For abelian anyons, the splitting state in \eqref{splittingstateabc} is a quasiparticle string ${\bf c}$ that splits into ${\bf a}$ and ${\bf b}$.

Eq.~\eqref{M2cochain} dictates the projective symmetry action on splitting states: \begin{align}&\widehat{MN}\left[\vcenter{\hbox{\includegraphics[height=0.3in]{splittingabc}}}\right]\widehat{MN}^\dagger=e^{i\Phi_{M,N}({\bf a},{\bf b},{\bf c})}\hat{M}\hat{N}\left[\vcenter{\hbox{\includegraphics[height=0.3in]{splittingabc}}}\right]\hat{N}^\dagger\hat{M}^\dagger\label{splittingMN}\\&e^{i\Phi_{M,N}({\bf a},{\bf b},{\bf c})}=e^{i[\phi_{M,N}({\bf a})+\phi_{M,N}({\bf b})-\phi_{M,N}({\bf c})]}.\label{splittingMNsol}\end{align} Unlike the $\phi_{M,N}$ in Eq.~\eqref{M2cochain}, the projective phase $\Phi_{M,N}$ between splitting states (or local string operators) in \eqref{splittingMN} is actually a fixed physical quantity. Associativity of symmetries \begin{gather}\widehat{(LM)N}=\widehat{L(MN)},\quad\begin{diagram}\widehat{LMN}&\rTo&\widehat{LM}\widehat{N}\\\dTo&&\dTo\\\hat{L}\widehat{MN}&\rTo&\widehat{L}\widehat{M}\widehat{N}\end{diagram}\label{quantumassociativity}\end{gather} when applied to splitting states requires the triviality of \begin{align}d\Phi_{L,M,N}=N\cdot\Phi_{L,M}-\Phi_{L,MN}+\Phi_{LM,N}-\Phi_{M,N}=0\label{projassociativity}\end{align} mod $2\pi\mathbb{Z}$, where $N\cdot\Phi_{L,M}({\bf a},{\bf b},{\bf c})\equiv\Phi_{L,M}(N{\bf a},N{\bf b},N{\bf c})$. Hence, quantum anyonic symmetries are specified by (i) the anyon relabeling group $G,$ and (ii) the physical projective phases $\Phi_{M,N}$ obeying the associativity condition $d\Phi_{L,M,N}=0$ (mod $2\pi\mathbb{Z}$).

The phases $\phi_{M,N}$ defined in \eqref{M2cochain} are, however, representation-dependent and unphysical. In particular, a new representation would change $\phi_{M,N}\to\phi_{M,N}+\delta\phi_{M,N}$, but since the physical phase in \eqref{splittingMN} is fixed, the {\em change} of phase must respect the fusion rule \begin{align}\delta\phi_{M,N}({\bf c})=\delta\phi_{M,N}({\bf a})+\delta\phi_{M,N}({\bf b})\label{fusionphasesum}\end{align} modulo an integer multiple of $2\pi$, given ${\bf c}={\bf a}\times{\bf b}$. 
For all examples we will consider later, the phases $\Phi_{M,N}$ are trivial and thus the $\phi_{M,N}$ themselves respect the fusion rules \eqref{fusionphasesum}. However we will see later that there are non-trivial cases where this is not true.

In mathematical terms, a $\delta\phi_{M,N}$ respecting \eqref{fusionphasesum} is a group homomorphism from the group of {\em Abelian} anyons $\mathcal{B}^\times$ to the group of phases $U(1)$, i.e. $\delta\phi_{M,N}\in\mbox{Hom}(\mathcal{B}^\times,U(1))\cong\mathcal{B}^\times$. If the parent topological phase is Abelian,  then one can make a correspondence between $\delta\phi_{M,N}$ and a unique Abelian anyon ${\bf p}_{M,N}$ so that the phase difference \begin{align}e^{i\delta\phi_{M,N}({\bf a})}=\vcenter{\hbox{\includegraphics[width=0.1\textwidth]{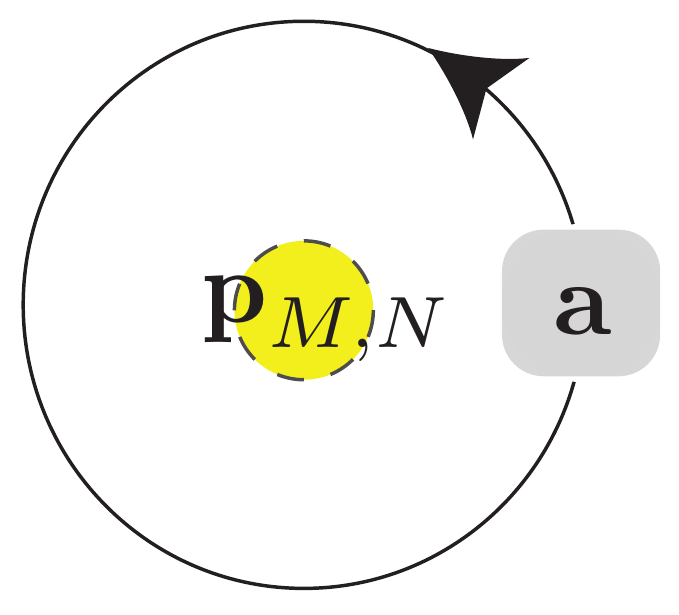}}}=e^{2\pi i{\bf a}^TK^{-1}{\bf p}_{M,N}}\label{PMNp}\end{align} 
represents a braiding phase. 


Now, to continue with our classification of the quantum anyonic symmetries we note that there are further constraints on the phases $\phi_{M,N}$, as well as some redundancies. First, associativity of the symmetries \eqref{quantumassociativity} when applied to local anyon states $|{\bf a}\rangle$ requires the (mod $2\pi\mathbb{Z}$) triviality of \begin{align}d\phi_{L,M,N}=N\cdot\phi_{L,M}-\phi_{L,MN}+\phi_{LM,N}-\phi_{M,N}\label{M3cochain}\end{align}  where $N\cdot\phi_{L,M}({\bf a})\equiv\phi_{L,M}(N{\bf a})$. $d\phi_{L,M,N}=0$ (mod $2\pi\mathbb{Z}$) is known as a {\em 2-cocycle} condition (also see \eqref{bootstrapsymm} below). 
Second, one can redefine the state basis by a phase \begin{align}|M{\bf a}\rangle'=e^{i\vartheta_M({\bf a})}\hat{M}|{\bf a}\rangle'.\label{symm2coboundary}\end{align}  A set of basis redefinitions is consistent if it leaves the splitting states (i.e.~string operators) in \eqref{splittingstateabc} invariant. This means that, like $\delta\phi_{M,N}$ in \eqref{fusionphasesum}, $\vartheta_M$ also respects the fusion rule \begin{align}\vartheta_M({\bf c})=\vartheta_M({\bf a})+\vartheta_M({\bf b})\end{align} mod $2\pi\mathbb{Z}$ if ${\bf a}\times{\bf b}={\bf c}$. This basis redefinition would modify $\phi_{M,N}\to\phi_{M,N}+d\vartheta_{M,N}$ by the {\em 2-coboundary} \begin{align}d\vartheta_{M,N}=N\cdot\vartheta_M-\vartheta_{MN}+\vartheta_N\label{M2coboundary}\end{align} where $N\cdot\vartheta_M({\bf a})\equiv\vartheta_M(N{\bf a})$. Real physical properties should not be sensitive to basis transformations, and therefore different phase characterizations $\phi_{M,N}$ should be identified if they differ by coboundaries. With these considerations in hand we are ready to classify the possible anyonic symmetries. 

\paragraph{Classification of quantum symmetries} 
The physical projective phases $\Phi_{M,N}$ on splitting states (see \eqref{splittingMN} and \eqref{projassociativity}) can be generated by inequivalent projective phases $\phi_{M,N}$ on local anyon states (see \eqref{M2cochain} and \eqref{splittingMNsol}). We will now classify these inequivalent possibilities. 
Associativity of the quantum symmetries requires \eqref{M3cochain} to be trivial, 
and the difference between any two phases $\delta\phi_{M,N}=\phi_{M,N}-\phi'_{M,N}$ now also satisfies associativity, i.e. $d\delta\phi_{L,M,N}=0$ mod $2\pi$ where $d\delta\phi_{L,M,N}$ is defined by replacing the $\phi$'s in \eqref{M3cochain} by $\delta\phi$'s. The collection of such phase differences forms the group of 2-cocycles $Z^2(G,\mathcal{B}^\times)$, where $G$ is the symmetry group and $\mathcal{B}^\times$ is the group of Abelian anyons.\cite{footnoteHomB}

The basis ambiguity/redundancy \eqref{M2coboundary} can be removed by identifying cocycles that differ by basis redefinitions \begin{align}\left[\delta\phi\right]=\left[\delta\phi+d\vartheta\right].\label{H2equivalentclass}\end{align} The change of basis $d\vartheta$ forms the group of coboundaries $B^2(G,\mathcal{B}^\times)$. The equivalence classes \eqref{H2equivalentclass} belong to the {\em second cohomology group},\cite{Cohomologybook,ChenGuLiuWen11} which is defined to be the quotient \begin{align}H^2(G,\mathcal{B}^\times)=\frac{Z^2(G,\mathcal{B}^\times)}{B^2(G,\mathcal{B}^\times)}.\label{H2GB}\end{align} (Notice, in general,  $G$ acts non-trivially on $\mathcal{B}^\times$ by anyon re-labeling.) This classifies quantum symmetries. Or to be precise, quantum symmetries are {\em torsors} over the cohomology so that the {\em difference} between any two quantum phase characterizations $\delta\phi=\phi-\phi'$ is an element in $H^2(G,\mathcal{B}^\times)$. To avoid begin overly pedantic, we here will assume $\Phi$ in \eqref{splittingMN} is trivial and treat $\phi$ as if it {\em is} a cohomological element itself instead of worrying about the distinction between $\phi$ and $\delta\phi$. 

Interestingly the presence of the Abelian anyons $\mathcal{B}^\times$ themselves actually extends the global symmetry group such that \begin{align}G=\widehat{G}/\mathcal{B}^\times.\label{Gextension}\end{align} This, in many ways, mimics the extension of a rotation point group by the set of translations to form a space group. The anyon relabeling symmetry $G$ acts as the ``rotations" ${\bf a}\to M{\bf a}$, while the set of Abelian anyons can act on themselves by ``translations" ${\bf a}\to{\bf a}\times{\bf b}$. The full quantum symmetry group incorporates both.

The simplest case of this type of extension is a {\em semi-direct product} $\widehat{G}=G\ltimes\mathcal{B}^\times$ whose elements are of the form of a 2-tuple $(\hat{M},{\bf a})$ so that group multiplication is given by $(M,{\bf a})\cdot(N,{\bf b})=(MN,{\bf a}+M{\bf b})$.  
While this obeys the definition \eqref{Gextension} of a group extension, the assumption that the relabeling group $G$ is an honest subgroup of the quantum symmetry $\widehat{G}$ is too strong. In general, group multiplication is modified by  \begin{align}(M,{\bf a})\cdot(N,{\bf b})=(MN,{\bf a}+M{\bf b}+{\bf p}_{M,N})\end{align} where ${\bf p}_{M,N}$ corresponds uniquely to the projective phase $\phi_{M,N}$ (see \eqref{PMNp}). The second cohomology $H^2(G,\mathcal{B}^\times)$ measures the failure of $G$ being a subgroup, and thus $\widehat{G}$ being a semi-direct product as well.

In the special case when the symmetry group $G$ acts trivially on the anyon labels in $\mathcal{B}^\times$, that is, when we just have a conventional global symmetry for example, it extends to a projective symmetry group (PSG).\cite{Wenspinliquid02} In this case it is known as a {\em central} extension of $G$ by $\mathcal{B}^\times$ -- or in Wen's original description the invariant gauge group (IGG)\cite{Wenspinliquid02} -- so that the set of Abelian anyons $\mathcal{B}^\times$ sit in the center of $\widehat{G}$ and commute with all group elements. In this paper we are focused instead on symmetries that permute anyons. The quantum symmetry group $\widehat{G}$ is now a {\em non-central} extension, and is not Abelian, even when $G$ is. Thus, we see that non-trivial quantum symmetries are the analog of non-symmorphic space groups,\cite{spacgroupbook} and we will call them non-symmorphic symmetry groups (NSG) to distinguish them from PSG.

In the examples listed in Table~\ref{tab:ASexamples}, only one symmetry group can be extended non-symmorphically into a quantum symmetry. This is because all of them, except one, has trivial cohomology, i.e.~$H^2(G,\mathcal{B}^\times)=0$, and thus, there is a unique way of quantizing the anyon relabeling symmetry. The conjugation symmetry ${\bf a}\to\overline{\bf a}=-{\bf a}$, however, supports multiple inequivalent quantum variations in general. We have demonstrated this by considering the bosonic Laughlin $\nu=1/2n$ state, where its conjugation symmetry necessarily squares to a possible sign choice in \eqref{Laughlinsigmasquare} due to the cocycle condition \eqref{M3cochain}. The non-trivial second group cohomology\cite{Cohomologybook,ChenGuLiuWen11} \begin{align}H^2(\mathbb{Z}_2,\mathbb{Z}_{2n})=\mathbb{Z}_2\label{H2Z2Z2n=Z2}\end{align} suggests there are two inequivalent NSG's and they are characterized by the sign in \eqref{Laughlinsigmasquare}.

The Laughlin $1/2$-state is a non-generic example in this series of states. It has anyon content $\{1,e\},$ but the $e$ quasiparticle is self-conjugate, and the anyon relabeling action is trivial. The quantum symmetry group is a PSG in this case, and there are two possibilities -- the trivial extension $\widehat{G}=\mathbb{Z}_2\times\mathbb{Z}_2$ and the projective one $\widehat{G}=\mathbb{Z}_4$. The difference between these symmetries leads to distinct ``twist liquids" after gauging. We will see the former is gauged to a $\mathbb{Z}_2$ gauge theory which is decoupled from the Laughlin state, while the latter forms the strong-paired state which is equivalent to the Laughlin $1/8$-state (or the $U(1)_4$-state). We will see this in Section~\ref{sec:4statePotts}. The quantum conjugation symmetries in all other filling fractions are not PSG's but non-Abelian NSG's. For instance, the trivial conjugation in the Laughlin $1/4$-state has the quantum symmetry group $\widehat{G}=\mathbb{Z}_2\ltimes\mathbb{Z}_4=D_4$, the dihedral group, while the non-symmorphic conjugation has the group structure $\widehat{G}=Q_8=\{\pm1,\pm i\sigma_x,\pm i\sigma_y,\pm i\sigma_z\}$ of the unit quaternions.

Finally, let us go back to  consider the richer $G=\mathbb{Z}_2\times\mathbb{Z}_2$ symmetry in the Laughlin $\nu=1/12$ state, where the first $\mathbb{Z}_2$ is the conjugation symmetry $\sigma:e^m\to e^{-m},$ and the second $\mathbb{Z}_2$ is the hidden symmetry $\tau:e^m\to e^{5m}$. The cohomology $H^2(\mathbb{Z}_2\times\mathbb{Z}_2,\mathbb{Z}_{12})=\mathbb{Z}_2^3$ suggests there are 8 inequivalent ways the symmetries can act on quantum states. They are specified by the three phases: \begin{gather}\hat\sigma\hat\sigma|e^m\rangle=(-1)^{s_1m}|e^m\rangle,\quad\hat\tau\hat\tau|e^m\rangle=i^{s_2m}|e^m\rangle\\\hat\sigma\hat\tau|e^m\rangle=(-1)^{s_3m}\hat\tau\hat\sigma|e^m\rangle\label{12sigmatau}\end{gather} where $s_1,s_2,s_3=0,1$. A sign in the $\hat\tau^2$ phase can be absorbed by a basis transformation \eqref{symm2coboundary} and therefore $i^{s_2m}$ and $(-i)^{s_2m}$ are equivalent representations. Eq.~\eqref{12sigmatau} demonstrates the possibility that quantum symmetries do not necessarily commute, even if they appear to do so at the level of the anyon relabeling operations.

\paragraph{Obstruction to quantum symmetries} 
Obstruction to a consistent quantum symmetry comes from the failure of associativity \eqref{quantumassociativity}.
This arises from non-trivial physical projective phases $\Phi_{M,N}$ on splitting states (see Eq.~\eqref{splittingMN}). Even if these phases are associative and obey $d\Phi=0$ (see Eq.~\eqref{projassociativity}), there might not exist consistent projective phases $\phi_{M,N}$ on local anyon states (see Eq.~\eqref{M2cochain}) to generate $\Phi_{M,N}$ by \eqref{splittingMNsol}. Here we classify these obstructions.

Let us take a look at associativity. There are multiple ways of boot-strapping the action of three symmetry operations $LMN$ on local anyon states: \begin{align}\begin{diagram}\widehat{LMN}&\rTo^{e^{i\phi_{LM,N}}}&\widehat{LM}\widehat{N}\\\dTo^{e^{i\phi_{L,MN}}}&&\dTo_{e^{iN\cdot\phi_{L,M}}}\\\hat{L}\widehat{MN}&\rTo^{e^{i\phi_{M,N}}}&\widehat{L}\widehat{M}\widehat{N}.\end{diagram}\label{bootstrapsymm}\end{align} A consistent theory should not depend on the path taken in \eqref{bootstrapsymm}, i.e.~$d\phi$ defined in \eqref{M3cochain} should be trivial. This, in general, might not be achievable because of the fixed non-trivial physical phase factor on splitting states (or local string operators): \begin{align}\Phi_{M,N}({\bf a},{\bf b},{\bf c})=\phi_{M,N}({\bf a})+\phi_{M,N}({\bf b})-\phi_{M,N}({\bf c})\label{Phi=phiphiphi}\end{align} mod $2\pi\mathbb{Z}$ for ${\bf a}\times{\bf b}={\bf c}$.

For example, let us take the $U(1)_n$-state with $2n$ Abelian anyons $\mathcal{A}=\{[e^{p}]:[p]\in\mathbb{Z}_{2n}\}$, where $[p]=[p+2n]$ and $1=e^0$ is the vacuum. This topological state is identical to the Laughlin $\nu=1/2n$ FQH state with quasiparticles $e^p$, for $-\infty<p<\infty$, so that $e^{2n}$ is the fundamental local boson, and $e^p$, $e^{p+2n}$ belong to the same anyon type $[e^p]$. The $\mathbb{Z}_2=\{1,\sigma\}$ conjugation symmetry switches $\sigma:e^p\leftrightarrow e^{-p}$. This symmetry could seemingly be projectively represented at the splitting state level \eqref{splittingMN} with $\Phi_{1,1}=\Phi_{1,\sigma}=\Phi_{\sigma,1}=0$ and  \begin{align}e^{i\Phi_{\sigma,\sigma}(e^p,e^q,e^r)}=(-1)^{\frac{p+q-r}{2n}}\in\mathbb{Z}_2\label{U(1)nobstruction}\end{align} for $p+q\equiv r$ mod $2n$ so that $[e^p]\times[e^q]=[e^r]$. For this choice, the physical phases $\Phi_{M,N}$ are associative, i.e.~they obey \eqref{projassociativity}, and are generated by \eqref{Phi=phiphiphi} using the projective phases $e^{i\phi_{1,1}}=e^{i\phi_{1,\sigma}}=e^{i\phi_{\sigma,1}}=1$ and \begin{align}e^{i\phi_{\sigma,\sigma}}|e^p\rangle=e^{2\pi i\frac{p}{4n}}|e^p\rangle\label{U(1)nphi}\end{align} on local anyon states \eqref{M2cochain}. For instance, the symmetry differentiate the fundamental local boson from the vacuum so that $\hat\sigma^2|e^{2n}\rangle=-|e^{2n}\rangle$. However, these $\phi$ phases violate associativity since \begin{align}d\phi_{\sigma,\sigma,\sigma}(e^p)=-2\pi\frac{p}{2n}\neq0\label{U(1)ndphi}\end{align} mod $2\pi\mathbb{Z}$ (see Eq.~\eqref{M3cochain}). 


In general, given a set of projective phases $\phi_{M,N}$ that generate the physical $\Phi_{M,N}$ by \eqref{Phi=phiphiphi}, the {\em obstruction} $d\phi_{L,M,N}$ defined in \eqref{M3cochain} must respect fusion rules although $\phi_{M,N}$ in general does not. [For instance in \eqref{U(1)nphi}, $\phi_{\sigma,\sigma}(e^n)+\phi_{\sigma,\sigma}(e^n)=\pi$ is not identical to $\phi_{\sigma,\sigma}(e^n\times e^n)=\phi_{\sigma,\sigma}(e^0)=0$.] This is because \begin{align}&d\phi_{L,M,N}({\bf a})+d\phi_{L,M,N}({\bf b})-d\phi_{L,M,N}({\bf c})\nonumber\\&=d\Phi_{L,M,N}({\bf a},{\bf b},{\bf c})=0\label{dphifusion}\end{align} mod $2\pi\mathbb{Z}$ for ${\bf a}\times{\bf b}={\bf c}$ (see \eqref{projassociativity}). Moreover, it is straightforward to show $d\phi_{L,M,N}$ is $d$-closed \begin{align}dd\phi_{K,L,M,N}=&N\cdot d\phi_{K,L,M}-d\phi_{K,L,MN}\nonumber\\&+d\phi_{K,LM,N}-d\phi_{KL,M,N}+d\phi_{L,M,N}=0\label{M3cocycle}\end{align} mod $2\pi\mathbb{Z}$, where $N\cdot d\phi_{K,L,M}({\bf a})=d\phi_{K,L,M}(N{\bf a})$. The collection of obstructions $d\phi$ that respect fusion \eqref{dphifusion} and the $d$-closed condition \eqref{M3cocycle} forms the group of {\em 3-cocycles} $Z^3(G,\mathcal{B}^\times)$.\cite{footnoteHomB}

Now, the projective symmetry will have a consistent quantum representation if the phases $\phi_{M,N}$ in \eqref{M2cochain} can be redefined via $\phi'_{M,N}=\phi_{M,N}+\delta\phi_{M,N}$ (without modifying the overall physical phase $\Phi_{M,N}$ in \eqref{splittingMN} and \eqref{Phi=phiphiphi}), such that $d\phi'_{L,M,N}=0$ mod $2\pi\mathbb{Z}$. 
The difference $\delta\phi_{M,N}$ must respect fusion as discussed in \eqref{fusionphasesum}. In general it modifies the obstruction by a {\em 3-coboundary} $d\phi'_{L,M,N}=d\phi_{L,M,N}+d\delta\phi_{L,M,N}$ where $d\delta\phi\in B^3(G,\mathcal{B}^\times)$ is defined by replacing $\phi$ in \eqref{M3cochain} by $\delta\phi$. Similar to the classification of quantum symmetries by the second cohomology \eqref{H2GB}, obstructions are therefore classified by equivalence classes \begin{align}[d\phi]=[d\phi+d\delta\phi]\end{align} in the {\em third cohomology group}\cite{Cohomologybook,ChenGuLiuWen11} \begin{align}H^3(G,\mathcal{B}^\times)=\frac{Z^3(G,\mathcal{B}^\times)}{B^3(G,\mathcal{B}^\times)}.\end{align} Again, $G$ acts non-trivially on $\mathcal{B}^\times$ if it relabels anyons.

The cohomology class $[d\phi]$ is entirely a property of the physical phase $\Phi_{M,N}({\bf a},{\bf b},{\bf c})$ on splitting states \eqref{splittingMN}, and is independent of the solution for $\phi$ to \eqref{Phi=phiphiphi}. When it is trivial, the obstruction $d\phi_{L,M,N}$ is identical to some coboundary $d\delta\phi,$ and can be removed by a redefinition $\phi'=\phi+\delta\phi$ such that $d\phi'=0,$ and hence an associative projective representation (see Eq.~\eqref{quantumassociativity}) exists, even at the local anyon state level \eqref{M2cochain}. On the contrary, if $[d\phi]$ is cohomologically non-trivial, then the obstruction is irremovable. In this case, projective symmetries cannot be associatively defined at the local anyon state level \eqref{M2cochain}, despite the fact that they are associative at the splitting state level \eqref{splittingMN}. For example, the projective $\mathbb{Z}_2$-symmetry \eqref{U(1)nobstruction} for the $U(1)_n$-state {\em cannot} be extended to local anyon states. This is because the obstruction in \eqref{U(1)ndphi} belongs to the non-trivial cohomology class in $H^3(\mathbb{Z}_2,\mathbb{Z}_{2n})=\mathbb{Z}_2$.\cite{Cohomologybook,ChenGuLiuWen11}

Beyond this discussion of classification, we will not further investigate anyonic symmetries that exhibit obstructions to quantum representations in this article. The absence of an obstruction can be ensured by requiring that the phases $\phi_{M,N}$ themselves respect fusion, or equivalently, that the quantum operators $\widehat{MN}$ and $\widehat{M}\widehat{N}$ act identically on splitting states so that $\Phi_{M,N}({\bf a},{\bf b},{\bf c})=0$. 
This requirement applies to all symmetries listed in Table~\ref{tab:ASexamples}. We suspect this requirement could fail on the surface of certain 3D SPTs, where surface topological order could give rise to a non-trivial obstruction to the quantum symmetry. We leave discussions of this to future work.
 




\subsection{Defect fusion category}
\label{sec:twistdefects}
\begin{figure}[htbp]
\includegraphics[width=0.3\textwidth]{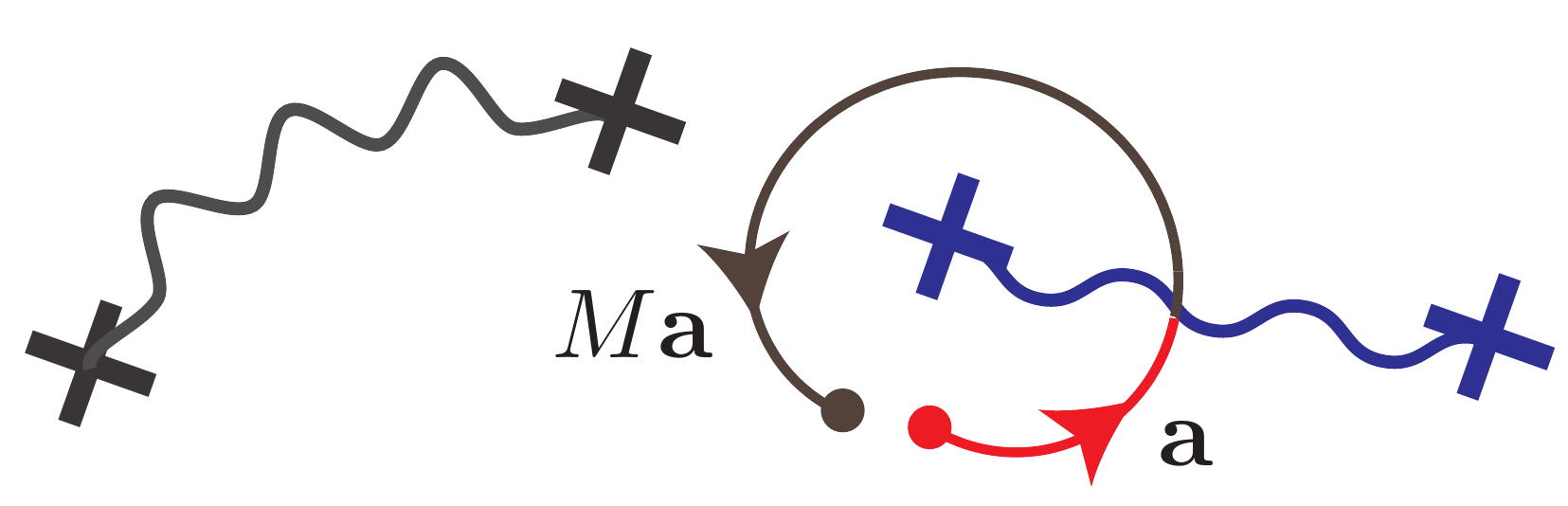}
\caption{Twist defects (crosses) connected by arbitrary branch cuts (curvy lines) where a passing anyon changes type ${\bf a}\to M{\bf a}$ according to an anyonic symmetry $M$.}\label{fig:defectanyoncuts}
\end{figure}

Now that we have classified anyonic symmetries we will proceed to the construction of the defect fusion category, i.e. the object that serves as the input for the twist liquid state.
Let us start by reviewing some recent developments in the theory of twist defects in topological phases.\cite{Kitaev06, EtingofNikshychOstrik10, barkeshli2010, Bombin, Bombin11, KitaevKong12, kong2012A, YouWen, YouJianWen, PetrovaMelladoTchernyshyov14, BarkeshliQi, BarkeshliQi13, BarkeshliJianQi, MesarosKimRan13, TeoRoyXiao13long, teo2013braiding, khan2014, BarkeshliBondersonChengWang14} A twist defect is a semiclassical topological point defect (with an attached branch-cut) labeled by an anyonic symmetry $M$. It is characterized by its action on anyons that encircle the defect. A quasiparticle will change type according to the symmetry operation $M:{\bf a}\to M{\bf a}$ when it travels once, counter-clockwise,  around the defect (see Fig.~\ref{fig:defectanyoncuts}). 
In a system with a finite number of defects, there exists a quasi-global definition of anyon labels that covers the system almost everywhere in space except along certain {\em branch cuts} between defects where the anyon label definition changes (see Fig.~\ref{fig:defectanyoncuts}).

Unlike anyonic quasiparticles, topological twist defects are not dynamical excitations of a quantum Hamiltonian. They are classical configurations or textures that vary slowly away from the defect points/cores. For example, in the absence of vortices the phase of an s-wave superconductor order parameter is locally uniform, but the phase winds by $2\pi$ around a flux vortex. A distortion in the defect texture in two dimensions usually generates  a confining potential between defect partners that grows at least logarithmically in their separation. Therefore, one would not expect unbound defect pairs to appear spontaneously at long length scales.

In many example models, topological order and discrete spatial order are intertwined, especially in lattice spin or rotor models with topological order. In these cases, twist defects can manifest themselves as lattice defects such as dislocations and disclinations. For example, a dislocation in the toric code\cite{Kitaev06, Bombin, KitaevKong12} (see Section~\ref{sec:toriccodetwistdefects}) switches the anyon type $e\leftrightarrow m$ of a quasiparticle after it travels once around the dislocation\cite{Bombin}. Other examples include the defects in the $\mathbb{Z}_k$ plaquette model of Wen\cite{YouWen, YouJianWen, teo2013braiding}, the Kitaev honeycomb model\cite{Kitaev06, PetrovaMelladoTchernyshyov14} and the color code model\cite{Bombin11, TeoRoyXiao13long}. Twist defects can also capture the fusion properties of parafermionic zero modes trapped at domain walls of fractional quantum spin Hall edges\cite{ClarkeAliceaKirill,LindnerBergRefaelStern,MChen,mongg2,Vaezi}. We will not focus on the realization of such defects, but instead we are interested in their fusion characteristics.\cite{TeoRoyXiao13long, teo2013braiding, khan2014, BarkeshliBondersonChengWang14}

Now we will move on to define and discuss the general defect fusion category structure. The fundamental objects in this semiclassical description are defect-quasiparticle composites denoted by $M_\lambda$, where $M$ is an anyonic symmetry element in $G$ associated to the defect, and $\lambda$ is a {\em species label} representing the equivalence class of the anyon charge bound at the defect-quasiparticle composite. For example, $\lambda$ could specify the fractional electric charge carried by a twist defect in a FQH state, or more general anyonic charges. Additionally, a species label can change, or \emph{mutate}, when the defect is fused with a quasiparticle. For example, \begin{align}{\bf a}\times M_\lambda=M_\lambda\times{\bf a}=M_{\lambda'}.\label{generalspeciesmutation}\end{align} In general, the species label can be statistically distinguished by a Wilson loop measurement via dragging a quasiparticle, ${\bf a}$, $p$ times around the defect, where $M^p{\bf a}={\bf a}$. For instance, the two species of dislocations in the toric code give distinct phase factors under the double Wilson loop operator $\Theta$ in Fig.~\ref{fig:defect1}(b), which is well defined since the anyonic symmetry is twofold.

In general, the globally (anyonic) symmetric parent topological state is equipped with a {\em braided fusion category}\cite{Kitaev06, Walkernotes91, Turaevbook, FreedmanLarsenWang00, BakalovKirillovlecturenotes, Wangbook} $\mathcal{B}$ which contains all the fusion and braiding data of the quasiparticles. A {\em defect fusion category}\cite{EtingofNikshychOstrik10,TeoRoyXiao13long,teo2013braiding,BarkeshliBondersonChengWang14} is a {\em $G$-graded} extension \begin{align}\mathcal{C}=\bigoplus_{M\in G}\mathcal{C}_M\label{Ggradedfusion}\end{align} containing $\mathcal{C}_1$, i.e. the fusion data of $\mathcal{B}$ itself, which corresponds to 1, i.e. the identity group element. Each other sector $\mathcal{C}_M$ is generated by twist defects associated to the anyonic symmetry $M$ with different species labels. A quasiparticle encircling two defects associated to symmetries $M$ and $N$ is relabeled by the combination $MN$. Group multiplication is therefore carried over to defect fusion \begin{align}\mathcal{C}_{M}\times\mathcal{C}_{N}\longrightarrow\mathcal{C}_{MN}\label{Ggradedfusionf}\end{align} and does not in general commute when $G$ is non-Abelian. 

As an example, the defect fusion category of the toric code has $\mathcal{C}_1=\langle1,e,m,\psi\rangle$ and $\mathcal{C}_\sigma=\langle\sigma_0,\sigma_1\rangle$, where $\mathbb{Z}_2=\{1,\sigma\}$ is the electric-magnetic anyonic symmetry group. Additionally, defect fusion respects group multiplication. For instance, the Ising fusion rule $\sigma_0\times\sigma_0=1+\psi$ is consistent with the requirement that $\mathcal{C}_\sigma\times\mathcal{C}_\sigma\to\mathcal{C}_1$ according to the twofold group structure $\sigma^2=1$. Moreover, the quasiparticle sector $\mathcal{C}_1$ acts transitively on the defect sectors $\mathcal{C}_\sigma$ and gives rise to distinct species labels $e\times\sigma_\lambda=m\times\sigma_\lambda=\sigma_{\lambda+1}$. In a general defect theory, the quasiparticle sector $\mathcal{C}_1$ is always closed under fusion, $\mathcal{C}_1\times\mathcal{C}_1\to\mathcal{C}_1,$ and it acts on individual defect sectors \begin{align}\mathcal{C}_1\times\mathcal{C}_M\longrightarrow\mathcal{C}_M,\quad\mathcal{C}_M\times\mathcal{C}_1\longrightarrow\mathcal{C}_M\end{align} by combining with the defects to form defect-quasiparticle composites. Mathematically, each defect sector $\mathcal{C}_M$ is known as a {\em $\mathcal{C}_1$-bimodule} (like a $\mathcal{C}_1$-vector space) and is equipped with associative ``vector addition" and ``scalar multiplication" operations such as $\sigma_0+\sigma_1=(1+e)\times\sigma_0$ and $e\times(m\times\sigma_0)=(e\times m)\times\sigma_0=\sigma_0$.

\subsubsection{Defects in an Abelian parent state}
Defect species can be described more explicitly if the globally (anyonic) symmetric parent topological state is Abelian, and we will focus on this case for now. In a $K$-matrix description \eqref{CSaction}, Abelian quasiparticles are labeled by an anyon lattice $\mathcal{A}=\mathbb{Z}^N/K\mathbb{Z}^N$ with lattice addition reflecting the fusion rule $\psi^{\bf a}\times\psi^{\bf b}=\psi^{{\bf a}+{\bf b}}$. Species of twist defects associated to an anyonic symmetry operation $M$ can be labeled by the quotient lattice\cite{teo2013braiding, khan2014} \begin{align}\mathcal{A}_M=\frac{\mathcal{A}}{(1-M)\mathcal{A}}.\label{defectsectorquotient}\end{align} This is because combining a defect with quasiparticles that are related by the symmetry should give the identical defect-quasiparticle composite: \begin{align}M_\lambda=\psi^{\bf l}\times M_0=\psi^{{\bf l}+(M-1){\bf b}}\times M_0.\label{defectQPcompositeeq}\end{align} This can be diagrammatically explained by comparing topologically equivalent quasiparticle string patterns of the composite object: 
\begin{align}\vcenter{\hbox{\includegraphics[width=0.2\textwidth]{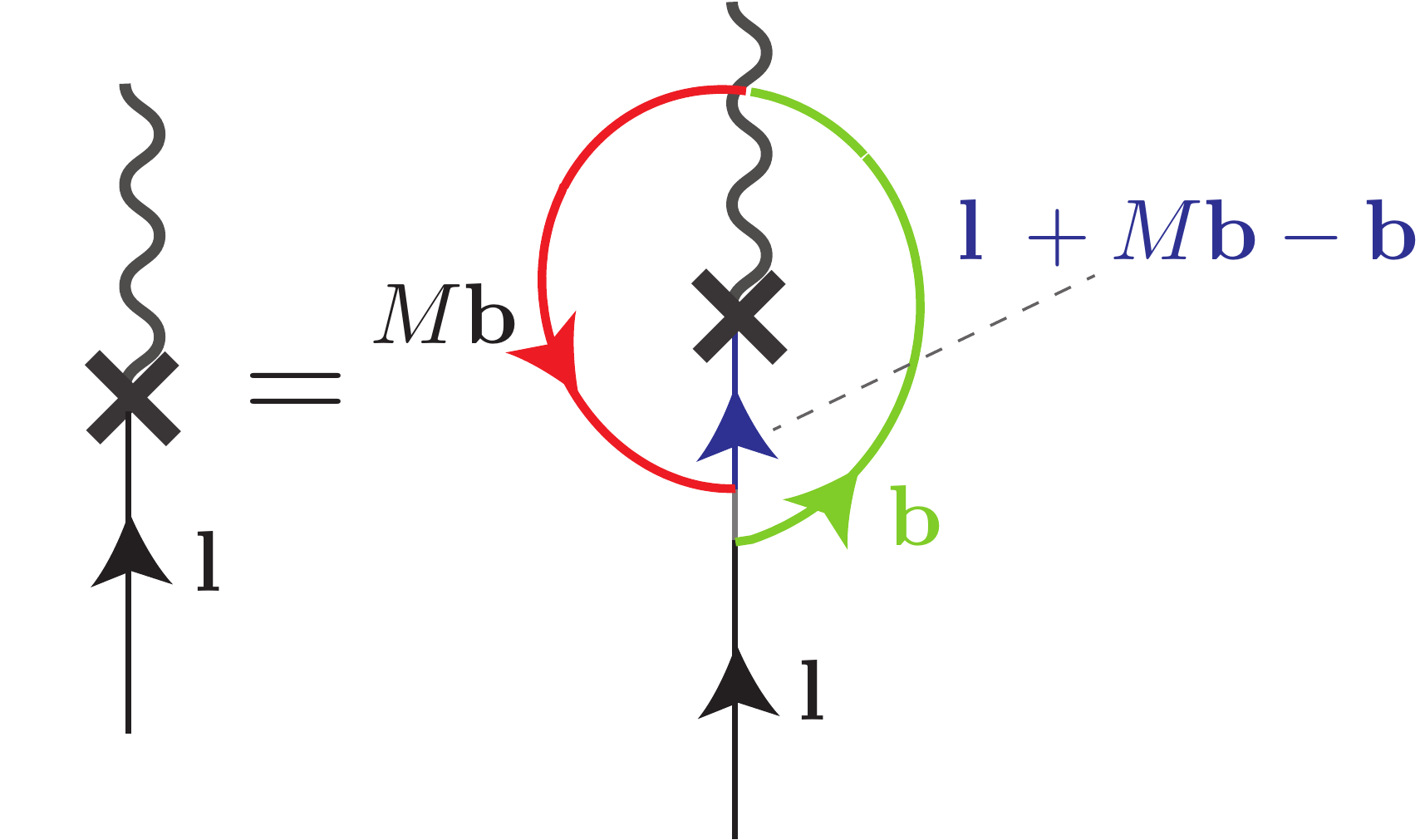}}}\label{defectQPcomposite}\end{align} where the ${\bf l}$ QP string attached to the defect $M_\lambda$ can be modified to ${\bf l}+M{\bf b}-{\bf b}$ by splitting off a ${\bf b}$ QP and letting it orbit once around the defect. This process cannot change the defect specices $\lambda$ as it cannot be detected by any Wilson measurements. 
For the toric code, $\mathcal{A}=\{1,e,m,\psi\}=\mathbb{Z}_2\oplus\mathbb{Z}_2$ and $(1-\sigma)\mathcal{A}=\{1,\psi\}=\mathbb{Z}_2$. The quotient is thus $\mathcal{C}_\sigma=\mathbb{Z}_2\oplus\mathbb{Z}_2/\mathbb{Z}_2=\mathbb{Z}_2$ and contains the two species of twist defects $\sigma_0,\sigma_1$.

\begin{figure}[htbp]
\centering\includegraphics[width=0.4\textwidth]{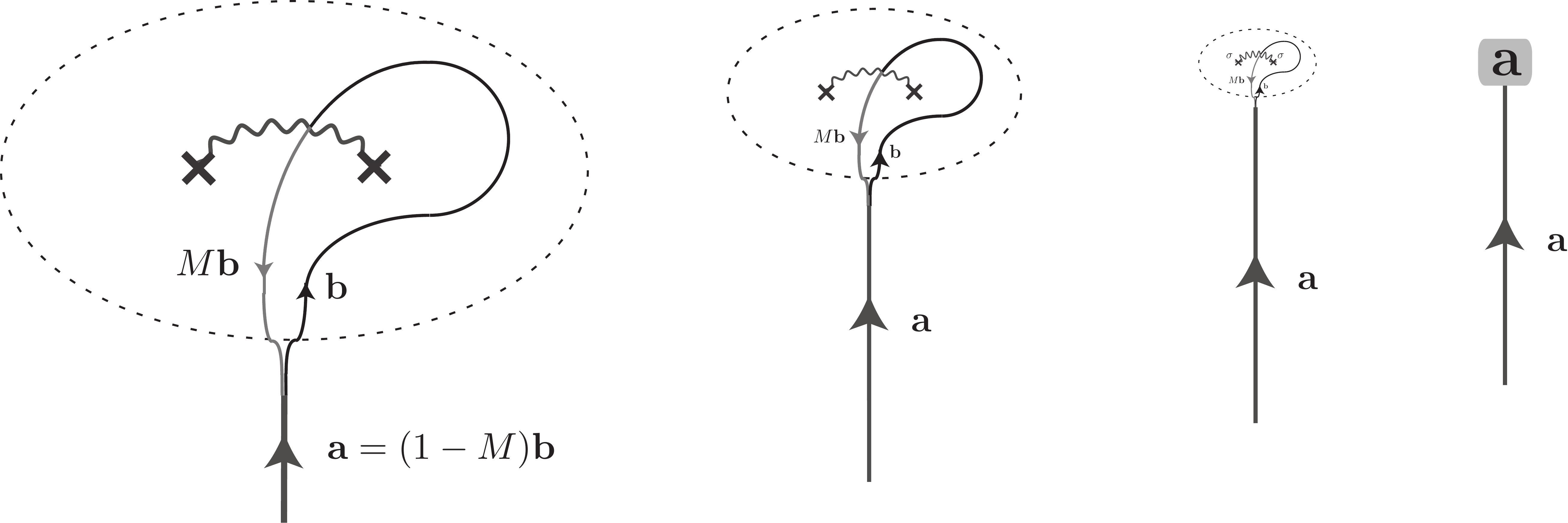}
\caption{Fusion of a pair of (bare) defects associated to opposite anyonic symmetries $M$ and $M^{-1}$.}\label{fig:splittingstategeneraldefect1}
\end{figure}

Next we consider the fusion of a conjugate defect pair, but allowing for the possibility of different species labels on each defect. They can be diagrammatically represented by point defects with canceling branch cuts (see Fig.~\ref{fig:splittingstategeneraldefect1}). Their fusion outcome must be a trivial defect, i.e. an Abelian quasiparticle (in $\mathcal{C}_1$), because of their trivial overall relabeling action to anyons encircling the pair of defects. The overall fusion channel depends on how quasiparticle strings are hung between the defects. Fig.~\ref{fig:splittingstategeneraldefect1} shows the general admissible string configurations that are irremovable when the defects fuse together. The overall open string contributes ${\bf a}=(1-M){\bf b}$ to the fusion channel of the defect pair. In other words, there is a one-to-one correspondence between the admissible fusion channels and the anyon sublattice $(1-M)\mathcal{A}.$ This leads to the fusion rule structure \begin{align}M_\lambda\times\overline{M}_{\lambda'}={\bf e}\times\sum_{{\bf a}\in(1-M)\mathcal{A}}{\bf a}\label{conjugatedefectfusion}\end{align} where $\overline{M}_{\lambda'}$ is a defect conjugate to $M_\lambda$ associated to the inverse symmetry $M^{-1}$, and ${\bf e}$ is some Abelian quasiparticle that depends on the species labels $\lambda,\lambda'$ as well as the projective phase $\phi_{M,M^{-1}}$ in Eq.~\eqref{M2cochain}. Each defect, in general, is attached to a quasiparticle string ${\bf l}$ that reflects its species label $\lambda$ (see Eqs.~\eqref{defectQPcompositeeq} and \eqref{defectQPcomposite}). Hence, the quasiparticle strings ${\bf l}$ and ${\bf l}'$ of the defects $M_\lambda$ and $\overline{M}_{\lambda'}$ reflect their species labels. Both strings contribute to the fusion outcome in \eqref{conjugatedefectfusion}, and are encapsulated by the anyon label ${\bf e}$. Given a fixed species label, the choice of ${\bf l}$ and ${\bf l}'$ is not unique, however the choice does not affect the overall outcome since all anyons in $(1-M)\mathcal{A}$ are summed over in \eqref{conjugatedefectfusion}. The effect of the projective phase $\phi_{M,M^{-1}}$ from possible non-symmorphic symmetries will be discussed in Section~\ref{sec:defectclassificationobstruction}. Hence the precise form of the QP ${\bf e}$ will be presented later in Eq.~\eqref{eanyon}.

Since Abelian quasiparticles have unit quantum dimension, and defects with different species labels are related by absorbing or emitting Abelian quasiparticles \eqref{generalspeciesmutation}, all defects in the same sector $\mathcal{C}_M$ must share identical quantum dimension $d_M$. [This is {\em not} true if the globally symmetry parent state is non-Abelian. An example is shown for a chiral state we call the ``4-Potts" state in Section~\ref{sec:4statePotts}.] Moreover, since conjugate objects in a fusion theory must carry identical quantum dimension, we have $d_M=d_{\overline{M}}$. By equating the quantum dimensions on both sides of Eq.~\eqref{conjugatedefectfusion}, each defect carries a dimension of \begin{align}d_M=\sqrt{\left|(1-M)\mathcal{A}\right|}.\label{defectdimensionabelian}\end{align} This number determines the ground state degeneracy of a system of $\mathcal{N}$ defects in the thermodynamic limit, $G.S.D.\propto (d_M)^{\mathcal{N}}$ for $\mathcal{N}\to\infty$. For example, twist defects in the toric code satisfy $\sigma_\lambda\times\sigma_\lambda=1+\psi$ and have quantum dimension $d_\sigma=\sqrt{2}.$

The total quantum dimension of the defect fusion category is defined so that it squares to the sum of the squares of the dimensions of all the simple objects in $\mathcal{C}$: \begin{align}\mathcal{D}_{\mbox{\small defect}}\equiv\sqrt{\sum_{M\in G}\sum_{\lambda\in\mathcal{A}_M}d^2_{M}}=\mathcal{D}_0\sqrt{|G|}\label{defecttotaldimension}\end{align} where we note that sum also contains the the trivial identity element of $G,$ $\mathcal{D}_0=\sqrt{|\mathcal{A}|}$ is the total quantum dimension of the globally symmetric Abelian parent state without considering defects, and $|\ast|$ is the number of elements in $\ast$. Eq.~\eqref{defecttotaldimension} can be proven by seeing that all defect sectors $\mathcal{C}_M$ have identical total dimension. \begin{align}\sum_{\lambda\in\mathcal{A}_M}d^2_M=\left|\frac{\mathcal{A}}{(1-M)\mathcal{A}}\right|\times\left|(1-M)\mathcal{A}\right|=\left|\mathcal{A}\right|\end{align} where the quotient accounts for the number of defect species $\lambda$ (see Eq.~\eqref{defectsectorquotient}) and its denominator cancels $d_M^2$ (see Eq.~\eqref{defectdimensionabelian}). 
A proof of Eq.\eqref{defecttotaldimension} by a general argument that applies even to non-Abelian parent states can be found in Ref.~[\onlinecite{BarkeshliBondersonChengWang14}].

\subsubsection{Basis transformations for defect states}\label{sec:basis-transformations-defects}

\begin{figure}[htbp]
\centering\includegraphics[width=0.4\textwidth]{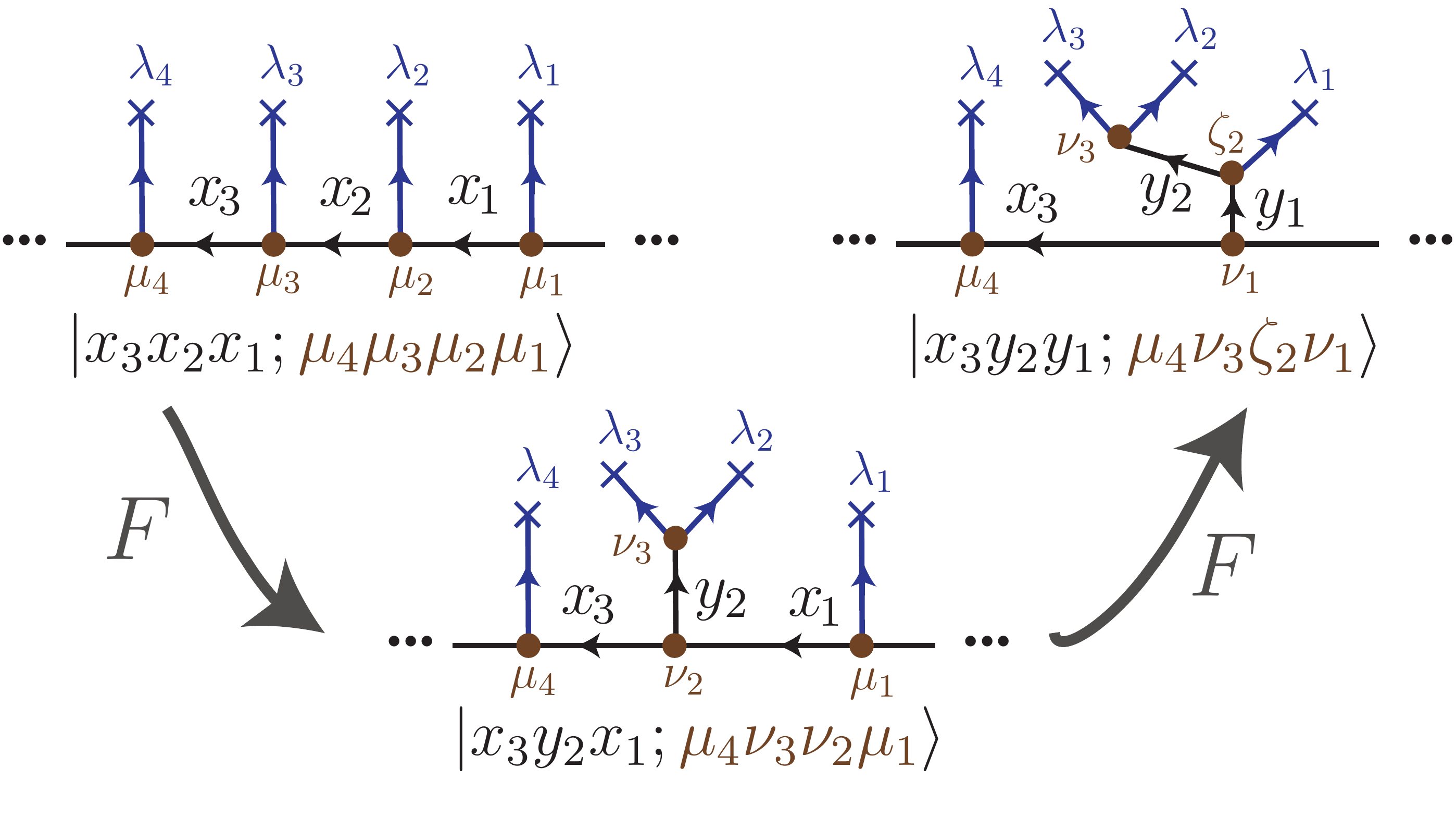}
\caption{Basis transformation of defect states by composition of fundamental $F$-moves. $\lambda_i$ are defect objects, $x_i,y_i$ are admissible internal fusion channels, and $\mu_i,\nu_i,\zeta_i$ label vertex degeneracies.}\label{fig:Fmoves}
\end{figure}

Just as the case when non-Abelian anyons are present, the (ground) states of a system of defects can be labeled by the quantum numbers of a maximal set of commuting Wilson-line observables, whose eigenvalues are associated to the internal fusion channels and vertex degeneracies of a fusion tree (see Fig.~\ref{fig:Fmoves}). Examples of this type of representation were demonstrated for the defect states in the toric code above (see Section~\ref{sec:toriccodetwistdefects} and Appendix~\ref{app:toricdefect}). 
For instance, quantum states in a system of four Ising defects (or zero energy Majorana bound states) $\sigma_1,\sigma_2,\sigma_3,\sigma_4$ can be labeled by the commuting local fermion parities $\{(-1)^{F_{12}},(-1)^{F_{34}}\}$ of pairs of zero modes. They label the fusion channels $1$ or $\psi$ of $\sigma_1\times\sigma_2$ and $\sigma_3\times\sigma_4$.
Different fusion trees correspond to distinct sets of Wilson operators that may not commute. Each tree defines a complete basis for the degenerate ground states, and basis transformations between different fusion trees are generated by the $F$-symbols (see Fig.~\ref{fig:Fmoves}). They are defined in Eq.~\eqref{Fsymboldef} and \eqref{Fdefinition}.
Continuing with our example, quantum states of the four Ising defects in the toric code can also be labeled by a different set of local fermion parities $\{(-1)^{F_{23}},(-1)^{F_{41}}\}$, which does not commute with the original. Basis states with respect to these two sets of local observables are related by the transformation $F^{\sigma\sigma\sigma}_\sigma$ in Table~\ref{tab:Fsymbolstoriccode}. 

\begin{figure}[htbp]
\centering\includegraphics[width=0.4\textwidth]{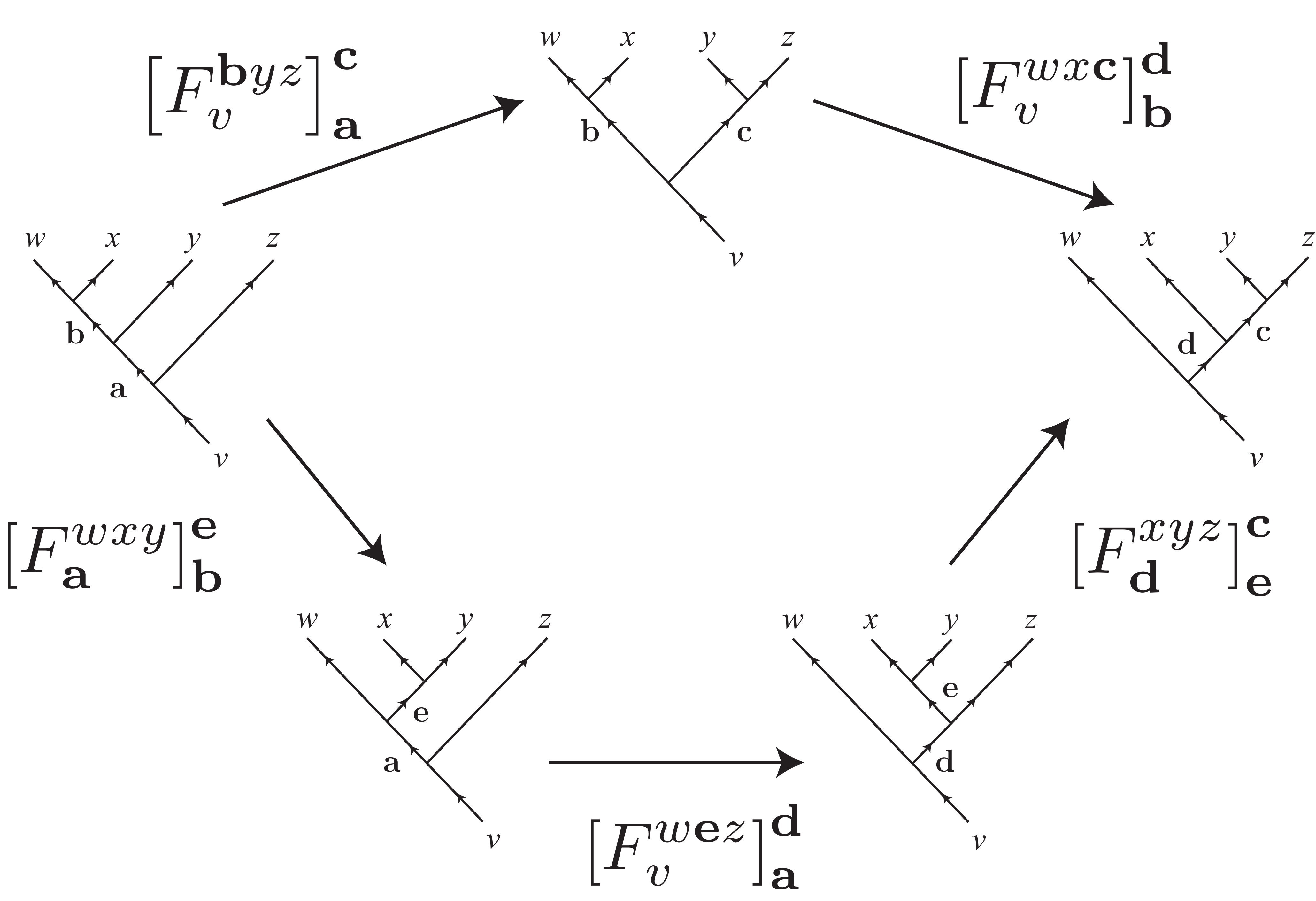}
\caption{The pentagon equation.\cite{Kitaev06} Summation is taken over the internal anyon label ${\bf e}$. Vertex labels are suppressed.}\label{fig:Fpentagon}
\end{figure}

The overall basis transformation between any two fusion trees, or maximally commuting sets of observables, is independent of the sequence of $F$-moves in between. This cocycle consistency condition is ensured by the pentagon equation ``$FF=FFF$" (see Fig.~\ref{fig:Fpentagon}) and MacLane's coherence theorem (see Refs. \onlinecite{Kitaev06, MacLanebook}). As mentioned earlier, instead of solving the algebraic pentagon equation, the $F$-matrices can be more efficiently computed directly from their definition.\cite{TeoRoyXiao13long,teo2013braiding} This is done by choosing a particular set of splitting states $[L^{xy}_z]$ for an admissible fusion process $x\times y\to z$. Each state can be diagrammatically represented by quasiparticle Wilson strings around $x$ and $y$ with fixed boundary conditions associated to the fusion outcome $z$. An example of this is given for defects in the toric code (see Fig.~\ref{fig:splittingstatestoriccode} in Appendix~\ref{app:toricdefect}). In fact the string configuration in Fig.~\ref{fig:splittingstategeneraldefect1} can also be treated as a choice of splitting state for a conjugate pair of bare defects in a general Abelian system. 


In general, different fusion trees (see Fig.~\ref{fig:Fmoves}) will associate different quasiparticle string patterns around defects.
From these trees, the $F$-symbols can be evaluated from their definition by overlapping quasiparticle strings between two fusion trees, keeping track of crossing phases when strings intersect, and condensing trivial quasiparticle loops into the condensate. An illustration for the transformation matrix $F^{\sigma\sigma\sigma}_\sigma$ can be found in Appendix~\ref{app:toricdefect}. In general, \emph{defect} $F$-symbols can be derived based on the known quasiparticle braiding information  and the $F$-symbols of the globally symmetric parent topological state. More examples will be considered in Section~\ref{sec:gauging-em-bilayer-su3}, \ref{sec:so(N)}, \ref{sec:so(8)symmetry} and \ref{sec:4statePotts}.

\subsubsection{Classification and obstruction of defect fusion categories}
\label{sec:defectclassificationobstruction}

In Section~\ref{sec:globalquantumsymmetries}, we have discussed the phase ambiguities in the representations of quantum symmetries that operate {\em projectively} on quantum states. The phase difference $\phi_{M,N}({\bf a})$ between the individual symmetry operations $\hat{M}\hat{N}$ and the combined operation $\widehat{MN}$ on the state $|{\bf a}\rangle$ (see Eq.~\eqref{M2cochain}) leads to {\em non-symmorphic} quantum symmetries which are classified by the cohomology group $H^2(G,\mathcal{B}^\times)$. One reason we were so careful with this classification is that, as hinted in Eq.~\eqref{conjugatedefectfusion}, the projective phases $\phi$ affect defect fusion, i.e. they affect the result of taking two independent anyonic symmetry operations  $\hat{M},\hat{N}$ and then fusing them to form $\widehat{MN}.$ 

\begin{figure}[htbp]
\centering\includegraphics[width=0.4\textwidth]{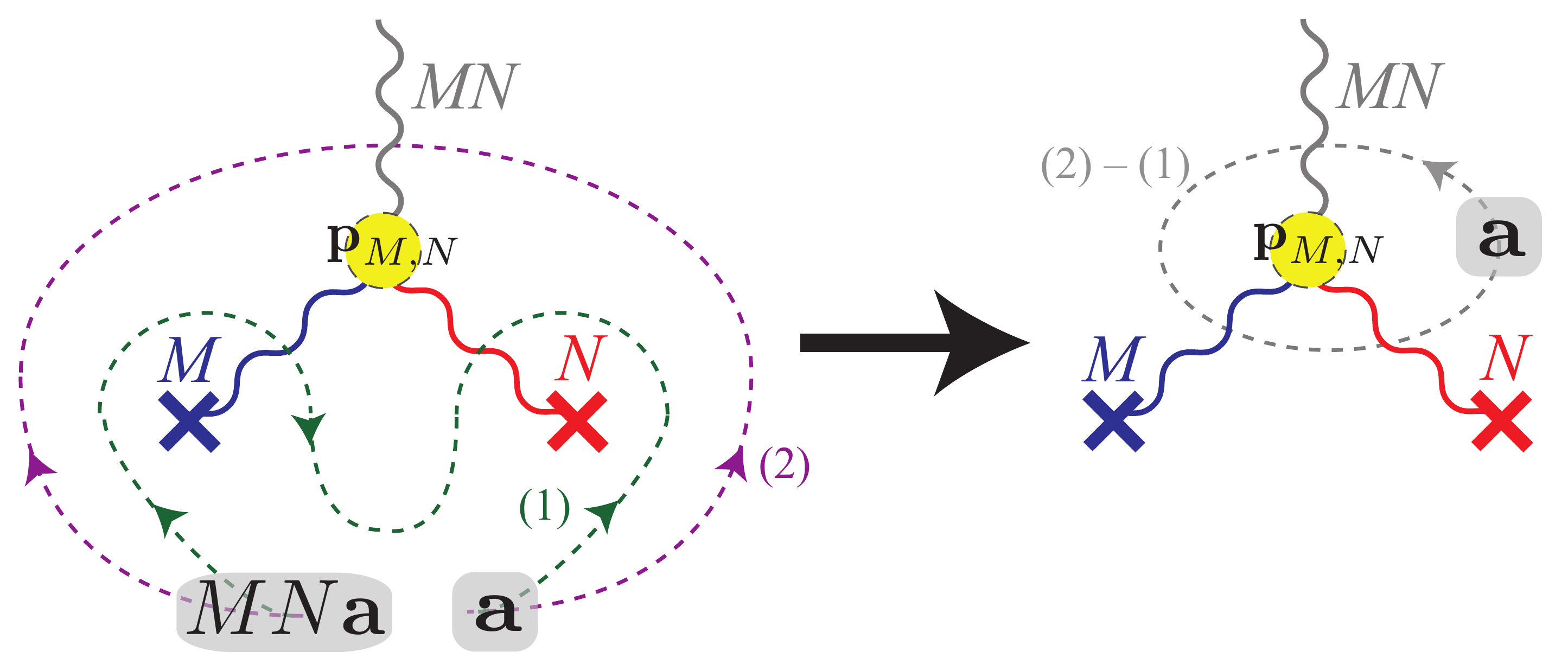}
\caption{(1) and (2) are trajectories of an Abelian anyon ${\bf a}$ around two defects associated to the anyonic symmetries $M$ and $N$. The difference between the two paths $(2)-(1)$ brings ${\bf a}$ around the tri-junction of branch cuts.}\label{fig:NSfusion}
\end{figure}
Let us assume there is no obstruction to a consistent representation of the quantum symmetries, and for simplicity we assume that the projective phases $\phi_{M,N}$ themselves respect the fusion of Abelian anyons, i.e.  $\phi_{M,N}({\bf c})=\phi_{M,N}({\bf a})+\phi_{M,N}({\bf b})$ mod $2\pi\mathbb{Z}$ if ${\bf a}\times{\bf b}={\bf c}.$ Now,  let us consider a pair of defects associated with symmetries $M$ and $N$ respectively. Under fusion $M\times N\to MN$, and during this process their branch cuts join at a tri-junction represented by the orange blob in Fig.~\ref{fig:NSfusion}. The projective phase $\phi_{M,N}({\bf a})$ can be physically interpreted as the phase difference between two different trajectories of an anyon ${\bf a}$: (1) a path that encloses the defects individually and (2) a path that encloses the full pair after fusion. 
As an abelian anyon ${\bf a}$ is dragged along the path, the quantum state $|{\bf a}\rangle$ is acted upon by a quantum symmetry operation every time it passes across a branch cut. As a result, there is a phase difference of $e^{i\phi_{M,N}({\bf a})}$ between the two as seen in Eq.~\eqref{M2cochain}. Since the \emph{difference} between path (1) and (2) is a loop around the tri-junction where the branch cuts meet (see Fig.~\ref{fig:NSfusion}), the phase difference can be regarded as the braiding phase \begin{align}e^{i\phi_{M,N}({\bf a})}=\vcenter{\hbox{\includegraphics[width=0.1\textwidth]{NSbraiding}}}=e^{2\pi i{\bf a}^TK^{-1}{\bf p}_{M,N}}\label{PMN}\end{align} 
between the quasi-particle ${\bf a}$ and another Abelian quasiparticle ${\bf p}_{M,N}$ trapped at the tri-junction (see Fig.~\ref{fig:NSfusion}). This physical interpretation of the projective phase always holds when the globally symmetric parent state is Abelian and the projective phase $\phi$ respects fusion. A more careful treatment might be required for general scenarios, but will not be presented in this article.

\begin{figure}[htbp]
\centering\includegraphics[width=0.45\textwidth]{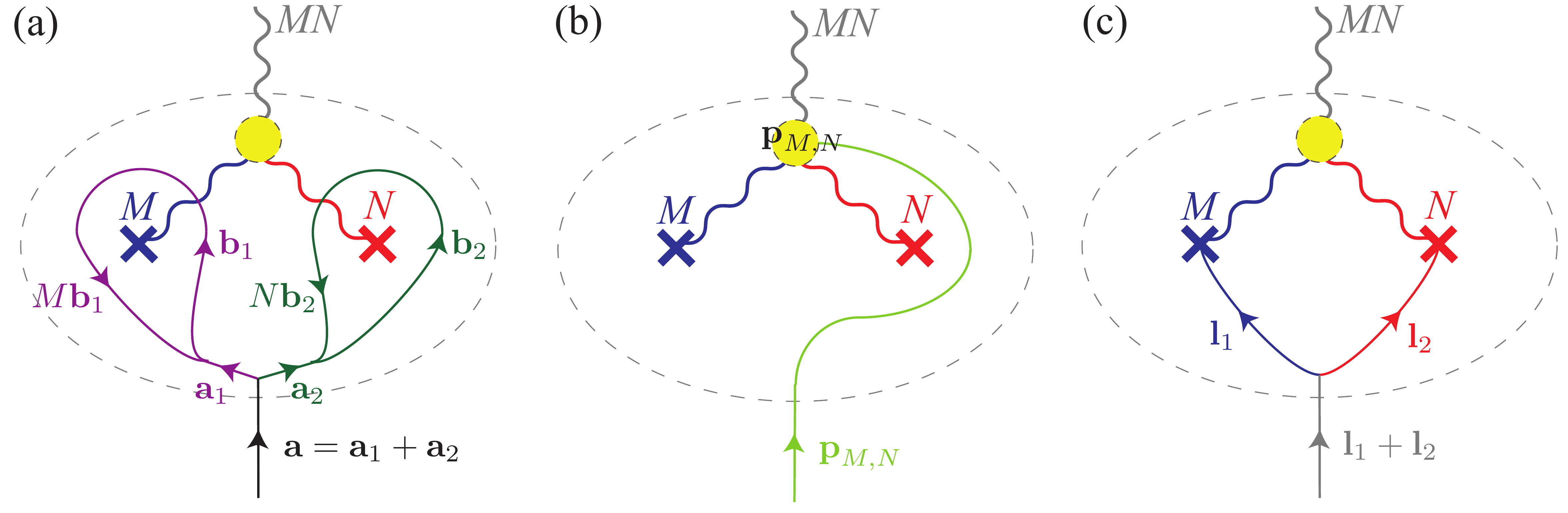}
\caption{The three contributions to the overall species label after the fusion of two defects $M\times N\to MN$.}\label{fig:MNfusion}
\end{figure}
Defect fusion rules $\mathcal{C}_M\times\mathcal{C}_N\to\mathcal{C}_{L}$, for $L=MN$, in an Abelian topological parent state $\mathcal{A}$ take the following general form \begin{align}M_{\lambda_1}\times N_{\lambda_2}={\bf e}^{M,N}_{\lambda_1,\lambda_2}\times\sum_{\lambda\in\mathcal{A}^{M,N}_{L}}\mathcal{N}^{M,N}_LL_\lambda.\label{generaldefectfusion}\end{align} Here $M_{\lambda_1},N_{\lambda_2},L_\lambda$ are defects associated to the symmetries $M,N,L$ each with their own species labels $\lambda_1, \lambda_2, \lambda_3$ living in the quotient lattices $\mathcal{A}_M,\mathcal{A}_N,\mathcal{A}_L$ defined in \eqref{defectsectorquotient} respectively. The quantity ${\bf e}^{M,N}_{\lambda_1,\lambda_2}$ is an Abelian quasiparticle to be discussed below. The admissible fusion channels range over a restricted set $\mathcal{A}^{M,N}_L$ of species labels $\lambda.$ Additionally,  a fusion channel can be degenerate so that there are multiple ways for the defects $M_{\lambda_1}\times N_{\lambda_2}$ to fuse into ${\bf e}^{M,N}_{\lambda_1,\lambda_2}\times L_\lambda.$ This degeneracy is accounted for through the non-negative integer $\mathcal{N}^{M,N}_L$, which does not depend on defect species. 

As we will now describe, the allowed fusion channels can be understood diagrammatically as in Fig.~\ref{fig:MNfusion}.
Figure~\ref{fig:MNfusion}(a) shows all possible quasiparticles ${\bf a}$ that can be attached to the fused defect $MN$ without modifying the initial species labels of its components $M$ and $N$.  This means that we can write ${\bf a}={\bf a}_1+{\bf a}_2=(1-M){\bf b}_1+(1-N){\bf b}_2$ for some anyon lattice vectors ${\bf b}_1,{\bf b}_2$ in $\mathcal{A}.$ Thus,  the quasiparticle ${\bf a}$ belongs to the sublattice $(1-M)\mathcal{A}+(1-N)\mathcal{A}$. The admissible species labels $\lambda$ in \eqref{generaldefectfusion} are restricted to the quotient \begin{align}\mathcal{A}^{M,N}_L=\frac{(1-M)\mathcal{A}+(1-N)\mathcal{A}}{(1-MN)\mathcal{A}}\label{AMNL}\end{align} since ${\bf a}\equiv{\bf a}+(1-MN){\bf b}$ corresponds to an identical species label for the combined defect $MN$ (see Eq.~\eqref{defectsectorquotient}).

The fusion degeneracy $\mathcal{N}^{M,N}_L$ of each admissible channel $M_{\lambda_1}\times N_{\lambda_2}\to{\bf e}^{M,N}_{\lambda_1,\lambda_2}\times L_\lambda$ is illustrated through the Wilson-loop algebra of closed quasiparticle strings between the defects. These closed Wilson loops are generated by the set of strings in Fig.~\ref{fig:MNfusion}(a), where 
${\bf a}={\bf a}_1+{\bf a}_2\equiv 0$. 
The set of Wilson operators do not necessarily commute due to possible non-trivial crossing phases, and they can form a non-Abelian Wilson algebra. An irreducible representation of this algebra corresponds to a $\mathcal{N}^{M,N}_L$-fold degenerate ground state. The total degeneracy can be evaluated by equating the quantum dimensions of both sides of Eq.~\eqref{generaldefectfusion}: \begin{align}d_Md_N=\mathcal{N}^{M,N}_Ld_L\left|\mathcal{A}_L^{M,N}\right|\end{align} where $\mathcal{A}^{M,N}_L$, the group defined in \eqref{AMNL}, has order \begin{align}\left|\mathcal{A}_L^{M,N}\right|&=\left|\frac{(1-M)\mathcal{A}\oplus(1-N)\mathcal{A}}{(1-M)\mathcal{A}\cap(1-N)\mathcal{A}}\right|\times\frac{1}{\left|(1-MN)\mathcal{A}\right|}\nonumber\\&=\frac{d_M^2d_N^2}{d_{L}^2}\times\frac{1}{\left|(1-M)\mathcal{A}\cap(1-N)\mathcal{A}\right|}.\end{align} Thus, the fusion degeneracy is \begin{align}\mathcal{N}^{M,N}_L=\frac{d_L}{d_Md_N}\left|(1-M)\mathcal{A}\cap(1-N)\mathcal{A}\right|.\end{align} Such defect fusion degeneracies have been observed for $\mathbb{Z}_3$ symmetric states\cite{teo2013braiding,khan2014}. Other examples will be presented for the $SO(8)_1$ state and the chiral ``4-Potts" state in Sections~\ref{sec:so(8)symmetry} and \ref{sec:4statePotts} respectively.

Next we want to determine the Abelian anyon attached to each fusion channel. Fig.~\ref{fig:MNfusion}(b) shows the contribution to the overall species label from non-symmorphic quantum symmetries. As seen in Eq.~\eqref{PMN}, and Fig.~\ref{fig:NSfusion}, the projective phase difference $e^{i\phi_{M,N}}$ between  individual and combined quantum symmetries $\hat{M}\hat{N}$ and $\widehat{MN}$ corresponds to an Abelian anyon ${\bf p}_{M,N}$ at the tri-junction of fused branch cuts. By convention, we arrange this quasiparticle string on the right side so that the braiding phase \eqref{PMN} can be recovered by intersecting strings to the right of defect $N$. This quasi-particle combines with the quasiparticles ${\bf l}_1$ and ${\bf l}_2$ that are attached to the defects $M$ and $N$ according to their species labels $\lambda_1$ and $\lambda_2$ (see Fig.~\ref{fig:MNfusion}(c)), to form the resulting Abelian quasiparticle \begin{align}{\bf e}^{M,N}_{\lambda_1,\lambda_2}={\bf p}_{M,N}+{\bf l}_1+{\bf l}_2.\label{eanyon}\end{align}

Given a particular defect species label there is a choice for the possible Abelian anyon bound to the defect since the species labels represent an equivalence class of anyons. Let us show that the defect fusion rules are independent of the choice of ${\bf l}_1$ and ${\bf l}_2$. We could have picked any of the equivalent representations ${\bf l}_1+(1-M){\bf b}_1$ and ${\bf l}_2+(1-N){\bf b}_2.$ This will alter \eqref{eanyon}, but the difference is just an anyon ${\bf a}$ in $(1-M)\mathcal{A}+(1-N)\mathcal{A}$ that is automatically absorbed by the sum in \eqref{generaldefectfusion}.

\begin{figure}[htbp]
\centering\includegraphics[width=0.4\textwidth]{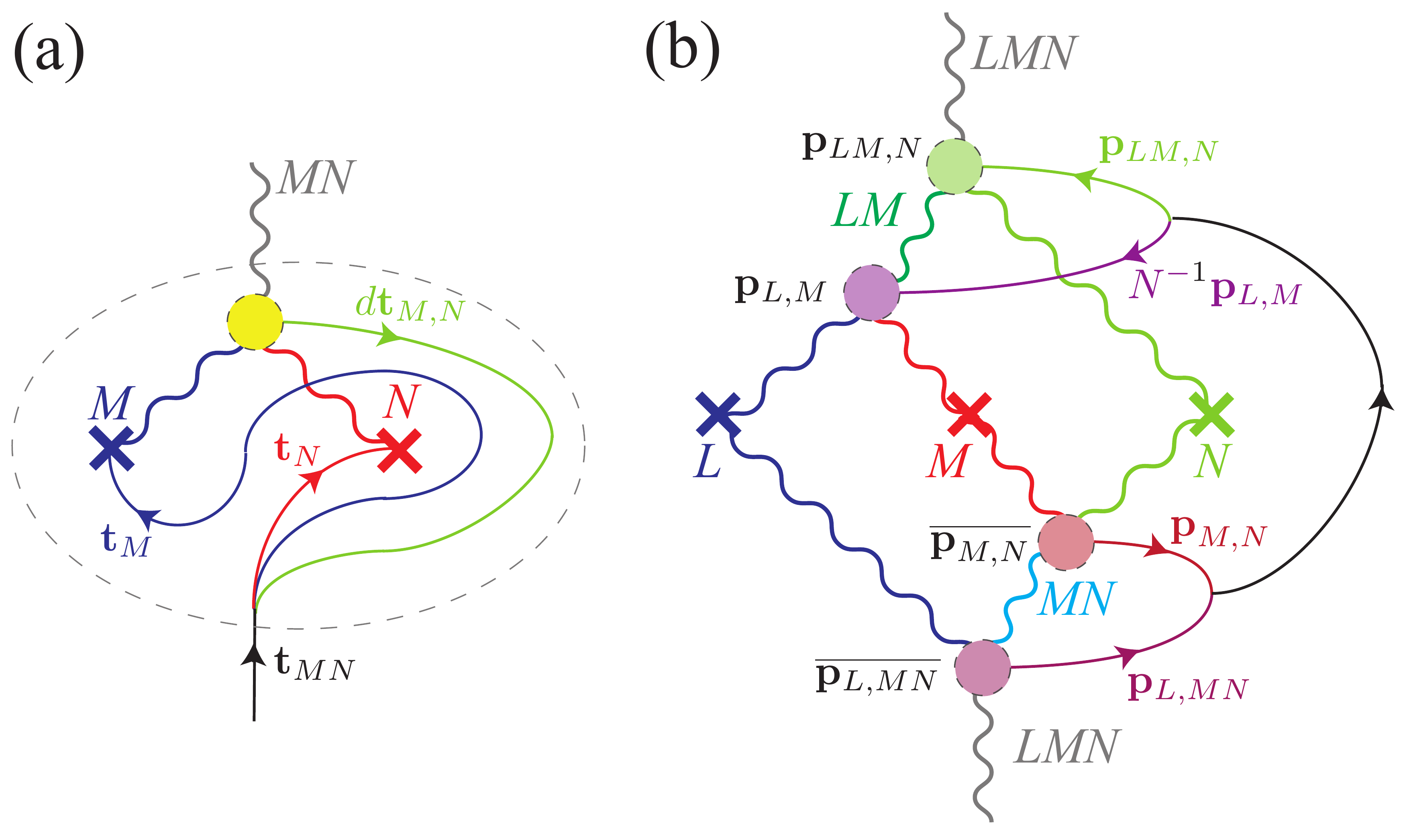}
\caption{(a) Absorbing the 2-coboundary $d{\bf t}_{M,N}$ by species redefinition \eqref{speciesredefinition}. (b) Cocycle condition for a consistent set of ${\bf p}_{M,N}$.}\label{fig:speciesredefinition}
\end{figure}

Next, we recall that the projective phase $\phi_{M,N}$ is not physical and can be modified by a 2-coboundary $\phi_{M,N}\to\phi_{M,N}+d\vartheta_{M,N}$ in \eqref{M2coboundary}. Thus, the corresponding anyon ${\bf p}_{M,N}$ is only well defined up to ${\bf p}_{M,N}\to{\bf p}_{M,N}+d{\bf t}_{M,N}$ for \begin{align}d{\bf t}_{M,N}=N^{-1}{\bf t}_M-{\bf t}_{MN}+{\bf t}_N\label{speciesredefinition}\end{align} where $\{{\bf t}_M:M\in G\}$ is an arbitrary set of Abelian anyons that relates to a set of phases $\{\vartheta_M:M\in G\}$ in \eqref{symm2coboundary} by the 1-1 correspondence \begin{align}e^{i\vartheta_M({\bf a})}=\mathcal{D}S_{{\bf a},{\bf t}_M}=e^{2\pi i{\bf a}^TK^{-1}{\bf t}_M}.\end{align}
\noindent Eq.~\eqref{speciesredefinition} will alter a particular fusion rule since the Abelian anyon in \eqref{eanyon} will be modified. However, the entire defect fusion structure (i.e., the full set of defect fusion rules) will still remain unchanged because the modification \eqref{speciesredefinition} can be absorbed by a redefinition of species labels \begin{align}M_{\lambda}\to M_{\lambda'}={\bf t}_M\times M_{\lambda}.\end{align} Fig.~\ref{fig:speciesredefinition}(a) explicitly shows how the 2-coboundary in \eqref{speciesredefinition} can be split into three parts and independently be absorbed into individual defects in the fusion $M\times N\to MN$.

Finally, if the fusion rules are associative $\left(L\times M\right)\times N=L\times\left(M\times N\right)$ the coboundary \begin{align}d{\bf p}_{L,M,N}=N^{-1}{\bf p}_{L,M}-{\bf p}_{L,MN}+{\bf p}_{LM,N}-{\bf p}_{M,N}=0,\label{M3cochain2}\end{align} i.e. there is a constraint on the relationship between the quasiparticles that are bound at different branch cut tri-junctions (see Fig.~\ref{fig:speciesredefinition}(b)). This matches the cocycle condition in \eqref{M3cochain} for the corresponding projective phase $\phi_{M,N}$. From \eqref{speciesredefinition} and \eqref{M3cochain2}, we see that consistent defect fusion rules are classified by 2-cocycles ${\bf p}$ up to 2-coboundaries $d{\bf t}$. In other words, they are classified by the same cohomology in $H^2(G,\mathcal{A})$ that characterized the quantum symmetry (see Eq.~\eqref{H2GB}). Similarly, a violation of the cocycle condition \eqref{M3cochain2} implies an inconsistency of fusion, which is classified by $H^3(G,\mathcal{A})$.

Let us take a moment to consider an example. A non-trivial example of the effect of non-symmorphic quantum symmetries on defect fusion rules can be given in an Abelian parent state\cite{MooreSeiberg89,Bondersonthesis,BarkeshliBondersonChengWang14} $\mathbb{Z}_{2n}^{(q+1/2)}$, where $q$ is an integer. This family of Abelian topological phase is parameterized by $q$ and contains  $2n$ quasiparticles $\langle1,e,e^2,\ldots,e^{2n-1}\rangle$, where the generator has $q$-dependent exchange statistics $\theta_e=e^{2\pi i(q+1/2)/(2n)},$ and obeys the fusion rule $e^{2n}=1$. This family of phases includes some familiar examples. For instance, when $q=0$, this Abelian state is identical to the Laughlin $\nu=1/2n$ FQH state (or the $U(1)_n$-state); and when $q=n-1$, this is identical to the $SU(2n)_1$ state.
For all values of $q$ the non-trivial anyonic symmetry operation on $\mathbb{Z}_{2n}^{(q+1/2)}$ is just conjugation $\sigma$ that switches $e^m\leftrightarrow\overline{e^m}=e^{2n-m}.$ As described above, the quantum symmetry of this type is classified by $H^2(\mathbb{Z}_2,\mathbb{Z}_{2n})=\mathbb{Z}_2$ (see Eq.~\eqref{H2Z2Z2n=Z2}). The two distinct classes are characterized by the projective phase $e^{i\phi_{\sigma,\sigma}(e^m)}=(-1)^{sm}$, for $s=0,1$, (see Eq.~\eqref{Laughlinsigmasquare}) which corresponds to a bound branch-cut tri-junction quasiparticle ${\bf p}_{\sigma,\sigma}=e^{sn}$ (see Eq.~\eqref{PMN}). When $n$ is odd, this gives rise to two possible inequivalent defect fusion rules \begin{align}\sigma_0\times\sigma_0&=\left\{\begin{array}{*{20}c}1+e^2+\ldots+e^{2n-2},&\mbox{for $s=0$}\\e+e^3+\ldots+e^{2n-1},&\mbox{for $s=1$.}\end{array}\right.\label{U(1)NSGfusion}\end{align} Gauging the latter fusion structure for the Laughlin $1/2n$ state would lead to a $U(1)_n/\mathbb{Z}_2$ orbifold theory.\cite{Ginsparg88,DijkgraafVafaVerlindeVerlinde99,bigyellowbook} 
For the simplest case, when $n=1$, the trivial fusion rule $\sigma_0\times\sigma_0=1$ leads to the twist liquid $U(1)_1\otimes(\mbox{Toric code})$, where $\sigma_0$ becomes the flux $m$ of the toric code. On the other hand, the non-trivial fusion rule $\sigma_0\times\sigma_0=e$ acts as a fractionalization of the Laughlin quasiparticle, and leads to a different gauging result, i.e. $U(1)_4$, with 8 quasiparticles generated by $\sigma_0$. This will reappear later in Section~\ref{sec:4statePotts}.

Other than the fusion rules, a defect theory is also characterized by its basis transformations, which are generated by the $F$-symbols (see Appendix~\ref{app:reviewTQFT} and Section~\ref{sec:basis-transformations-defects}). The defect $F$-matrices can be evaluated diagrammatically by specifying a particular set of splitting states like those in Fig.~\ref{fig:splittingstatestoriccode} for the toric code, or Fig.~\ref{fig:MNfusion} in general. However, this diagrammatic evaluation leaves the $F$-symbols undetermined by certain Abelian phase factors that arise when branch cuts are rearranged: \begin{align}\left\langle\vcenter{\hbox{\includegraphics[width=0.1\textwidth]{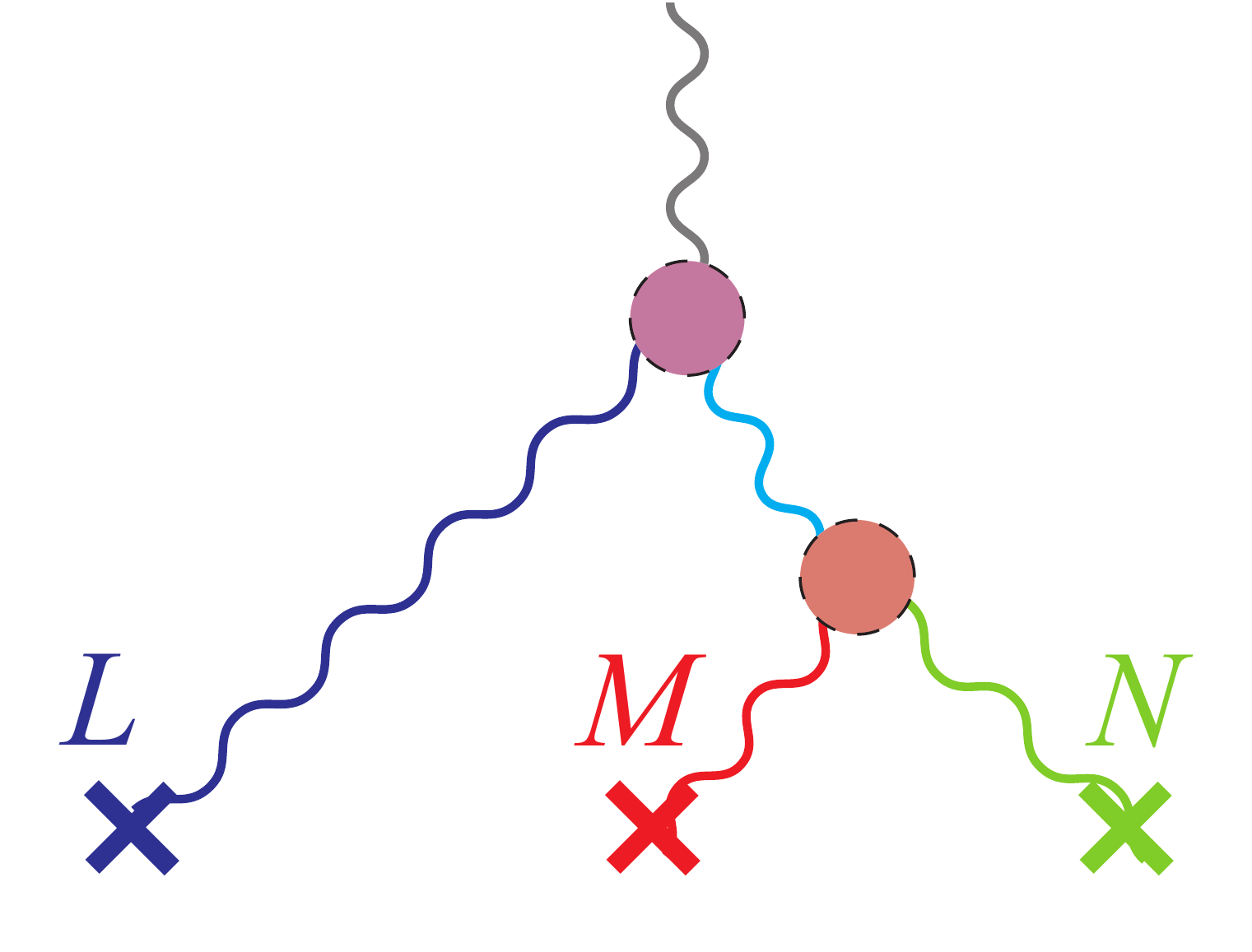}}}\right|\left.\vcenter{\hbox{\includegraphics[width=0.1\textwidth]{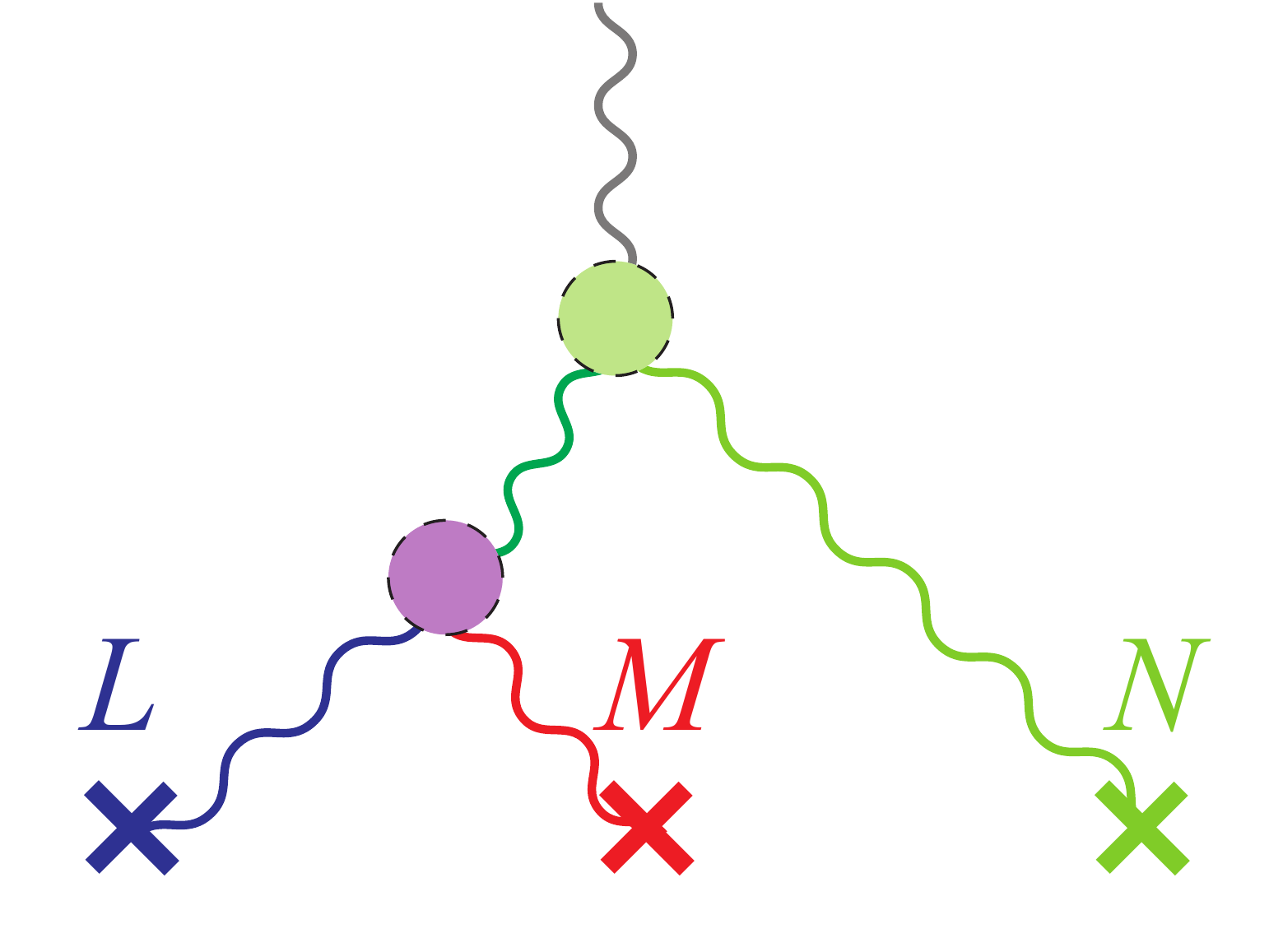}}}\right\rangle=\frac{1}{\sqrt{\mathcal{N}}}e^{i\chi_{L,M,N}}\label{Fphase}\end{align} where $\mathcal{N}$ is some normalization so that the $F$-symbols are unitary. The phase $\chi_{L,M,N}$ in general depends on the details of the parent state, and, for example, it could be trivial as it is in the toric code lattice model.

A change of these phase factors $\delta\chi_{L,M,N}$ is restricted by the pentagon identity (see Fig.~\ref{fig:Fpentagon}) so that \begin{align}d\delta\chi_{K,L,M,N}&=\delta\chi_{K,L,M}-\delta\chi_{KL,M,N}+\delta\chi_{K,LM,N}\nonumber\\&\;\;\;-\delta\chi_{K,L,MN}+\delta\chi_{L,M,N}\label{4cochain}\end{align} must vanish modulo $2\pi$. This is known as a {\em 3-cocycle condition}. Some choices of the F-symbol phases can simply be absorbed by a redefinition of splitting states \begin{align}\left[\vcenter{\hbox{\includegraphics[width=0.075\textwidth]{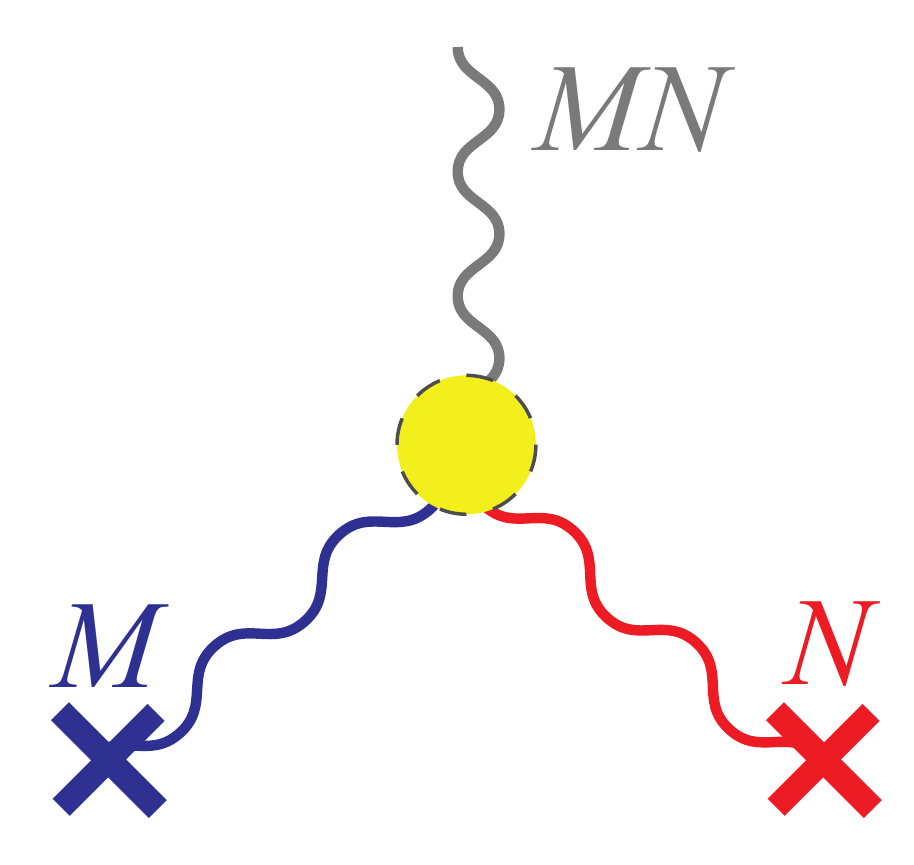}}}\right]\to e^{i\eta_{M,N}}\left[\vcenter{\hbox{\includegraphics[width=0.075\textwidth]{branchcuteta}}}\right].\end{align} 
By adding these phases to each vertex in \eqref{Fphase}, the phase can be modified $\delta\chi\to\delta\chi+d\eta$ by the {\em 3-coboudary} \begin{align}d\eta_{L,M,N}=\eta_{L,M}-\eta_{L,MN}+\eta_{LM,N}-\eta_{M,N}\label{3coboundary}\end{align} modulo $2\pi$. Consistent defect $F$-symbols are therefore classified by the equivalence classes $[\delta\chi]=[\delta\chi+d\eta]$ in the third cohomology group\cite{Cohomologybook,ChenGuLiuWen11} \begin{align}H^3(G,U(1))=\frac{Z^3(G,U(1))}{B^3(G,U(1))}\label{H3GU1}\end{align} where $Z^3$ contains all 3-cocycles, i.e. phase choices $\delta\chi$ with vanishing \eqref{4cochain}, and $B^3$ consists of 3-coboundaries \eqref{3coboundary}.

One can provide a nice physical explanation of the possible non-trivial phase factor choices. We can see this by noting that the cohomology group \eqref{H3GU1} also classifies symmetry protected topological (SPT) phases with symmetry $G$ in two dimensions\cite{ChenGuLiuWen12, LuVishwanathE8, MesarosRan12, EssinHermele13, BiRasmussenSlagleXu14, Kapustin14}. For example, $H^3(\mathbb{Z}_2,U(1))=\mathbb{Z}_2$ implies that there are two topologically distinct short-range entangled 2D phases with symmetry $\mathbb{Z}_2=\{1,\sigma\}$. Defect theories in these parent SPT phases have identical fusion rules $\sigma\times\sigma=1$ but different basis transformations \begin{align}F^{\sigma\sigma\sigma}_\sigma=\pm1\end{align}where all other $F$-symbols are trivial. Upon gauging, these choices lead to different theories: the positive sign leads to a $\mathbb{Z}_2$ gauge theory while the negative sign leads to the double-semion theory\cite{LevinWen05}.

\begin{figure}[htbp]
\centering\includegraphics[width=0.4\textwidth]{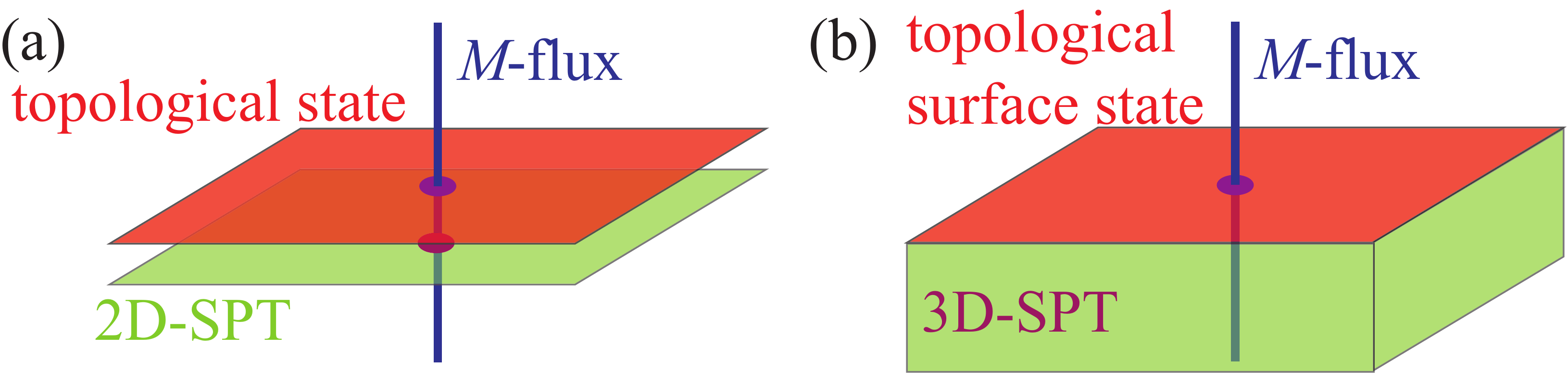}
\caption{(a) A composite defect of a globally symmetry topological parent state with a 2D-SPT. (b) A defect of the topologically ordered surface state of a 3D-SPT.}\label{fig:SPT}
\end{figure}
From this result one can see that a change of the phase factors $\delta\chi_{L,M,N}$ in the defect $F$-symbols (see Eq.~\eqref{Fphase}) \begin{align}F^{LMN}_K\to e^{i\delta\chi_{L,M,N}}F^{LMN}_K\label{Fphasemodification}\end{align} can be generated by an additional layer of an SPT phase on top of the original globally symmetric topological parent state. Since an SPT is short-ranged entangled, it will not alter the anyon structure of the parent state. Its presence however is reflected in the defect theory by the modification \eqref{Fphasemodification}. This is because the new twist defect is now a combination of a defect in the topological parent state and the SPT \begin{align}M_\lambda\to M_\lambda\otimes M_{SPT}\end{align} (see also Fig.~\ref{fig:SPT}(a)). The additional phase factor in $e^{i\delta\chi}$ comes from the defect $F$-symbols in the SPT. When $\delta\chi$ represents a non-trivial cohomology class, the modification \eqref{Fphasemodification} cannot be {\em gauged away} by a basis redefinition of splitting states. Hence $F$ and $e^{i\delta\chi}F$ correspond to two inequivalent defect fusion theories, although their fusion \emph{rules} are identical. For example, the addition of an extra $\mathbb{Z}_2$-SPT to the toric code would contribute an overall minus sign to the defect $F$-symbol $F^{\sigma\sigma\sigma}_\sigma$ in \eqref{Ising0Fsymbols}. This will alter the resulting twist liquid after gauging the anyonic electric-magnetic symmetry so that the four Ising quasiparticles no longer carry spins $h_\sigma=\pm1/16,\pm7/16,$ but $h_\sigma=\pm3/16,\pm5/16$ instead. We will discuss more about these issues in Section~\ref{sec:gaugingSPT}.

We should also make one final point about the defect $F$-symbols. As shown in Ref.~\onlinecite{ChenBurnellVishwanathFidkowski14}, if the globally symmetric parent state lives on the anomalous surface of a 3D-SPT protected by the same symmetry (see Fig.~\ref{fig:SPT}(b)), this will lead to obstructions to a consistent defect fusion theory. Three dimensional $G$-SPT's are classified by the cohomology group $H^4(G,U(1))$, and this also classifies the obstruction to the pentagon equation \eqref{fig:Fpentagon} for the defect $F$-symbols. Surface defects in these systems do not in general admit defect fusion theories with consistent basis transformations. As a consequence, there is an obstruction to gauging the $G$-symmetry on the surface topological state. Explicit examples can be found in Ref.~\onlinecite{ChoTeoRyu14} for decomposable Abelian symmetry group $G$. The obstruction to extending a given $G$-symmetric parent state to a $G$-crossed theory, which encodes fusion and $G$-braiding of classical defects, is explained in more detail in Ref.\onlinecite{BarkeshliBondersonChengWang14}.

In Table~\ref{tab:classificationobstruction} we provide a summary of the classification and possible obstruction of defect fusion theories of globally symmetric topological parent states using group cohomology. We will not consider much about cases with obstructions as that is not our focus in this article. In fact, all examples listed in Table~\ref{tab:ASexamples} do not admit obstructions to defect fusion or the $F$-symbols. This is because the cohomology groups $H^3(G,\mathcal{B}^\times)$ and $H^4(G,U(1))$ are all trivial except for a $\mathbb{Z}_{2n}^{(q+1/2)}$ theory with a $\mathbb{Z}_2$ conjugation symmetry, where $H^3(\mathbb{Z}_2,\mathbb{Z}_{2n})=\mathbb{Z}_2$. However, a typical conjugation symmetry in such a theory would still correspond to a trivial cohomology class and would not lead to inconsistent defect fusion rules. On the other hand, we will need to pay attention to inequivalent quantum symmetries as they lead to different twist liquids after gauging, even though the anyon relabeling operation naively seems identical. The classification of defect fusion theories in different topological phases is summarized in Table~\ref{tab:H23summary}.

\begin{table}[htbp]
\centering
\begin{tabular}{l|ll}
&Classification&Obstruction\\\hline
Fusion Rules&$H^2(G,\mathcal{B}^\times)$&$H^3(G,\mathcal{B}^\times)$\\
$F$-symbols&$H^3(G,U(1))$&$H^4(G,U(1))$
\end{tabular}
\caption{Summary of cohomological classification and obstruction of defect fusion categories. $G$ is the anyonic symmetry group. $\mathcal{B}^\times$ is the group of Abelian quasiparticles in the globally $G$-symmetric topological parent phase $\mathcal{B}$. $G$ acts on $\mathcal{B}^\times$ by anyon relabeling.}\label{tab:classificationobstruction}
\end{table}

\begin{table}[htbp]
\centering
\begin{tabular}{llcc}
Topological&Symmetry&$H^2(G,\mathcal{B}^\times)$&$H^3(G,U(1))$\\
phase $\mathcal{B}$&$G$&&\\\hline
$U(1)_n$&$\mathbb{Z}_2$ conjugation&$\mathbb{Z}_2$&$\mathbb{Z}_2$\\\noalign{\smallskip}
bilayer system&$\mathbb{Z}_2$ bilayer&\multirow{2}{*}{$0$}&\multirow{2}{*}{$\mathbb{Z}_2$}\\
$\mathcal{B}^\uparrow\otimes\mathcal{B}^\downarrow$&symmetry&&\\\noalign{\smallskip}
$\mathbb{Z}_k$ gauge&$\mathbb{Z}_2$ e-m&\multirow{2}{*}{$0$}&\multirow{2}{*}{$\mathbb{Z}_2$}\\
theory&symmetry&&\\\noalign{\smallskip}
$SO(8)_1$&triality&\multirow{2}{*}{$0$}&\multirow{2}{*}{$\mathbb{Z}_6$}\\
&symmetry $S_3$&&\\\noalign{\smallskip}
``4-Potts" Phase &$S_3$&$0$&$\mathbb{Z}_6$
\end{tabular}
\caption{Summary of the classification of fusion rules and $F$-symbols of the defect theory in a globally $G$-symmetric topological parent phase $\mathcal{B}$.}\label{tab:H23summary}
\end{table}

\subsection{Quasiparticle structure of twist liquids: a generalization of discrete gauge theories}
\label{sec:QPstructuretwistliquid}

In the previous section we provided the general classification and interpretation of defect fusion structures in parent states with a global anyonic symmetry. Our goal is to understand the theory of the twist liquid that arises once the twist defects are promoted to fully dynamical quantum fluxes when the anyonic symmetry is gauged. In general, anyonic excitations of a twist liquid are a composition of gauge flux and charge, and quasiparticle super-sectors from the parent state. This naturally generalizes the structure of discrete gauge theories, where its parent state would be a trivial boson condensate and its anyonic excitations, after gauging a conventional global symmetry, are gauge flux-charge composites, or dyons. Before providing a systematic way to count and characterize the excitations of twist liquids, we begin by reviewing the simpler calculation for discrete gauge theories.\cite{BaisDrielPropitius92,Bais-2007,Propitius-1995,PropitiusBais96,Preskilllecturenotes,Freedman-2004,Mochon04}

\subsubsection{Review of Quasiparticle Accounting in Discrete Gauge Theories}
In a two dimensional gauge theory with a finite, discrete gauge group $G$, the anyon excitations are labeled by the 2-tuple $\chi=\left([M],\rho\right)$. The flux component is characterized by a {\em conjugacy class} \begin{align}[M]=\left\{M'\in G:M'=NMN^{-1}\mbox{ for some }N\in G\right\}\label{conjugacyclass}\end{align} of the gauge group. Given a particular conjugacy class for the flux, the possible charge components are characterized by an irreducible representation $\rho:Z_M\to U(\mathcal{N}_\rho)$ of the {\em centralizer} of $M$ (or any representative of $[M]$) defined by \begin{align}Z_M=\left\{N\in G:NM=MN\right\}.\label{centralizer}\end{align}

\begin{figure}[htbp]
\centering\includegraphics[width=0.27\textwidth]{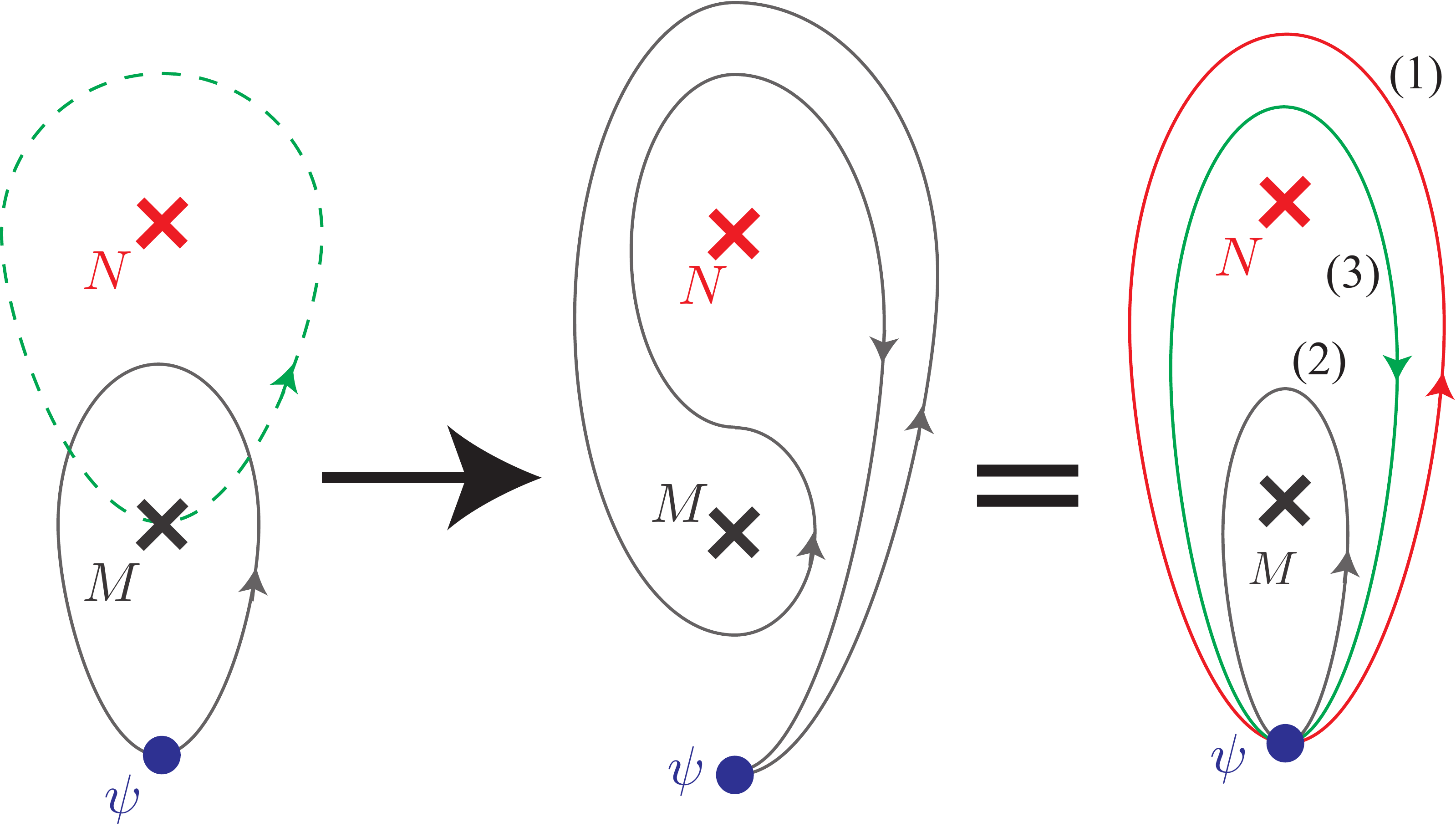}
\caption{Conjugation of holonomy $M\to N^{-1}MN$.}\label{fig:fluxconjugation}
\end{figure}
To understand the rules for anyon construction, we can begin with pure flux (i.e. the charge is in the trivial representation) and pure charge excitations (the flux conjugacy class is that of the identity element). A charge $\psi^a$ will acquire a non-Abelian holonomy $\psi^a\to\rho(M)_b^a\psi^b$ when encircling a flux $M$ according to the representation $\rho$ which characterizes the charge. The holonomy, however, would change to $\rho(N^{-1}MN)$ if the flux $M$ is also braided around another flux $N$ (see Fig.~\ref{fig:fluxconjugation}). This means the holonomy measurement is defined only up to conjugacy. In fact, a projective measurement can only read off the character of the representation $\mbox{Tr}[\rho(M)]$, and therefore fluxes are naturally characterized by conjugacy classes rather than the individual elements of the class. Additionally, a fundamental charge should be described by an irreducible representation because if $\rho$ is decomposable, then the charge can be split-off into simpler components. Equivalent representations describe identical charges because they are merely related by basis transformation $\psi^a\to U_b^a\psi^b$.

Now that we understand why inequivalent fluxes are described by conjugacy classes, let us motivate why charges only take the representations of the centralizer group of a given class. When a charge $\psi^a$ is combined with a flux $M$ to form a dyon, the holonomy it acquires $\psi^a\to\rho(N)_b^a\psi^b$ when encircling another flux $N$ is only well-defined when $M$ and $N$ commute. If this were not the case, then the flux would change by conjugation $M\to NMN^{-1}.$ Although this operation respects the conjugacy class it can permute the individual flux elements within the class. If this occurs then the evolution would not be cyclic, since the initial and final dyon configuration spaces could be different, and the holonomy could be gauged away by basis redefinition for $\psi^a$ instead. As a result, the charge component of a dyon is characterized by the representation of the centralizer \eqref{centralizer} rather than the whole group. For pure charges the flux conjugacy class is that of the identity element, and thus, in that case, the centralizer is the entire group. This is expected since pure charges are not dyonic and therefore do not run into the same consistency issue. 

Using the theory of finite groups we can develop a systematic procedure to count the number of inequivalent anyon excitations. The quantum dimension of the dyon $\chi=\left([M],\rho\right)$ is the product \begin{align}d_\chi=\left|[M]\right|\dim(\rho)\end{align} where $|[M]|$ is the number group elements in the conjugacy class $[M]$. By choosing a set of conjugating representatives $N_\mu\in G$ ($\mu=1\ldots |[M]|$) such that $[M]=\{N_1MN_1^{-1},\ldots,N_nMN_n^{-1}\}$, it is straightforward to see that \begin{align}|G|=\left|[M]\right||Z_M|\label{cosetdecomposition1}\end{align} for any choice of $M\in G,$ since each group element can be uniquely be expressed as $N_\mu z$ for $z\in Z_M$ and $1\leq\mu\leq |[M]|$. A useful theorem in the representation theory of finite groups\cite{grouptheorybook} relates the order of any finite group $H$ to the dimensions of its irreducible representations by \begin{align}|H|=\sum_{\mbox{\tiny irred. rep.}}\dim(\rho)^2.\label{repsumrule}\end{align} Eq.~\eqref{cosetdecomposition1} and \eqref{repsumrule} lead to the identification of the total quantum dimension $\mathcal{D}$ of the discrete gauge theory and the order of the gauge group $|G|$: \begin{align}\mathcal{D}^2&=\sum_\chi d_\chi^2=\sum_{[M]}\left[|[M]|^2\mathop{\sum_{\mbox{\tiny irred. rep.}}}_{\mbox{\tiny of $Z_M$}}\dim(\rho)^2\right]\nonumber\\&=\sum_{[M]}|[M]|^2|Z_M|=\sum_{[M]}|[M]||G|=|G|^2.\label{DGTdimension}\end{align}

To illustrate this general structure we can use the simple example of $G=\mathbb{Z}_2=\{1,\sigma\}.$ This group is Abelian and has two conjugacy classes $[1], [\sigma],$ and the full group has two representations $\rho_{+}, \rho_{-},$ both of which are one-dimensional. Since the group is Abelian there is a nice simplification because the centralizer of every conjugacy class is just the entire group. Thus we can see that we should have four anyon excitations $\chi_1=([1],\rho_{+}), \chi_{e}=([1],\rho_{-}), \chi_{m}=([\sigma],\rho_{+}),$ and $\chi_{\psi=em}([\sigma],\rho_{-}).$ Each of the anyons has quantum dimension $d_{i}=1$ and the total quantum dimension is $\mathcal{D}=\sqrt{1^2+1^2+1^2+1^2}=2=\vert G\vert.$

\subsubsection{Structure of Quasi-Particles in Twist Liquids}
We now generalize to twist liquids. For simplicity, we assume that the globally $G$-symmetric parent state is an Abelian topological phase with a group of Abelian quasiparticles $\mathcal{A}$.  A full treatment of non-Abelian parent states will not be presented here, although we expect a similar anyon structure in the general case. In fact, we do give one example of a non-Abelian parent state in Section~\ref{sec:4statePotts} which discusses the chiral ``4-Potts" state. 
We present an explicit approach that targets general Abelian states; a more abstract description which applies even to non-Abelian states and can be found in Ref.~[\onlinecite{BarkeshliBondersonChengWang14}].

The anyon excitations in the $G$-twist liquid are flux-quasiparticle-charge composites, i.e.~composites of the flux and charge from the gauged $G$-symmetry, and quasiparticles from the parent Abelian theory. The anyons can be labeled by the 3-tuple \begin{align}\chi=\left(\left[M\right],\boldsymbol\lambda,\rho\right)\label{fqccomposite}\end{align} which is a natural generalization of the 2-tuple label in discrete gauge theories. 

Here $[M]$ is a conjugacy class, as defined in \eqref{conjugacyclass}, that describes the flux component of the gauged $G$-symmetry. 
Just as in the case for the discrete gauge theory, the choice of a flux class restricts the allowed values for the remaining anyon structure. In the 3-tuple, $\boldsymbol\lambda$ is a super-sector of defect species labels drawn from the quotient group $\mathcal{A}_M=\mathcal{A}/(1-M)\mathcal{A}$, which depends on the equivalence class $[M]$ (see Eq.~\eqref{defectsectorquotient}). However, the anyonic symmetry may still act on the elements of $\mathcal{A}_M$ and permute them. This forces elements in $\mathcal{A}_M$ to form into super-sectors, i.e.~ {\em irreducible} sets of species labels of the form:
\begin{align}\boldsymbol\lambda=\lambda_1+\ldots+\lambda_l,\quad\lambda_i\in\mathcal{A}_M\label{super-sector}\end{align} which corresponds to the non-simple object $M_{\lambda_1}+\ldots+M_{\lambda_l}$ in the defect fusion category. The set of species labels in $\boldsymbol\lambda,$ i.e.~$\{\lambda_1,\ldots,\lambda_l\}\subseteq\mathcal{A}_M,$ must be chosen such that the centralizer $Z_M$ (defined in \eqref{centralizer}) acts {\em transitively} on it. The transitive action of $Z_M$ means that (i) the labels in $\boldsymbol\lambda$ are permuted by any symmetries $N$ that commute with the flux $M$ according to the $Z_M$-action \begin{align}\lambda_i={\bf a}+(1-M)\mathcal{A}\longrightarrow\lambda_j=N{\bf a}+(1-M)\mathcal{A},\label{speciesrelabeling}\end{align} (notice $N(1-M)\mathcal{A}=(1-M)\mathcal{A}$ as $\mathcal{A}$ is closed under $N$); 
and (ii) that any two labels are related by some permutation in the centralizer, i.e. $\lambda_j=N\lambda_i$ for some $N\in Z_M.$ In other words, the irreducible sets $\boldsymbol{\lambda}$ are exactly the {\em $Z_M$-orbits} in $\mathcal{A}_M$. The reason it is the $Z_M$-orbits, and not full $G$-orbits, is that the quotient group $\mathcal{A}_M$, which contains the anyons/species that form the super-sectors, only remains unchanged under $Z_M.$

The final element of the 3-tuple is the $G$-charge component $\rho.$  The charge is characterized by a $\mathcal{N}_\rho$-dimensional irreducible representation \begin{align}\rho:Z_M^{\boldsymbol\lambda}\longrightarrow U(\mathcal{N}_\rho)\label{chargerepresentation}\end{align} of a \emph{restricted} centralizer of the flux $M$ \emph{and} super-sector $\boldsymbol\lambda$ \begin{align}Z_M^{\boldsymbol\lambda}=\{N\in Z_M:N\lambda_1=\lambda_1\},\label{centralizer2}\end{align} i.e. $Z_M^{\boldsymbol\lambda}$ is the subgroup in the centralizer $Z_M$ that fixes a particular choice of species $\lambda_1$ in $\boldsymbol\lambda$. Thus, Eqs.~\eqref{super-sector} and \eqref{centralizer2} depend on the choice of a particular representative $M$ in $[M]$ (other representatives can change the species labels entering $\boldsymbol\lambda$) and a species in the super-sector, for which we have arbitrarily chosen $\lambda_1.$ However, different choices are related by adjoint isomorphisms and are therefore equivalent, and lead to identical anyon content for the twist liquid.

Before we show why the general structure takes this form, let us illustrate the construction using the example of the toric code with a gauged electric-magnetic symmetry. In this case $G=\mathbb{Z}_2,$ which, as indicated above, has two conjugacy classes $[1], [\sigma],$ and two one-dimensional representations $\rho_{+}$ and $\rho_{-}.$ The quasiparticles in the Abelian parent state are those of the toric code $\{1, e, m, \psi\}.$ If we choose the flux class $[1]$ the full set of species labels $\mathcal{A}_1$ is just the set of Abelian anyons $\{1, e, m, \psi\}.$ However, what we need for our construction is the set of $\mathbb{Z}_2$ orbits of this set which is $\{1, e+m, \psi\}.$ Thus, if we pick the flux class $[1,]$ the only options for the quasiparticle super-sectors are $\boldsymbol\lambda = 1, e+m,$ or $\psi.$ Before fixing the charge representation possibilities, let us consider the other flux class $[\sigma].$ The set of species labels is $\mathcal{A}_\sigma=\{0,1\}$ where $0$ effectively represents (the vacuum) $1$ or $\psi$ and $1$ represents $e$ or $m.$ Both of the elements of $\mathcal{A}_\sigma$ are invariant under the $\mathbb{Z}_2$ action (e.g. even though the group action takes $e$ to $m$ they are both represented by $\lambda=1$), and thus the $\mathbb{Z}_2$ orbits are just $\{0\}$ and $\{1\}$ themselves.

To find the allowed charge representations we need to calculate the restricted centralizer groups according to Eq.~\eqref{chargerepresentation}. These groups are \begin{gather}Z_{[1]}^1=\mathbb{Z}_2,\quad Z_{[1]}^{e+m}=0,\quad Z_{[1]}^{\psi}=\mathbb{Z}_2,\nonumber\\
Z_{[\sigma ]}^{0}=\mathbb{Z}_2,\quad Z_{[\sigma ]}^{1}=\mathbb{Z}_2.\nonumber \end{gather} The only unusual example is $Z_{[1]}^{e+m}$ which is trivial, i.e. it only consists of the identity element, because the non-trivial element of $G (=Z_{[1]})$ does not preserve the first species label $e.$ The trivial group only has a trivial, one-dimensional representation which we will denote by $\rho_0.$ From this data we are ready to identify the anyon quasiparticles in the gauged electric-magnetic twist liquid. The anyons are 
\begin{gather}
1=([1],1,\rho_{+}),\quad z=([1],1,\rho_{-}),\;\;\; \mathcal{E}=([1],e+m,\rho_0)\nonumber\\
\psi=([1],\psi,\rho_{+}),\quad\overline{\psi}=([1],\psi,\rho_{-}),\nonumber\\
\sigma=([\sigma],1,\rho_{+}),\quad\sigma'=([\sigma],1,\rho_{-}),\nonumber\\ 
\overline{\sigma}=([\sigma],e,\rho_{+}),\quad\overline{\sigma}'=([\sigma],e,\rho_{-}),\nonumber
\end{gather} where for the super-sectors we have used $1,e$ to label the species instead of $0,1$ to help avoid any confusion (c.f.~Eqs.~(\ref{TCDrinfeldsigma1}-\ref{TCDrinfeldsigma4}) in Section~\ref{sec:Zkgaugetheory}). The quantum dimensions can be easily read-off from these $3$-tuples by multiplying the dimensions of the three separate components, e.g. $d_z=1\times 1\times 1 =1$, $d_{\mathcal{E}}=1\times 2\times 1$, and $d_{\sigma}=\sqrt{2}\times 1\times 1.$ See Eq.~\eqref{FQCcompositedimension} for the calculation of the quantum dimension of a general $3$-tuple.

Now that we have shown an example, let us consider the general construction. To understand the anyon structure, we begin with a pure quasiparticle super-sector, i.e. we choose the flux class $[1].$ When an anyonic symmetry group $G$ is gauged, quasiparticles from the parent state are indistinguishable if they are related by the symmetries in $G.$ They become components of a super-sector ${\bf a}_1+\ldots+{\bf a}_l$, i.e. a $G$-orbit in $\mathcal{A}_{[1]}$, which is now a simple, indecomposable object in the twist liquid. Its components form a $G$-orbit so that all ${\bf a}_i$ in a super-sector are equivalent since they transform into one another through symmetries. Alternatively, we note that a super-sector can also be treated as an irreducible representation of a set, i.e.~a homomorphism $G\to S_l$, (where $S_l$ is the group of permutations of $l$ elements) such that the image does not sit in any splitting $S_m\times S_{l-m}$ for $0<m<l$.  
When the super-sector object is dragged around a non-trivial flux (of $G$), it picks up a holonomy $P\in S_l$ that permutes its components. However, since the components of the super-sector are indistinguishable, they can be rearranged by a permutation $Q$ and the holonomy would be altered by conjugation $P\to QPQ^{-1}$. Thus we expect that a flux is physically characterized by a conjugacy class, rather than an individual group element, as stated above.

Next, to further constrain the theory, we can have a quasiparticle super-sector combined with a flux $M$ to form a composite object. 
Another way to see that only the conjugacy class of $M$ is physical meaningful, is to consider choosing a different representative of the  class $[M].$ If the flux is represented by a different symmetry $M',$ which is necessarily related to $M$ by conjugation if they are in the same class, then the basis of the quasiparticle super-sector could be rearranged ${\bf a}_i\to{\bf a}_{P(i)}$ according to a permutation $P$  represented by some symmetry $N\in G$ such that that $M'=NMN^{-1}.$ Physical properties cannot depend on this basis choice, and thus only flux classes are physically distinguishable. 

Now, let us further imagine that the composite is braided around another flux $L$; the super-sector would pick up a holonomy under this operation. However, if the flux is altered by conjugation $M\to M_L=LML^{-1}$, then the braiding evolution would not be cyclic since the basis of quasiparticles would need to be rearranged by some permutation to compare with the original basis. This means that the holonomy is not well-defined unless $M_L=M$, i.e.~the two fluxes $L$ and $M$ commute. As a result, quasiparticle super-sectors of a flux $M$ are characterized by a $Z_M$-orbit rather than a full $G$-orbit, where $Z_M$ the centralizer of $M$. Note that for the flux class  $[1],$ $Z_{[1]}=G$ so our conclusions for the pure quasiparticle sector are not altered. 
 
 Moreover, as shown in Section~\ref{sec:twistdefects}, quasiparticles attached to a flux $M$ are well defined only up to $(1-M)\mathcal{A}$ because they can go around the flux and change anyon type (see Eq.~\eqref{defectQPcomposite}). The components of the super-sector for non-trivial fluxes are thus species labels $\lambda_i$ in the quotient $\mathcal{A}_M=\mathcal{A}/(1-M)\mathcal{A}$ instead of the original group $\mathcal{A}$ of Abelian quasiparticles. Notice that the quotient $\mathcal{A}_M$ only has a reduced symmetry $Z_M$ because the invariance of the denominator $(1-M)\mathcal{A}$ in \eqref{speciesrelabeling} requires $N$ to commute with $M$.

The final consistency conditions require the most general anyon in a twist liquid: a flux-quasiparticle-charge composite $([M],\boldsymbol\lambda,\rho)$. Let us now introduce the appropriate notation so we can write the fully general form of this composite object. To be explicit, let $[M]=\{M_\mu=N_\mu MN_\mu^{-1}\}_{\mu=1,\ldots,n}$ be the conjugacy class that characterizes the flux, where $N_\mu$ ($\mu=1\ldots n$) are some (arbitrary) choice of conjugating symmetries, $n=|[M]|,$ and $N_1=1$. We can express the components of the super-sector $\boldsymbol\lambda=\lambda_1+\ldots+\lambda_l$ by $\lambda_i=z_i\lambda_1$ where $z_i$ are elements of the  centralizer $Z_M,$ and $z_1=1$. For a different representative, $M_\mu,$ of the flux class, the basis of the super-sector will be changed. To be explicit let us pick this isomorphism to be $N_\mu:\mathcal{A}_M\to\mathcal{A}_{M_\mu}$ which maps between the cosets \begin{align}{\bf a}+(1-M)\mathcal{A}\longrightarrow N_\mu{\bf a}+(1-M_\mu)\mathcal{A}.\end{align}  Notice $N_\mu(1-M)\mathcal{A}=(1-M_\mu)\mathcal{A}$ as $N_\mu M N_\mu^{-1}=M_\mu,$ and $\mathcal{A}$ is closed under $N_\mu$. 
From this we define the new super-sector basis $\boldsymbol\lambda_\mu=\lambda_{\mu,1}+\ldots+\lambda_{\mu,l}$ where $\lambda_{\mu,i}=N_\mu\lambda_i=N_\mu z_i\lambda_1$. 

With the notation now fixed, the composite object can be expressed as the combination \begin{align}\left([M],\boldsymbol\lambda,\rho\right)=\sum_{\mu=1}^{n}\sum_{i=1}^l\left(M_\mu\right)_{\lambda_{\mu,i}}\times\boldsymbol\psi_{(\mu,i)}\label{FQCcompositesum}\end{align} where $(M_\mu)_{\lambda_{\mu,i}}$ is a defect $M_\mu$ with species label $\lambda_{\mu,i}$ and $\boldsymbol\psi_{(\mu,i)}$ is the charge sector spanned by the vectors $\psi_{(\mu,i)}^a$. The sum over $\mu$ runs over the elements of the conjugacy class $[M]$, and the sum over $i$ runs over the Abelian anyon elements of a super-sector. When $G$ is Abelian, as in our toric code example above, each conjugacy class only has a single element, which simplifies all of the results. 

Conceptually, we want to understand if there are any physically distinct anyons with non-trivial charge. Hence, we need to understand what restrictions are placed on the charge sector to have the complete set of physically distinguishable objects. Like a discrete gauge theory, a pure charge corresponds to a Hilbert space spanned by $\psi^a$ that transforms holonomically by $\psi^a\to\rho(L)_b^a\psi^b$ according to an irreducible representation $\rho$ of the gauge group when the charge is braided around a flux $L$. Now, in the composite object \eqref{FQCcompositesum}, 
there are multiple Hilbert space sectors $\mathcal{H}_{(\mu,i)}=\mbox{span}\{\psi_{(\mu,i)}^a\},$ which are each associated to a defect $(M_\mu)_{\lambda_{\mu,i}}$. When braiding \eqref{FQCcompositesum} around a flux $L$, these Hilbert spaces will mix $\mathcal{H}_{(\mu,i)}\to\mathcal{H}_{(\nu,j)}$ since $L$ acts via \begin{align}LM_\mu L^{-1}=M_\nu\quad\mbox{and}\quad L\lambda_{\mu,i}=\lambda_{\nu,j}.\label{Ltransformmui}\end{align}
Since the charge sector is fused with a defect, the braiding evolution is cyclic only when the Hilbert space goes back to itself after a cycle, i.e.~$(M_\mu)_{\lambda_{\mu,i}}$ is unaltered by $L$. Otherwise the holonomy would be gauge dependent as it would be affected by the arbitrary choice of the $N_\mu$ and $z_i$. Let us consider the consequences of this constraint. Since different charge sectors of a single composite anyon are related by isomorphisms, it is most convenient to focus on the sector $\boldsymbol\psi_{(1,1)}$ where $N_1=z_1=1$. The defect $M_{\lambda_1}$ is unchanged when the composite object encircles flux $L_0$ if $M$ and $L_0$ commute \emph{and} the species label $\lambda_1$ is invariant under $L_0$. This requires the charge sector to be characterized by an irreducible representation $\rho:Z_M^{\boldsymbol\lambda}\to U(\mathcal{N}_\rho)$ of the restricted centralizer subgroup $Z^{\boldsymbol\lambda}_M$ defined in Eq.~\eqref{centralizer2}.

To see how the charges $\boldsymbol\psi_{(\mu,i)}$ in the composite object \eqref{FQCcompositesum} transform when braided around a general flux $L$, we need to extend the representation $\rho:Z_M^{\boldsymbol\lambda}\to U(\mathcal{N}_\rho)$ to the whole symmetry group $G$. This can be done by an {\em induced} representation \begin{align}L:\psi^a_{(\mu,i)}\longrightarrow\rho\left((N_\nu z_j)^{-1}L(N_\mu z_i)\right)_b^a\psi^b_{(\nu,j)}\label{inducedrep}\end{align} where the indices change $(\mu,i)\to(\nu,j)$ according to \eqref{Ltransformmui}. Notice that the operator $L_0=(N_\nu z_j)^{-1}L(N_\mu z_i)$ in \eqref{inducedrep} leaves $M$ and $\lambda_1$ unaltered, and therefore it belongs in the restricted centralizer $Z_M^{\boldsymbol\lambda}$ so that $\rho(L_0)$ is well-defined. In other words, the conjugation $L\to(N_\nu z_j)^{-1}L(N_\mu z_i)$ is required to compare the basis between defect $(M_\nu)_{\lambda_{\nu,j}}$ and $(M_\mu)_{\lambda_{\mu,i}}$, i.e.~defects which can be different representatives of the same flux class and carry species labels which may also need to be basis transformed for comparison. However, since the induced representation depends on the arbitrary choice of the representatives $N_\mu,N_\nu$ and $z_i,z_j$, it is gauge {\em dependent}. The only gauge independent part is the restriction $\rho:Z_M^{\boldsymbol\lambda}\to U(\mathcal{N}_\rho)$ on the centralizer, and hence these restricted representations offer the only physically distinguishable options for the charge sector.

For example, later in Section~\ref{sec:so(8)symmetry} we will consider the $SO(8)_1$-state, with an anyon content of three mutually semionic fermions $f_1,f_2,f_3,$ with the fusion rules $f_i^2=1$ and $f_1\times f_2=f_3$. It has a global $S_3$-symmetry that permutes the three fermions and, when gauged, forces them into a super-sector $\Psi=f_1+f_2+f_3.$  The restricted centralizer subgroup $Z_{[1]}^{f_1+f_2+f_3}$ that fixes $f_1$ is $\mathbb{Z}_2=\{1,\sigma_1\}$, where $\sigma_1$ switches $f_2\leftrightarrow f_3$. It has two one-dimensional representations $\rho_\pm:\sigma_1\to\pm1$. The full super-sector-charge composite is of the form $f_1\psi_1+f_2\psi_2+f_3\psi_3$, where the three-dimensional charge sector $\boldsymbol\psi=(\psi_1,\psi_2,\psi_3)$ transforms according to  \begin{align}\sigma_1:\left(\begin{array}{*{20}c}\psi_1\\\psi_2\\\psi_3\end{array}\right)\to\left(\begin{array}{*{20}c}\pm1&0&0\\0&0&1\\0&1&0\end{array}\right)\left(\begin{array}{*{20}c}\psi_1\\\psi_2\\\psi_3\end{array}\right).\label{sigma1inducedrep}\end{align} Here the induced representation in \eqref{inducedrep} is defined by choosing $z_1=N=1$, $z_2=\theta$ and $z_3=\theta^2$ , for $\theta:f_i\to f_{i+1}$ the cyclic threefold element in $S_3$. The character of the representation $\chi(\sigma_1)$, i.e.~its trace, depends only on the $\psi_1$ block and is gauge independent. The off-diagonal lower block that mixes $\psi_2\leftrightarrow\psi_3$ is however gauge dependent. For instance, a gauge transformation $(\psi_2,\psi_3)\to(\psi_2,-\psi_3)$ would change the lower $2\times2$ block in \eqref{sigma1inducedrep} from $\sigma_x$ to $-\sigma_x$. Hence, it is only the representation of the restricted centralizer, which provides a physically distinguishable charge.

Eq.~\eqref{inducedrep} extends the representation $\rho_+$ (resp. $\rho_-$) to the three-dimensional representation $A_1\oplus E$ (resp.~$A_2\oplus E$) of the full $S_3$ group (see the $S_3$ character table in Table~\ref{tab:S3character}). For instance, the induced representation of $\rho_\pm$ also determines the representation of another restricted centralizer $\mathbb{Z}_2=\{1,\sigma_2\}$, where $\sigma_2:f_1\leftrightarrow f_3$ now fixes $f_2$ instead of $f_1$. This is because the group operations $\sigma_1,\sigma_2$ are related through conjugation by $\theta,$ and thus have an identical character $\chi(\sigma_2)=\chi(\theta\sigma_1\theta^{-1})=\chi(\sigma_1)=\pm1$. The two representations $\rho_\pm$ are associated to two distinct anyons $\Psi$ and $z_2\Psi$, which differ by fusion with the $\mathbb{Z}_2$ charge $z_2,$ and can hence be distinguished by braiding operation around a $\mathbb{Z}_2$ flux.

The full construction can be distilled into a relatively simple mathematical algorithm. To summarize, the quasiparticles in a twist liquid where the anyonic symmetry group $G$ has been gauged are represented by a $3$-tuple $([M],\boldsymbol{\lambda},\rho)$ which is a flux-quasiparticle-charge composite defined via the rules:
\begin{itemize}
\item $[M]$ represents a flux of $G$ and is a conjugacy class of $G.$
\item Given a choice of $[M]$ and a representative of the class, construct the group of species labels $\mathcal{A}_M=\mathcal{A}/(1-M)\mathcal{A}$ and the centralizer of $[M]$: $Z_{[M]}.$ The values that the quasiparticle component $\boldsymbol{\lambda}$ can take are the $Z_{[M]}$ orbits in $\mathcal{A}_M.$
\item Given a choice of $[M]$ and $\boldsymbol{\lambda},$ construct the restricted centralizer group $Z_{[M]}^{\boldsymbol\lambda}$ which is the group of elements of $G$ that commute with $M$ and fix the (say) first species label of $\boldsymbol\lambda.$ The $G$-charges are given by the representations $\rho$ of $Z_{[M]}^{\boldsymbol\lambda}.$
\end{itemize}

\subsubsection{Total Quantum Dimension and Chiral Central Charge of the Twist Liquid} 
We are now ready to prove some of the main results of the article, namely the relationship between total quantum dimension of the twist liquid and the quantum dimension of the initial parent state, and the invariance of the central charge during the gauging procedure.
Recall from \eqref{defectdimensionabelian} that a defect has quantum dimension $d_M=\sqrt{|(1-M)\mathcal{A}|}$, and the dimension of the charge sector is given by the dimension of the representation $\mathcal{N}_\rho=\dim(\rho)$. The quantum dimension of the composite $\chi=([M],\boldsymbol\lambda,\rho)$ can be evaluated by summing over the dimensions of its components in \eqref{FQCcompositesum} which is equivalent to the following product: \begin{align}d_{\chi}=|[M]||\boldsymbol\lambda|\dim(\rho)\sqrt{|(1-M)\mathcal{A}|}\label{FQCcompositedimension}\end{align} where $|[M]|=n$ is the number of elements in the conjugacy class and $|\boldsymbol\lambda|=l$ is the number of Abelian components in the super-sector $\boldsymbol\lambda$. 

Hence, the total quantum dimension $\mathcal{D}_{TL}$ of the twist liquid is calculated by a sum \begin{align}\mathcal{D}_{TL}^2=\sum_\chi d_\chi^2=&\sum_{[M]}\left(|[M]|^2|(1-M)\mathcal{A}|\right.\nonumber\\&\;\;\;\times\left.\mathop{\sum_{\mbox{\tiny $Z_M$-orbits}}}_{\mbox{\tiny $\boldsymbol\lambda$ in $\mathcal{A}_M$}}\left\{|\boldsymbol\lambda|^2\mathop{\sum_{\mbox{\tiny irred. rep.}}}_{\mbox{\tiny of $Z_M^{\boldsymbol\lambda}$}}\dim(\rho)^2\right\}\right).\label{TLdimension1}\end{align} To simplify this, we first see from Eq.~\eqref{repsumrule} that $\sum\dim(\rho)^2=|Z_M^{\boldsymbol\lambda}|$. So the second line of \eqref{TLdimension1} becomes a sum of $|\boldsymbol\lambda|^2|Z_M^{\boldsymbol\lambda}|$ over species super-sectors $\boldsymbol\lambda$. Next we see that \begin{align}|Z_M|=|\boldsymbol\lambda||Z_M^{\boldsymbol\lambda}|.\label{ZM=lZMl}\end{align} 
This is because each $z$ in $Z_M$ have a unique decomposition $z=z_i\zeta$, where $\lambda_1$ is transformed into $z\lambda_1=\lambda_i=z_i\lambda_1$ under $z$, and $\zeta=z_i^{-1}z$ fixes $\lambda_1$ and, hence belongs in $Z_M^{\boldsymbol\lambda}$. The one-to-one correspondence between $Z_M$ and $\{\lambda_i\}\times Z_M^{\boldsymbol\lambda}$ gives \eqref{ZM=lZMl}.

Using this result, the second line of \eqref{TLdimension1} then becomes \begin{align}\mathop{\sum_{\mbox{\tiny $Z_M$-orbits}}}_{\mbox{\tiny $\boldsymbol\lambda$ in $\mathcal{A}_M$}}|\boldsymbol\lambda|^2|Z_M^{\boldsymbol\lambda}|=\mathop{\sum_{\mbox{\tiny $Z_M$-orbits}}}_{\mbox{\tiny $\boldsymbol\lambda$ in $\mathcal{A}_M$}}|\boldsymbol\lambda||Z_M|=|\mathcal{A}_M||Z_M|\label{TLdimension2}\end{align} where we have used the simple result that the sum of elements in all orbits is just the number of elements in the whole group $\mathcal{A}_M=\mathcal{A}/(1-M)\mathcal{A}$. Combining \eqref{TLdimension2} and \eqref{TLdimension1}, \begin{align}\mathcal{D}_{TL}^2&=\sum_{[M]}|[M]|^2|(1-M)\mathcal{A}|\left|\frac{\mathcal{A}}{(1-M)\mathcal{A}}\right||Z_M|\nonumber\\&=|\mathcal{A}|\sum_{[M]}|[M]|^2|Z_M|=\mathcal{D}_0^2|G|^2\label{TLdimension3}\end{align} where $\mathcal{D}_0=\sqrt{|\mathcal{A}|}$ is the total quantum dimension of the parent state $\mathcal{A},$ and the final step was already proven in \eqref{DGTdimension} for the case of a discrete gauge theory. Eq.~\eqref{TLdimension3} is one of the central results of this article: the total quantum dimension is increased by $|G|$ after gauging an anyonic symmetry $G$. This result holds true even when the globally symmetric parent state is non-Abelian. A proof is available in Ref.~[\onlinecite{BarkeshliBondersonChengWang14}].

The chiral central charge $c_-=c_R-c_L$ along the gapless boundary of a topological phase is related to the anyon content in the bulk by the Gauss-Milgram formula\cite{FrohlichGabbiani90,MooreRead,ReadRezayi,Rehren89} \begin{align}\exp\left(2\pi i\frac{c_-}{8}\right)=\frac{1}{\mathcal{D}}\sum_{\bf a}d_{\bf a}^2\theta_{\bf a},\label{GaussMilgram}\end{align} where the sum is taken over all quasiparticles with exchange phases $\theta_{\bf a}=e^{2\pi ih_{\bf a}}$. Remarkably Eq.~\eqref{GaussMilgram} is unchanged by gauging.\cite{BarkeshliBondersonChengWang14} Here we specifically show this for Abelian parent states.  

First, just like a discrete gauge theory, 
the braiding phase of a pure charge $\psi$ orbiting around a flux $[M]$ is given by $\mbox{Tr}[\rho(M)]/\dim(\rho)$, where $\rho$ is the irreducible representation that describes $\psi$. As the twist liquid anyon $\chi=([M],\boldsymbol\lambda,\rho)$ is a composition of the flux $\chi_0=([M],\boldsymbol\lambda,1)$ and the pure charge $\psi=(1,1,\rho)$, its twist phase -- which is also its exchange phase -- differs from that of the chargeless $\chi_0$ by a braiding phase, i.e. \begin{align}\theta_{([M],\boldsymbol\lambda,\rho)}=\theta_{([M],\boldsymbol\lambda,1)}\frac{\mbox{Tr}[\rho(M)]}{\dim(\rho)}.\label{twistfluxcharge}\end{align} 

Hence, the Gauss-Milgram sum \eqref{GaussMilgram} in a twist liquid decomposes into \begin{align}e^{\frac{i\pi c_-}{4}}&=\frac{1}{\mathcal{D}_{TL}}\sum_{\chi=([M],\boldsymbol\lambda,\rho)}d_\chi^2\theta_\chi\nonumber\\&=\frac{1}{\mathcal{D}_0|G|}\sum_{\chi_0=([M],\boldsymbol\lambda,1)}d_{\chi_0}^2\theta_{\chi_0}\nonumber\\&\quad\quad\times\sum_{\rho}\dim(\rho)^2\frac{\mbox{Tr}[\rho(M)]}{\dim(\rho)}\end{align} where $d_\chi=d_{\chi_0}\dim(\rho)$ (see \eqref{FQCcompositedimension}), and $\mathcal{D}_{TL}=\mathcal{D}_0|G|$. Next, a useful formula from the representation theory of finite groups\cite{grouptheorybook} requires \begin{align}\mathop{\sum_{\mbox{\tiny irred. rep.}}}_{\mbox{\tiny $\rho$ of $H$}}\dim(\rho)\mbox{Tr}\left[\rho(M)\right]=0\label{sumchM=0}\end{align} for any non-trivial group operation $M$ in a finite group $H$. Thus, the only contribution to the Gauss-Milgram sum is from the trivial flux sector $M=1$. For instance, the $\mathbb{Z}_2$ defect fluxes $\sigma,\sigma',\overline\sigma,\overline\sigma'$ of the  toric code have canceling exchange phases $e^{i\pi/8},-e^{i\pi/8},e^{-i\pi/8},-e^{-i\pi/8}$ respectively (see Section~\ref{sec:Zkgaugetheory}). 

Consequently, the sum becomes \begin{align}e^{\frac{i\pi c_-}{4}}&=\frac{1}{\mathcal{D}_0|G|}\sum_{\boldsymbol\lambda}d_{\boldsymbol\lambda}^2\theta_{\boldsymbol\lambda}\mathop{\sum_{\mbox{\tiny irred. rep.}}}_{\mbox{\tiny $\rho$ of $Z_1^{\boldsymbol\lambda}$}}\dim(\rho)^2\nonumber\\&=\frac{1}{\mathcal{D}_0|G|}\sum_{\boldsymbol\lambda}d_{\boldsymbol\lambda}^2\theta_{\boldsymbol\lambda}|Z_1^{\boldsymbol\lambda}|\end{align} after using \eqref{repsumrule}, and noting that the quantum dimension of the pure super-sector is $d_{\boldsymbol\lambda}=|\boldsymbol\lambda|$, i.e. the number of components. From \eqref{ZM=lZMl}, $d_{\boldsymbol\lambda}|Z_1^{\boldsymbol\lambda}|=|Z_1|=|G|$ and therefore \begin{align}e^{\frac{i\pi c_-}{4}}=\frac{1}{\mathcal{D}_0}\sum_{\boldsymbol\lambda}d_{\boldsymbol\lambda}\theta_{\boldsymbol\lambda}=\frac{1}{\mathcal{D}_0}\sum_{\lambda\in\mathcal{A}}\theta_\lambda\label{chiralcinvariance}\end{align} which is identical to the Gauss-Milgram sum of the globally symmetric Abelian parent phase $\mathcal{A}$. This is because the components of each super-sector $\boldsymbol\lambda=\lambda_1+\ldots+\lambda_l$ originate from Abelian anyons $\lambda_i$ that are related by anyonic symmetries and have the same spins, i.e.~$\theta_{\boldsymbol\lambda}=\theta_{\lambda_i}$. The invariance of \eqref{GaussMilgram} resembles the fact that the chiral central charge $c_-$ of a $(1+1)$D CFT is unaltered by an orbifold construction.~\cite{Ginsparg88,bigyellowbook}

\subsection{Exactly-solvable String-net Models of Twist Liquids}
\label{sec:stringnet}

The phase transition of Eq.~\eqref{gaugingtransition} between a globally anyonic symmetric parent state and its corresponding locally-symmetric, gauged twist liquid can be captured by a string-net model.\cite{LevinWen05} These types of lattice models are built using the fusion data from a defect fusion category described in Section~\ref{sec:twistdefects}. The transition between globally and locally symmetric states is driven by tuning string tensions that confine twist defects. In the absence of string tensions for the defects, the lattice model is exactly-solvable and provides a local, Hamiltonian description of a gapped topological phase describing the twist liquid. The transition essentially occurs when twist defects, i.e.~anyonic symmetry fluxes, become deconfined. As we have seen Section~\ref{sec:QPstructuretwistliquid}, general anyonic excitations in twist liquids are composite objects of fluxes, quasiparticle super-sectors, and gauge charges. Their fusion and braiding properties can be derived using the {\em Drinfeld construction}.\cite{LevinWen05,Kasselbook,BakalovKirillovlecturenotes} This procedure involves solving the hexagon equations \eqref{hexagoneq}, i.e.~Fig.~\ref{fig:hexagon1}. We have demonstrated this construction explicitly by gauging the electric-magnetic symmetry of the toric code in Section~\ref{Drinfeld-cosntruction}, where the Drinfeld construction led to the $\mbox{Ising}\times\overline{\mbox{Ising}}$ anyon structure. Here we present string-net Hamiltonians for general twist liquids, and the procedure for deriving their anyonic fusion and braiding properties.

\begin{figure}[htbp]
\centering\includegraphics[width=0.2\textwidth]{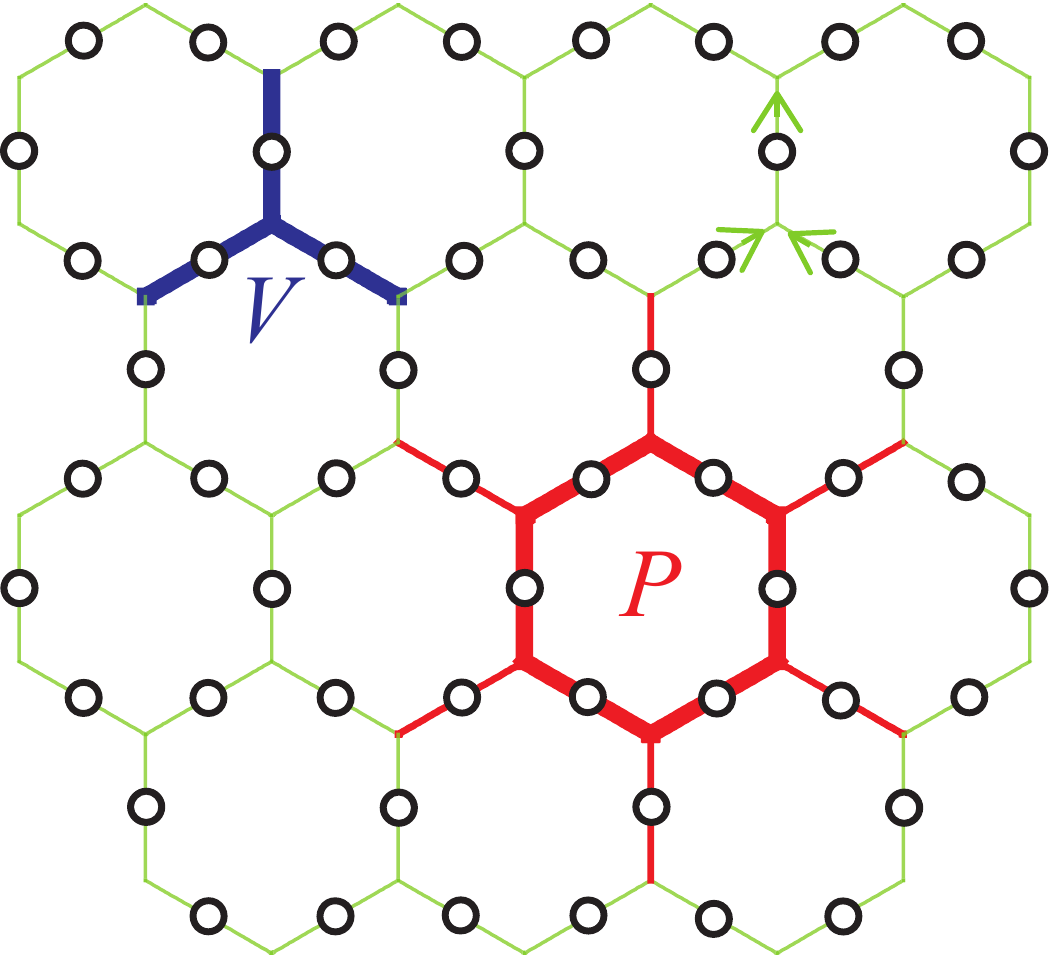}
\caption{String-net model on a honeycomb lattice. The $\circ$'s represent degrees of freedom on links. $V$ and $P$ are vertex and plaquette operators.}\label{fig:stringnet}
\end{figure}
The string-net model construction proposed by Levin and Wen is defined on a honeycomb lattice where the degrees of freedom live on the links\cite{LevinWen05}. In some cases the links are oriented, and the orientations are chosen so that they align as much as possible to the vertical direction (see Fig.~\ref{fig:stringnet}). To construct the model we begin with the defect fusion category $\mathcal{C}=\oplus_{M\in G}\mathcal{C}_M$ in \eqref{Ggradedfusion}, and associate a 1-dimensional space for each simple defect object $M_\lambda$. Thus, there is a $D$-dimensional space on each link, where $D$ is the number simple objects in $\mathcal{C}$. For example, the defect category $\langle 1,e,m,\psi\rangle\oplus\langle\sigma_0,\sigma_1\rangle$ of the toric code would give a 6-dimensional space on an edge, i.e. six possible string types on each link. The twist liquid Hamiltonian itself consists of a sum of local stabilizing operators over all vertices $V$ and plaquettes $P$ of the honeycomb lattice \begin{align}\hat{H}_{TL}=-\sum_V\hat{V}-\sum_P\hat{P}.\end{align} 

Each vertex operator acts to enforce the admissible fusion and splitting rules:  \begin{align}\hat{V}\left|\vcenter{\hbox{\includegraphics[height=0.25in]{splittingabc}}}\right\rangle=\delta_{\bf abc}\left|\vcenter{\hbox{\includegraphics[height=0.25in]{splittingabc}}}\right\rangle,\quad\hat{V}\left|\vcenter{\hbox{\includegraphics[height=0.25in]{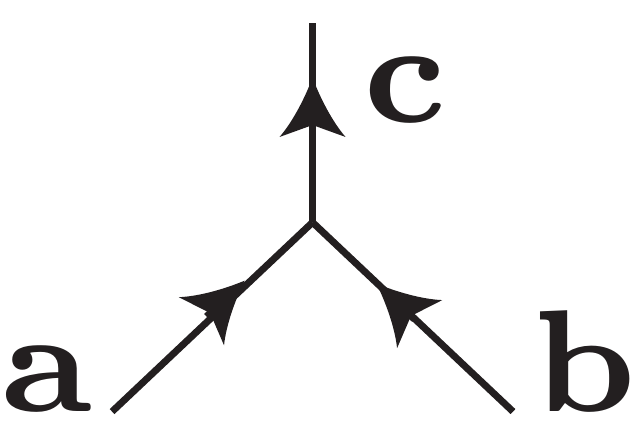}}}\right\rangle=\delta_{\bf abc}\left|\vcenter{\hbox{\includegraphics[height=0.25in]{splittingcab}}}\right\rangle\end{align} where $\delta_{\bf abc}=1$ if ${\bf c}$ is an admissible fusion channel of ${\bf a}\times{\bf b},$ and $\delta_{\bf abc}=0$ otherwise. Hence a vertex operator is a projection operator that acts locally on its three adjacent links. 

A plaquette operator is a local operator that switches the labels along the 6 sides of the hexagon plaquette and depends on, but does not change, the labels on its external legs. Its action is: \begin{align}&\hat{P}\vcenter{\hbox{\includegraphics[width=0.1\textwidth]{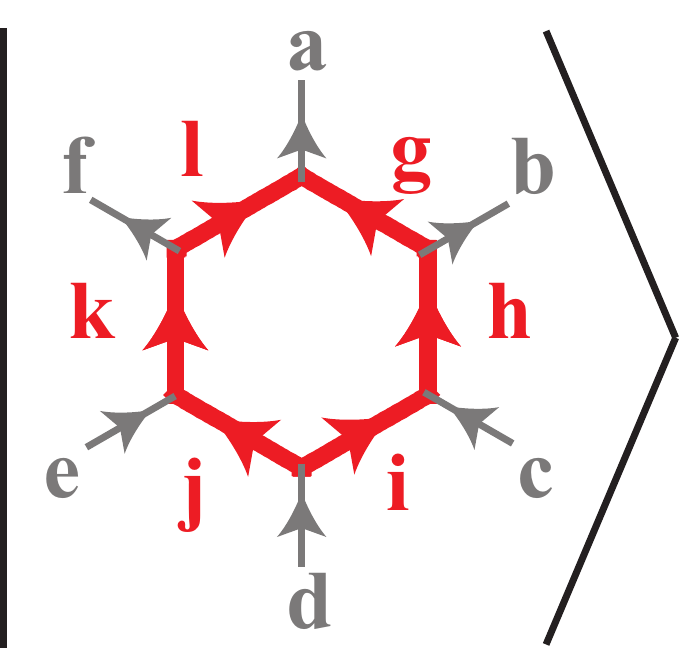}}}=\sum_{\bf s}d_{\bf s}\vcenter{\hbox{\includegraphics[width=0.12\textwidth]{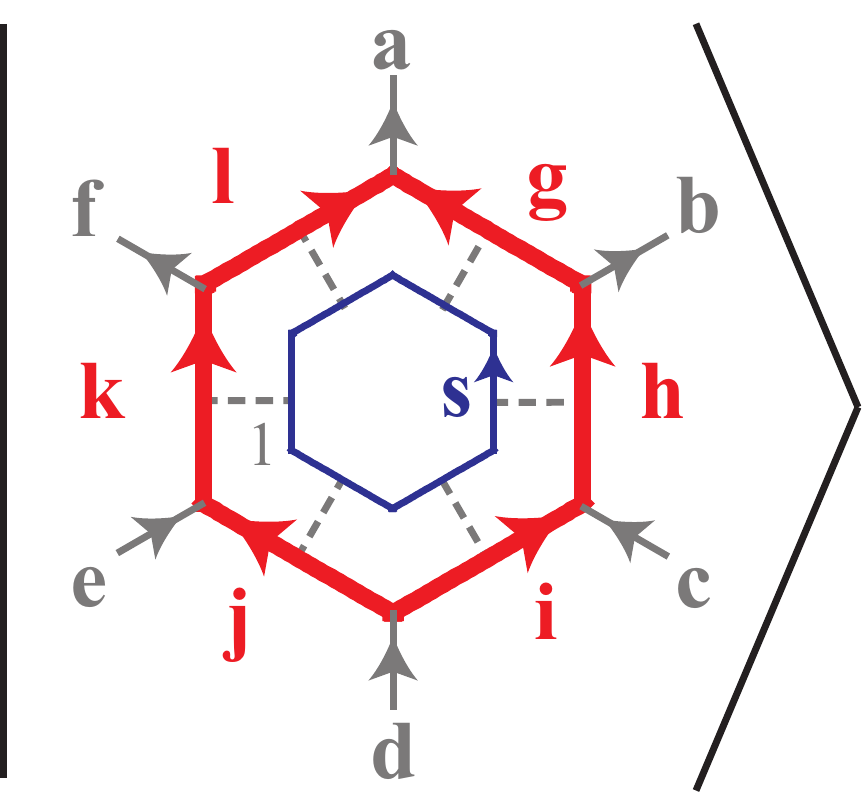}}}\label{plaquetteop1}\\&=\sum_{\bf s}d_{\bf s}\sum_{\bf h'g'i'j'k'l'}P^{\bf s}\vcenter{\hbox{\includegraphics[width=0.1\textwidth]{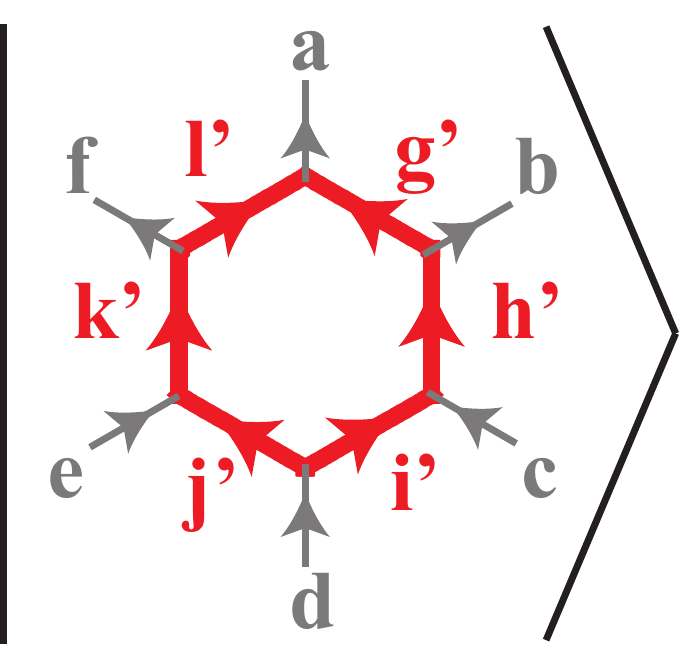}}}\label{plaquetteop2}\end{align} where $d_{\bf s}$ is the quantum dimension of ${\bf s}$, and the summation is taken over all admissible fusion channels \begin{align}\begin{array}{*{20}c}{\bf s}\times{\bf g}\to{\bf g'},&{\bf s}\times{\bf h}\to{\bf h'},&{\bf s}\times{\bf i}\to{\bf i'}\\\bar{\bf s}\times{\bf j}\to{\bf j'},&\bar{\bf s}\times{\bf k}\to{\bf k'},&\bar{\bf s}\times{\bf l}\to{\bf l'}.\end{array}\end{align} Eq.~\eqref{plaquetteop2} can be derived from \eqref{plaquetteop1} by applying a series of $F$-moves (see Fig.~\ref{fig:Fmoves} or Eq.~\eqref{Fsymboldef}) so that the inner ${\bf s}$-loop (which is created in the interior the plaquette, off the lattice) is eventually absorbed on the honeycomb lattice. The coefficients in \eqref{plaquetteop2} are given by a product of $F$-symbols \begin{align}P^{\bf s}=&\beta_{\bf g'j'}^{\bf sgj}\left[F^{\bf j'si}_{\bf d}\right]_{\bf j}^{\bf i'}\left[\left(F^{\bf sgb}_{\bf h'}\right)^{-1}\right]_{\bf h}^{\bf g'}\left[F^{{\bf fl}\bar{\bf s}}_{\bf k'}\right]_{\bf k}^{\bf l'}\nonumber\\&\times\left[\left(F^{\overline{\bf l}{\bf s}\overline{\bf g'}}_{\overline{\bf a}}\right)^{-1}\right]_{\overline{\bf g}}^{\overline{\bf l'}}\left[F^{\bar{\bf l}\bar{\bf j}\bf s}_{\overline{\bf k'}}\right]_{\overline{\bf k}}^{\overline{\bf j'}}\left[\left(F^{\overline{\bf s}\overline{\bf i}\overline{\bf c}}_{\overline{\bf h'}}\right)^{-1}\right]_{\overline{\bf h}}^{\overline{\bf i'}}\end{align} where the $\beta$ factor comes from particle-antiparticle duality (or bending): \begin{align}\beta_{\bf g'j'}^{\bf sgj}=\frac{1}{d_{\bf s}}\sqrt{\frac{d_{\bf g}d_{\bf j}}{d_{\bf g'}d_{\bf j'}}}\left(\left[F^{\bar{\bf s}\bf s\bar{\bf s}}_{\bf s}\right]_1^1\right)^2\left[F^{\bar{\bf s}\bf sg}_{\bf g}\right]_1^{\bf g'}\left[F^{\bar{\bf j}\bf s\bar{\bf s}}_{\bar{\bf j}}\right]^{\overline{\bf j'}}_1,\end{align} and the labels for vertex degeneracy have been suppressed.

The vertex and plaquette operators mutually commute essentially because of the enforcement of associative fusion rules and consistent $F$-symbols that obey the pentagon identity (see Fig.~\ref{fig:Fpentagon}). They therefore share simultaneous eigenstates and the model can be exactly solved. The resulting  ground state is a linear combination of all possible link/string configurations such that fusion at each vertex is admissible. It can be visualized as a superposition of all possible contractible loops with a normalization of $d_{\bf s}$ for each simple ${\bf s}$-loop. 

The simplest string-net model is the (globally symmetric) toric code on a honeycomb lattice, where the input data for the model is the ``defect category" $\langle1\rangle\oplus\langle\sigma\rangle$ with the fusion rule $\sigma\times\sigma=1,$ and basis transformation $F^{\sigma\sigma\sigma}_\sigma=1$. This string-net model gives a $\mathbb{Z}_2$ gauge theory, i.e.~the twist liquid of a trivial $\mathbb{Z}_2$ symmetric boson condensate. However, if the parent state is a topological $\mathbb{Z}_2$-SPT, or equivalently, stacked on top of one, then the basis transformation would be non-trivial $F^{\sigma\sigma\sigma}_\sigma=-1,$ which corresponds the non-trivial cohomology class in $H^3(\mathbb{Z}_2,U(1))=\mathbb{Z}_2$. This would lead to a distinct twist liquid as the corresponding string-net model yields a double semion theory\cite{LevinWen05} described by the 2 component Chern-Simons theory with $K=2\sigma_z$.

\begin{figure}[htbp]
\centering\includegraphics[width=0.45\textwidth]{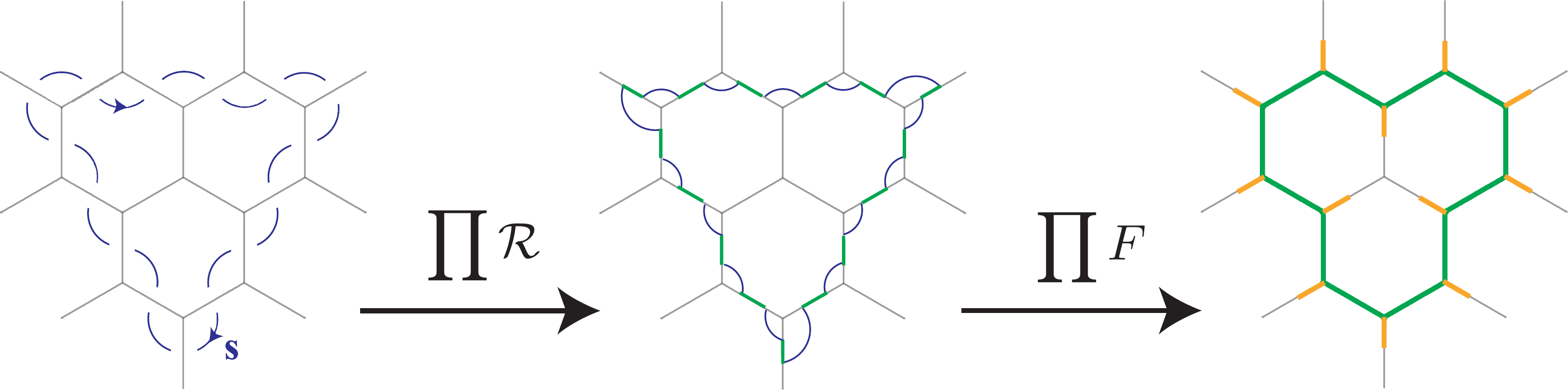}
\caption{A general Wilson loop constructed by crossing ($\mathcal{R}$-symbols) and bootstrapping ($F$-symbols).}\label{fig:stringnetloop}
\end{figure}

Just like the toric code model, the anyon excitations of the string-net model appear at the ends of open Wilson-line string operators.  To construct the allowed anyon excitations we can first consider how to construct generic closed Wilson loops by extending the plaquette operators in \eqref{plaquetteop1} (see Fig.~\ref{fig:stringnetloop}). The construction starts with a fictitious loop with label ${\bf s}$ in the background (underneath the lattice) and then manipulating the loop to eventually absorb it into the honeycomb lattice. This can be done by introducing the crossing operation ($\mathcal{R}$-move): \begin{align}\vcenter{\hbox{\includegraphics[width=0.05\textwidth]{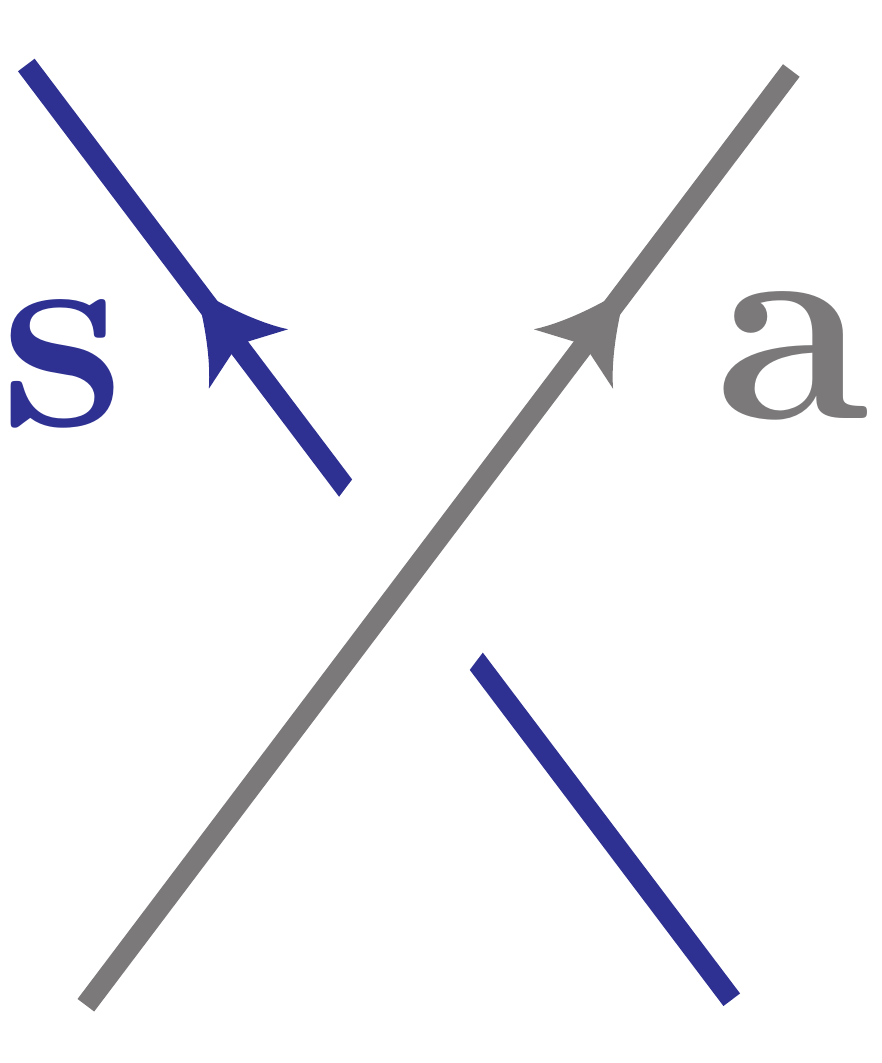}}}=\sum_{\bf c}\sqrt{\frac{d_{\bf c}}{d_{\bf a}d_{\bf s}}}\mathcal{R}^{{\bf a}{\bf s}}_{\bf c}\vcenter{\hbox{\includegraphics[width=0.05\textwidth]{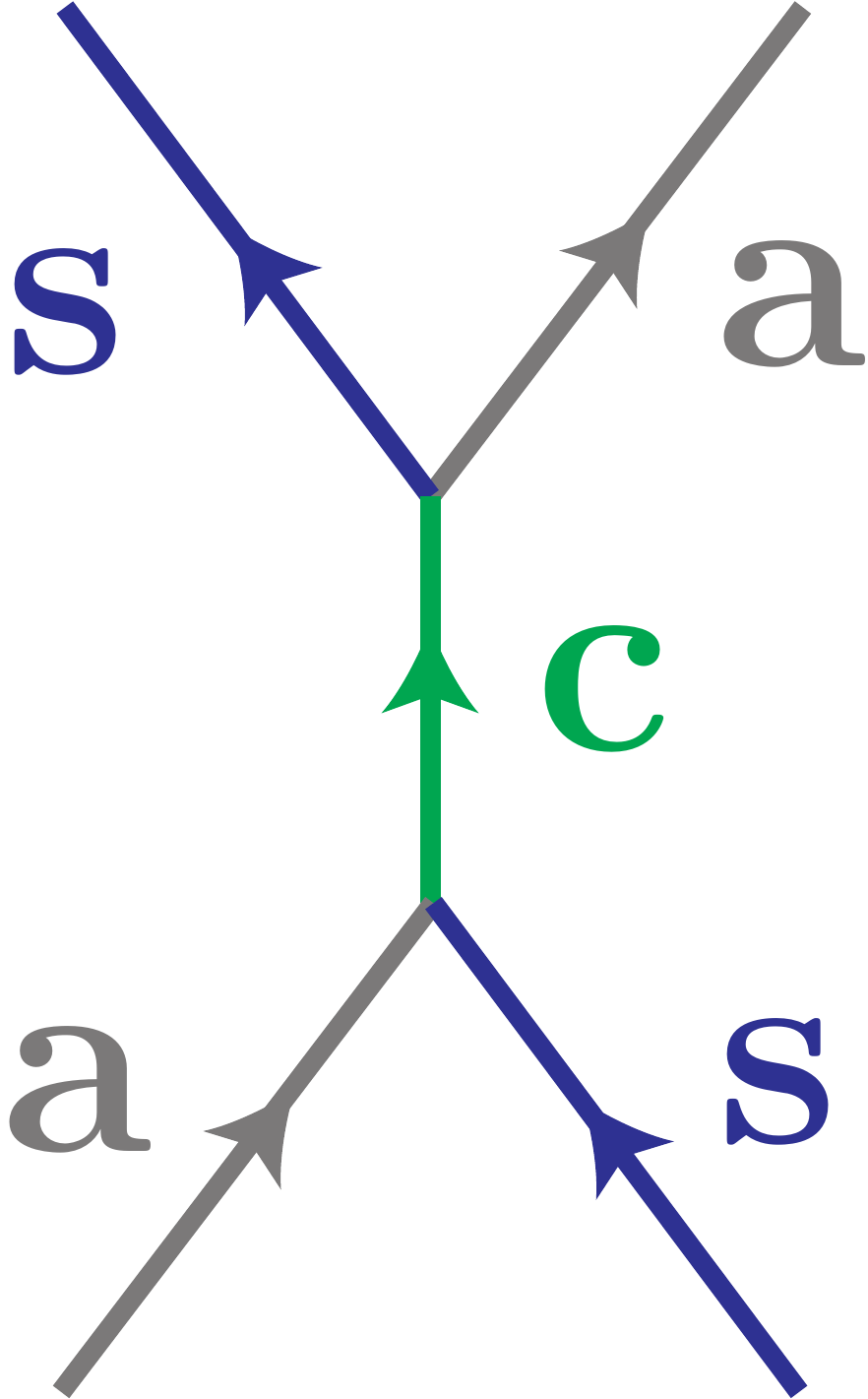}}}\end{align} where the summation is taken over admissible fusion channels, ${\bf a}\times{\bf s}\to{\bf c},$ where vertex labels for fusion degeneracy have been suppressed. The ${\bf s}$-string then can be put on the same ``level" as the honeycomb lattice by applying an $\mathcal{R}$-move for each intersection between the background loop and the honeycomb lattice. Finally, the rest can be put onto  the honeycomb by bootstrapping using the $F$-symbols. The result swaps link labels along the loop (green edges) with local phases depending on the adjacent links (orange edges). 

While the $F$-symbols are inputs of the string-net model, the crossing $\mathcal{R}$-symbols are not. Thus we need a way to specify these symbols in order to perform an $\mathcal{R}$-move for the above algorithm.
Importantly, not all choices of $\mathcal{R}^{\bf as}_{\bf c}$ produce an admissible Wilson loop. To be admissible, a loop operator must commute with all the plaquette operators, commute with the other Wilson loops, and condense to the ground state. These requirements are strict and impose a set of local consistency relations on $\mathcal{R}$ -- precisely the hexagon identity \eqref{hexagoneq} (see also Fig.~\ref{fig:hexagon1}) -- to ensure that local crossings of strings are compatible with the fusion rules. In short, this implies that $\mathcal{R}^{\bf as}\mathcal{R}^{\bf bs}\sim\mathcal{R}^{({\bf a}\times{\bf b}){\bf s}}$ up to bootstrapping by $F$-moves. A Wilson loop is therefore specified by the following information \begin{align}\chi=\left({\bf s},\{\mathcal{R}^{{\bf s}\ast}_\ast\}\right)\label{Drinfeld2tuple}\end{align} where $\mathcal{R}$ is a particular solution to the \hexeq's.  Since an open Wilson string realizes localized anyonic excitations at the ends of the string, Eq.~\eqref{Drinfeld2tuple} therefore prescribes a particular excitation type known as a {\em Drinfeld anyon}.\cite{LevinWen05,Kasselbook,BakalovKirillovlecturenotes} The pentagon identities Fig.~\ref{fig:Fpentagon} and the \hexeq's guarantee that  fusion and braiding between the Drinfeld anyons is consistent. 
 
To summarize, the string-net model of a fusion category $\mathcal{C}$ gives a topological field theory, whose anyon structure  consists of the Drinfeld anyons, and encodes both fusion and braiding information. It is mathematically known as the {\em Drinfeld center}~\cite{Kasselbook,BakalovKirillovlecturenotes} and is denoted by $Z(\mathcal{C})$.  In a general scenario, one begins by extending the globally $G$-symmetric parent state $\mathcal{B}$ to a defect fusion category $\mathcal{C}=\oplus_{M\in G}\mathcal{C}_M$. This involves specifying the defect fusion rules and $F$-symbols described in Section~\ref{sec:twistdefects}. The string-net construction, in which the initial anyon types and the defects themselves become string-types on the links, then provides a model for a topological state where defects become dynamical excitations of a quantum Hamiltonian. This fits the definition of a twist liquid where anyonic symmetries are gauged and symmetry fluxes are excitations. 

The string-net construction also provides a controlled gauging phase transition. This can be achieved by adding string tension terms \begin{align}\delta H=+h\sum_{\mbox{\small links $l$}}\mathcal{T}_l\end{align} where $\mathcal{T}_l$ acts locally at the edge $l,$ and is $\mathcal{T}_l=1$ if the link label is a non-trivial defect $M\neq1,$ or  $\mathcal{T}_l=0$ if the link label is a quasiparticle from the parent state. The string tension essentially gives an energy cost to a defect. These terms do not respect the fusion rules, and therefore do not commute with vertex and plaquette operators. For weak tension $h$, the perturbation would lower the bulk excitation energy gap but the topological state would remain in the same topological phase as the twist liquid. For strong tension, however, defects are confined as they are not energetically favorable, and the string-net model transitions back to the globally symmetric topological state. The gauging phase transition can therefore be driven by tuning $h$ between 0 and $\infty$ 
\begin{align}\begin{diagram}
\stackrel{\mbox{Globally Symmetric}}{\mbox{State ($h\to\infty$)}}&\pile{\rTo^{\mbox{\small Gauging}}\\\lTo_{\mbox{\small Condensation}}}&\stackrel{\mbox{Twist Liquid}}{(h=0)}
\end{diagram}
\label{gaugingtransition2}
\end{align} 
where a gap closing quantum phase transition occurs at certain finite $h>0$.

The transition from $h=0$ to $\infty$ can also be viewed through a complementary anyon condensation picture. Previously, in Section~\ref{sec:QPstructuretwistliquid}, we saw that a twist liquid contains pure gauge charges. They are bosonic excitations that acquire a holonomy when encircling a gauge flux. In the string-net model, they are Drinfeld anyons $({\bf s};\mathcal{R}^{{\bf s}\ast}_\ast)$ originating from the trivial fusion object ${\bf s}=1$. Condensing these bosonic gauge charges confines all the gauge fluxes with respect to which they are non-local. The string-net model therefore reduces to a topological state with only global (anyonic) symmetry when $h\to\infty$.

One subtle source of confusion with this construction is that, for our purposes, the string-net construction always contains a redundant time-reversal partner. For instance, the model in the limit $h\to\infty$ is the time-reversal symmetric doubled parent state \begin{align}Z(\mathcal{B})=\mathcal{B}\otimes\overline{\mathcal{B}}\label{TRdouble}\end{align} where $\mathcal{B}$ is the globally symmetric parent state in which we are interested, and which could be chiral, and $\overline{\mathcal{B}}$ is its time-reversed partner.\cite{EtingofOstrik04} The doubling appears because the string-net Hamiltonian, as well as the \hexeq's themselves, are time reversal symmetric. We have already illustrated the construction of the Drinfeld Center for the reduced defect fusion category $\mathcal{C}_{\mbox{\tiny toric code}}^{(0)}=\langle1,\psi\rangle\oplus\langle\sigma_0\rangle$ of the toric code in Section~\ref{Drinfeld-cosntruction}, and the resulting twist liquid has an $\mbox{Ising}\times\overline{\mbox{Ising}}$ anyon structure. However, we should have really applied the Drinfeld construction, i.e. solving the \hexeq's, on the entire defect fusion category $\mathcal{C}_{\mbox{\tiny toric code}}=\langle1,e,m,\psi\rangle\oplus\langle\sigma_0,\sigma_1\rangle.$ Had we done this the resulting twist liquid would have the anyon content of \begin{align}Z(\mathcal{C}_{\mbox{\tiny toric code}})=\left(\mbox{Ising}\otimes\overline{\mbox{Ising}}\right)\otimes\mbox{Toric code}.\label{ZCtoriccode}\end{align} This result can be  understood by noting that the full defect fusion category has another $\mathbb{Z}_2$-grading (which is not related to the $\mathbb{Z}_2$ e-m symmetry group) as we recall $\sigma_1=e\times\sigma_0$ and $m=e\times\psi$: \begin{align}\mathcal{C}_{\mbox{\tiny toric code}}&=\mathcal{C}_{\mbox{\tiny toric code}}^{(0)}\oplus\mathcal{C}_{\mbox{\tiny toric code}}^{(1)}=\langle1,\psi,\sigma_0\rangle\oplus\langle e,m,\sigma_1\rangle\nonumber\\&=\underline{\mathbb{Z}_2}\otimes\mathcal{C}_{\mbox{\tiny toric code}}^{(0)}=\langle1,e\rangle\otimes\langle1,\psi,\sigma_0\rangle.\end{align} Usefully, the Drinfeld construction decomposes naturally via the tensor product because the fusion rules, as well as the $F$-symbols, are simple products. This gives \eqref{ZCtoriccode} since $Z(\mathcal{C}_{\mbox{\tiny toric code}}^{(0)})=\mbox{Ising}\times\overline{\mbox{Ising}}$ as proven in Section~\ref{Drinfeld-cosntruction} and $Z(\underline{\mathbb{Z}_2})=\mbox{Toric code}$ as shown in Ref.[\onlinecite{LevinWen05}]. We are not interested in the extra redundant copy, and unfortunately it is not always possible to extract a defect subcategory which gives the correct twist liquid result, thus we need a method to remove the redundancy.

Let us explore this a bit more. Unlike the topological parent state $\mathcal{B},$ which has a modular braiding structure, its defect theory $\mathcal{C}$ only specifies fusion information. Thus, the corresponding string-net model, or the Drinfeld center $Z(\mathcal{C})$ in general, can no longer be written as a time reversal pair \eqref{TRdouble}. 
 For instance, \eqref{ZCtoriccode} does not split into a time-reversed pair.
However, the total quantum dimension of the Drinfeld center/string-net model is still related to that of the initial defect fusion category by\cite{MugerI03,MugerII03} \begin{align}\mathcal{D}_{Z(\mathcal{C})}=\mathcal{D}_{\mathcal{C}}^2.\end{align} On the other hand, we have already proved in Section~\ref{sec:twistdefects} that the total dimension of the initial defect fusion category is $\mathcal{D}_C=\mathcal{D}_0\sqrt{|G|}$, where $\mathcal{D}_0$ is the dimension of the parent state $\mathcal{B}$. Thus, this shows that the string-net model construction (Drinfeld center) built on such a defect fusion category yields a total dimension of \begin{align}\mathcal{D}_{Z(\mathcal{C})}=\mathcal{D}_0^2|G|.\end{align} However, this exceeds the total quantum dimension of the twist liquid that we derived in Eq.~\eqref{TLdimension3} precisely by a factor of $\mathcal{D}_0$. This is because, as mentioned above, the string-net model contains not only the twist liquid but also a redundant time-reversed partner of the original parent state to ensure an overall vanishing chirality. This leads to the identification \begin{align}Z(\mathcal{C})=(\mbox{Twist Liquid})\otimes\overline{\mathcal{B}}.\label{Drinfeld=TL}\end{align} For example, when one uses the full defect theory of the toric code, i.e. $\mathcal{C}=\langle1,e,m,\psi\rangle\oplus\langle\sigma_0,\sigma_1\rangle,$ in the string-net construction, the resulting theory \eqref{ZCtoriccode} includes a redundant copy of the toric code parent state (which is time reversal symmetric) in addition to the $\mbox{Ising}\times\overline{\mbox{Ising}}$ twist liquid.

If we want to extract the relevant anyon content for the twist liquid alone, we must work a bit harder and refine the Drinfeld construction by solving a conditioned set of \hexeq's. First, we note that the defect fusion category $\mathcal{C}$ is, in general, only {\em partially braided}. This is because while the trivial defect sector $\mathcal{C}_1$ is identical to the parent state $\mathcal{B}$ which has a modular braiding structure, the non-trivial defect pieces do not. The initial modular braiding structure specifies the $\mathcal{R}$-symbols that only involve exchanging quasiparticles from the parent state with each other. The redundant time reversal sector can be removed from the final string-net result by reinstating these \emph{known} $\mathcal{R}$-symbols in the \hexeq's that come from the original braiding structure of the parent state. The solutions to these restricted \hexeq's precisely correspond to the anyonic excitations in the twist liquid alone. This procedure is known as the {\em relative} Drinfeld construction, and we will demonstrate this in examples presented in Section~\ref{sec:gauging-em-bilayer-su3}, \ref{sec:so(N)}, \ref{sec:so(8)symmetry} and \ref{sec:4statePotts}.

\subsection{Gauging symmetry protected topological phases}\label{sec:gaugingSPT}
We will conclude the discussion of the general gauging procedure by elaborating on the brief earlier discussions of combining topologically ordered phases with an anyonic symmetry $G$ with a symmetry protected topological phase protected by the same symmetry group.
Generically, symmetry protected topological phases (SPTs) are short-range entangled states in $(2+1)$ dimensions.\cite{ChenGuLiuWen12, LuVishwanathE8, MesarosRan12, EssinHermele13, WangPotterSenthil13, BiRasmussenSlagleXu14, Kapustin14} They do not have bulk topological order, and do not support topological ground state degeneracy or anyonic excitations. They do, however, usually carry anomalous boundary theories protected by symmetries. Their bulk short-range entangled topologies are classified cohomologically by $H^3(G,U(1))$ where $G$ is the symmetry group\cite{ChenGuLiuWen11,ChenGuLiuWen12}. If we want to gauge the symmetry, then the cohomology class enters via a  set of consistent defect $F$-symbols $F^{LMN}_K$ where $L,M,N,K$ are group elements so that $LMN=K$. These were discussed in more detail in Section~\ref{sec:twistdefects} (in particular see Eq.~\eqref{Fphasemodification}). The relationship between $\mathbb{Z}_2$-SPT and gauging was first noticed by Levin and Gu in Ref.\onlinecite{LevinGu12}. The relation was later extended to general finite Abelian symmetries, fermion systems\cite{ChengGu14AbSPT} and higher dimensions\cite{HungWen12}, and realized in exact solvable models\cite{LinLevin14}. Here we demonstrate the gauging of $\mathbb{Z}_2$ and $\mathbb{Z}_3$ SPT's, and show how they modify general $\mathbb{Z}_2$ and $\mathbb{Z}_3$ twist liquids when combined with topologically ordered phases with global anyonic symmetries. 

\subsubsection{Gauging a \texorpdfstring{$\mathbb{Z}_2$}{Z2} SPT}\label{sec:gaugingZ2SPT}
There are two inequivalent short-range entangled phases with $\mathbb{Z}_2=\{1,m\}$ symmetry. They are classified by $H^3(\mathbb{Z}_2,U(1))=\mathbb{Z}_2$ and characterized by the two inequivalent defect $F$-symbols \begin{align}\varkappa_m=F^{mmm}_m=\pm1\end{align} where $m$ is the twofold defect with the defect fusion rule $m\times m=1$. It was shown by Levin and Wen in Ref. [\onlinecite{LevinWen05}] that these two inequivalent fusion theories give rise to two inequivalent topologically ordered phases -- the toric code (or a $\mathbb{Z}_2$ gauge theory) when $\varkappa_\sigma=+1$ and the double-semion model when $\varkappa_\sigma=-1$. The toric code was studied in detail in Section~\ref{sec:Zkgaugetheory}. The double-semion model is an Abelian topological state described by the $K$-matrix $K=2\sigma_z$ in a 2-component Chern-Simons action \eqref{toriccodeCSaction}. 

The anyon contents of both the toric code and double-semion theory contain the vacuum $1$, the $\mathbb{Z}_2$ charge $e$, the $\mathbb{Z}_2$ flux $m,$ and a flux-charge composite $\psi=e\times m$. The two theories have identical fusion rules, but different exchange statistics. While the $\mathbb{Z}_2$ charge is always a boson, the $\mathbb{Z}_2$ flux $m$ is bosonic in the toric code but semionic with spin $h_m=1/4$ in the double-semion theory. Additionally, $\psi$ is fermionic for the toric code but has spin $h_\psi=-1/4$ in the double-semion theory.

\subsubsection{Gauging a \texorpdfstring{$\mathbb{Z}_3$}{Z3} SPT}\label{sec:gaugingZ3SPT}
There are three inequivalent short-range entangled phases with $\mathbb{Z}_3=\{1,\rho,\overline\rho\}$ symmetry since $H^3(\mathbb{Z}_3,U(1))=\mathbb{Z}_3$. The three cohomology classes can be represented by the three inequivalent sets of defect $F$-symbols \begin{align}\begin{array}{*{20}l}F^{\rho\overline\rho\rho}_\rho=e^{2\pi mi/3},& F^{\overline\rho\rho\overline\rho}_{\overline\rho}=e^{-2\pi mi/3},\\F^{\rho\bar\rho\bar\rho}_{\overline\rho}=e^{-2\pi mi/3},& F^{\overline\rho\rho\rho}_\rho=e^{2\pi mi/3},\\F^{\bar\rho\bar\rho\rho}_{\overline\rho}=e^{2\pi mi/3},& F^{\bar\rho\bar\rho\bar\rho}_1=e^{-2\pi mi/3}\end{array}\label{Z3SPTF}\end{align} for $m=0,1,2$ modulo 3, where all unspecified $F$-symbols are trivial.

Gauging the $\mathbb{Z}_3$ symmetry leads to three distinct topological phases. They each contain nine anyons $\{1,z,\overline{z}\}\times\{1,\rho,\overline\rho\}$ expressed as composition of flux $\rho$ and charge $z,$ such that the braiding phase between $\rho$ and $z$ is $e^{2\pi i/3}$. The three topological phases each have different fusion rules and exchange statistics. These can be evaluated by the Drinfeld construction presented in Section~\ref{sec:stringnet}. The choice of defect $F$-symbols in \eqref{Z3SPTF}  does not affect the spins of the $\mathbb{Z}_3$ charges, and therefore they are always bosonic. The \hexeq~that determines the spin of a $\mathbb{Z}_3$ flux is given by \begin{align}\theta_\rho^2=\mathcal{R}^{\rho\rho}_{\overline\rho}\mathcal{R}^{\rho\rho}_{\overline\rho}=\mathcal{R}^{\rho\overline\rho}_1=\varkappa_\rho\theta_\rho^\ast\end{align} where the Frobenius-Schur indicator is given by $\varkappa_\rho=F^{\rho\overline\rho\rho}_\rho=e^{2\pi mi/3}$. There are three solutions to the \hexeq,~and they correspond to the exchange phases of the flux-charge composites $\{\rho_0,\rho_1,\rho_2\}=\{\rho,z\rho,\overline{z}\rho\}$. There are three possible sets of spins \begin{align}h_\rho=\left\{\begin{array}{*{20}c}0,1/3,2/3,&\mbox{for $m=0$}\\1/9,4/9,7/9,&\mbox{for $m=1$}\\2/9,5/9,8/9,&\mbox{for $m=2$}\end{array}.\right.\end{align}

Unlike the double-semion theory, which has identical fusion rules to the toric code, the fusion rules of the three $\mathbb{Z}_3$ topological phases are not the same. They are distinguished by \begin{align}\rho\times\rho\times\rho=\left\{\begin{array}{*{20}c}1,&\mbox{for $m=0$}\\\overline{z},&\mbox{for $m=1$}\\z,&\mbox{for $m=2$}\end{array}.\right.\end{align} This can be proven using the ribbon identity \eqref{ribbon}. The three topological phases are Abelian and can be characterized by 2-component Chern-Simons effective actions \eqref{toriccodeCSaction} with the $K$-matrices \begin{align}K=\left(\begin{array}{*{20}c}0&3\\3&0\end{array}\right),\quad\left(\begin{array}{*{20}c}4&5\\5&4\end{array}\right),\quad\left(\begin{array}{*{20}c}-4&5\\5&-4\end{array}\right)\end{align} for $m=0,1,2$ respectively. The $m=0$ state is the conventional $\mathbb{Z}_3$ gauge theory, and the other two have different properties. For example when $m=1$, the gauge charges $1,z,\overline{z}$ are represented by the lattice vectors $0,6{\bf e}_1,3{\bf e}_1$, and the fluxes $\rho_0,\rho_1,\rho_2$ are represented $7{\bf e}_1,4{\bf e}_1,{\bf e}_1$ respectively. The fusion group $\mathcal{A}=\mathbb{Z}^2/K\mathbb{Z}^2\cong\mathbb{Z}_9$ differs from that of the $\mathbb{Z}_3$-gauge theory, which has $\mathcal{A}=\mathbb{Z}_3\times\mathbb{Z}_3$.

\subsubsection{Modifying a twist liquid by a SPT}
The defect $F$-symbols of a globally anyonic $G$-symmetric topological state will be modified by $U(1)$ phases \begin{align}F^{LMN}_K\to e^{i\phi_{LMN}}F^{LMN}_K\end{align} if it is stacked with a copy of a $(2+1)D$ $G$-SPT. Here the set of phases $\{e^{i\phi_{LMN}}:L,M,N\in G\}$ is a representative of the cohomology class in $H^3(G,U(1))$ that classifies the SPT. This changes the \hexeq's in the Drinfeld construction (see Section~\ref{sec:stringnet}) that determine the spin and braiding of anyons in the twist liquid. 

When a $(2+1)D$-SPT $\mathcal{S}$ is sitting on top of a globally symmetric topological state $\mathcal{B}$, a symmetry flux must pass through both systems (see Fig.~\ref{fig:SPT}). When gauging the symmetry, the flux across the SPT is bound to the flux through the twist liquid. This confinement can be captured nicely using an artificially extended construction and then condensing  gauge charges to reproduce our system of interest. To illustrate, we first gauge the symmetries of  $\mathcal{S}$ and $\mathcal{B}$ independently. Their pure gauge charges are both characterized by irreducible representations $z^1,\ldots,z^r$ of the symmetry group $G$. We denote the charges by by $z^1_{\mathcal{S}},\ldots,z^r_{\mathcal{S}}$ and $z^1_{\mathcal{B}},\ldots,z^r_{\mathcal{B}}$. They form a closed fusion algebra built from tensor products of representations. Additionally, all pure charges are bosonic as they do not have internal structure. 

Next we put the two twist liquids together by a tensor product, and then condense the set of boson pairs \begin{align}\mathcal{Z}=\{z^1_{\mathcal{B}}\otimes\overline{z^1_{\mathcal{S}}},\ldots,z^1_{\mathcal{B}}\otimes\overline{z^1_{\mathcal{S}}}\}\label{GCcondensate}\end{align} where $\overline{z^i_{\mathcal{S}}}$ is the anti-partner so that the trivial anyon 1 is an admissible fusion channel of $z^i_{\mathcal{S}}\times\overline{z^i_{\mathcal{S}}}$. This condensation effectively identifies the pure charges \begin{align}z^i\equiv z^i_{\mathcal{B}}\otimes1\equiv1\otimes z^i_{\mathcal{S}},\end{align} and confines all single-layer fluxes $M_{\mathcal{B}}\otimes1$ and $1\otimes M_{\mathcal{S}}$ into flux pairs $M=M_{\mathcal{B}}\otimes M_{\mathcal{S}}$, for $M$ a non-trivial element in $G$. This is because each component $M_{\mathcal{B}}$ or $M_{\mathcal{S}}$ alone is non-local with respect to the condensate \eqref{GCcondensate} due to non-trivial braiding phases. The resulting state is a new twist liquid with modified exchange statistics and fusion rules. SPT's therefore act as a transformation/permutation of the category of twist liquids \begin{align}\{\mbox{SPT's}\}\boxtimes\left\{\mbox{Twist liquids}\right\}\longrightarrow\left\{\mbox{Twist liquids}\right\}\end{align} where $\boxtimes$ stands for tensor product with the boson pair condensation \eqref{GCcondensate}.


\section{Gauging Conjugation or Bilayer Symmetry in \texorpdfstring{$SU(3)_1$}{SU(3)}}\label{sec:gauging-em-bilayer-su3}
Our first example demonstrates how the non-chiral string-net model can also capture the gauging transition of chiral topological states.
The $SU(3)_1$ topological phase is a prototype bosonic bilayer quantum Hall state with the $K$-matrix \begin{align}K_{SU(3)_1}=\left(\begin{array}{*{20}c}2&-1\\-1&2\end{array}\right)\end{align} chosen to be the Cartan matrix of the Lie algebra $su(3)$.\cite{khan2014} As a result the system carries a chiral $su(3)$ Kac-Moody CFT at level 1\cite{bigyellowbook} on the edge, as illustrated by the conventional bulk-boundary correspondence. This is an Abelian topological state with Abelian quasiparticles $1,\psi_1,\psi_2,$ and the fusion rules $\psi_1\times\psi_2=1$ and $\psi_1\times\psi_1=\psi_2.$ The topological spins are \begin{align}\theta_{\psi_1}=\theta_{\psi_2}=e^{2\pi ih_\psi},\quad h_\psi=\frac{1}{3}.\end{align} This system has an effective $\mathbb{Z}_2$ bilayer symmetry which is represented by the matrix $M=\sigma_x,$ and interchanges $\psi_1\leftrightarrow\psi_2$. This symmetry is related to the non-trivial reflection symmetry of the Dynkin diagram of the Lie algebra; a symmetry which is also known as an outer automorphism: $\mbox{Outer}(su(3))=\mathbb{Z}_2$ as defined in \eqref{outer}.\cite{khan2014}

If we include the time reversed counterpart for $SU(3)_1$, then the resulting non-chiral theory has the same topological order (or is at least stably equivalent\cite{cano2013bulk}) as a $\mathbb{Z}_3$ discrete gauge theory\cite{BaisDrielPropitius92,Bais-2007,Propitius-1995,PropitiusBais96,Preskilllecturenotes,Freedman-2004,Mochon04} (or the $\mathbb{Z}_3$ plaquette rotor model\cite{Wenplaquettemodel}). For all of these cases the topological order is equivalent to the quantum double of $\mathbb{Z}_3$: \begin{align}D(\mathbb{Z}_3)\approx SU(3)_1\otimes\overline{SU(3)_1}\end{align} with $K$-matrix $3\sigma_x$. The electric charges $e,$ and magnetic fluxes $m,$ of the $\mathbb{Z}_3$ gauge theory are identified with the $SU(3)_1$ quasiparticles via $\psi_1=em^2$, $\psi_2=e^2m$, $\overline{\psi_1}=em$ and $\overline{\psi_2}=e^2m^2$. We choose the convention here that there is a braiding phase of $e^{-2\pi i/3}$ between $e$ and $m$. 

The electric-magnetic symmetry switches $e\leftrightarrow m$. If we translate this symmetry into the non-chiral $SU(3)_1\otimes\overline{SU(3)_1}$ theory, it matches the action of the $\mathbb{Z}_2$ bilayer symmetry, but it only acts  on the chiral sector $\psi_1\leftrightarrow\psi_2,$ while leaving the time reversed sector $\overline{\psi_1},\overline{\psi_2}$ unchanged. This is important since we are really interested in gauging the chiral $SU(3)_1$ state, and will eventually want to remove the vestigial $\overline{SU(3)_1}$ partner. Thus, even though the string-net construction is inherently time-reversal invariant, we can still gauge the anyonic symmetries of chiral states since the anti-chiral partner is inert under the anyonic symmetry action. At the end of the process he $\overline{SU(3)_1}$ sector  can be easily excised from the theory since it is both unaltered during the procedure, and is not coupled with its time reversed partner.

Now let us focus on gauging this $\mathbb{Z}_2$ bilayer symmetry. As mentioned, the non-chiral version of the theory has the advantage of having an exactly solvable string-net model describing the twist liquid, which we will prove is identified with $SU(2)_4\otimes\overline{SU(3)_1}$. 
The gauging/condensation relationships are illustrated in the following diagram: \begin{align}\begin{diagram}SU(2)_4\otimes\overline{SU(3)_1}&\pile{\rTo^{\mbox{\small Condensation}}\\\lTo_{\mbox{\small Gauging}}}&D(\mathbb{Z}_3)\\\uInto&&\uInto\\SU(2)_4&\pile{\rTo^{\mbox{\small Condensation}}\\\lTo_{\mbox{\small Gauging}}}&SU(3)_1\label{su(2)4su(3)1commdiagram}\end{diagram}\end{align} The chiral twist liquid $SU(2)_4$ of the bilayer symmetric $SU(3)_1$ will be extracted using the {\em relative} Drinfeld construction described in Section~\ref{sec:stringnet}. 

Before we carry out the gauging procedure,  it will be helpful if we first describe the taxonomy of quasiparticles in the resulting phase. From the bulk-boundary correspondence, the anyon structure of the $SU(2)_4$ bulk phase is given by the primary fields of the CFT along the boundary\cite{bigyellowbook}. They are labeled by $j={\bf 0},{\bf 1/2},{\bf 1},{\bf 3/2},{\bf 2}$ with spins (or conformal dimensions) \begin{align}\theta_j=e^{2\pi ih_j},\quad h_j=\frac{j(j+1)}{6}=0,\frac{1}{8},\frac{1}{3},\frac{5}{8},1,\label{su(2)4spins}\end{align} and the braiding (or modular) $S$-matrix is\eqref{braidingS} \begin{align}S_{j_1,j_2}=\frac{1}{\sqrt{3}}\sin\left[\frac{\pi(2j_1+1)(2j_2+1)}{6}\right].\end{align} The anyon ${\bf 0}$ (which we would usually call $1$) serves as the vacuum, and the rest obey the fusion rules \begin{gather}{\bf 2}\times{\bf 2}={\bf 0},\quad{\bf 2}\times{\bf 1}={\bf 1},\quad{\bf 2}\times{\bf \frac{1}{2}}={\bf \frac{3}{2}}\\{\bf \frac{1}{2}}\times{\bf \frac{1}{2}}={\bf 0}+{\bf 1},\quad{\bf \frac{1}{2}}\times{\bf 1}={\bf \frac{1}{2}}+{\bf \frac{3}{2}}\\{\bf 1}\times{\bf 1}={\bf 0}+{\bf 1}+{\bf 2}.\label{su(2)4fusion1}\end{gather} The quasiparticle ${\bf 2}$ is an Abelian boson, and will be identified as the $\mathbb{Z}_2$ charge for the gauged anyonic bilayer symmetry. The objects ${\bf 1/2}$ and ${\bf 3/2}$ are non-Abelian anyons, and they have quantum dimensions $d_{\bf 1/2}=d_{\bf 3/2}=\sqrt{3}$. They will be identified as the gauged $\mathbb{Z}_2$ fluxes. Finally, ${\bf 1}$ has integral quantum dimension $d_{\bf 1}=2,$ and will arise as the super-selection sector $\psi_1+\psi_2$ after gauging.

Beginning with the gauged theory we might ask how we can condense\cite{BaisSlingerlandCondensation, Kong14} quasiparticles to return to the un-gauged $SU(3)_1$ theory. We can do this by condensing the $j={\bf 2}$ boson. Hence, the non-Abelian quasiparticles ${\bf 1/2}$ and ${\bf 3/2}$ are confined because of the non-trivial $-1$ braiding phase with this boson. 
The dimension 2 quasiparticle must split into the Abelian ones ${\bf 1}=\psi_1+\psi_2$ so that the fusion rule \eqref{su(2)4fusion1} becomes \begin{align}(\psi_1+\psi_2)\times(\psi_1+\psi_2)=1+1+\psi_1+\psi_2\label{su(2)4fusionsuper}\end{align} when ${\bf 2}$ is identified with the vacuum ${\bf 0} (=1)$. Both $\psi$'s must have the same spin as their parents, $\theta_{\psi_i}=\theta_{\bf 1}=e^{2\pi i/3}$, which, from the spin-statistics theorem, is also the $180^\circ$ exchange phase between a pair of identical $\psi_i$. The full $360^\circ$ braid is its square $\mathcal{D}S_{\psi_i\psi_i}=e^{4\pi i/3}$. From the ribbon identity \eqref{ribbon}, the spin of the fused quasiparticle is \begin{align}\theta_{\psi_i\times\psi_i}=\mathcal{D}S_{\psi_i\psi_i}\theta_{\psi_i}\theta_{\psi_i}=e^{2\pi i/3},\end{align} and this forces $\psi_1\times\psi_1=\psi_2$ and $\psi_2\times\psi_2=\psi_1.$ The trivial fusion channels of \eqref{su(2)4fusionsuper} then require $\psi_1\times\psi_2=1$. The resulting condensed phase thus matches with $SU(3)_1=\langle1,\psi_1,\psi_2\rangle$. As a consistency check, the braiding phase between ${\bf 1}$ (which recall is not the vacuum) and itself matches that between $\psi_i$ and $\psi_j$: \begin{align}\vcenter{\hbox{\includegraphics[height=0.6in]{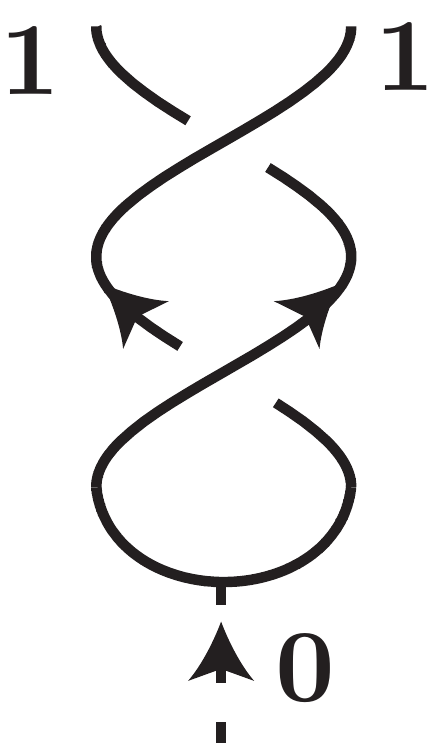}}}&=\frac{1}{\theta_{\bf 1}\theta_{\bf 1}}=\vcenter{\hbox{\includegraphics[height=0.6in]{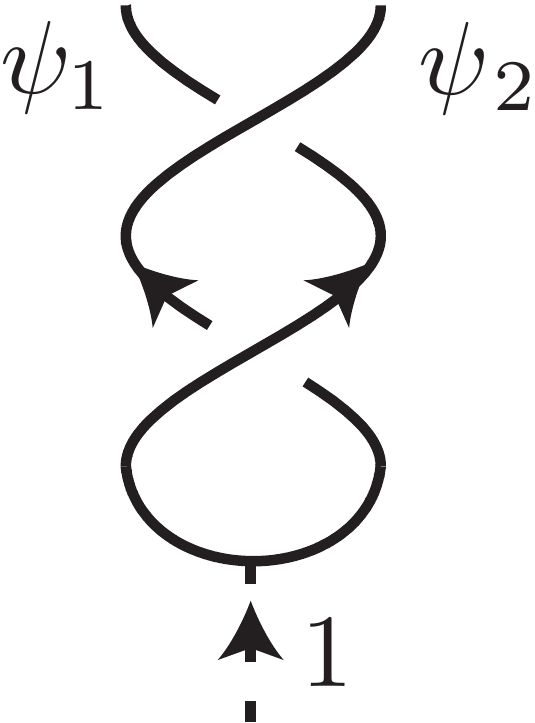}}}=\frac{1}{\theta_{\psi_i}\theta_{\psi_i}}=e^{2\pi i/3}\\\vcenter{\hbox{\includegraphics[height=0.6in]{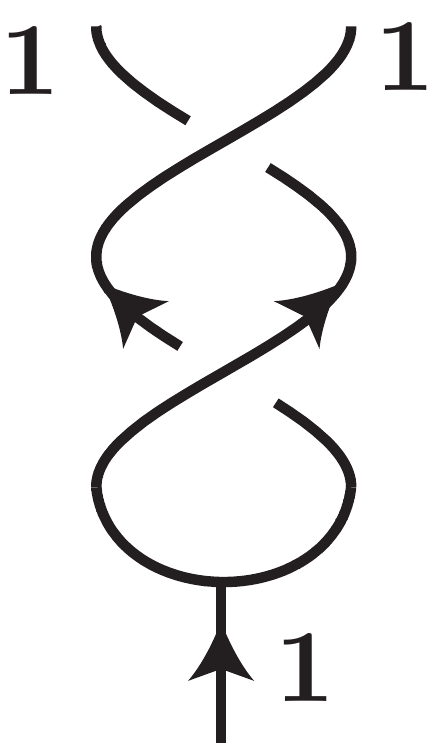}}}&=\frac{\theta_{\bf 1}}{\theta_{\bf 1}\theta_{\bf 1}}=\vcenter{\hbox{\includegraphics[height=0.6in]{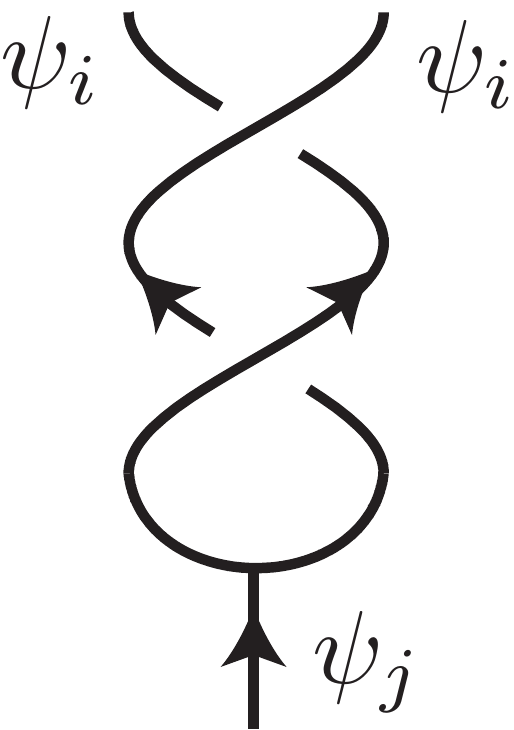}}}=\frac{\theta_{\psi_j}}{\theta_{\psi_i}\theta_{\psi_i}}=e^{-2\pi i/3}.\end{align} 
Moreover, the opposing braiding phases  \begin{align}\vcenter{\hbox{\includegraphics[height=0.6in]{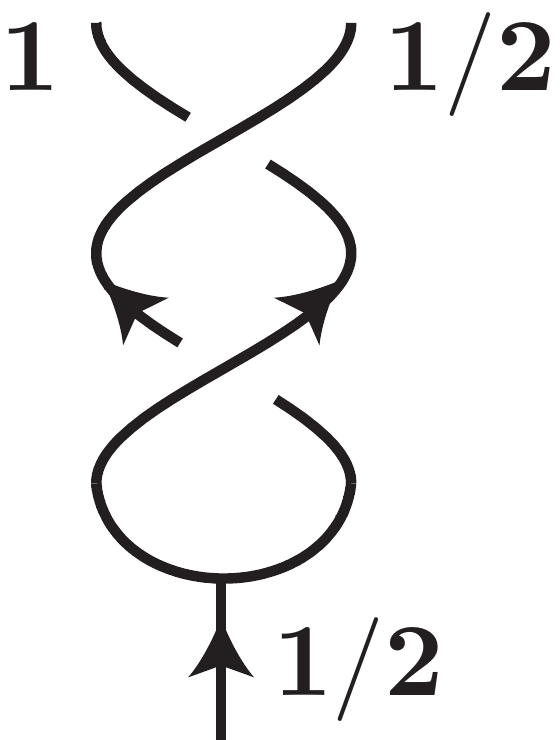}}}=-\vcenter{\hbox{\includegraphics[height=0.6in]{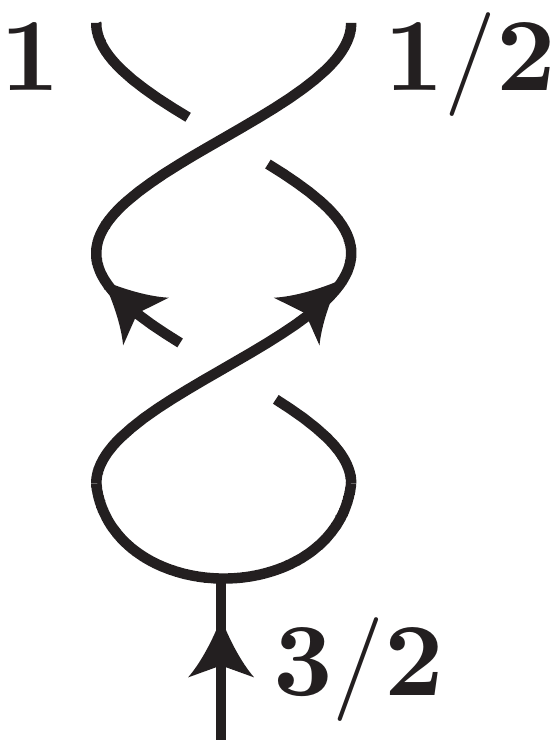}}}=\frac{\theta_{\bf 1/2}}{\theta_{\bf 1/2}\theta_{\bf 1}}=e^{-2\pi i/3}\label{su(2)4superfluxbraid}\end{align} ensure the vanishing of the $S$-matrix element \begin{align}\mathcal{D}S_{{\bf 1},{\bf 1/2}}=\vcenter{\hbox{\includegraphics[width=0.5in]{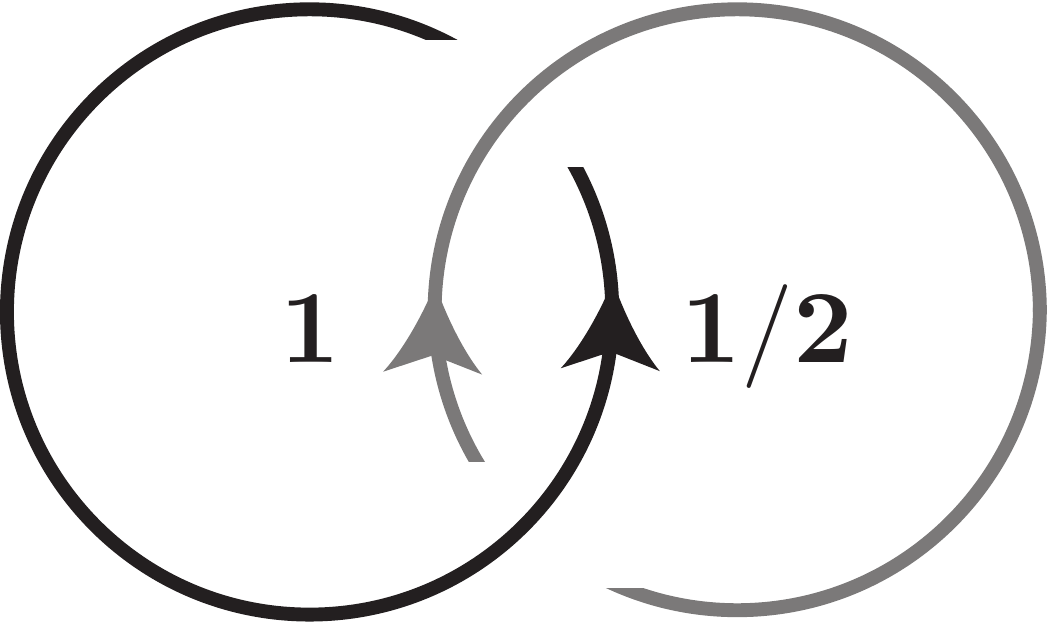}}}=0.\end{align} This indicates that the components of the super-sector ${\bf 1}=\psi_1+\psi_2$ must switch when cycling around the $\mathbb{Z}_2$ flux ${\bf 1/2}$ as expected. The square of the braiding phases in \eqref{su(2)4superfluxbraid} is $e^{2\pi i/3}$, which matches with the eigenvalue of the double Wilson loop (see Fig.~\ref{fig:defect1}(b) and Ref.~[\onlinecite{teo2013braiding}]) in the defect theory.

\subsection{Defect Fusion Category and Twist Liquid Anyon Content}
\label{sec:defect-fusion-category}

These arguments prove the forward horizontal arrows of \eqref{su(2)4su(3)1commdiagram}. To show the reverse directions, we actually need to gauge the $\mathbb{Z}_2$ anyonic symmetry of $SU(3)_1$ by constructing the Drinfeld center  \begin{align}Z(\mathcal{C}_{SU(3)_1})=SU(2)_4\otimes\overline{SU(3)_1}\label{Z(SU(3)1)}\end{align} of the fusion category $\mathcal{C}_{SU(3)_1}=\langle1,\psi_1,\psi_2\rangle\oplus\langle\sigma\rangle$, where $\sigma$ is the twist defect of the bilayer symmetry. The defect object $\sigma$ satisfies the fusion rules \begin{align}\sigma\times\psi_i=\sigma,\quad\sigma\times\sigma=1+\psi_1+\psi_2.\label{su(3)defectfusioncategory}\end{align}  This fusion category can be treated as a sub-category of the full defect theory $\langle1,\psi_1,\psi_2\rangle\otimes\langle1,\overline{\psi_1},\overline{\psi_2}\rangle\oplus\langle\sigma_0,\sigma_1,\sigma_2\rangle$ of the quantum double $D(\mathbb{Z}_3)$, where $\sigma=\sigma_0$ is the (bare) twist defect associated with the e-m/bilayer symmetry. This is similar to the sub-category $\langle1,\psi,\sigma_0\rangle$ of the full defect theory $\langle1,e,m,\psi,\sigma_0,\sigma_1\rangle$ of the toric code in Section~\ref{Drinfeld-cosntruction}.

The non-trivial, admissible $F$-symbols are given by\cite{teo2013braiding} \begin{align}F^{\sigma{\bf a}\sigma}_{\bf b}=F^{{\bf a}\sigma{\bf b}}_\sigma=e^{-\frac{2\pi i}{3}ab},\quad\left[F^{\sigma\sigma\sigma}_\sigma\right]_{\bf a}^{\bf b}=-\frac{1}{\sqrt{3}}e^{\frac{2\pi i}{3}ab}\label{su(3)defectFsymbols}\end{align} where ${\bf a}=e^{-a}m^{a}$, ${\bf b}=e^{-b}m^{b}$ run over the Abelian subset $1$, $\psi_1=em^2,$ and $\psi_2=e^2m$, where the un-bolded labels $a,b\in\mathbb{Z}_3$. Here the sign of $F^{\sigma\sigma\sigma}_\sigma$ gives rise to a non-trivial Frobenius-Schur indicator \begin{align}\varkappa_\sigma=\frac{[F^{\sigma\sigma\sigma}_\sigma]_1^1}{\left|[F^{\sigma\sigma\sigma}_\sigma]_1^1\right|}=-1,\label{su(3)FS}\end{align} which cannot be determined purely by the quasiparticle string manipulation methods explained in Section~\ref{sec:twistdefects}. The FS indicator in general affects particle-antiparticle duality (or bending) of a defect object $\chi$ \begin{align}\vcenter{\hbox{\includegraphics[height=0.8in]{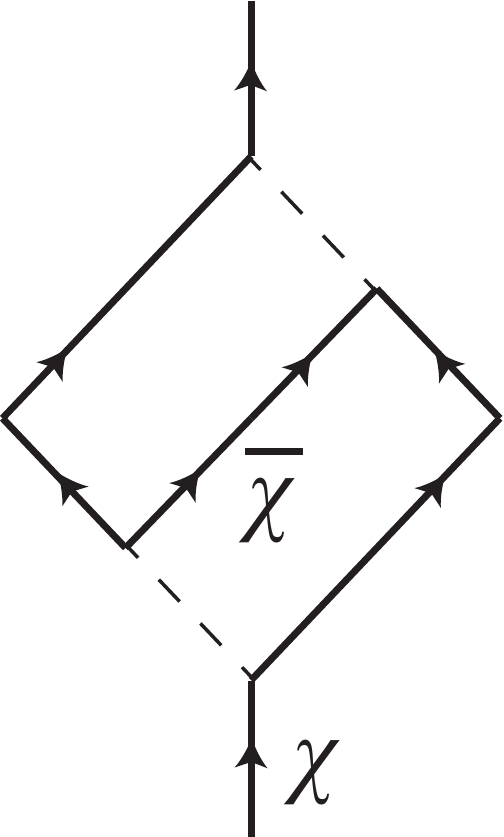}}}=d_\chi \left[F^{\chi\overline\chi\chi}_\chi\right]_1^1\;\vcenter{\hbox{\includegraphics[height=0.8in]{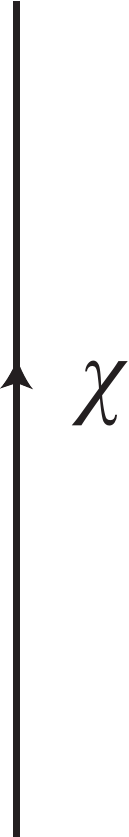}}}=\varkappa_\chi\;\vcenter{\hbox{\includegraphics[height=0.8in]{bending2}}}\label{bendingdef}\end{align} where $d_\chi=\left|[F^{\sigma\sigma\sigma}_\sigma]_1^1\right|^{-1}$. The sign \eqref{su(3)FS} depends on the details of the parent topological state and can change when an extra layer of a non-trivial $\mathbb{Z}_2$ SPT is added (see Section~\ref{sec:twistdefects}). The particular sign convention in Eq.~\eqref{su(3)defectFsymbols} is chosen to match the $F$-symbols of the $SU(2)_4$ state, and also the $\mathbb{Z}_2$-orbifold of the $SU(3)_1$ edge CFT.\cite{chenroyteoinprogress}

Now that we have given the properties of the parent defect fusion category, we are ready to calculate the anyon content of the twist liquid. As shown above, anyons in a general twist liquid can be understood as composition of flux, quasiparticle super-sector, and charge (see Sections~\ref{sec:QPstructuretwistliquid} and \ref{sec:gaugingSPT}). First, for the case of trivial flux, we have either the quasiparticle super-sectors $1$ or $\mathcal{E}=\psi_1+\psi_2$. For the vacuum super-sector $1$, its centralizer is the full $\mathbb{Z}_2$ symmetry group which has two irreducible representations for the charge. The two representations produce either the $1$ or $z$ twist liquid anyons that correspond to the vacuum and pure $\mathbb{Z}_2$ charge respectively. On the other hand, for $\mathcal{E}=\psi_1+\psi_2$, it has the trivial group as its centralizer since neither $\psi_1$ nor $\psi_2$ are fixed under the symmetry. This shows that this super-sector cannot carry non-trivial $\mathbb{Z}_2$ charge, i.e.~$z\times\mathcal{E}=\mathcal{E}$. Next, for non-trivial flux $\sigma$, there are no non-trivial species labels, and the resulting reduced centralizer is the full $\mathbb{Z}_2$ group, whose non-trivial representation corresponds to non-trivial gauge charge. The two fluxes are denoted by $\sigma$ and $z\sigma=z\times\sigma$. This set forms a collection of five anyons and matches the content of the $SU(2)_4$ state: \begin{align}\begin{array}{*{20}c}1={\bf 0},\quad z={\bf 2},\quad\mathcal{E}=\psi_1+\psi_2={\bf 1}\\\sigma={\bf 1/2},\quad z\sigma={\bf 3/2}.\end{array}\end{align}

\subsection{Explicit Drinfeld Construction of the Twist Liquid}
\label{sec:drinfeld-construction-twist}

Now we can explicitly derive the twist liquid theory through the Drinfeld construction defined in Section~\ref{sec:stringnet}. The Drinfeld anyons are specified by $\chi=(x;\mathcal{R}^{x\ast}_\ast)$, where $x$ belongs to the original defect fusion category with fusion rules \eqref{su(3)defectfusioncategory}, and can be any linear combination of the labels $1,\psi_1,\psi_2,\sigma.$ The matrix/symbol $\mathcal{R}^{x\ast}_\ast$ encodes the braiding information, and is a solution to the \hexeq's in Eq.~\ref{hexagoneq} (see also Fig.~\ref{fig:hexagon1}).

We begin with $x=1$. The exchange between $1$ and a pair of $\sigma$'s gives the \hexeq's \begin{align}\mathcal{R}^{1\sigma}_\sigma\mathcal{R}^{1\sigma}_\sigma=\mathcal{R}^{11}_1=\mathcal{R}^{1\psi_i}_{\psi_i}=1\end{align} for $i=1,2$, and there are two solutions \begin{align}1&=\left(1;\mathcal{R}^{1\psi_i}_{\psi_i}=\mathcal{R}^{1\sigma}_\sigma=1\right)\\z&=\left(1;\mathcal{R}^{1\psi_i}_{\psi_i}=-\mathcal{R}^{1\sigma}_\sigma=1\right).\end{align} The trivial solution represents the true vacuum, and the non-trivial $z$ boson is the $\mathbb{Z}_2$ charge. It has a $-1$ braiding phase around the $\mathbb{Z}_2$ flux since $\mathcal{R}^{1\sigma}_\sigma=-1.$

Next we solve the \hexeq's for $\mathcal{R}^{\sigma\ast}_\ast$. The exchange of $\sigma$ with a pair of Abelian objects ${\bf a},{\bf b}$ gives the equation \begin{align}e^{-\frac{2\pi i}{3}ab}\mathcal{R}^{\sigma{\bf a}}_\sigma\mathcal{R}^{\sigma{\bf b}}_\sigma=\mathcal{R}^{\sigma({\bf a}\times{\bf b})}_\sigma.\label{su(3)1hexeq1}\end{align} The exchange of $\sigma$ with another $\sigma$ and an Abelian object ${\bf a}$ gives the \hexeq's \begin{align}\mathcal{R}^{\sigma\sigma}_{\bf b}\mathcal{R}^{\sigma{\bf a}}_\sigma=\mathcal{R}^{\sigma\sigma}_{{\bf a}\times{\bf b}}e^{-\frac{2\pi i}{3}a(a+b)}.\label{su(3)1hexeq2}\end{align} Also, the exchange of $\sigma$ with a pair of $\sigma's$ gives \begin{align}\frac{1}{\sqrt{3}}e^{\frac{2\pi i}{3}ab}\mathcal{R}^{\sigma\sigma}_{\bf a}\mathcal{R}^{\sigma\sigma}_{\bf b}=-\frac{1}{3}\sum_{{\bf c}}e^{\frac{2\pi i}{3}(a+b)c}\mathcal{R}^{\sigma{\bf c}}_\sigma,\label{su(3)1hexeq3}\end{align} where the three intermediate channels ${\bf c}$ run through $1,\psi_1,\psi_2$ for $c=0,-1,1$ respectively.

We can see that Eq.~\eqref{su(3)1hexeq1} has three solutions \begin{align}\mathcal{R}^{\sigma{\bf a}}_\sigma=e^{\frac{2\pi i}{3}a(a+r)},\quad\mbox{for $r=0,\pm1$}.\end{align} Now, for each $r$, \eqref{su(3)1hexeq2} and \eqref{su(3)1hexeq3} have two solutions \begin{align}\mathcal{R}^{\sigma\sigma}_{\bf a}=\pm e^{-\pi i/4}e^{\frac{2\pi i}{3}\left(r^2+ar-a^2\right)}.\end{align} These solutions correspond to six anyons in the Drinfeld center, and they all represent the possible $\mathbb{Z}_2$ fluxes: \begin{align}\sigma_r&=\left(\sigma;\mathcal{R}^{\sigma\sigma}_{\bf a}=-e^{-\pi i/4}e^{\frac{2\pi i}{3}\left(r^2+ar-a^2\right)}\right)\\z\sigma_r&=\left(\sigma;\mathcal{R}^{\sigma\sigma}_{\bf a}=e^{-\pi i/4}e^{\frac{2\pi i}{3}\left(r^2+ar-a^2\right)}\right).\end{align} The topological spins of the fluxes are the average of the exchange phases over the admissible fusion channels \begin{gather}\theta_\sigma=\frac{1}{d_\sigma}\left(\mathcal{R}^{\sigma\sigma}_1+\mathcal{R}^{\sigma\sigma}_{\psi_1}+\mathcal{R}^{\sigma\sigma}_{\psi_2}\right)=\varkappa_\sigma(\mathcal{R}^{\sigma\sigma}_1)^\ast\\\theta_{\sigma_r}=e^{\pi i/4}e^{-\frac{2\pi i}{3}r^2},\quad\theta_{z\theta_{\sigma_r}}=-\theta_{\sigma_r},\end{gather} where $\varkappa_\sigma=-1$ is the non-trivial Frobenius-Schur indicator in \eqref{su(3)FS}. Comparing with the conformal dimensions in \eqref{su(2)4spins}, $\sigma_0$ (where the subscript means $r=0$) is identified with $j={\bf 1/2}$ in $SU(2)_4$. The flux-charge composite $z\sigma_0=z\times\sigma_0$ is identified with $j={\bf 3/2}$. The other $\mathbb{Z}_2$ fluxes are the composition of the bare flux with quasiparticles in the time reversed $\overline{SU(3)_1}$ sector in \eqref{Z(SU(3)1)} (to be discussed below), \begin{align}\sigma_r=\sigma_0\times\overline{\psi_r}\end{align} for $r=0,\pm1$ mod 3 and $\overline{\psi_0}=1$.

The \hexeq's due to  exchanging a $\psi_i$ with a $\sigma$ and a $\psi_{i/j}$ are
\begin{align}\mathcal{R}^{\psi_i\sigma}_\sigma\mathcal{R}^{\psi_i\psi_i}_{\psi_j}&=e^{-2\pi i/3}\mathcal{R}^{\psi_i\sigma}_\sigma\label{su(3)Rpsisigma1}\\\mathcal{R}^{\psi_i\sigma}_\sigma\mathcal{R}^{\psi_i\psi_j}_{1}&=e^{2\pi i/3}\mathcal{R}^{\psi_i\sigma}_\sigma.\label{su(3)Rpsisigma2}\end{align} These require $\mathcal{R}^{\psi_i\psi_i}_{\psi_j}=e^{-2\pi i/3}$ and $\mathcal{R}^{\psi_i\psi_j}_1=e^{2\pi i/3}$. The quadratic \hexeq's from exchanging $\psi_i$ with a pair of $\sigma$'s \begin{align}\mathcal{R}^{\psi_i\sigma}_\sigma\mathcal{R}^{\psi_i\sigma}_\sigma&=\mathcal{R}^{\psi_i\psi_j}_1\end{align} require $\mathcal{R}^{\psi_i\sigma}_\sigma=\pm e^{\pi i/3}$. There are two solutions for $\mathcal{R}^{\psi_i\ast}_\ast$: \begin{align}\overline{\psi_i}&=\left(\psi_i;{\mathcal{R}^{\psi_i\psi_i}_{\psi_j}}^\ast=\mathcal{R}^{\psi_i\psi_j}_1=e^{\frac{2\pi i}{3}},\mathcal{R}^{\psi_i\sigma}_\sigma=-e^{\frac{\pi i}{3}}\right)\label{Rsu(3)psisol1}\\z\overline{\psi_i}&=\left(\psi_i;{\mathcal{R}^{\psi_i\psi_i}_{\psi_j}}^\ast=\mathcal{R}^{\psi_i\psi_j}_1=e^{\frac{2\pi i}{3}},\mathcal{R}^{\psi_i\sigma}_\sigma=e^{\frac{\pi i}{3}}\right)\label{Rsu(3)psisol2}.\end{align} Since $\overline{\psi_i}$ generates the time reversal $\overline{SU(3)_1}$ sector in \eqref{Z(SU(3)1)}, these anyons are in the time-reversed sector.

Because of the non-trivial action of the $\mathbb{Z}_2$ bilayer symmetry on the chiral sector,  we find explicitly that \eqref{su(3)Rpsisigma1} and \eqref{su(3)Rpsisigma2} forbid the original $\psi_i$ from the parent $SU(3)_1$-state from appearing in the twist liquid. Similar to the gauging in the toric code example, they form a super-sector $\psi_1+\psi_2$ that corresponds to irreducible matrix solutions to the \hexeq's. 
The \hexeq's involving exchanging the super-sector with a pair of $\psi$'s resulting in \begin{align}\mathcal{R}^{(\psi_1+\psi_2)\psi_i}_1\mathcal{R}^{(\psi_1+\psi_2)\psi_i}_1&=\mathcal{R}^{(\psi_1+\psi_2)\psi_j}_{\psi_i}\\\mathcal{R}^{(\psi_1+\psi_2)\psi_i}_{\psi_j}\mathcal{R}^{(\psi_1+\psi_2)\psi_i}_{\psi_j}&=\mathcal{R}^{(\psi_1+\psi_2)\psi_j}_1\\\mathcal{R}^{(\psi_1+\psi_2)\psi_i}_1\mathcal{R}^{(\psi_1+\psi_2)\psi_j}_{\psi_i}&=\mathcal{R}^{(\psi_1+\psi_2)1}_{\psi_j}=1,\end{align} which require $\mathcal{R}^{(\psi_1+\psi_2)\psi_i}_1={\mathcal{R}^{(\psi_1+\psi_2)\psi_j}_{\psi_i}}^\ast$ be cubes of unity, for $i\neq j$. $\mathcal{R}^{(\psi_1+\psi_2)\sigma}_\sigma$ is a $2\times2$ matrix acting on the vertex degeneracy for $(\psi_1+\psi_2)\times\sigma=2\sigma$. The \hexeq's \begin{align}&\mathcal{R}^{(\psi_1+\psi_2)\sigma}_\sigma e^{-\frac{2\pi i}{3}\mu\sigma_z}\mathcal{R}^{(\psi_1+\psi_2)\sigma}_\sigma\nonumber\\&=\left(\begin{array}{*{20}c}\mathcal{R}^{(\psi_1+\psi_2)\psi_{\mu-1}}_{\psi_\mu}&0\\0&\mathcal{R}^{(\psi_1+\psi_2)\psi_{\mu+1}}_{\psi_\mu}\end{array}\right)\label{Rsu(3)supersigma}\end{align} for $\mu=0,1,2$ mod 3 and $\psi_0=1$,
have solutions only when \begin{align}&\mathcal{R}^{(\psi_1+\psi_2)\psi_1}_1=\mathcal{R}^{(\psi_1+\psi_2)\psi_2}_1\label{Rsu(3)superpsi}\\&=\left\{\begin{array}{*{20}c}e^{+2\pi i/3}&\to&\mbox{decomposable $\mathcal{R}^{(\psi_1+\psi_2)\sigma}_\sigma$}\\e^{-2\pi i/3}&\to&\mbox{irreducible $\mathcal{R}^{(\psi_1+\psi_2)\sigma}_\sigma$}\end{array}\right. .\nonumber\end{align} As labeled, in the former case, Eq.~\eqref{Rsu(3)superpsi} leads to decomposable (diagonal) solutions for $\mathcal{R}^{(\psi_1+\psi_2)\sigma}_\sigma=\pm e^{\pi i/3},\pm e^{\pi i\sigma_z/3}$. These are a trivial decomposition of $\overline{\psi_1}+\overline{\psi_2}$ into Abelian components in \eqref{Rsu(3)psisol1} and \eqref{Rsu(3)psisol2}. When Eq.~\eqref{Rsu(3)superpsi} is $e^{-2\pi i/3}$, then the \hexeq's \eqref{Rsu(3)supersigma} have off diagonal solutions \begin{align}\mathcal{R}^{(\psi_1+\psi_2)\sigma}_\sigma=e^{2\pi i/3}\left(\begin{array}{*{20}c}0&e^{i\phi}\\e^{-i\phi}&0\end{array}\right)\label{Rsu(2)supersigmasol}\end{align} where $\phi$ is a vertex gauge degree of freedom and can be set to 0. The resulting anyon from the Drinfeld construction is \begin{align}\mathcal{E}=\left(\psi_1+\psi_2;\begin{array}{*{20}c}\mathcal{R}^{(\psi_1+\psi_2)\psi_i}_1={\mathcal{R}^{(\psi_1+\psi_2)\psi_i}_{\psi_j}}^\ast=e^{-2\pi i/3}\\\mathcal{R}^{(\psi_1+\psi_2)\sigma}_\sigma=e^{2\pi i/3}\sigma_x\end{array}\right),\label{Rsu(2)4Esol}\end{align} and corresponds to the quasiparticle $j={\bf 1}$ in $SU(2)_4.$ This quasiparticle also has the correct spin \begin{align}\theta_{\mathcal{E}}=\left(R^{\mathcal{E}\mathcal{E}}_1\right)^\ast=\left(\mathcal{R}^{\mathcal{E}\psi_i}_1\right)^\ast=e^{2\pi i/3}.\end{align} 

The fusion $z\times\mathcal{E}$ changes the $\mathcal{R}$-symbol \eqref{Rsu(2)supersigmasol} by a minus sign, $\mathcal{R}^{z\sigma}_\sigma=-1$. This however can be absorbed into the gauge degree of freedom $\phi,$  and corresponds to the same solution to the \hexeq's. Thus we have the fusion rule \begin{align}z\times\mathcal{E}=\mathcal{E}.\label{su(2)4zEfusion}\end{align} The original degenerate fusion data $\sigma\times\mathcal{E}=2\sigma$ must split into \begin{align}\sigma_0\times\mathcal{E}=\sigma_0+z\sigma_0\end{align} since the flux-charge composite $z\sigma_0$ must also be admissible due to \eqref{su(2)4zEfusion}  and fusion associativity: $\sigma_0\times\mathcal{E}=\sigma_0\times(\mathcal{E}\times z)=(\sigma_0\times\mathcal{E})\times z$. 
Finally, the fusion channels of a pair of super-sectors $\mathcal{E}$ must respect the original data \eqref{su(2)4fusionsuper} and \eqref{su(2)4zEfusion}, so that both the vacuum 1 and $\mathbb{Z}_2$ charge $z$ are admissible. The distinct channels of \begin{align}\mathcal{E}\times\mathcal{E}=1+z+\mathcal{E}\end{align} can be distinguished by their braiding phases with the $\mathbb{Z}_2$ flux $\sigma$.

We notice that, because the bilayer symmetry does not act on the time-reversed sector, the time reversed quasiparticles $\overline{\psi}_i$ are relatively local with respect to the $\mathbb{Z}_2$ fluxes $\sigma_0,z\sigma_0,$ as well as the super-sector $\mathcal{E}$. From the $\mathcal{R}$-symbols in \eqref{Rsu(3)psisol1} and \eqref{Rsu(2)4Esol} we can see explicitly that \begin{align}\vcenter{\hbox{\includegraphics[height=0.6in]{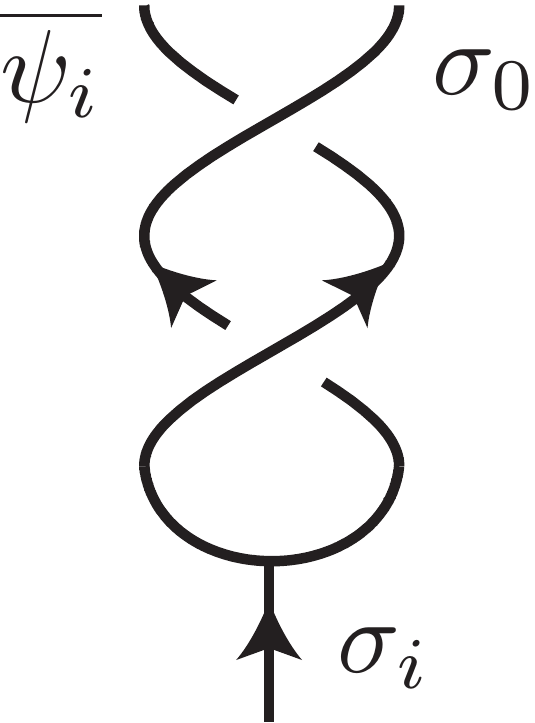}}}&=\mathcal{R}^{\sigma\psi_i}_\sigma\mathcal{R}^{\overline{\psi_i}\sigma}_\sigma=1\\\vcenter{\hbox{\includegraphics[height=0.6in]{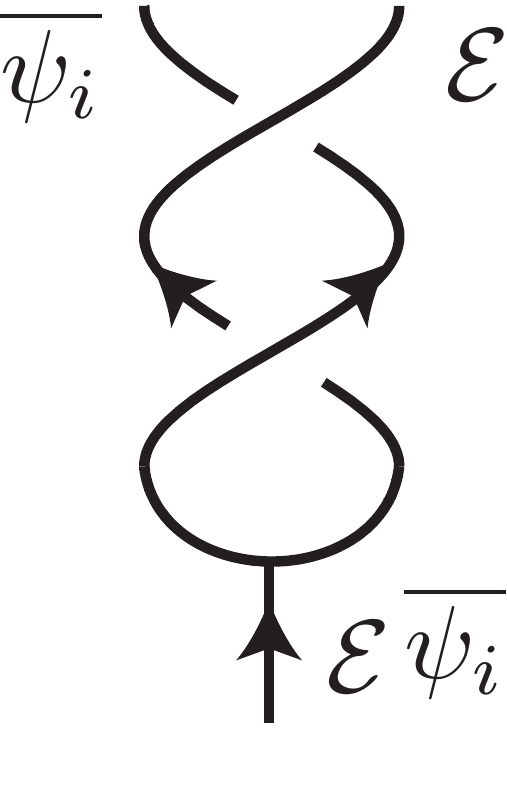}}}&=\mathcal{R}^{\mathcal{E}\psi_i}_{\psi_j}\mathcal{R}^{\overline{\psi_i}\psi_i}_{\psi_j}=\mathcal{R}^{\mathcal{E}\psi_i}_{1}\mathcal{R}^{\overline{\psi_i}\psi_j}_{1}=1.\end{align} The time reversed sector $\overline{SU(3)_1}=\{1,\overline{\psi_1},\overline{\psi_2}\},$ thus completely decouples from $SU(2)_4=\{1,\sigma_0,\mathcal{E},z\sigma_0,z\}$. The super-sector $\mathcal{E}$ can fuse with quasiparticles $\overline{\psi_i}$ in the anti-chiral sector to give two additional super-sectors $\mathcal{E}\overline{\psi_i}$. These last anyons complete the set of 15 anyons in $SU(2)_4\otimes\overline{SU(3)_1},$ and completes the proof of the reverse directions of \eqref{su(2)4su(3)1commdiagram}.

At this point we have completed the conventional Drinfeld construction and we are poised to excise the unwanted anti-chiral sector. The time reversed $\overline{SU(3)_1}$ sector can be removed by refining the Drinfeld construction relative to the parent $SU(3)_1$-state. This refinement simply means that we take into account the initial specified braiding information of the parent state in the defect fusion category, which requires, for instance, $R^{\psi_i\psi_i}_1=\theta_{\psi_i}^\ast=e^{-2\pi i/3}$. This forbids the  solutions to the \hexeq's \eqref{su(3)Rpsisigma1} and \eqref{su(3)Rpsisigma2}, which would lead to the time-reversed partners $\overline{\psi_i}$. This relative Drinfeld construction also does not allow the fluxes $\sigma_{\pm1}$ and $z\sigma_{\pm1}$. This is because their fusion rules, for instance $\sigma_1\times\sigma_1=\overline{\psi_2}\times(1+\psi_1+\psi_2)$, necessarily produce time-reversed quasiparticles. Hence, the end result is the removal of $\overline{SU(3)_1}$, leaving behind the twist liquid $SU(2)_4.$

 
Finally, we notice that the non-trivial Frobenius-Schur indicator $\varkappa_\sigma=-1$ in \eqref{su(3)FS} stems from the negative sign of the $F$-symbol $F^{\sigma\sigma\sigma}_\sigma$ in \eqref{su(3)defectFsymbols}. A switch of the sign will lead to a slightly different theory with identical fusion rules but conjugate spins for the $\mathbb{Z}_2$ fluxes. This however does not affect the overall chiral central charge (mod 8) of the twist liquid as the spins of the fluxes $\theta_\sigma$ and $\theta_{z\sigma}$ cancel in the Gauss-Milgram formula \eqref{GaussMilgram}. We will denote these two theories by $SU(2)_4^\pm$, i.e.~these are the  twist liquids of $SU(3)_1$ with $\varkappa_\sigma=\mp1,$ and $h_\sigma=\pm1/8,$ respectively. Hence, $SU(2)_4^+$ refers the conventional $SU(2)$ Kac-Moody theory at level 4.

\subsection{Connections between the \texorpdfstring{$\mathbb{Z}_3$}{Z3}-parafermions, 3-state Potts model, and the \texorpdfstring{$\mathcal{M}(6,5)$}{M(6,5)} model}\label{sec:connections}
We have just proved that gauging the bilayer symmetry of of the FQH state $SU(3)_1$ provides a possible route to the desirable $SU(2)_4$ state, and since experimental proposals for realizing twist defects in bilayer FQH states are already available\cite{BarkeshliQi13, BarkeshliOregQi14}, it is plausible that this state could be realized in experiments in the future. 
This particular twist liquid may be of practical interest in the context of quantum computing. Although the $SU(2)_4$ state does not contain a Fibonacci anyon\cite{SlingerlandBais01, ChetanSimonSternFreedmanDasSarma}, it may still support universal quantum computing\cite{CuiWang14,LevaillantBondersonBauerFreedmanunpublished} if interferometric measurements\cite{Bonderson13} are incorporated with braiding operations. 

As we will now argue, it turns out that there is also a close connection between $SU(3)_1$ and theories containing Fibonacci anyons.  The  $SU(3)_1$ theory appears as the Abelian sector of the $\mathbb{Z}_3$ parafermion theory\cite{ZamolodchikovFateev85, bigyellowbook} given by the coset \begin{align}\frac{SU(2)_3}{U(1)_3}=\left(G_2\right)_1\times\overline{SU(3)_1}=\langle1,\tau\rangle\times\langle1,\overline{\psi_1},\overline{\psi_2}\rangle,\label{Z3parafermioncoset}\end{align} which supports a Fibonacci anyon $\tau$ in $(G_2)_1=\langle1,\tau\rangle$ with spin $h_\tau=2/5$ and quantum dimension $d_\tau=(1+\sqrt{5})/2$. The notation $(G_2)_1$ refers to the exceptional Lie group $G_2$ at level 1.\cite{bigyellowbook} Since $G_2$ has a non-simply-laced Lie algebra, it represents a non-Abelian theory even at level 1\cite{mongg2, KhanTeoHughesoappearsoon}. As a CFT, the parafermion theory was shown to have 6 primary fields with conformal dimensions summarized in Table~\ref{tab:Z3parafermion}.\cite{ZamolodchikovFateev85, bigyellowbook} 
\begin{table}[htbp]
\centering
\begin{tabular}{|c|c|c|}
\multicolumn{3}{c}{$\mathbb{Z}_3$-parafermion}\\\hline
0&$2/3$&$2/3$\\\hline
$2/5$&$1/15$&$1/15$\\\hline
\end{tabular}
\begin{tabular}{|c|c|c|}
\multicolumn{3}{c}{3-state Potts}\\\hline
0&$2/3$&$3$\\\hline
$2/5$&$1/15$&$7/5$\\\hline
\end{tabular}
\begin{tabular}{|c|c|c|c|c|}
\multicolumn{5}{c}{$\mathcal{M}(6,5)$ minimal model}\\\hline
0&$1/8$&$2/3$&$13/8$&$3$\\\hline
$2/5$&$21/40$&$1/15$&$1/40$&$7/5$\\\hline
\end{tabular}
\caption{Conformal dimensions of the primary fields of three $c=4/5$ CFT's. The two rows are associated to the Fibonacci sectors $1,\tau$ coming from  $(G_2)_1.$ The columns in each separate sub-table are associated to the anyon content of the $\overline{SU(3)_1}$, $\overline{SO(3)_2}$ and $\overline{SU(2)_4^-}$ sectors respectively  (see \eqref{Z3parafermioncoset} and \eqref{3statePotts}).}\label{tab:Z3parafermion}
\end{table}

Interestingly, there are two other theories that are closely related to the coset theory \eqref{Z3parafermioncoset}, and all share central charge $c=4/5.$ These are the 3-state Potts model\cite{Potts52,Dotsenko84,Baxterbook,bigyellowbook} and the $\mathcal{M}(6,5)$ minimal model\cite{bigyellowbook} (also known as the tetracritical Ising model\cite{FisherNelson74, chaikin-1995}). The field contents of these theories are related to the parafermion theory by the gauging of an anyonic symmetry. This is easily seen by writing the latter two theories as \begin{align}\mathcal{M}(6,5)&=\left(G_2\right)_1\times\overline{SU(2)_4^-}=\langle1,\tau\rangle\times\langle1,\sigma,\overline{\mathcal{E}},z\sigma,z\rangle\nonumber\\\mbox{3-Potts}&=\left(G_2\right)_1\times\overline{SO(3)_4}=\langle1,\tau\rangle\times\langle1,\overline{\mathcal{E}},z\rangle.\label{3statePotts}\end{align} We can easily see the connection between the $\mathbb{Z}_3$-parafermion and the $\mathcal{M}(6,5)$ minimal model, i.e. the $(G_2)_1$ part remains fixed, but the bilayer anyonic symmetry of the $SU(3)_1$ sector of the parafermion theory is gauged to generate the $SU(2)_4^-$ factor appearing in the $\mathcal{M}(6,5)$ model. Additionally, since it  is $SU(2)_4^-$ appearing, this represents the twist liquid of the $SU(3)_1$ factor with an additional ${\mathbb{Z}}_2$ SPT added in such that $\varkappa_\sigma=+1$ and $h_\sigma=-1/8.$ From this we can see that the $SO(3)_4$ factor in the $3$-state Potts model is just a closed sub-theory of $SU(2)_4$ generated only by ${\bf 0}$, $\mathcal{E}={\bf 1}$ and $z={\bf 2}$.  However, $SO(3)_4$ loses modularity as the non-trivial boson $z$ leads to braiding degeneracy. 
As CFT's, the conformal dimensions of the primary fields of these theories are summarized in Table~\ref{tab:Z3parafermion} as well. The quantum dimensions match up with the spins of the ten anyons in $\langle1,\tau\rangle\times\langle1,\sigma,\overline{\mathcal{E}},z\sigma,z\rangle$ (modulo $\mathbb{Z}$). For example the $\mathbb{Z}_2$ charge $z$ corresponds to the bosonic field with conformal dimension $3$, and indicates a $W_3$-algebra structure.\cite{ZamolodchikovWalgebra, Blumenhagenbook}

\section{The sixteenfold \texorpdfstring{$SO(N)_1$}{SO(N)} series}\label{sec:so(N)}
For our second example we will present a noteworthy family of interconnected topological phases.
We begin with the real orthogonal Lie groups $SO(N)$ which have simple Lie algebras that fall into two series that are separated into even and odd dimensional series: \begin{align}B_r=so(2r+1),\quad D_r=so(2r). \label{eq:banddseries}\end{align} At the boundaries of the topological quantum field theories associated to the $SO(N)$ groups there will be a sequence of $(1+1)D$ affine Kac-Moody edge CFT's at level 1,\cite{bigyellowbook} and the interlaced combination of the CFTs of both series has a sixteen-fold classification. Consequently, through the bulk-boundary correspondence, these define a sixteenfold classification of bulk topological phases in $(2+1)D$.\cite{Kitaev06} 

The (2+1)D topological phases arising from the $B_r$ series are Ising non-Abelian states. On the other hand, the $D_r$-states are Abelian and have  $K$-matrix Chern-Simons descriptions (see Eq.~\eqref{toriccodeCSaction}), where the $K$-matrix is identified with the symmetric Cartan matrix of the simply-laced Lie algebra $so(2r)$.\cite{bigyellowbook,khan2014} The $\mathbb{Z}_{16}$ classification is periodic because the $B_r$-state (or $D_r$-state) is {\em stably equivalent}\cite{cano2013bulk} to the $B_{r+8}$-state (resp.~$D_{r+8}$-state) up to the addition of a ``trivial" $E_8$-state. The $E_8$ state is a  bosonic Abelian integer quantum Hall state with a $K$-matrix consisting of the unimodular Cartan matrix of $E_8$. It has trivial topological order, but carries a chiral edge central charge of $c_-=8$\cite{Kitaev06, LuVishwanathE8, khan2014}. The resulting periodicity can be written as\footnote{This equality is a {\em stable equivalence}\cite{cano2013bulk} that indicates the two $(2+1)$D bulk theories carry identical fusion, spin and braiding information as well as chiral central charge along their edges. It is not an equality in the affine Lie algebra sense.}  \begin{align}SO(N+16)_1\approx SO(N)_1\otimes(E_8)_1\end{align} where the subscripts indicate level 1. Thus, the $SO(N+8)_1$ topological phase is said to be equivalent to the $SO(N)_1$ topological phase up to a ``trivial" $(E_8)_1$ integer quantum Hall state which has no topological order. In the following, the number $N$ is to be understood as being defined modulo 16, and the rank $r$ (c.f. \ref{eq:banddseries}) is defined modulo 8. We will also use the fact that, generally, an $SO(N)_1$ state has chiral central charge $c_-=N/2$. 

Let us now consider the anyon content of these theories. The non-Abelian $B_r$ series has an Ising fusion structure with quasiparticles $1,\psi,\sigma$, where \begin{align}\psi\times\psi=1,\quad\psi\times\sigma=\sigma,\quad\sigma\times\sigma=1+\psi.\end{align} The anyon $\psi$ is always a fermion for all $r$, i.e. $\theta_\psi=-1$, while the spin for the Ising anyon $\sigma$ depends on the rank $r$ of $B_r=so(2r+1)$ as \begin{align}\theta_\sigma=e^{2\pi ih_\sigma},\quad h_\sigma=\frac{2r+1}{16}. \end{align} or small values of $r$ we take $SO(1)_1$ to be the conventional Ising theory with $c_-=1/2$ and $h_\sigma=1/16$; and $SO(3)_1$ to be $SU(2)_2=\langle1,\psi,\sigma\rangle$ with $c_-=3/2$ and $h_\sigma=3/16$. These are chosen to match the structure found in $SO(17)_1$ and $SO(19)_1$ respectively, up to a trivial $E_8$ integer quantum Hall state.

The Abelian $D_r$ series has an Abelian fusion structure with four quasiparticles.  When $r$ is even, $D_r$ has an identical fusion content to that of the toric code with quasiparticles $1,e,m,\psi$ where  $e^2=m^2=\psi^2=em\psi=1$. Again, $\psi$ is always a fermion, but $e,m$ are not always bosons, and generally they have the $r$-dependent spins \begin{align}\theta_{e/m}=e^{2\pi ih_{e/m}},\quad h_e=h_m=\frac{r}{8}.\end{align} When $r$ is odd, $D_r$ has similar fusion content to the bosonic Laughlin $\nu=1/4$ state with quasiparticles $1,m,m^2=\psi,m^3,$ such that $m^4=1$. The anyon $\psi$ is again aways a fermion while the spins of the remaining quasiparticles  are \begin{align}\theta_{m/m^3}=e^{2\pi ih_{m/m^3}},\quad h_m=h_{m^3}=\frac{r}{8}.\end{align}
 For small values of $r$ we take $SO(0)_1$ to be the toric code with $c_-=0$; and $SO(2)_1$ to be the Abelian $U(1)_2$-state (or the Laughlin $\nu=1/4$ FQH state) with the $K$-matrix $K=4$. These are chosen to match the structure found in $SO(16)_1$ and $SO(18)_1$ respectively, up to a trivial $E_8$ integer quantum Hall state.

Let us consider some additional properties of each series, beginning with the Abelian case. A $D_r$-state can be described by an $r$-component Abelian Chern-Simons theory \eqref{toriccodeCSaction} with a $K$-matrix identical to the Cartan matrix of $D_r=so(2r)$: \begin{align}K=\left(\begin{array}{*{20}c}2&-1&&&\\-1&\ddots&\ddots&&\\&\ddots&2&-1&-1\\&&-1&2&0\\&&-1&0&2\end{array}\right)_{r\times r},\end{align} where the ellipsis indicates a continuation of the pattern, and all other non-specified entries are 0. 

Physically, these states can, in principle, be realized by strongly proximity-coupling topological Chern insulators\cite{Haldane1988} to an $s$-wave superconductor as shown in Ref.~[\onlinecite{qhz2010}]. A Chern insulator is an electronic insulator that realizes the quantum anomalous Hall effect, i.e. the integer quantum Hall effect in the absence of an external magnetic field. A Chern insulator is characterized by an integer Chern invariant that both counts the number of gapless chiral edge modes,\cite{TKNN}  and specifies the Hall conductance, e.g. it determines the amount of electron-charge attached to each unit of quantum flux $hc/q_e$ (where $q_e$ is the electron charge) that is inserted\cite{Haldane1988}. Starting with a Chern insulator with Chern number $r$ (note this $r$ will be connected to the rank $r$ above), the proximity-coupling to the  superconductor breaks charge conservation and condenses Cooper pairs $\sim\psi\psi$. If the fermion parity symmetry is gauged then this establishes the set of anyon excitations. First, the fermionic electron quasiparticle $\psi$ is now non-local with respect to the superconducting flux/vortex $m=hc/2q_e$, as it acquires a minus sign when encircling the flux. There are also Caroli-de Gennes-Matricon bound states\cite{CarolideGennesMatricon} in the core of the superconducting flux vortex $m$. Finally, the $m\times\psi=e$ or $m^3$ quasiparticle -- for even or odd $r$ respectively -- corresponds to an excited vortex with an opposite fermion parity compared to the un-excited state.\cite{KhanTeoVishveshwaraappearsoon} It is only after gauging the fermion parity that the $hc/2q_e$ vortices become deconfined quantum excitations, and the resulting state is the bosonic topological state $D_r$. This also can be corroborated by observing that the chiral edge CFT  of the Chern insulator has a Kac-Moody structure  $(u(1)_{1/2})^r,$ (i.e. $r$ copies of chiral fermions) which is effectively extended to $so(2r)_1$ in the presence of the superconductivity after decomposing the $r$ complex chiral fermions into real and imaginary chiral Majorana modes.

The topological states in the $D_r$-series are equipped with a global $\mathbb{Z}_2$ anyonic symmetry, which acts to switch the parity of the $hc/2q_e$ vortex. For even $r$, this is the same as an electric-magnetic symmetry that interchanges $e\leftrightarrow m$, while for odd $r$, there is effectively a conjugation symmetry that switches $m\leftrightarrow m^3$. For example, the $D_8=SO(16)_1$ state has an anyon content identical to that of the conventional toric code, and thus exhibits and electric-magnetic anyonic symmetry. In fact, the two theories, i.e. $SO(16)_1$ and our convention for $SO(0)_1,$ are even {\em stably equivalent} up to an $E_8$-state, which has a trivial anyon content (and simply accounts for the difference in chirality). Thus, as we have already seen the twist liquid arising from the toric code with a gauged e-m symmetry, we expect that  gauging the $\mathbb{Z}_2$ symmetry of the $D_r$-series will therefore give rise to a series of twist liquids that generalizes the $\mbox{Ising}\times\overline{\mbox{Ising}}$ state presented in Section~\ref{sec:Zkgaugetheory}.

Let us now work out the details of this gauging procedure more precisely. We will show that gauging the $\mathbb{Z}_2$ symmetry of $SO(2r)_1$ leads to two possible twist liquids: \begin{align}\begin{diagram}SO(2r)_1&\pile{\rTo^{\mbox{\small Gauging}}\\\lTo_{\mbox{\small Condensation}}}&SO(N_0)_1\otimes SO(N_1)_1\end{diagram}\label{so(N1)xso(N2)}\end{align} where $N_0$ and $N_1$ are odd integers satisfying \begin{align}N_0=2r-N_1=\left\{\begin{array}{*{20}c}r+1&\mbox{or}&r+3,&\mbox{for even $r$}\\r&\mbox{or}&r+4,&\mbox{for odd $r$}.\end{array}\right.\label{N0N1values}\end{align} We immediately see that the fusion content of the resulting twist liquid always has an $(\mbox{Ising})^2$ structure since the gauging takes us from the $D$-series to two states in the $B$-series. We denote the anyons in each sector by \begin{align}\{1,\psi,\sigma_0\}\times\{1,\psi',\sigma_1\}\label{IsingxIsing}\end{align} where the Ising anyons have spins $h_{\sigma_0}=N_0/16$ and $h_{\sigma_1}=N_1/16$ respectively. This structure can be derived by the relative Drinfeld construction described in Section~\ref{sec:stringnet}. The detailed calculations are similar to that of the toric code example considered earlier, and we will only highlight the main features. 

We begin proving \eqref{so(N1)xso(N2)} starting with the defect fusion category of the parent $SO(2r)_1$ state given by $\langle1,e,m,\psi\rangle\oplus\langle\sigma_0,\sigma_1\rangle$ for $r$ even, or $\langle1,e,\psi,e^3\rangle\oplus\langle\sigma_0,\sigma_1\rangle$ for $r$ odd, with the fusion rules $\sigma_0\times\sigma_0=1+\psi$ and $\sigma_1=e\times\sigma_0$. We notice for the odd case, although there are two inequivalent quantum anyonic symmetries, i.e.~$H^2(\mathbb{Z}_2,\mathbb{Z}_4)=\mathbb{Z}_2$ (see \eqref{H2GB}), they correspond to identical defect fusion rules. Just like the toric code, the flux-super-sector-charge characterization defined in Section~\ref{sec:QPstructuretwistliquid} leads to the $(\mbox{Ising})^2$ fusion structure in the twist liquid given in \eqref{IsingxIsing}. For instance, \begin{align}\sigma_0\sigma_1=\sigma_0\times\sigma_1=\left\{\begin{array}{*{20}c}e+m,&\mbox{for even $r$}\\m+m^3,&\mbox{for odd $r$}\end{array}\right.,\quad z=\psi\times\psi'\end{align} are the quasiparticle super-sector and $\mathbb{Z}_2$ charge respectively.

Now we can derive the spins of the fluxes by solving the \hexeq's involving $\sigma$ (see Eq.~\eqref{hexagoneq} or Fig.~\ref{fig:hexagon1}). They are identical to those in the toric code \begin{align}\left(\theta_\sigma^\ast\right)^2=\left(\mathcal{R}^{\sigma\sigma}_1\right)^2=\varkappa_\sigma\left(\frac{1\pm i}{\sqrt{2}}\right)=\varkappa_\sigma e^{\pm\pi i/4}\end{align} (see Eq.~\eqref{TCRsigmasigma}) except for the possible extra sign in the Frobenius-Schur (FS) indicator\cite{FredenhagenRehrenSchroer92,Kitaev06} \begin{align}\varkappa_\sigma=\frac{[F^{\sigma\sigma\sigma}_\sigma]_1^1}{\left|[F^{\sigma\sigma\sigma}_\sigma]_1^1\right|}=\pm1.\label{IsingFSindicator}\end{align}  
This possible choice of sign \eqref{IsingFSindicator} appears due to the non-trivial defect classification $H^3(\mathbb{Z}_2,U(1))=\mathbb{Z}_2$ (see Eq.\eqref{H3GU1}). This leads to the eight possible solutions $\theta_\sigma=e^{2\pi i N/16}$, for $N$ odd, and proves that the anyon decomposition \eqref{IsingxIsing} must match the form $SO(N_0)_1\otimes SO(N_1)_1$ for \emph{odd-integer} $N_0,N_1$. 

Next we notice that the edge of a twist liquid must have the same chiral central charge $c_-=c_R-c_L$ as its globally symmetric parent state. This was shown in Eq.~\eqref{chiralcinvariance}. It can also be shown by the string-net construction (see Eq.~\eqref{Drinfeld=TL}) because the non-chiral string-net model contains the twist liquid as well as the time reversed copy of the parent state. Alternatively, we believe the edge theory of the twist liquid can be derived from that of its parent state by taking a $G$-orbifold,\cite{Ginsparg88,bigyellowbook} which does not affect the chiral central charge. For example, the $D_1$ state has the same anyon content as $U(1)_2$ with the $K$-matrix $K=4,$ and it is well known that its $\mathbb{Z}_2$ orbifold theory has a double Ising structure: \begin{align}U(1)_2/\mathbb{Z}_2=\mbox{Ising}\otimes\mbox{Ising}=B_1\otimes B_1.\end{align} 

In general one can compute the central charge modulo 8 by using the Gauss-Milgram formula \eqref{GaussMilgram}, which applies to any modular symmetric topological phase. In the case of \eqref{IsingxIsing}, the contributions to this formula from the Abelian sector $\{1,\psi,\psi',z=\psi\psi'\}$ cancel. This is because $\psi$ is the fermion from the parent state, and $z$ is the pure $\mathbb{Z}_2$ charge which must be bosonic. The contributions from the twist-defect sector $\{\sigma_0,\sigma_1,z\sigma_0,z\sigma_1\}$ must also cancel due to the opposite spins \begin{align}\theta_\sigma=-\theta_{z\sigma},\end{align} which can be proven by the ribbon identity \eqref{ribbon}, and the fact that the braiding phase between a $\mathbb{Z}_2$ charge $z$ and a flux $\sigma$ is minus one. Hence the central charge only depends on the super-sector $\sigma_0\sigma_1=e+m$ or $m+m^3$, which gives \begin{align}\frac{c_-}{8}=h_{\sigma_0\sigma_1}=h_{\sigma_0}+h_{\sigma_1}=h_m=\frac{r}{8}\label{chiralitySO(2r)}\end{align} modulo 1. This enforces the constraint $N_0+N_1=2r$ if the twist liquid is of the form of $SO(N_0)_1\otimes SO(N_1)_1$.

Finally, we notice that the Frobenius-Schur indicator \eqref{IsingFSindicator} for the fluxes $\sigma_0$ and $\sigma_1$ must be identical since they originate from the same defect fusion category. This rules out cases such as $N_0=2r-N_1=r\pm2$ for odd $r$ since $\varkappa_\sigma=(-1)^{(N_0^2-1)/8}=(-1)^{(N_1^2-1)/8}$.\cite{Kitaev06} Hence, the possible decompositions $SO(N_0)_1\otimes SO(N_1)_1$ of the $\mathbb{Z}_2$ twist liquid are therefore only those given in \eqref{N0N1values}. Now let us try to give a more physical understanding of this result. 

\subsection{Topological Transitions and Gauging Fermion Parity in Chern Insulators}
\label{sec:gauging-TRSB}

\begin{figure}[htbp]
\centering\includegraphics[width=0.48\textwidth]{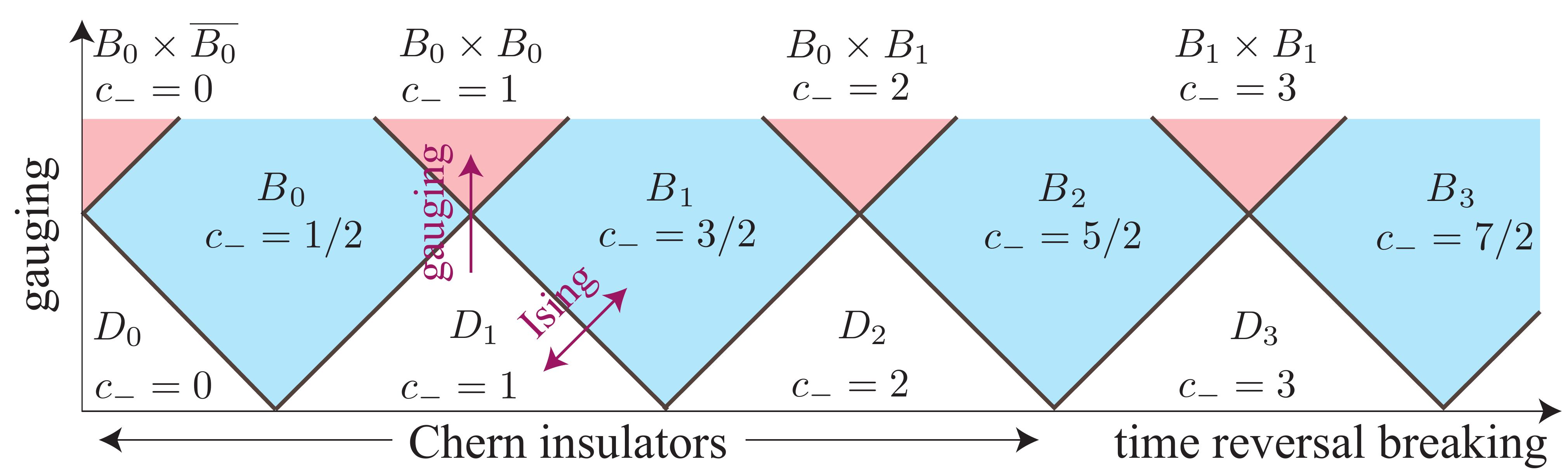}
\caption{Phase diagram of Chern insulators with gauged fermion parity. Vertical axis describes gauging transition \eqref{so(N1)xso(N2)} and chirality/Chern number increases along horizontal axis. Phase boundaries are $(2+1)$D quantum Ising transitions.
}\label{fig:so(N)phasediagram}
\end{figure}
It is known that, when proximity coupled to an $s$-wave superconductor, the quantum phase transition between a trivial insulator ($c_-=0$) and a Chern insulator ($c_-=1$) can be split into two separate Ising transitions that sandwich an intermediate, gapped chiral $p_x+ip_y$-like superconducting phase ($c_-=1/2$).\cite{qhz2010} Using our analysis, we can expand the idea behind this set of transitions to construct a family of topological phase transitions connecting the $B$ and $D$ series. Ref.~[\onlinecite{qhz2010}] presented a phase diagram for the proximity-coupled Chern insulator as a function of two competing parameters: (i) a time-reversal breaking parameter on the horizontal axis which drives the trivial-to-topological insulator transition by changing the Chern invariant, and (ii) the superconducting proximity coupling on the vertical axis, which drives the insulator-to-superconductor transition. 

In order to connect to that work we present a qualitatively similar phase diagram in Fig. \ref{fig:so(N)phasediagram}. In our setting, we should replace the parameter axis representing the superconductor proximity-coupling by a ``gauging" axis, i.e.~a parameter axis under which the system becomes unstable toward superconductivity and the subsequent gauging of fermion parity. This is necessary if we are to produce a state with topological order. For example, we have seen that gauging the fermion parity of a Chern insulator with Chern number $r$ leads to the bosonic $D_r=SO(2r)_1$ states. A direct transition between the $D_r$ and $D_{r+1}$ phases could be driven through a change in the Chern number along the horizontal direction of Fig.\ref{fig:so(N)phasediagram}. However, a transition where the Chern number changes by one will generically be separated into two Ising transitions in the presence of superconductivity. The chiral central charge is modified by $\delta c_-=1/2+1/2=1$ through these two Ising transitions. In fact, the two Ising transitions will sandwich an intervening non-Abelian $B_r=so(2r+1)_1$ phase (i.e.~a $(p_x+ip_y)$-like chiral phase, but with gauged fermion parity). The phase boundaries eventually would meet at high enough superconducting/fermion-parity gauging coupling, after which the $B_{r-1}$ and $B_r$ phases will no longer be separated by an intermediate Abelian $D_r$ phase but instead by its non-Abelian twist liquid $B_{r_1}\otimes B_{r_2}$ (see Fig.~\ref{fig:so(N)phasediagram} for a phase diagram). 
In other words, the gauging transition in \eqref{so(N1)xso(N2)} -- the vertical direction in Fig.\ref{fig:so(N)phasediagram} -- would generically also be separated into two Ising transitions sandwiching some non-Abelian $B_r$ state. However, in this case the two Ising transitions are opposing and  have opposite chiralities so that $\delta c_-=1/2-1/2=0$. Indeed this matches our expectation that the twist liquid should have a matching central charge with the parent state. 
Although we do not have a model which exhibits the full phase diagram, this illustration is useful to show the interconnected relationships between the $B$ and $D$ series states, and also that the appearance of the twist liquid phase of the $D$-series is quite natural in this context.

\section{Anyonic Triality Symmetry in \texorpdfstring{$SO(8)_1$}{SO(8)}}\label{sec:so(8)symmetry}

Now we move onto an Abelian topological state that carries a non-Abelian anyonic symmetry. 
The $SO(8)_1$ fractional quantum Hall state is described by a four-component Chern-Simons effective action \eqref{CSaction} with \begin{align}K_{SO(8)_1}=\left(\begin{array}{*{20}c}2&-1&-1&-1\\-1&2&0&0\\-1&0&2&0\\-1&0&0&2\end{array}\right).\label{so(8)Kmatrix}\end{align} The $K$-matrix is identical to the Cartan matrix of the Lie algebra $so(8),$ and as a result, the edge CFT carries a chiral $SO(8)$ Kac-Moody structure at level 1\cite{bigyellowbook, khan2014}. This topological phase has four quasiparticle types $1,\psi_1,\psi_2,\psi_3$, where $1$ is the local bosonic vacuum, and the $\psi_i$ are all fermions with mutual semionic statistics: $\mathcal{D}S_{\psi_i\psi_j}=-1$ for $i\neq j$. The fermions obey the fusion properties \begin{align}\psi_i^2=1,\quad\psi_1\times\psi_2\times\psi_3=1.\label{so(8)fusion}\end{align}

Since all non-trivial quasiparticles have the same spin, and the fusion rules are also invariant under permutation, the $SO(8)_1$ state has a triality anyonic symmetry $S_3$ that permutes the three fermions $\psi_i$ (see Ref. \onlinecite{khan2014}). The permutation group $S_3=\mathbb{Z}_2\ltimes\mathbb{Z}_3$ is generated by a twofold reflection $\sigma_1$ and a threefold rotation $\rho$ represented by \begin{align}\sigma_1=\left(\begin{array}{*{20}c}1&0&0&0\\0&1&0&0\\0&0&0&1\\0&0&1&0\end{array}\right),\quad \rho=\left(\begin{array}{*{20}c}1&0&0&0\\0&0&0&1\\0&1&0&0\\0&0&1&0\end{array}\right).\end{align} The reflection generator $\sigma_1$ interchanges $\psi_2\leftrightarrow\psi_3$ while fixing $\psi_1$, and $\rho$ (or $\rho^2$) cyclicly rotates $\psi_i\to\psi_{i+1}$ (resp.~$\psi_i\to\psi_{i-1}$). The other two reflections are defined as \begin{align}\sigma_2=\sigma_1\rho,\quad\sigma_3=\sigma_1\rho^2,\end{align} and they fix $\psi_2$ and $\psi_3$ respectively, while interchanging the remaining two fermions.

The non-Abelian symmetry group $S_3$ contains an Abelian {\em normal} subgroup $\mathbb{Z}_3$ and three {\em abnormal} $\mathbb{Z}_2$ subgroups. 
Gauging different symmetries will lead to different twist liquids as summarized by the following diagram: \begin{align}\vcenter{\hbox{\includegraphics[width=0.3\textwidth]{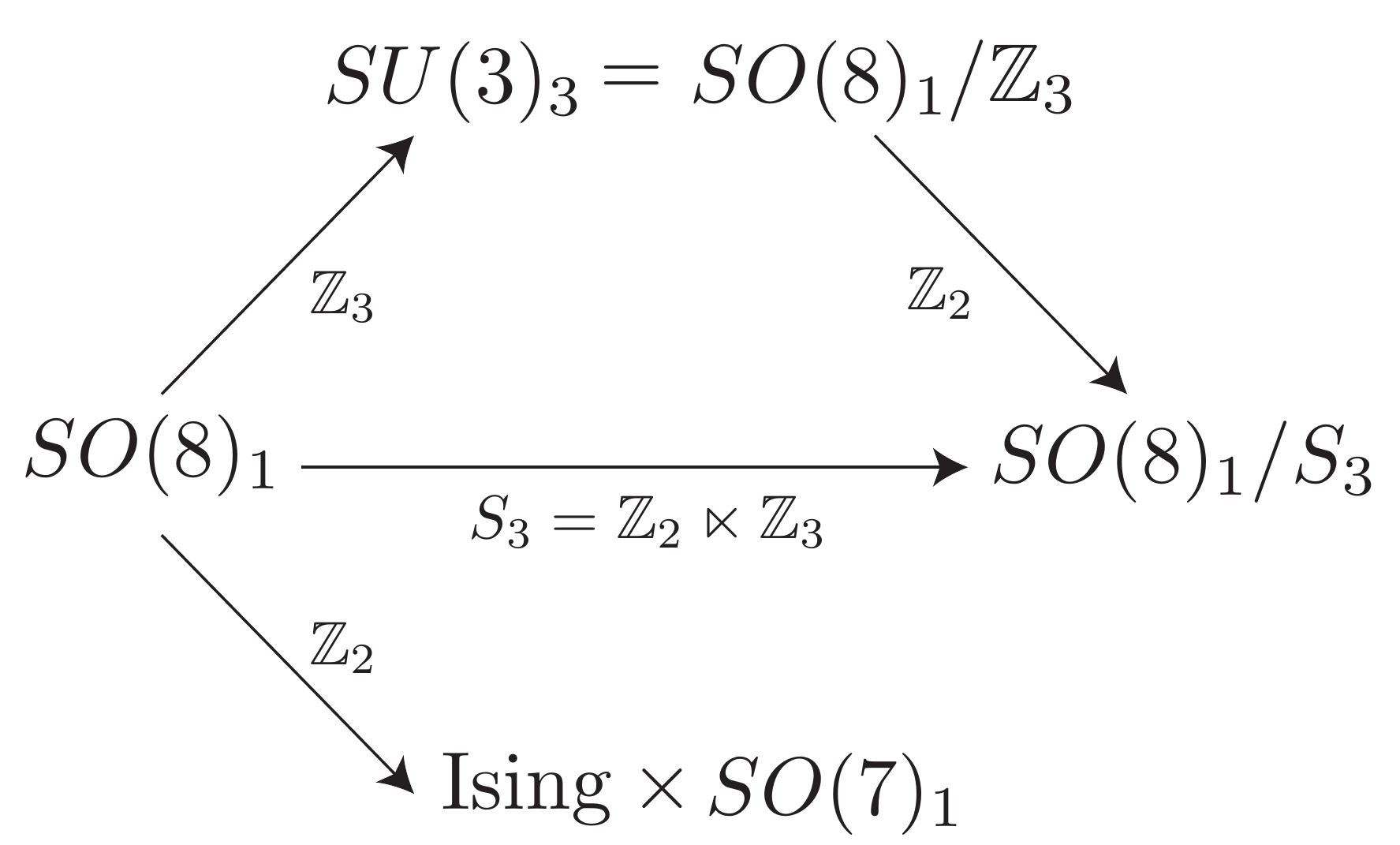}}}\label{so(8)gaugingdiagram}\end{align} Each arrow represents a gauging phase transition of a particular symmetry. The quotients here refer to the twist liquids obtained after gauging the symmetry in the denominator. This notation is adopted from the corresponding orbifolds of the $(1+1)$D edge CFT,\cite{Ginsparg88, bigyellowbook, Alimohammadi94, chenroyteoinprogress} whose primary fields match the $(2+1)$D bulk anyons of the twist liquid. The bulk-boundary correspondence, however, will not be presented in this article as it is not a necessary tool.


\subsection{Twofold and Threefold Defects and the Defect Fusion Category}\label{sec:so(8)defectcategory}
We will label twist defects according to their symmetries. The three reflections correspond to twofold defects $(\sigma_i)_\lambda$ which can have species labels $\lambda=0,1.$ The two rotations correspond to threefold defects $\rho$ and $\overline\rho$ which have no non-trivial species labels\cite{khan2014}. The fusion structure of the twofold defects is similar to the $\mathbb{Z}_2$-symmetric toric code, or the more general $SO(2r)_1$ models, discussed in Section~\ref{sec:so(N)}: \begin{align*}&(\sigma_i)_\lambda\times\psi_i=(\sigma_i)_\lambda,\quad(\sigma_i)_\lambda\times\psi_{i\pm1}=(\sigma_i)_{\lambda+1}\\&(\sigma_i)_0\times(\sigma_i)_0=1+\psi_i,\quad(\sigma_i)_0\times(\sigma_i)_1=\psi_{i-1}+\psi_{i+1},\end{align*} where the subscripts $i$ refer to which reflection operation each twofold defect corresponds, and the $\lambda$ subscripts refer to the species labels. Each twofold defect has quantum dimension $d_\sigma=\sqrt{2}$. 

The threefold defects obey the fusion rules \begin{align}\rho\times\rho=2\overline{\rho},\quad\rho\times\overline{\rho}=1+\psi_1+\psi_2+\psi_3\label{Z3defectfusion}\end{align} where $\overline{\rho}$ is the anti-defect of $\rho,$ and is thus associated to the symmetry $\rho^2$, and $N^{\overline\rho}_{\rho\rho}=2$ indicates a fusion degeneracy. The fusion degeneracy implies that there are two distinct ways to split the anti-partner $\overline{\rho}$ into a pair of $\rho$'s. These splittings can be distinguished by a Wilson loop measurement,\cite{TeoRoyXiao13long} and are represented by orthogonal splitting states (see Eq.~\eqref{so(8)splittingrhorho}). By equating the quantum dimensions on both sides of the fusion equation, the fusion rule for $\rho\times\overline{\rho}$ indicates that threefold defects should have a quantum dimension of $d_\rho=2$. The fusion degeneracy is therefore necessary to balance the quantum dimensions of the fusion equation $\rho\times\rho=2\overline{\rho}$. 

Finally, we note that the twofold and threefold defects satisfy non-commutative fusion rules\cite{TeoRoyXiao13long, khan2014}: \begin{align*}\begin{array}{*{20}c}(\sigma_i)_\lambda\times(\sigma_{i+1})_{\lambda'}=\rho,\\(\sigma_i)_\lambda\times(\sigma_{i-1})_{\lambda'}=\overline{\rho},\end{array}\quad\begin{array}{*{20}c}\sigma_i\times\rho=(\sigma_{i+1})_0+(\sigma_{i+1})_1\\\sigma_i\times\overline{\rho}=(\sigma_{i-1})_0+(\sigma_{i-1})_1\end{array}.\end{align*} 
From the cohomology classification we see that there is not a non-trivial non-symmorphic quantum $S_3$-symmetry, i.e.~$H^2(S_3,\mathcal{A}_{SO(8)_1})=0$,\cite{ChenGuLiuWen11} where $\mathcal{A}_{SO(8)_1}\cong\mathbb{Z}_2^2$ is the group of Abelian quasiparticles in $SO(8)_1$. Hence, the above set of fusion rules is the only consistent one. We will see that fter gauging the $S_3$ symmetry, the three twofold defects group into a single anyon by forming a super-sector, and similarly for the two threefold defects (see \eqref{so(8)twofoldfluxsector} and table~\ref{tab:anyonsS3} later). Fusion commutativity is then restored in the twist liquid.

The full defect fusion category is \begin{align}\mathcal{C}_{S_3}=\mathcal{C}_1\oplus\mathcal{C}_\rho\oplus\mathcal{C}_{\overline\rho}\oplus\mathcal{C}_{\sigma_1}\oplus\mathcal{C}_{\sigma_2}\oplus\mathcal{C}_{\sigma_3}\end{align} where $\mathcal{C}_1=\langle1,\psi_1,\psi_2,\psi_3\rangle,$ as usual, contains quasiparticles in the parent state and \begin{align}\mathcal{C}_\rho=\langle\rho\rangle,\quad\mathcal{C}_{\overline{\rho}}=\langle\overline{\rho}\rangle,\quad\mathcal{C}_{\sigma_i}=\left\langle(\sigma_i)_0,(\sigma_i)_1\right\rangle\end{align} are the non-trivial defect sectors. We also find that \begin{align}\mathcal{C}_{\mathbb{Z}_3}=\mathcal{C}_1\oplus\mathcal{C}_\rho\oplus\mathcal{C}_{\overline\rho},\quad\mathcal{C}_{\mathbb{Z}_2}^i=\mathcal{C}_1\oplus\mathcal{C}_{\sigma_i}\label{so(8)defectcategory}\end{align} are closed defect subcategories involving only threefold or twofold defects respectively.

To complete the defect fusion category structure we need to calculate the $F$-symbols. To express the $F$-symbol basis transformations in a simple notation, it is convenient to represent the Abelian quasiparticles by two dimensional $\mathbb{Z}_2$-valued vectors ${\bf a}=(0,0)=1$, $(1,0)=\psi_1$, $(0,1)=\psi_2,$ and $(1,1)=\psi_3$. In this notation, the threefold symmetry, for example, is represented by the $\mathbb{Z}_2$-valued matrix \begin{align}\Lambda_3=\left(\begin{array}{*{20}c}0&-1\\1&-1\end{array}\right)\equiv\left(\begin{array}{*{20}c}0&1\\1&1\end{array}\right).\end{align} The $R$-symbols between Abelian anyons in $SO(8)_1$ can be chosen to be \begin{align}R^{{\bf a}{\bf b}}=(-1)^{{\bf a}^T\sigma_x\Lambda_3^2{\bf b}}\label{Rso8abelian}\end{align} so that the braiding phase, $\mathcal{D}S_{{\bf a}{\bf b}}=R^{{\bf a}{\bf b}}R^{{\bf b}{\bf a}}=(-1)^{{\bf a}^T\sigma_x{\bf b}},$ agrees with that from the conventional $K$-matrix description. 

Non-trivial $F$-symbols can be calculated in a similar fashion to that of a $S_3$-symmetric lattice model in Ref.~[\onlinecite{TeoRoyXiao13long}]. We sketch the calculation Appendix~\ref{app:so8fsymbols}. The $F$-symbols for the threefold defects in $SO(8)_1$ are listed in Table~\ref{tab:so(8)Fsymbols}. They are evaluated by matching the Wilson strings of the splitting states $x\times(y\times z)$ and $(x\times y)\times z,$ as highlighted in Section~\ref{sec:twistdefects}. As discussed in Appendix~\ref{app:so8fsymbols}, the $F$-transformations can be understood as a representation of the double cover of $A_4$ which is the group of even permutations of 4 elements, or equivalently, the rotation symmetries of a tetrahedron. For instance the $\pi$-rotations are represented by \begin{align}\mathcal{A}_{\psi_1}=i\sigma_x,\quad\mathcal{A}_{\psi_2}=-i\sigma_y,\quad\mathcal{A}_{\psi_3}=i\sigma_z\label{so(8)Arep}\end{align} where $\sigma_{x,y,z}$ are the Pauli matrices; and $2\pi/3$-rotation, say about the 111-axis, takes the form: $\exp\left[\frac{\pi}{3}\left(\frac{\mathcal{A}_{\psi_1}+\mathcal{A}_{\psi_2}+\mathcal{A}_{\psi_3}}{\sqrt{3}}\right)\right]$. And indeed, these matrices show up in the defect $F$-symbols (see Table~\ref{tab:so(8)Fsymbols} and Appendix~\ref{app:so8fsymbols}).


\begin{table}[ht]
\centering
\begin{tabular}{ll}
\multicolumn{2}{c}{Threefold defect $F$-symbols for $SO(8)_1$}\\\hline\noalign{\smallskip}
$F^{{\bf a}{\bf b}{\bf c}}_{{\bf a}+{\bf b}+{\bf c}}$ & $1$\\
$F^{\rho{\bf a}{\bf b}}_\rho$ & $R^{{\bf a}(\Lambda_3{\bf b})}$\\
$F^{{\bf a}{\bf b}\rho}_\rho$ & $R^{{\bf b}(\Lambda_3^2{\bf a})}$\\
$F^{{\bf a}\rho{\bf b}}_\rho$ & $\mathcal{D}S_{(\Lambda_3{\bf a}){\bf b}}$\\
$F^{\rho\overline\rho{\bf a}}_{\bf b}$ & $R^{{\bf a}(\Lambda_3^2{\bf b})}$\\
$F^{{\bf a}\rho\overline\rho}_{\bf b}$ & $R^{({\bf a}+{\bf b})(\Lambda_3^2{\bf a})}$\\
$F^{\rho{\bf a}\overline\rho}_{\bf b}$ & $\mathcal{D}S_{{\bf a}(\Lambda_3{\bf b})}R^{{\bf a}(\Lambda_3{\bf a})}$\\
$F^{\rho\rho{\bf a}}_{\overline\rho}$ & $R^{{\bf a}{\bf a}}\mathcal{A}_{\bf a}$\\
$F^{{\bf a}\rho\rho}_{\overline\rho}$ & $\mathcal{A}_{\Lambda_3^2\bf a}$\\
$F^{\rho{\bf a}\rho}_{\overline\rho}$ & $R^{{\bf a}(\Lambda_3^2{\bf a})}\mathcal{A}_{\Lambda_3\bf a}$\\
$F^{\rho\rho\rho}_{\bf a}$ & $\mathcal{A}_{\bf a}\exp\left[\frac{\pi}{3}\left(\frac{\mathcal{A}_{\psi_1}+\mathcal{A}_{\psi_2}+\mathcal{A}_{\psi_3}}{\sqrt{3}}\right)\right]$\\
$\left[F^{\rho\overline\rho\rho}_{\rho}\right]^{\bf b}_{\bf a}$ & $\frac{1}{2}\mathcal{D}S_{(\Lambda_3{\bf a}){\bf b}}R^{{\bf b}(\Lambda_3{\bf b})}$\\
$\left[F^{\overline\rho\rho\rho}_\rho\right]_{\bf a}$ & $\frac{1}{\sqrt{2}}R^{{\bf a}{\bf a}}\mathcal{A}_{\Lambda_3{\bf a}}\mathcal{A}_{\psi_2}$\\
$\left[F^{\rho\rho\overline\rho}_\rho\right]^{\bf b}$ & $\frac{1}{\sqrt{2}}R^{{\bf b}{\bf b}}\exp\left[\frac{\pi}{3}\left(\frac{\mathcal{A}_{\psi_1}+\mathcal{A}_{\psi_2}+\mathcal{A}_{\psi_3}}{\sqrt{3}}\right)\right]\mathcal{A}_{\psi_1}\mathcal{A}_{\bf b}$\\
\end{tabular}
\caption{The $R$-symbols are defined in \eqref{Rso8abelian}. $\mathcal{D}S_{{\bf a}{\bf b}}=R^{{\bf a}{\bf b}}R^{{\bf b}{\bf a}}$ is the braiding phase of ${\bf a}$ around ${\bf b}$. The $2\times2$ matrices $\mathcal{A}_{\bf a}$ act on splitting degeneracy $\overline\rho\to\rho\times\rho$ or $\rho\to\overline\rho\times\overline\rho,$ and can be represented by Pauli matrices \eqref{so(8)Arep}. $F$-symbols with interchanged $\rho\leftrightarrow\overline\rho$ are not listed explicitly but can obtained by replacing $\Lambda_3\leftrightarrow\Lambda_3^2$.}\label{tab:so(8)Fsymbols}
\end{table}

The $F$-matrices in Table~\ref{tab:so(8)Fsymbols} also come with a certain phase ambiguity. They can be modified by the $\mathbb{Z}_3$ phases 
\begin{align}\begin{array}{*{20}c}F^{\rho\overline\rho\rho}_\rho\to e^{2\pi mi/3}F^{\rho\overline\rho\rho}_\rho,\hfill& F^{\overline\rho\rho\overline\rho}_{\overline\rho}\to e^{-2\pi mi/3}F^{\overline\rho\rho\overline\rho}_{\overline\rho},\hfill\\F^{\rho\bar\rho\bar\rho}_{\overline\rho}\to e^{-2\pi mi/3}F^{\rho\bar\rho\bar\rho}_{\overline\rho},\hfill& F^{\overline\rho\rho\rho}_\rho\to e^{2\pi mi/3}F^{\overline\rho\rho\rho}_\rho,\hfill\\F^{\bar\rho\bar\rho\rho}_{\overline\rho}\to e^{2\pi mi/3}F^{\bar\rho\bar\rho\rho}_{\overline\rho},\hfill& F^{\bar\rho\bar\rho\bar\rho}_{\bf a}\to e^{-2\pi mi/3}F^{\bar\rho\bar\rho\bar\rho}_{\bf a},\hfill\end{array}\label{so(8)Fsymbolsmod}\end{align}
 while keeping all others unchanged. These modifications preserve the pentagon identities (Fig.~\ref{fig:Fpentagon}), and give rise to three topologically inequivalent sets of $F$-symbols, for $m=0,1,2$, that cannot be related by gauge transformations on the splitting states. Their differences are classified cohomologically by $H^3(\mathbb{Z}_3,U(1))=\mathbb{Z}_3$ (see also Section \ref{sec:gaugingZ3SPT}). This affects particle-antiparticle duality (see \eqref{bendingdef}) in threefold defects 
as the phase contributes to the Frobenius-Schur indicator\cite{FredenhagenRehrenSchroer92,Kitaev06} \begin{align}\varkappa_\rho=\varkappa_{\overline{\rho}}^\ast=\left[F^{\rho\overline\rho\rho}_{\overline\rho}\right]_1^1\left/\left|\left[F^{\rho\overline\rho\rho}_{\overline\rho}\right]_1^1\right|\right.=e^{2\pi im/3}.\label{SU3FSindicator}\end{align} The three distinct collections of $F$-matrices are all legal sets of defect basis transformations. Physically, the non-trivial phase modification can be understood by the addition of a $(2+1)$D $\mathbb{Z}_3$ SPT, which is classified by the same group cohomology and contributes the $\mathbb{Z}_3$ phases \eqref{Z3SPTF}, to the $SO(8)_1$ state. Consequently, these choices correspond to three distinct non-Abelian twist liquid states that can arise after gauging the $\mathbb{Z}_3$ subgroup of the $S_3$ symmetry. 
 
 The $K$-matrix \eqref{so(8)Kmatrix} for the Chern-Simons description of the parent state provides more information than just the bulk anyon structure. We can also determine a possible edge theory. The edge state in this case consists of four chiral $U(1)$-bosons, and by performing a $\mathbb{Z}_3$-orbifold of the chiral CFT,\cite{chenroyteoinprogress} the phase ambiguity of the $F$-symbols can be fixed to $m=2$ modulo $3\mathbb{Z}$. However, in this article we consider the more general situation with all possible $m$. 

Table~\ref{tab:so(8)Fsymbols} only specifies the $F$-symbols for the $\mathcal{C}_{\mathbb{Z}_3}$ subcategory of threefold defects. The full set of $F$-symbols for the entire $\mathcal{C}_{S_3}$ defect category will not be presented in this article. Although they would be necessary for constructing the string-net model, one can derive the anyon properties of the full $S_3$-twist liquid from just a subset of $F$-symbols. In addition to the symbols in Table~\ref{tab:so(8)Fsymbols}, we will only need the ones from a single $\mathcal{C}_{\mathbb{Z}_2}$ sector:\begin{align}\left[F^{\sigma_3\sigma_3\sigma_3}_{\sigma_3}\right]_{\bf a}^{\bf b}=\frac{(-1)^{s+a_2b_2}}{\sqrt{2}},\quad F^{{\bf a}\sigma_3{\bf b}}_{\sigma_3}=F^{\sigma_3{\bf a}\sigma_3}_{\bf b}=(-1)^{a_2b_2},\end{align} where ${\bf a}=(a_1,a_2)$ is a $\mathbb{Z}_2$-valued vector labeling the four $SO(8)_1$ quasiparticles. We have already discussed the derivation of these $F$-symbols in our discussion of the $\mathbb{Z}_2$ defect fusion category of the toric code. We remind the reader that there is a phase ambiguity represented by the $\mathbb{Z}_2$ valued quantity $s=0,1$ that relates to the Frobenius-Schur indicator $\varkappa_\sigma=(-1)^s$.

With the structure of the defect fusion category in place, we will now discuss the gauging procedure. We will consider three separate procedures: (i) we will briefly comment on gauging a $\mathbb{Z}_2$ sector, (ii) we will discuss gauging of the $\mathbb{Z}_3$ sector, and (iii) we will gauge the full $S_3$ symmetry. 


\subsection{Gauging \texorpdfstring{$\mathbb{Z}_2$}{Z2}}\label{sec:gaugingZ2}

Gauging the $\mathbb{Z}_2$ symmetry would lead to one of the Ising-like theories as described in Section~\ref{sec:so(N)} Eq.~\eqref{so(N1)xso(N2)} \begin{align}\begin{diagram}SO(8)_1&\rTo^{\mbox{\small Gauging $\mathbb{Z}_2$}}\left\{\begin{array}{*{20}c}SO(3)_1\otimes SO(5)_1&\mbox{if $\varkappa_\sigma=-1$}\\\mbox{Ising}\otimes SO(7)_1&\mbox{if $\varkappa_\sigma=+1$}\end{array}\right.\end{diagram}\label{gaugeZ2so(8)}\end{align} where the $SO(N)_1$ theory for odd $N$ contains quasiparticles $1,\psi,\sigma$ with spins $h_1=0$, $h_\psi=1/2$, and $h_\sigma=N/16$. 
The two inequivalent theories in \eqref{gaugeZ2so(8)} differ from each other by the Frobenius-Schur indicator $\varkappa_\sigma=\pm1$, which is the sign of $[F^{\sigma\sigma\sigma}_\sigma]^1_1,$ and is classified cohomologically by $H^3(\mathbb{Z}_2,U(1))=\mathbb{Z}_2$. These two theories are connected by attaching the parent $SO(8)_1$ state with a $\mathbb{Z}_2$-SPT (see Section~\ref{sec:gaugingSPT}). Interestingly,  since the $\mathbb{Z}_2$ symmetry is not a normal subgroup of the full $S_3$ triality symmetry, the $\mathbb{Z}_3$ symmetry is \emph{broken} by the gauging process, and it is no longer an anyonic symmetry of the non-Abelian $\mathbb{Z}_2$ twist liquid state. Thus, if we gauge the $\mathbb{Z}_2$ symmetry initially this is as far as we can take it.

\subsection{Gauging \texorpdfstring{$\mathbb{Z}_3$}{Z3}}\label{sec:gaugingZ3}

Now let us consider gauging the threefold symmetry of $SO(8)_1$. \begin{align}\begin{diagram}SO(8)_1&\pile{\rTo^{\mbox{\small Gauging $\mathbb{Z}_3$}}\\\lTo_{\mbox{\small Condensation}}}&SU(3)_3\end{diagram}.\label{so(8)su(3)diagram}\end{align} We will show that the result is a non-Abelian topological phase very similar to the $SU(3)_3$ state. Let us review the properties of the $SU(3)_3$ topological phase.  From the bulk boundary correspondence, the anyons of the bulk non-Abelian state are in one-to-one correspondence to the primary fields of the edge chiral affine Kac-Moody $su(3)$ algebra at level 3.\cite{bigyellowbook, Alimohammadi94} There are 10 primary fields in this CFT, and they are labeled by the dimensions of the truncated irreducible representations of $su(3)$. Their conformal dimensions -- the spins of the corresponding anyons -- and their quantum dimensions are listed in Table~\ref{tab:su(3)3spindim}. They obey the following fusion rules:
\begin{gather}1\times x=x,\quad{\bf 10}\times{\bf 10}=\overline{\bf 10},\quad{\bf 10}\times\overline{\bf 10}=1\nonumber\\{\bf 3}\times{\bf 10}={\bf 6},\quad{\bf 3}\times\overline{\bf 10}={\bf 15},\quad{\bf 8}\times{\bf 10}={\bf 8}\nonumber\\{\bf 3}\times{\bf 3}=\overline{\bf 3}+\overline{\bf 6},\quad{\bf 3}\times\overline{\bf 3}=1+{\bf 8}\label{su(3)3fusion}\\{\bf 8}\times{\bf 8}=1+{\bf 10}+\overline{\bf 10}+{\bf 8}+{\bf 8}\nonumber\\{\bf 8}\times{\bf 3}={\bf 3}+{\bf 6}+{\bf 15}.\nonumber\end{gather}
\begin{table}[t!]
\centering
\begin{tabular}{lcccccccccc}
Anyons&$1$&${\bf 3}$&$\overline{\bf 3}$&${\bf 6}$&$\overline{\bf 6}$&${\bf 8}$&${\bf 10}$&$\overline{\bf 10}$&${\bf 15}$&$\overline{\bf 15}$\\\hline
$h_x$&$0$&$\frac{2}{9}$&$\frac{2}{9}$&$\frac{5}{9}$&$\frac{5}{9}$&$\frac{1}{2}$&$1$&$1$&$\frac{8}{9}$&$\frac{8}{9}$\\
$d_x$&$1$&$2$&$2$&$2$&$2$&$3$&$1$&$1$&$2$&$2$\\
\end{tabular}
\caption{The conformal dimensions $h_x$ (and therefore the resulting spins $\theta_x=e^{2\pi ih_x}$) and quantum dimensions $d_x$ of the ten anyons of the non-Abelian $SU(3)_3$ state.}\label{tab:su(3)3spindim}
\end{table}

The Abelian bosons ${\bf 10}$ and $\overline{\bf 10}$ will be identified as $\mathbb{Z}_3$ charges because of their threefold braiding phases \begin{align}\vcenter{\hbox{\includegraphics[height=0.6in]{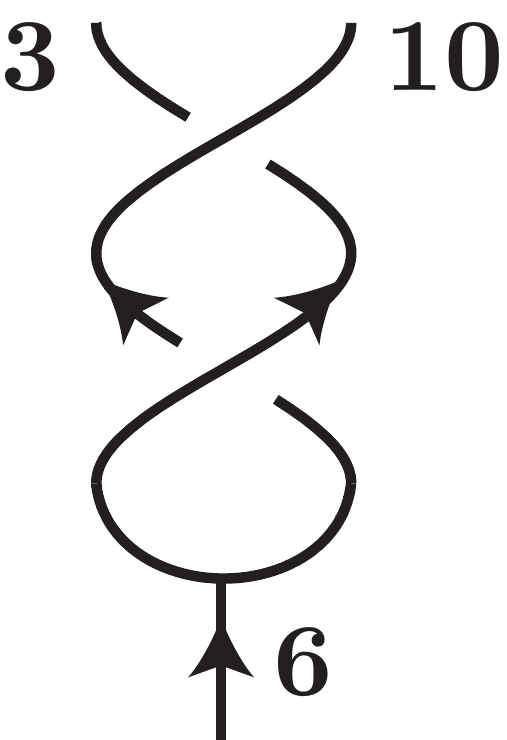}}}=\frac{\theta_{\bf 6}}{\theta_{\bf 3}\theta_{\bf 10}}=e^{2\pi i/3},\quad\vcenter{\hbox{\includegraphics[height=0.6in]{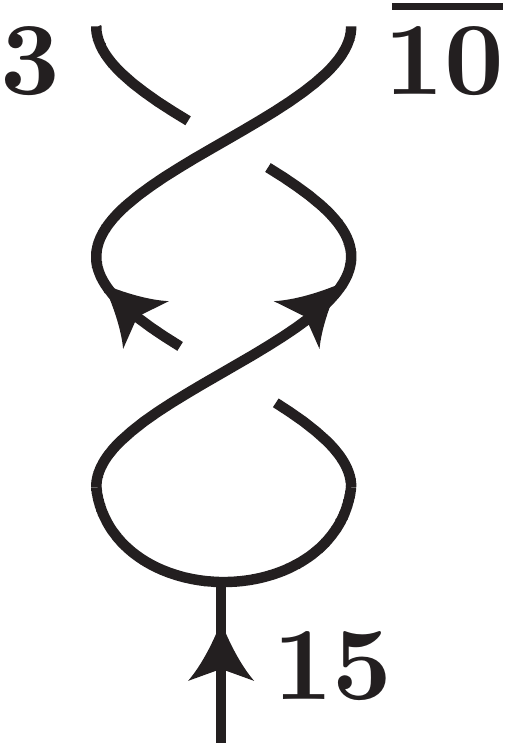}}}=\frac{\theta_{\bf 15}}{\theta_{\bf 3}\theta_{\overline{\bf 10}}}=e^{-2\pi i/3}\end{align} around the non-Abelian anyon ${\bf 3}$, which, along with ${\bf 6}$ and ${\bf 15},$ will be identified  as $\mathbb{Z}_3$ fluxes. The threefold nature of the fluxes is shown by the fusion rule \begin{align}{\bf 3}\times{\bf 3}\times{\bf 3}=1+\ldots\end{align} obtained by combining the fusion equations \eqref{su(3)3fusion} with the reverse fluxes $\overline{\bf 3}$, $\overline{\bf 6},$ and $\overline{\bf 15}$.

The non-Abelian $SU(3)_3$ state can be reduced back to the Abelian $SO(8)_1$ state by condensing the boson charges ${\bf 10}$ and $\overline{\bf 10}$. All six threefold fluxes are confined during this process, as they are non-local with respect to the charges. The dimension 3 fermion will be identified as a super-sector ${\bf 8}=\psi_1+\psi_2+\psi_3$ with three Abelian components. Upon condensation, the fusion rule for ${\bf 8}\times{\bf 8}$ in \eqref{su(3)3fusion} becomes \begin{align}(\psi_1+\psi_2+\psi_3)\times(\psi_1+\psi_2+\psi_3)\nonumber\\=1+1+1+2\psi_1+2\psi_2+2\psi_3\label{Z3super-sectorfusion1}\end{align} and ${\bf 8}$ must therefore decompose into the three distinct Abelian fermions $\psi_i$ that form $SO(8)_1$.

\begin{table}[htbp]
\centering
\begin{tabular}{lll}
Anyons $\chi$ & Dimensions $d_\chi$ & Spins $\theta_\chi$\\\hline
$1$ & $1$ & $1$\\
$z_3$ & $1$ & $1$\\
$\overline{z}_3$ & $1$ & $1$\\
$\Psi=\psi_1+\psi_2+\psi_3$ & $3$ & $-1$\\
$\rho_0$ & $2$ & $e^{\frac{2\pi im}{9}}$\\
$\rho_1$ & $2$ & $e^{\frac{2\pi i(m+3)}{9}}$\\
$\rho_2$ & $2$ & $e^{\frac{2\pi i(m-3)}{9}}$\\
$\overline\rho_0$ & $2$ & $e^{\frac{2\pi im}{9}}$\\
$\overline\rho_1$ & $2$ & $e^{\frac{2\pi i(m+3)}{9}}$\\
$\overline\rho_2$ & $2$ & $e^{\frac{2\pi i(m-3)}{9}}$\\
\end{tabular}
\caption{The quantum dimensions $d_\chi$ and spin/statistics $\theta_\chi=e^{2\pi ih_\chi}$ of anyonic excitations in the twist liquid $SO(8)_1/\mathbb{Z}_3$ derived by gauging the $\mathbb{Z}_3$ symmetry of $SO(8)_1$. $m=0,1,2$ is fixed by the choice of $F$-symbols (\ref{so(8)Fsymbolsmod},\ref{SU3FSindicator}) in the defect fusion category.}\label{tab:anyonsZ3}
\end{table}

Now we will explicitly gauge the $\mathbb{Z}_3$ symmetry of $SO(8)_1,$ and show that the twist liquid state must have an $SU(3)_3$-like structure. From the defect fusion category, we know there are three distinct (cohomologically) inequivalent possibilities, labeled by $m=0,1,2$, distinguished by the fusion and spin/statistics of $\mathbb{Z}_3$ fluxes. First, we summarize the resulting anyon structure in Table~\ref{tab:anyonsZ3}, and the braiding $S$-matrix \eqref{su(3)3Smatrix} can be determined by \eqref{braidingS} to be:
\begin{widetext}
\begin{align}S=\frac{1}{\mathcal{D}}
\left(
\begin{smallmatrix}
 1 & 1 & 1 & 3 & 2 & 2 & 2 & 2 & 2 & 2 \\
 1 & 1 & 1 & 3 & 2 e^{\frac{2\pi i}{3}} & 2 e^{\frac{2\pi i}{3}} & 2 e^{\frac{2\pi i}{3}} & 2 e^{-\frac{2\pi i}{3}} & 2 e^{-\frac{2\pi i}{3}} & 2 e^{-\frac{2\pi i}{3}} \\
 1 & 1 & 1 & 3 & 2 e^{-\frac{2\pi i}{3}} & 2 e^{-\frac{2\pi i}{3}} & 2 e^{-\frac{2\pi i}{3}} & 2 e^{\frac{2\pi i}{3}} & 2 e^{\frac{2\pi i}{3}} & 2 e^{\frac{2\pi i}{3}} \\
 3 & 3 & 3 & -3 & 0 & 0 & 0 & 0 & 0 & 0 \\
 2 & 2e^{\frac{2\pi i}{3}} & 2e^{-\frac{2\pi i}{3}} & 0 & -2e^{\frac{4\pi im}{9}} & -2e^{\frac{4\pi i(m-3)}{9}} & -2e^{\frac{4\pi i(m+3)}{9}} & -2e^{-\frac{4\pi im}{9}} & -2e^{-\frac{4\pi i(m-3)}{9}} & -2e^{-\frac{4\pi i(m+3)}{9}} \\
 2 & 2e^{\frac{2\pi i}{3}} & 2e^{-\frac{2\pi i}{3}} & 0 & -2e^{\frac{4\pi i(m-3)}{9}} & -2e^{\frac{4\pi i(m+3)}{9}} & -2e^{\frac{4\pi im}{9}} & -2e^{-\frac{4\pi i(m-3)}{9}} & -2e^{-\frac{4\pi i(m+3)}{9}} & -2e^{-\frac{4\pi im}{9}} \\
 2 & 2e^{\frac{2\pi i}{3}} & 2e^{-\frac{2\pi i}{3}} & 0 & -2e^{\frac{4\pi i(m+3)}{9}} & -2e^{\frac{4\pi im}{9}} & -2e^{\frac{4\pi i(m-3)}{9}} & -2e^{-\frac{4\pi i(m+3)}{9}} & -2e^{-\frac{4\pi im}{9}} & -2e^{-\frac{4\pi i(m-3)}{9}} \\
 2 & 2e^{-\frac{2\pi i}{3}} & 2e^{\frac{2\pi i}{3}} & 0 & -2e^{-\frac{4\pi im}{9}} & -2e^{-\frac{4\pi i(m-3)}{9}} & -2e^{-\frac{4\pi i(m+3)}{9}} & -2e^{\frac{4\pi im}{9}} & -2e^{\frac{4\pi i(m-3)}{9}} & -2e^{\frac{4\pi i(m+3)}{9}} \\
 2 & 2e^{-\frac{2\pi i}{3}} & 2e^{\frac{2\pi i}{3}} & 0 & -2e^{-\frac{4\pi i(m-3)}{9}} & -2 e^{-\frac{4\pi i(m+3)}{9}} & -2e^{-\frac{4\pi im}{9}} & -2e^{\frac{4\pi i(m-3)}{9}} & -2e^{\frac{4\pi i(m+3)}{9}} & -2 e^{\frac{4\pi im}{9}} \\
 2 & 2e^{-\frac{2\pi i}{3}} & 2e^{\frac{2\pi i}{3}} & 0 & -2e^{-\frac{4\pi i(m+3)}{9}} & -2e^{-\frac{4\pi im}{9}} & -2e^{-\frac{4\pi i(m-3)}{9}} & -2e^{\frac{4\pi i(m+3)}{9}} & -2e^{\frac{4\pi im}{9}} & -2e^{\frac{4\pi i(m-3)}{9}} \\
\end{smallmatrix}
\right)\label{su(3)3Smatrix}\end{align}
\end{widetext}
where the total quantum dimension is $\mathcal{D}=6$, and the entries of the $S$-matrix and anyon list in Table~\ref{tab:anyonsZ3} have the same ordering.

Now let us derive this structure. We first interpret the anyons as compositions of gauge flux, quasiparticle super-sector, and gauge charge as described in Section~\ref{sec:QPstructuretwistliquid}. The $\mathbb{Z}_3$ gauge group has conjugacy classes $[1],[\rho],[\rho^2]$. The trivial defect sector $\mathcal{C}_1=\langle1,\psi_1,\psi_2,\psi_3\rangle$ has two $\mathbb{Z}_3$-orbits $1$ and $\Psi=\psi_1+\psi_2+\psi_3$. These are the quasiparticle super-sectors for trivial gauge flux. As $1$ is fixed by the entire gauge group, it can carry any of the three $\mathbb{Z}_3$ charges $1,z_3,\overline{z}_3,$ which are distinguished by their braiding phases around a threefold flux $\rho$ according to the irreducible representations $1,e^{2\pi i/3},e^{-2\pi i/3}$ respectively. The super-sector $\Psi$ is incapable of carrying a gauge charge because the only symmetry that fixes a component of $\Psi$ is the trivial one, which only has a trivial representation. 

Now let us consider the non-trivial flux sectors. The threefold flux $\rho$ does not carry species labels because the quotient $\mathcal{A}_{SO(8)_1}/(1-\Lambda_3)\mathcal{A}_{SO(8)_1}$ is trivial. Hence, the centralizer is the whole group $\mathbb{Z}_3$, which has three irreducible representations. Thus there are three inequivalent fluxes $\rho_0,\rho_1,\rho_2$ which differ from each other by the gauge charge they carry. Similarly there are three reverse fluxes $\overline{\rho}_0,\overline{\rho}_1,\overline{\rho}_2$. This completes the full set of allowed anyons.

Just as in our previous discussions in Section~\ref{sec:stringnet}, the string-net model constructed from the defect subcategory $\mathcal{C}_{\mathbb{Z}_3}$ in \eqref{so(8)defectcategory} will be non-chiral. In this case it captures the gauging phase transition of one of its chiral sectors while leaving the other chiral sector untouched: \begin{align}\begin{diagram}SO(8)_1^L\otimes SO(8)_1^R&\pile{\rTo^{\mbox{\small Gauging $\mathbb{Z}_3$}}\\\lTo_{\mbox{\small Condensation}}}&SU(3)_3^L\otimes SO(8)_1^R\end{diagram}\label{so(8)su(3)diagram2}\end{align} where $L/R$ labels the sectors with opposite chiralities. We can work out more properties of the anyons by constructing the excitations of the string-net theory using the Drinfeld anyon construction: $\chi=\left(x;\mathcal{R}^{x\ast}_\ast\right)$, where $x$ is some simple object in $\mathcal{C}_{\mathbb{Z}_3},$ and the $R$-symbols $\mathcal{R}^{x\ast}_\ast$ are unitary solutions to the \hexeq's in \eqref{hexagoneq} (see also Fig.~\ref{fig:hexagon1}).

First we begin with $x=1$. The \hexeq's for exchanging $x$ with a pair of threefold defects are \begin{align}\mathcal{R}^{1\rho}_\rho\mathcal{R}^{1\rho}_\rho=\mathcal{R}^{1\overline{\rho}}_{\overline{\rho}},\quad\mathcal{R}^{1\overline{\rho}}_{\overline{\rho}}\mathcal{R}^{1\overline{\rho}}_{\overline{\rho}}=\mathcal{R}^{1\rho}_\rho.\end{align} They have three solutions corresponding to the true vacuum and two $\mathbb{Z}_3$ charges \begin{align}1&=\left(1;\mathcal{R}^{1\rho}_\rho=\left(\mathcal{R}^{1\overline{\rho}}_{\overline{\rho}}\right)^{-1}=1\right)\nonumber\\z_3&=\left(1;\mathcal{R}^{1\rho}_\rho=\left(\mathcal{R}^{1\overline{\rho}}_{\overline{\rho}}\right)^{-1}=e^{2\pi i/3}\right)\label{Z3charges}\\\overline{z}_3&=\left(1;\mathcal{R}^{1\rho}_\rho=\left(\mathcal{R}^{1\overline{\rho}}_{\overline{\rho}}\right)^{-1}=e^{4\pi i/3}\right).\nonumber\end{align} They are Abelian bosons that have non-trivial braiding phases $e^{2\pi ni/3}$ around a threefold flux $\rho$. Their fusion properties are fixed by multiplying the braiding phases. \begin{align}z_3\times z_3=\overline{z}_3,\quad z_3\times{\overline{z}_3}=1.\end{align} Therefore, they correspond the anyons $z_3={\bf 10}$ and $\overline{z}_3=\overline{\bf 10}$ of $SU(3)_3$.

Next we solve the \hexeq's that determine the Drinfeld anyons with $x=\rho$. They are associated with the threefold defect of $SO(8)_1$ and correspond to $\mathbb{Z}_3$ fluxes. Exchanging the defect $\rho$ with another pair of $\rho$'s requires \begin{align}\left[\mathcal{R}^{\rho\rho}_{\overline{\rho}}\right]\left[F^{\rho\rho\rho}_1\right]\left[\mathcal{R}^{\rho\rho}_{\overline{\rho}}\right]=\left[F^{\rho\rho\rho}_1\right]\mathcal{R}^{\rho\overline{\rho}}_1\left[F^{\rho\rho\rho}_1\right]\label{Z3defecthexeq}\end{align} when the triplet fuses to $\rho\times\rho\times\rho\to1.$ Note that the bracketed symbols are $2\times2$ matrices due to the fusion degeneracy $\rho\times\rho=2\overline{\rho}$. Since $F^{\rho\rho\rho}_1$ has unit determinant, \eqref{Z3defecthexeq} implies \begin{align}\det\left[\mathcal{R}^{\rho\rho}_{\overline{\rho}}\right]^2=\left(\mathcal{R}^{\rho\overline\rho}_1\right)^2.\label{Z3defecthexeqdet}\end{align} This condition determines the possible exchange statistics of the $\mathbb{Z}_3$ fluxes \begin{align}\theta_\rho=\mbox{Tr}\left[\mathcal{R}^{\rho\rho}_{\overline{\rho}}\right]=\left(\mathcal{R}^{\rho\overline{\rho}}_1\right)^\ast\varkappa_\rho\label{Z3fluxspin}\end{align} where $\varkappa_\rho$ is the Frobenius-Schur indicator \begin{align}\varkappa_\rho={\left[F^{\rho\overline{\rho}\rho}_\rho\right]_1^1}{\left|\left[F^{\rho\overline{\rho}\rho}_\rho\right]_1^1\right|}^{-1}.\end{align} Here the second equality of \eqref{Z3fluxspin} is the spin-statistics theorem that identifies exchange statistics with a $2\pi$ twist.\cite{Kitaev06} In order for $\mathcal{R}^{\rho\rho}_{\overline{\rho}}$ to be a unitary matrix and $\mbox{Tr}\left[\mathcal{R}^{\rho\rho}_{\overline{\rho}}\right]$ to be a $U(1)$-phase, the $R$-symbol must take the form of $\mathcal{R}^{\rho\rho}_{\overline{\rho}}=\theta_\rho e^{i(\pi/3)\hat{d}\cdot\vec\sigma}$ where $\hat{d}$ is a unit 3-vector, $\sigma_j$ are Pauli matrices, and by  combining \eqref{Z3defecthexeqdet} and \eqref{Z3fluxspin}, the phase $\theta_\rho$ must satisfy \begin{align}{\theta_\rho}^3=\pm\varkappa_\rho.\end{align} 

From Table~\ref{tab:so(8)Fsymbols} and \eqref{so(8)Fsymbolsmod}, there are three (cohomologically) inequivalent Frobenius-Schur indicators $\varkappa_\rho\in\mathbb{Z}_3$. They correspond to three distinct sets of solutions, and describe three different gauged phases: \begin{align}\theta_{\rho_n}=\pm e^{2\pi (m+3n)i/9},\quad\mbox{for }\varkappa_\rho=e^{2\pi mi/3}\label{Z3fluxspins}\end{align} where $m$ is fixed, and $n=0,1,2$ label three distinct $\mathbb{Z}_3$ flux-charge composites $\rho_n$ which satisfy \begin{align}\rho_n\times z_3=\rho_{n+1},\quad\rho_n\times\overline{z}_3=\rho_{n-1}.\end{align} Since the string-net model is non-chiral, and only the symmetry in the left sector is gauged, the right $SO(8)_1^R$ sector is left unchanged. Combining the threefold flux with a fermion in the opposite sector $\rho^L\times\psi^R$ will change the spin statistics relation by a minus sign in \eqref{Z3fluxspins}. 

We see the exchange statistics $\theta_{\rho_n}$ only matches the conformal dimensions of the primary fields ${\bf 3}$, ${\bf 6}$ and ${\bf 15}$ in $SU(3)_3$  for a particular $m$ (see Table~\ref{tab:su(3)3spindim}), which is fixed by the Frobenius-Schur indicator $\varkappa_\rho$. The three distinct theories originate from three different symmetry enriched topological phases of the underlying $SO(8)_1$ state, and each is separated from another by a $\mathbb{Z}_3$ SPT. Upon gauging, the threefold flux also passes through the additional SPT (see Fig.~\ref{fig:SPT}). This accounts for the difference of exchange statistics.

Next, we see there is a 1-parameter family of solutions to the hexagon equation \eqref{Z3defecthexeq}. Particular solutions can be chosen from the $A_4$ group structure of the $F$-symbols, for example the threefold rotation about the $11\overline{1}$-axis \begin{align}\mathcal{R}^{\rho\rho}_{\overline\rho}=\theta_{\rho_n}\exp\left[\frac{\pi}{3}\left(\frac{\mathcal{A}_{\psi_1}+\mathcal{A}_{\psi_2}-\mathcal{A}_{\psi_3}}{\sqrt{3}}\right)\right]\end{align} where $\mathcal{A}_{\psi_i}$ are given by \eqref{so(8)Arep}. The full $2\pi$ braid $\mathcal{R}^{\rho\rho}_{\overline\rho}\mathcal{R}^{\rho\rho}_{\overline\rho}$ of a pair of threefold fluxes has two distinct eigenvalues $\theta_\rho^2e^{\pm2\pi i/3}$. The ribbon identity (see \eqref{ribbon}) \begin{align}\theta_{\overline\rho}=\vcenter{\hbox{\includegraphics[width=0.5in]{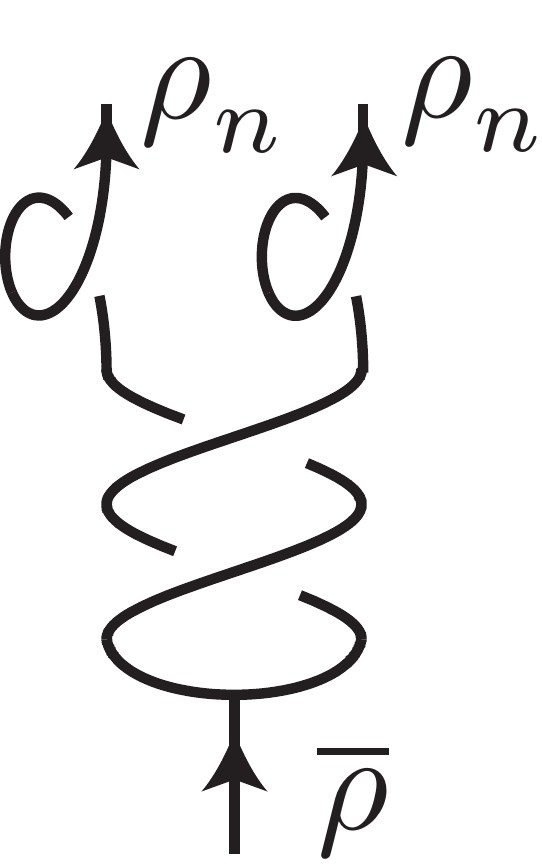}}}=R^{\rho\rho}_{\overline\rho}R^{\rho\rho}_{\overline\rho}\theta_{\rho_n}\theta_{\rho_n}=\theta_{\rho_n}^4e^{\pm2\pi i/3}\label{rhorhoribbon}\end{align} can be used to determine the fusion channels of $\rho\times\rho\to\overline\rho$. For example, for $m=0$, $\rho_0$ is a boson ($\theta_{\rho_0}=1$), as seen in \eqref{Z3fluxspins}, and \eqref{rhorhoribbon} requires the fusion outcome $\rho_0\times\rho_0\to\overline\rho$ to have spin $\theta_{\overline\rho}=e^{\pm2\pi i/3}$. There are therefore two equally possible outcomes $\overline\rho=\overline\rho_1$ or $\overline\rho_2$. In general, the ribbon identity \eqref{rhorhoribbon} forces the fusion rule \begin{align}\rho_n\times\rho_n=\left\{\begin{array}{*{20}c}\overline{\rho}_{n-1}+\overline{\rho}_{n+1}\hfill&\mbox{for $m=0$}\\\overline{\rho}_{n-1}+\overline{\rho}_{n}\hfill&\mbox{for $m=1$}\\\overline{\rho}_{n}+\overline{\rho}_{n+1}\hfill&\mbox{for $m=2$}\end{array}\right. .\label{Z3fluxfusion}\end{align} This resolves the fusion degeneracy $\rho\times\rho=2\overline\rho$ that was present in the defect theory.

Now let us consider the fermion sector. The \hexeq's for the triple $\psi_i\times\psi_j\times\rho$  enforce a requirement on the exchange phase $\mathcal{R}^{\rho\psi_i}_\rho\mathcal{R}^{\rho\psi_j}_\rho=(-1)^{1+\delta_{ij}}\mathcal{R}^{\rho(\psi_i\times\psi_j)}_\rho$. The $\mathbb{Z}_3$ fluxes, restricted to the left chiral sector, correspond to the threefold symmetric solution $\mathcal{R}^{\rho\psi_i}_\rho=-1$.  On the other hand, non-symmetric solutions, like $\mathcal{R}^{\rho\psi_1}_\rho=\mathcal{R}^{\rho\psi_2}_\rho=-\mathcal{R}^{\rho\psi_3}_\rho=-1$, correspond to $\mathbb{Z}_3$ fluxes attached with a fermion in the right (anti-chiral) sector, such as $\rho^L\times\psi_3^R$. 
Similar to the previous examples, the original Abelian anyons group together to form a threefold symmetric non-Abelian super-sector \begin{align}\Psi=\psi_1+\psi_2+\psi_3.\end{align} There is an irreducible solution to the \hexeq's involving exchanging $\Psi$ with a pair of threefold defects \begin{align}\mathcal{R}^{\Psi\rho}_\rho=\left(\mathcal{R}^{\Psi\overline\rho}_{\overline\rho}\right)^{-1}=\left(\begin{array}{*{20}c}0&1&0\\0&0&1\\1&0&0\end{array}\right)\label{Z3super-sectorR}\end{align} where the $3\times3$ matrix acts on the fusion degeneracy $\Psi\times\rho=3\rho$. Its eigenvalues are cube roots of unity, and therefore, if we multiply by an additional exchange phase $\mathcal{R}^{z_3\rho}_\rho=e^{2\pi i/3}$ it only modifies \eqref{Z3super-sectorR} up to a basis transformation. As a result, the super-sector is unchanged by an addition of a threefold charge: \begin{align}z_3\times\Psi=\overline{z}_3\times\Psi=\Psi.\label{Z3super-sectorfusion2}\end{align} The full $2\pi$ braid of the $\Psi$ with a threefold flux $R^{\Psi\rho}_\rho R^{\rho\Psi}_\rho$ has three distinct eigenvalues $-1,-e^{2\pi i/3},-e^{-2\pi i/3}$. They correspond to the three fusion channels \begin{align}\Psi\times\rho_n=\rho_0+\rho_1+\rho_2.\end{align} This can be verified by the ribbon identity and the fermionic statistics $\theta_\Psi=\theta_{\psi_i}=-1$. Finally, the original fusion rule \eqref{Z3super-sectorfusion1} must be modified into \begin{align}\Psi\times\Psi=1+z_3+\overline{z}_3+2\Psi\end{align} due to \eqref{Z3super-sectorfusion2}. This also shows that the super-sector is a simple self-conjugate object after gauging the $\mathbb{Z}_3$ symmetry. The non-Abelian fermion $\Psi$ has a quantum dimension of $d_\Psi=3,$ and matches the adjoint representation ${\bf 8}$ of $SU(3)_3$. This completes the properties of the twist liquid anyons and shows their correspondence to the content of $SU(3)_3.$


\subsection{Gauging the full \texorpdfstring{$S_3$}{S3} symmetry}
We have seen that gauging the $\mathbb{Z}_3$ symmetry that cyclicly rotates the fermions $\psi_1,\psi_2,\psi_3$ of $SO(8)_1$ leads to a non-Abelian $SU(3)_3$-like state. It has ten anyonic excitations including the trivial vacuum $1$, two threefold charges $z_3$ and $\overline{z}_3$, six threefold fluxes $\rho_n$, $\overline{\rho}_n$ for $n=0,1,2$, and a super-sector $\Psi=\psi_1+\psi_2+\psi_3$ originating from the three original fermions. The full anyonic symmetry of $SO(8)_1$ is the permutation group $S_3=\mathbb{Z}_2\ltimes\mathbb{Z}_3$. After gauging $\mathbb{Z}_3,$ the global anyonic $\mathbb{Z}_2$ symmetry remains  un-gauged  in the $su(3)_3$-like state. The remaining $\mathbb{Z}_2$ is identified with the conjugation symmetry that acts to switch $z_3\leftrightarrow\overline{z}_3,$ since the threefold and twofold symmetries do not commute. Additionally the symmetry switches $\rho_n\leftrightarrow\overline{\rho}_n$, which we can see from the adjoint action of the twofold operator, $\sigma\rho\sigma^{-1}=\rho^{-1}$. One can also see this from the fact that the threefold charges must be flipped to preserve the braiding phases $\mathcal{D}S_{z_3\rho_n}=\mathcal{D}S_{\overline{z}_3\overline{\rho}_n}=e^{2\pi i/3}$.

\begin{table}[htbp]
\centering
\begin{tabular}{lll}
Anyons $\chi$ & Dimensions $d_\chi$ & Spins $\theta_\chi$\\\hline
$1$ & $1$ & $1$\\
$z_2$ & $1$ & $1$\\
$z_3+\overline{z}_3$ & $2$ & $1$\\
$\rho_0+\overline{\rho}_0$ & 4 & $e^{2\pi im/9}$\\
$\rho_1+\overline{\rho}_1$ & 4 & $e^{2\pi i(m+3)/9}$\\
$\rho_2+\overline{\rho}_2$ & 4 & $e^{2\pi i(m-3)/9}$\\
$\Psi$ & $3$ & $-1$\\
$z_2\Psi$ & $3$ & $-1$\\
$\Sigma^0$ & $3\sqrt{2}$ & $e^{2\pi i(1+2s)/16}$\\
$z_2\Sigma^0$ & $3\sqrt{2}$ & $e^{2\pi i(9+2s)/16}$\\
$\Sigma^1$ & $3\sqrt{2}$ & $e^{2\pi i(7-2s)/16}$\\
$z_2\Sigma^1$ & $3\sqrt{2}$ & $e^{2\pi i(15-2s)/16}$\\
\end{tabular}
\caption{The quantum dimensions $d_\chi$ and spin/statistics $\theta_\chi=e^{2\pi ih_\chi}$ of deconfined fluxes, charges, and super-sectors of the twist liquid $SO(8)_1/S_3$ derived from gauging the $S_3$ symmetry of $SO(8)_1$. $m=0,1,2$ and $s=0,1$ correspond to six (cohomologically) distinct gauging theories.
}\label{tab:anyonsS3}
\end{table}

Gauging the full $S_3$ symmetry of $SO(8)_1$ is therefore equivalent to gauging the $\mathbb{Z}_2$ conjugation symmetry of the $SU(3)_3$-like state. We will not present the string-net (Drinfeld) construction below, but instead we will argue the anyonic structure from the general gauging principle. The quasiparticle excitations, their spin/statistics, and quantum dimensions are summarized in Table~\ref{tab:anyonsS3}. 

First, the non-self-conjugate anyons in $SU(3)_3$ must group into super-sectors. This includes the threefold charges $z_3+\overline{z}_3$ and fluxes $\rho_n+\overline{\rho}_n$. The two dimensional super-sector $z_3+\overline{z}_3$ can be interpreted as the pure $S_3$ charge corresponding to the 2D irreducible representation $E$ of $S_3$ (see the character table of $S_3$ in Table~\ref{tab:S3character}). The flux super-sector $\rho_n+\overline{\rho}_n$ is a combination of gauge flux $[\rho]=[\rho^2],$ and one of the three gauge charges that irreducibly represents the centralizer $Z_\rho=\mathbb{Z}_3$ of the threefold element of $S_3$. The pure $\mathbb{Z}_2$ charge $z_2$ is characterized by the 1D irreducible representation $A_2$ of $S_3$ that generates a $-1$ braiding phase around a twofold flux. From the tensor product rules for representations \begin{align}\begin{array}{*{20}c}A_2\otimes A_2=A_1,\quad A_2\otimes E=E\\E\otimes E=A_1\oplus A_2\oplus E\end{array}\end{align} we can determine the fusion rules for pure charges \begin{align}\begin{array}{*{20}c}z_2\times z_2=1,\quad z_2\times(z_3+\overline{z}_3)=(z_3+\overline{z}_3)\\(z_3+\overline{z}_3)\times(z_3+\overline{z}_3)=1+z_2+(z_3+\overline{z}_3).\end{array}\label{S3chargefusion}\end{align}

\begin{table}[htbp]
\centering
\begin{tabular}{c|ccc}
Tr&$[1]$&$[\rho]$&$[\sigma]$\\\hline
$A_1$&1&1&1\\
$A_2$&1&1&$-1$\\
$E$&2&$-1$&0\\
\end{tabular}
\caption{The character table of $S_3$. The rows are labeled by irreducible representations and the columns are labeled by conjugacy classes.}\label{tab:S3character}
\end{table}

Since the twofold elements of $S_3$ live outside of the centralizer $Z_\rho=\mathbb{Z}_3$ of the threefold element $\rho$, the fluxes $\rho_n+\overline\rho_n$ cannot carry twofold charges, and therefore \begin{align}z_2\times(\rho_n\times\overline\rho_n)=(\rho_n\times\overline\rho_n).\label{Z2xrho+rho}\end{align} Other fusion rules involving the threefold fluxes originate from the $SU(3)_3$ theory: \begin{align}(\rho_n+\overline{\rho}_n)\times(\rho_n+\overline{\rho}_n)=&1+z_2+\Psi+z_2\Psi\nonumber\\&+(\rho_{n-1+m}+\overline{\rho}_{n-1+m})\nonumber\\&+(\rho_{n+1+m}+\overline{\rho}_{n+1+m}),\nonumber\\ \nonumber\\(z_3+\overline{z}_3)\times(\rho_n+\overline{\rho}_n)=&(\rho_{n-1}+\overline{\rho}_{n-1})\times(\rho_{n+1}+\overline{\rho}_{n+1}),\nonumber\\\Psi\times(\rho_n+\overline\rho_n)&=\sum_{r=0}^2(\rho_r+\overline\rho_r),\label{S3fusionZ3}\end{align} where $m=0,1,2$ are fixed and (cohomologically) label three distinct possible $SU(3)_3$-like states. We note that the appearance of the twofold charge $z_2$ in the first equation is a consequence of \eqref{Z2xrho+rho}. Here the quasiparticle super-sector $\Psi=\psi_1+\psi_2+\psi_3$ can carry a twofold charge: \begin{align}z_2\Psi=z_2\times\Psi.\end{align} This is because each of its components $\psi_i$ are fixed by a twofold permutation $\sigma_i:\psi_{i-1}\leftrightarrow\psi_{i+1}$. The twofold charge therefore carries the non-trivial representation of the centralizer subgroup $Z_1^\Psi=\mathbb{Z}_2$ (see Section~\ref{sec:QPstructuretwistliquid} for the notation). The  fusion of $\Psi$ with the threefold charge is modified to \begin{align}(z_3+\overline{z}_3)\times\Psi=\Psi+z_2\Psi\end{align} required by fusion associativity and $z_2\times(z_3+\overline{z}_3)=z_3+\overline{z}_3$.

Now let us consider the properties of the twofold defects. We mentioned earlier that gauging a $\mathbb{Z}_2$ symmetry of $SO(8)_1$ would lead to the Ising-like theory $SO(N)_1\otimes SO(8-N)_1$ for some odd $N$. However, making a particular twofold defect $\sigma_i$ quantum dynamical breaks the threefold symmetry. This is because the defect twists orbiting fermions asymmetrically, $\psi_i\leftrightarrow\psi_i$ and $\psi_{i-1}\leftrightarrow\psi_{i+1}$. This is related to the fact that $\mathbb{Z}_2$ is not a normal subgroup of the full symmetry group $S_3$. However, the super-sectors \begin{align}\Sigma^0=\sigma_1^0+\sigma_2^0+\sigma_3^0,\quad\Sigma^1=\sigma_1^1+\sigma_2^1+\sigma_3^1\label{so(8)twofoldfluxsector}\end{align} are threefold symmetric. Here the subscripts on the $\sigma$ represent which fermion each twofold defect leaves \emph{fixed}, and the superscripts label the species of the defects so that $\sigma^1_i=\sigma^0_i\times\psi_{i\pm1},$ i.e. the non-trivial species label represents the attachment of a fermion which is not fixed. 
The full centralizer subgroup is $Z_{\sigma_i}=\{1,\sigma_i\}=\mathbb{Z}_2$. Since the symmetry acts trivially on the species labels $\lambda$, which live in the quotient $\mathcal{A}_{SO(8)_1}/(1-\sigma_i)\mathcal{A}_{SO(8)_1},$ then the reduced centralizer is also $\mathbb{Z}_2.$ As a result, the twofold flux can carry a non-trivial twofold charge $z_2$ which  carries the non-trivial representation of the centralizer $Z_{\sigma_i}^\lambda=\mathbb{Z}_2.$ However, it cannot carry a threefold charge. We thus have \begin{align}\begin{array}{*{20}c}z_2\Sigma^\lambda\equiv z_2\times\Sigma^\lambda\\(z_3+\overline{z}_3)\times\Sigma^\lambda=\Sigma^\lambda+z_2\Sigma^\lambda\end{array}\end{align} where the first equation is a notational definition. 

Other fusion rules for these defects can be deduced from what we already know about the rules for the Ising-like state $SO(N)_1\otimes SO(8-N)_1$ that arises from gauging the $\mathbb{Z}_2$ symmetry of the toric code: \begin{gather}\Sigma^\lambda\times\Sigma^\lambda=1+(z_3+\overline{z}_3)+\Psi+\sum_{n=0}^2(\rho_n+\overline{\rho}_n)\nonumber\\\Sigma^0\times\Sigma^1=\Psi+z_2\Psi+\sum_{n=0}^2(\rho_n+\overline{\rho}_n)\nonumber\\\Psi\times\Sigma^0=\Sigma^0+\Sigma^1+z_2\Sigma^1\label{S3fusionZ2}\\\Psi\times\Sigma^1=\Sigma^0+z_2\Sigma^0+z_2\Sigma^1.\nonumber\end{gather} For instance, for $\chi=z_3,\Psi,\rho_n$, the admissible vertex structures $\chi\times\chi\to\overline{\chi}$ in the splitting state \begin{align}\left[\vcenter{\hbox{\includegraphics[width=0.4in]{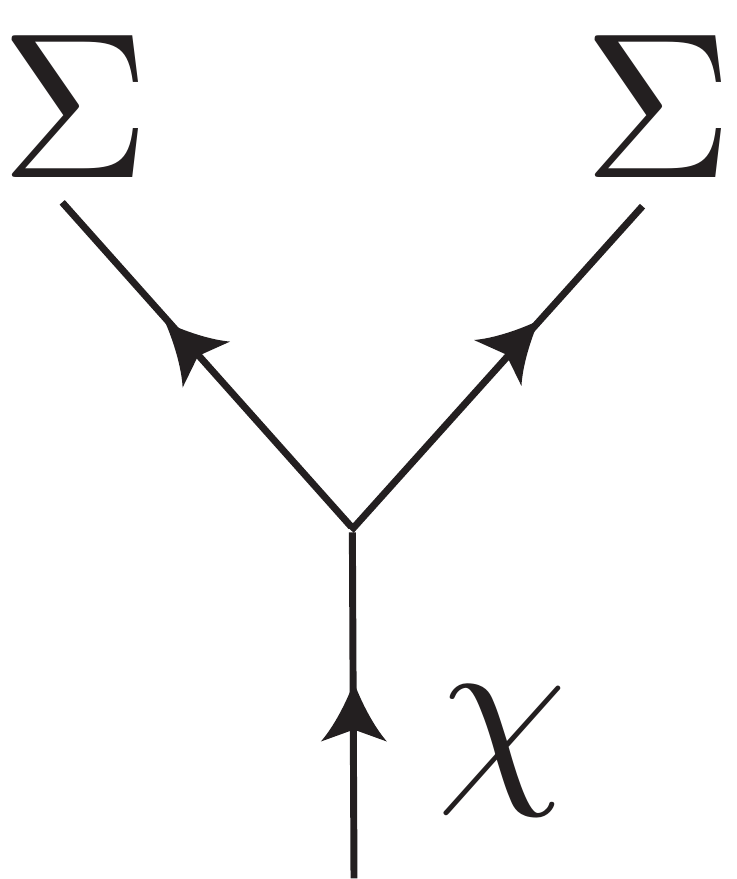}}}\right]=\left[\vcenter{\hbox{\includegraphics[width=0.6in]{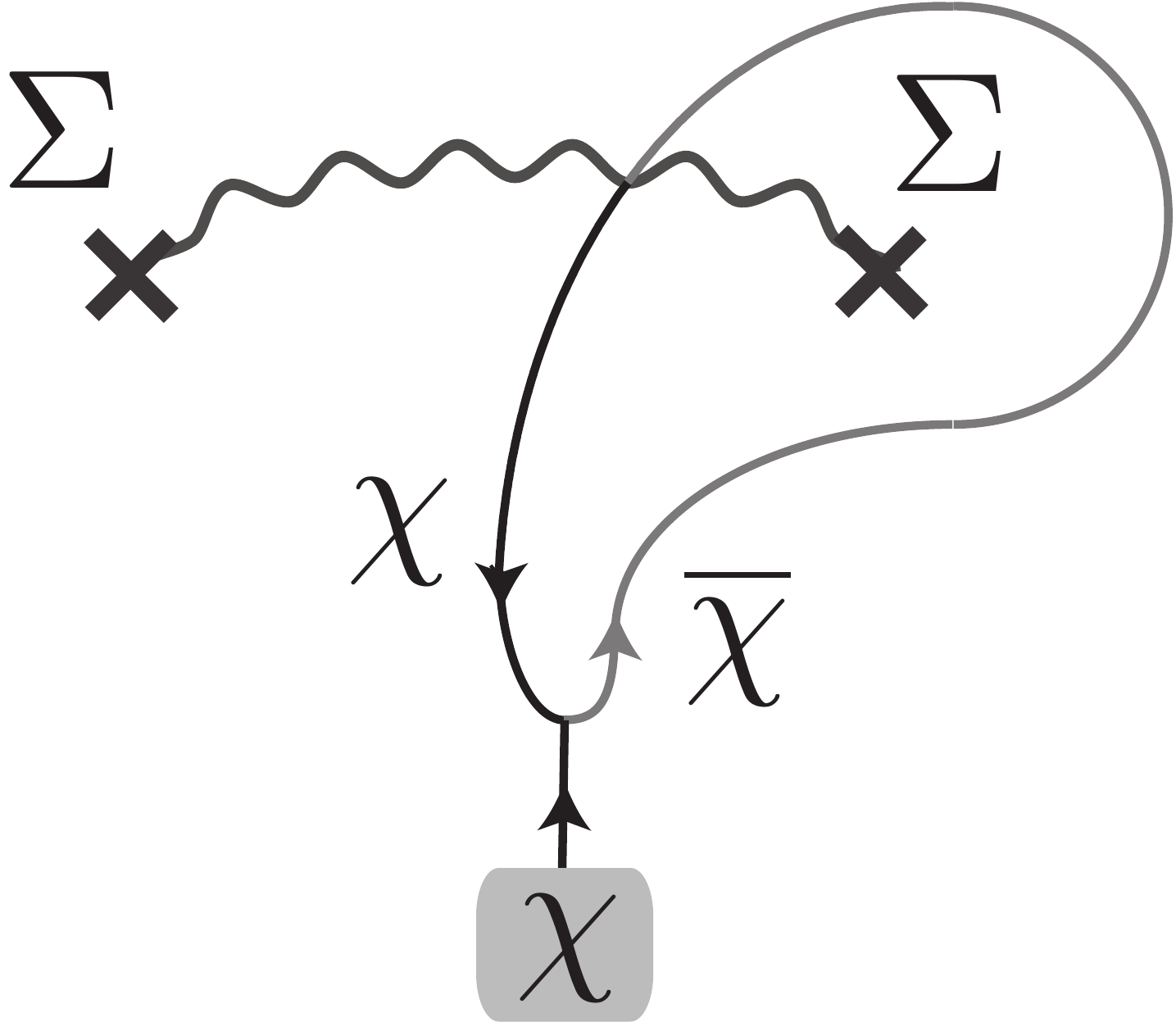}}}\right]\end{align} allow the fusion channels in the first equation of \eqref{S3fusionZ2}. Finally, the fusion of a twofold and a threefold flux must result in twofold fluxes because of the $S_3$ group structure, e.g.~the operation $\rho\sigma_i=\sigma_{i-1}$. We thus have the fusion rule \begin{align}(\rho_n+\overline{\rho}_n)\times\Sigma^\lambda=\Sigma^0+z_2\Sigma^0+\Sigma^1+z_2\Sigma^1.\end{align} Here both species $\lambda=0,1$ are admissible because  a threefold defect can always donate a fermion ($\rho=\psi_{i+1}\times\rho$) and change the species of a twofold defect $(\sigma_i)_{\lambda+1}=\psi_{i+1}\times(\sigma_i)_\lambda$ (see Eq.~\eqref{so(8)splittings} for the splitting state); 
and similarly the twofold charges can be generated by $z_2\times(\rho+\overline\rho)=\rho+\overline\rho$ (see \eqref{Z2xrho+rho}).
Hence, the $\mathbb{Z}_2$ fluxes have quantum dimension $d_\Sigma=3\sqrt{2}.$ Their spin/statistics are determined by their constituents, and are identical to the Ising anyons in the $SO(N)_1\otimes SO(8-N)_1$ state.

We notice that gauging the $S_3$ symmetry could lead to six distinct theories. They are distinguished by the two integer labels $m=0,1,2$ and $s=0,1$ that appear in the spin statistics of the $\mathbb{Z}_3$ and $\mathbb{Z}_2$ fluxes respectively (Table~\ref{tab:anyonsS3}). The integer $m$ also dictates the fusion rules of a pair of $\mathbb{Z}_3$ fluxes (see \eqref{Z3fluxfusion} and \eqref{S3fusionZ3}). These quantities originate from the Frobenius-Schur indicators\cite{FredenhagenRehrenSchroer92,Kitaev06} $\varkappa_\sigma=[F^{\sigma\sigma\sigma}_\sigma]_1^1\left|[F^{\sigma\sigma\sigma}_\sigma]_1^1\right|^{-1}=(-1)^s$ and $\varkappa_\rho=[F^{\rho\overline\rho\rho}_\rho]_1^1\left|[F^{\rho\overline\rho\rho}_\rho]_1^1\right|^{-1}=e^{2\pi im/3}$ (see eq.\eqref{IsingFSindicator} and \eqref{SU3FSindicator}) that govern the particle-antiparticle duality \eqref{bendingdef} of the twofold and threefold defects. These $U(1)$-phases in the $F$-symbols cannot be absorbed by basis transformations of fusion vertices, and they correspond to the six cohomology classes $(m,s)\in\mathbb{Z}_3\times\mathbb{Z}_2=H^3(S_3,U(1))$ that classify inequivalent defect fusion theories of $SO(8)_1$. Different theories are physically represented by adding a 2D bosonic $S_3$-SPT to the $SO(8)_1$ state. These SPT states are classified by the same cohomology group. The addition of the SPT state does not alter the topological order or anyon structure because SPT's are short range entangled states. However, the basis transformations ($F$-symbols) of defects are modified by non-trivial phases as twist defects now live on both the $SO(8)_1$ and the SPT ``layers" (see Fig.\ref{fig:SPT}).

We can summarize the anyon braiding via the modular $S$-matrix in \eqref{S3Smatrix} which is calculated by \eqref{braidingS} using the fusion and spin statistics information:\begin{widetext}
\begin{align}S=\frac{1}{\mathcal{D}}\left(
\begin{smallmatrix}
 1 & 1 & 2 & 4 & 4 & 4 & 3 & 3 & 3 \sqrt{2} & 3 \sqrt{2} & 3 \sqrt{2} & 3 \sqrt{2} \\
 1 & 1 & 2 & 4 & 4 & 4 & 3 & 3 & -3 \sqrt{2} & -3 \sqrt{2} & -3 \sqrt{2} & -3 \sqrt{2} \\
 2 & 2 & 4 & -4 & -4 & -4 & 6 & 6 & 0 & 0 & 0 & 0 \\
 4 & 4 & -4 & -8\cos\frac{4\pi m}{9} & -8\cos\frac{4\pi(m-3)}{9} & -8\cos\frac{4\pi(m+3)}{9} & 0 & 0 & 0 & 0 & 0 & 0 \\
 4 & 4 & -4 & -8\cos\frac{4\pi(m-3)}{9} & -8\cos\frac{4\pi(m+3)}{9} & -8\cos\frac{4\pi m}{9} & 0 & 0 & 0 & 0 & 0 & 0 \\
 4 & 4 & -4 & -8\cos\frac{4\pi(m+3)}{9} & -8\cos\frac{4\pi m}{9} & -8\cos\frac{4\pi(m-3)}{9} & 0 & 0 & 0 & 0 & 0 & 0 \\
 3 & 3 & 6 & 0 & 0 & 0 & -3 & -3 & -3 \sqrt{2} & -3 \sqrt{2} & 3 \sqrt{2} & 3 \sqrt{2} \\
 3 & 3 & 6 & 0 & 0 & 0 & -3 & -3 & 3 \sqrt{2} & 3 \sqrt{2} & -3 \sqrt{2} & -3 \sqrt{2} \\
 3 \sqrt{2} & -3 \sqrt{2} & 0 & 0 & 0 & 0 & -3 \sqrt{2} & 3 \sqrt{2} & 0 & 0 & 6 & -6 \\
 3 \sqrt{2} & -3 \sqrt{2} & 0 & 0 & 0 & 0 & -3 \sqrt{2} & 3 \sqrt{2} & 0 & 0 & -6 & 6 \\
 3 \sqrt{2} & -3 \sqrt{2} & 0 & 0 & 0 & 0 & 3 \sqrt{2} & -3 \sqrt{2} & 6 & -6 & 0 & 0 \\
 3 \sqrt{2} & -3 \sqrt{2} & 0 & 0 & 0 & 0 & 3 \sqrt{2} & -3 \sqrt{2} & -6 & 6 & 0 & 0 
\end{smallmatrix}
\right)\label{S3Smatrix}\end{align}
\end{widetext}
where the total quantum dimension is $\mathcal{D}=12$, and the entries are arranged to have the same order as the anyon listed in Table~\ref{tab:anyonsS3}. This matrix depends on the cohomological parameter $m$, but not $s$, however, generically one would expect the $S$-matrix to have a dependence on both. For this case, while the $S$-matrix is $s$-independent, the $T$-matrix, which contains the spin information, will depend on $s$ as seen above.


\section{Triality Symmetry in the chiral ``4-Potts" Phase}\label{sec:4statePotts}

Previously, we discussed the globally $S_3$-symmetric $SO(8)_1$ state, which led to the $\mathbb{Z}_2$, $\mathbb{Z}_3,$ and $S_3$-twist liquids in \eqref{so(8)gaugingdiagram}. That was the simplest Abelian state that carries a global triality anyonic symmetry. Now, as a lead-in to future work, we will explore a non-Abelian parent state that possesses a $S_3$-anyonic symmetry. Although the general gauging framework presented previously in Section~\ref{sec:gauginganyonicsymmetries} tackles only Abelian parent states in general, we will see that the main ideas also apply in this non-Abelian case. For instance, the anyons in the twist liquid can still be understood as compositions of gauge fluxes, quasiparticle super-sectors, and gauge charges. We also still find that the total quantum dimension increases by a factor of $|G|$, the order of the anyonic symmetry group, after gauging. This strongly suggests the gauging framework in Section~\ref{sec:gauginganyonicsymmetries} should be applicable to all globally symmetric parent states regardless of their character.

\begin{table}[htbp]
\centering
\begin{tabular}{lll}
Anyons $\chi$&Quantum Dim $d_\chi$& Conformal Dim $h_\chi$\\\hline
1&1&0\\
$j_1,j_2,j_3$&1&1\\
$\Phi$&2&1/4\\
$\sigma_1,\sigma_2,\sigma_3$&2&1/16\\
$\tau_1,\tau_2,\tau_3$&2&9/16\\
\end{tabular}
\caption{The spins (encoded as conformal dimensions)  $h_\chi$ and quantum dimensions $d_\chi$ of anyons in the $S_3$-symmetric parent state $``SU(2)_1/Dih_2"=\mbox{``4-Potts"}$.}\label{tab:4statePottsanyons}
\end{table}
The $S_3$-symmetric parent state under consideration here has an anyon content that corresponds to the 11 primary fields in one of the chiral sectors of the 4-state Potts model.\cite{DijkgraafVafaVerlindeVerlinde99, CappelliAppollonio02} The $(1+1)D$ boundary of the (2+1)D topological parent state we want to consider is characterized by the orbifold CFT $SU(2)_1/Dih_2$,\cite{Ginsparg88, Harris88} where $Dih_2$ is the (double cover) group of $\pi$-rotations about $x,y,z,$ which are a subgroup in the continuous 3D rotation group $SU(2)$. Since we lack a better name, and to abbreviate the terminology, we will refer to this $(2+1)D$ bulk topological state by its boundary theory, and will call it either  \begin{align}\mbox{``4-Potts"} \;\;{\rm{or\; equivalently}}\;\;``SU(2)_1/Dih_2"\end{align} where the quotation marks indicate that it is referring to the $(2+1)$D bulk. The spins (encoded as conformal dimensions) and the quantum dimensions of the anyons in this theory are listed in Table~\ref{tab:4statePottsanyons}. 

The fusion rules are generated by \begin{gather}j_a\times j_a=1,\quad j_a\times j_{a\pm1}=j_{a\mp1}\nonumber\\\Phi\times j_a=\Phi,\quad\Phi\times\Phi=1+j_1+j_2+j_3\nonumber\\\sigma_a\times\sigma_a=1+j_a+\Phi\label{4statePottsfusion}\\\sigma_a\times\sigma_{a\pm1}=\sigma_{a\mp1}+\tau_{a\mp1}\nonumber\\\sigma_a\times j_a=\sigma_a,\quad\sigma_a\times j_{a\pm1}=\tau_a,\quad\sigma_a\times\Phi=\sigma_a+\tau_a\nonumber\end{gather} where $a=1,2,3$ modulo $3\mathbb{Z}$. The $F$-symbols that generate basis transformations can be evaluated (up to gauge transformations) by solving the hexagon and pentagon equation (Eq.!\eqref{hexagoneq} and Fig.~\ref{fig:Fpentagon}). In particular we choose \begin{gather}F^{j_\mu j_\nu j_\lambda}_{j_\mu\times j_\nu\times j_\lambda}=F^{j_a\Phi j_a}_\phi=F^{\Phi j_a\Phi}_{j_a}=1\nonumber\\F_\Phi^{\Phi j_aj_a}=F_\Phi^{j_aj_a\Phi}=F_{j_a}^{j_a\Phi\Phi}=F_{j_a}^{\Phi\Phi j_a}=1\nonumber\\F^{j_a\Phi j_{a\pm1}}_\Phi=F^{\Phi j_a\Phi}_{j_{a\pm1}}=-1\label{4PottsFsymbols}\\F_\Phi^{\Phi\Phi\Phi}=\frac{1}{2}\left(\begin{smallmatrix}1&1&1&1\\1&1&-1&-1\\1&-1&1&-1\\1&-1&-1&1\end{smallmatrix}\right),\quad F^{\sigma_a\sigma_a\sigma_a}_{\sigma_a}=\frac{1}{2}\left(\begin{smallmatrix}1&1&\sqrt{2}\\1&1&-\sqrt{2}\\\sqrt{2}&-\sqrt{2}&0\end{smallmatrix}\right)\nonumber\\F_\Phi^{\Phi j_aj_{a\pm1}}=\left(F_\Phi^{j_aj_{a\pm1}\Phi}\right)^{-1}=F_{j_a}^{j_{a\pm1}\Phi\Phi}=\left(F_{j_{a\pm1}}^{\Phi\Phi j_a}\right)^{-1}=\pm i\nonumber\end{gather} where the rows and columns for $F^{\Phi\Phi\Phi}_\Phi$ are arranged according to the internal fusion channels $\{1,j_1,j_2,j_3\}$ of $\Phi\times\Phi,$ and those for $F^{\sigma_a\sigma_a\sigma_a}_{\sigma_a}$ are arranged according to $\{1,j_a,\Phi\}$. These will be useful in understanding the defect fusion category later.

The modular $S$-matrix that characterizes braiding can be generated from the spin and fusion properties via Eq.~\eqref{braidingS} and is given by:\cite{DijkgraafVafaVerlindeVerlinde99, CappelliAppollonio02} 
\begin{align}S=\frac{1}{\mathcal{D}_0}\left(
\begin{smallmatrix}
 1 & 1 & 1 & 1 & 2 & 2 & 2 & 2 & 2 & 2 & 2\\
 1 & 1 & 1 & 1 & 2 & 2 & -2 & -2 & 2 & -2 & -2\\
 1 & 1 & 1 & 1 & 2 & -2 & 2 & -2 & -2 & 2 & -2\\
 1 & 1 & 1 & 1 & 2 & -2 & -2 & 2 & -2 & -2 & 2\\
 2 & 2 & 2 & 2 & -4 & 0 & 0 & 0 & 0 & 0 & 0\\
 2 & 2 & -2 & -2 & 0 & \sqrt{8} & 0 & 0 & -\sqrt{8} & 0 & 0\\
 2 & -2 & 2 & -2 & 0 & 0 & \sqrt{8} & 0 & 0 & -\sqrt{8} & 0\\
 2 & -2 & -2 & 2 & 0 & 0 & 0 & \sqrt{8} & 0 & 0 & -\sqrt{8}\\
 2 & 2 & -2 & -2 & 0 & -\sqrt{8} & 0 & 0 & \sqrt{8} & 0 & 0\\
 2 & -2 & 2 & -2 & 0 & 0 & -\sqrt{8} & 0 & 0 & \sqrt{8} & 0\\
 2 & -2 & -2 & 2 & 0 & 0 & 0 & -\sqrt{8} & 0 & 0 & \sqrt{8}\\
\end{smallmatrix}
\right)\label{4statePottsSmatrix}\end{align}
where the total quantum dimension is $\mathcal{D}_0=4\sqrt{2}$, and the entries are arranged to have the same order as the anyon listed in Table~\ref{tab:4statePottsanyons}.

The chiral ``4-state Potts" phase is $S_3$-symmetric because the fusion, spin, and braiding properties are invariant under the simultaneous permutation of $\{j_1,j_2,j_3\}$, $\{\sigma_1,\sigma_2,\sigma_3\},$ and $\{\tau_1,\tau_2,\tau_3\}$. The threefold ($\theta$) and twofold ($\alpha_a$) generators of the group respectively relabel \begin{align}\begin{array}{*{20}c}\theta:(j_a,\sigma_a,\tau_a)\to(j_{a+1},\sigma_{a+1},\tau_{a+1})\\\alpha_a:(j_{a\pm1},\sigma_{a\pm1},\tau_{a\pm1})\to(j_{a\mp1},\sigma_{a\mp1},\tau_{a\mp1})\end{array}\label{S3operation4Potts}\end{align} while fixing the other anyons. 

\subsection{The symmetry tower \texorpdfstring{$SU(2)_1/\Gamma$}{SU(2)/G}}
Remarkably, the chiral ``4-state Potts" phase lies in the middle of an interesting series of topological states. The series begins with the conventional $SU(2)_1$ state, which has the same topological order as the Laughlin $\nu=1/2$ bosonic FQH state. It is described by the $(2+1)D$ Chern-Simons theory \eqref{toriccodeCSaction} with $K=2$, and this Abelian theory contains the anyon content $\{1,\phi\}$, where $\phi$ is a semion with spin $h_\phi=1/4$. The boundary theory of this state is realized by the $(1+1)D$ conformal field theory with an $SU(2)$ current algebra at level 1.\cite{bigyellowbook} This CFT also appears in one of the chiral sectors of the anti-ferromagnetic Heisenberg spin chain (also known as the XXX model).\cite{Baxterbook}

Beginning with this edge CFT,  Ginsparg, in Ref.~\onlinecite{Ginsparg88}, observed a series of CFT's with central charge $c=1$ generated by taking orbifolds of the $SU(2)_1$ CFT. This family of orbifolds is characterized as a set of $(1+1)D$ theories defined on periodic space-time where the $SU(2)_1$ currents $J_x,J_y,J_z$ transform across the periodic boundaries according to operations in a discrete point group  $\Gamma,$ which is a subgroup of $SU(2)$. This series can be classified, as it is known that three dimensional point groups have an $ADE$ classification.\cite{Slodowy83} The $\hat{A}_r$ series corresponds to $r$-fold cyclic rotations $C_r$ around a primary axis. The $\hat{D}_r$ series corresponds to dihedral groups $Dih_r$, each containing a twofold rotation with an axis perpendicular to that of a primary $r$-fold rotation axis. The exceptional groups $\hat{E}_6$, $\hat{E}_7,$ and $\hat{E}_8$  correspond to the tetrahedral group $T$, the octahedral group $O,$ and the icosahedral group $I$ respectively. For a physical interpretation, we can view the family of orbifold CFT's  $SU(2)_1/\Gamma$\cite{Ginsparg88, DijkgraafVafaVerlindeVerlinde99,CappelliAppollonio02} as arising at the self-dual critical points of certain 2D classical restricted solid-on-solid (RSOS) models where nearest neighbor ``heights" correspond to adjacent nodes of the Dynkin diagram of the extended $ADE$ algebra.\cite{Ginsparg88, Pasquier87,AndrewsBaxterForrester84,Baxterbook} However, we will not appeal to this physical interpretation here. 

From the bulk boundary correspondence, these orbifold CFT's are the chiral boundary theories of a series of $(2+1)D$ twist liquids ``$SU(2)_1/\Gamma$" that are related to each other by a sequence of gauging transitions. We note that here, the notation ``$SU(2)_1/\Gamma$" is abused to refer a $(2+1)D$ topological phase with a chiral boundary that carries a $\Gamma$-orbifold CFT of $SU(2)_1$. Specifically, we are interested in the tower of symmetry groups \begin{align}\begin{diagram}C_2&\rInto^{\mathbb{Z}_2}&Dih_2&\rInto^{\mathbb{Z}_3}&T&\rInto^{\mathbb{Z}_2}&O\end{diagram}\label{S4tower}\end{align} where it should be noted that each of these finite point groups here is actually a double cover in $SU(2)$. To be explicit $Dih_2$ is the double-cover of the group of $180^\circ$ rotations about the $x,y,z$-axes, the tetrahedral group $T$ extends $Dih_2$ by the threefold rotations about the $(111)$-axis, and the octahedral group $O$ extends $T$ by twofold rotations about the $(110)$-axis. The important result is that this tower \eqref{S4tower} will lead to a series of twist liquids obtained by sequentially gauging the symmetry groups on the arrows beginning from the $SU(2)_1$ state. 

By applying this idea, we immediately see the parent state we wish to study in this section -- the $(2+1)D$ topological state with the anyon content that corresponds to the chiral 4-state Potts model (see Table~\ref{tab:4statePottsanyons}) -- is actually already a twist liquid of another parent state \begin{align}\vcenter{\hbox{\includegraphics[width=0.3\textwidth]{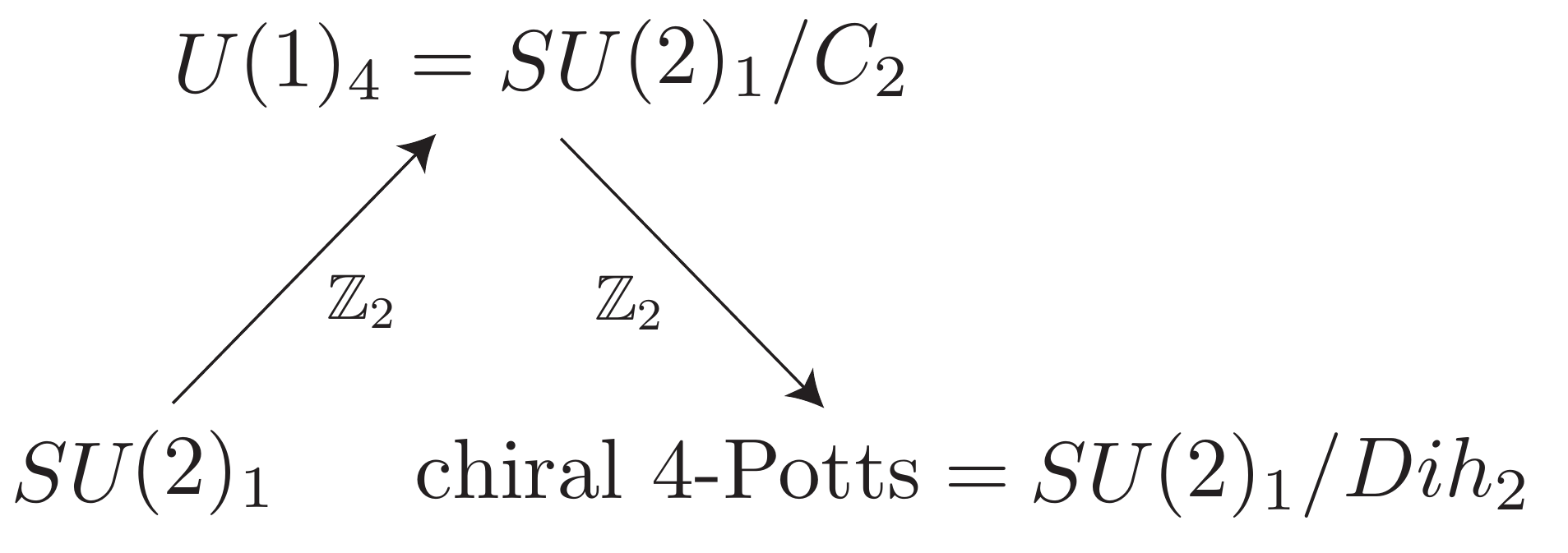}}}.\label{4statePottspregauge}\end{align}Here the first arrow signifies the gauging of a twofold symmetry of the bosonic Laughlin state $SU(2)_1$. As mentioned earlier, the $\mathbb{Z}_2=\{1,s\}$ symmetry does {\em not} relabel the anyons in this state as the semion $\phi$ is self-conjugate. We thus have a non-trivial {\em projected} quantum symmetry\cite{Wenspinliquid02} that is cohomologically classified by $H^2(\mathbb{Z}_2,\mathbb{Z}_2)=\mathbb{Z}_2$ (see Section~\ref{sec:anyonicsymmetries}). For the non-trivial cohomology class we will have the defect fusion rule \begin{align}s\times s=\phi\end{align} instead of the conventional $s\times s=1$, for the (bare) twofold flux $s$ (see \eqref{U(1)NSGfusion} in Section~\ref{sec:defectclassificationobstruction}). This difference is related to the fact that $360^\circ$ rotation in the double cover $SU(2)$ of $SO(3)$ is minus one. 
The resulting twist liquid ``$SU(2)_{1}/C_2$" is topologically identical to the strong paired state $U(1)_4,$ which can be represented in Abelian Chern-Simons theory \eqref{toriccodeCSaction} with $K=8$. The 8 quasiparticles $1,s,s^2,\ldots,s^7$ are generated by the fundamental flux $s$, which has spin $h_s=1/16$. The non-trivial boson $s^4$ is identified as the $\mathbb{Z}_2$ charge of the twist liquid. 

Interestingly, $U(1)_4$ also has a conjugation symmetry $s\leftrightarrow\overline{s}=s^7,$ which is an anyonic symmetry since it clearly changes the anyon types. If we proceed to gauge this twofold symmetry we will find the twist liquid ``$SU(2)_1/Dih_2$"$=$``$U(1)_4/\mathbb{Z}_2$", which  has exactly the anyon content of the ``4-Potts" phase listed in Table~\ref{tab:4statePottsanyons}. We can identify the anyons as follows: \begin{gather}j_1=\mbox{conjugation charge},\quad j_2=s^4,\quad j_3=j_1\times j_2\nonumber\\\Phi=s^2+\overline{s^2}
\\\sigma_1=s+\overline{s},\quad\sigma_2=\mbox{conjugation flux},\quad\sigma_3=s\times\sigma_2\nonumber\\\tau_1=s^3+\overline{s^3},\quad\tau_2=j_1\times\sigma_2,\quad\tau_3=s\times\tau_2.\nonumber\end{gather} The theory has a total quantum dimension of \begin{align}\mathcal{D}_{\mbox{\small 4-Potts}}=4\sqrt{2}\end{align} which is twice of that of the strong paired state $U(1)_4$, and four times that of the Laughlin state $SU(2)_1$. This is exactly what one would expect from our general analysis, and it shows that the ``4-Potts" state is already a twist liquid. 
There is a subtlety here which we have avoided by successively gauging the symmetries to arrive at ``4-Potts" instead of attempting to gauge the double cover of $Dih_2$ right from the start in the parent $SU(2)_1$ state. The trouble arises from the fact that this group would have non-trivial projective representations and, so far, our anyons have been carrying honest charge representations. This detail adds another layer of complication onto this process, and we will save a full discussion for future work.

As mentioned above, the ``4-Potts" state also has anyonic symmetry, and we want to continue our gauging process. Just like $SO(8)_1$, the $S_3$-symmetry of the chiral 4-state Potts phase can be successively gauged. This gives rise to the twist liquids \begin{align}\vcenter{\hbox{\includegraphics[width=0.3\textwidth]{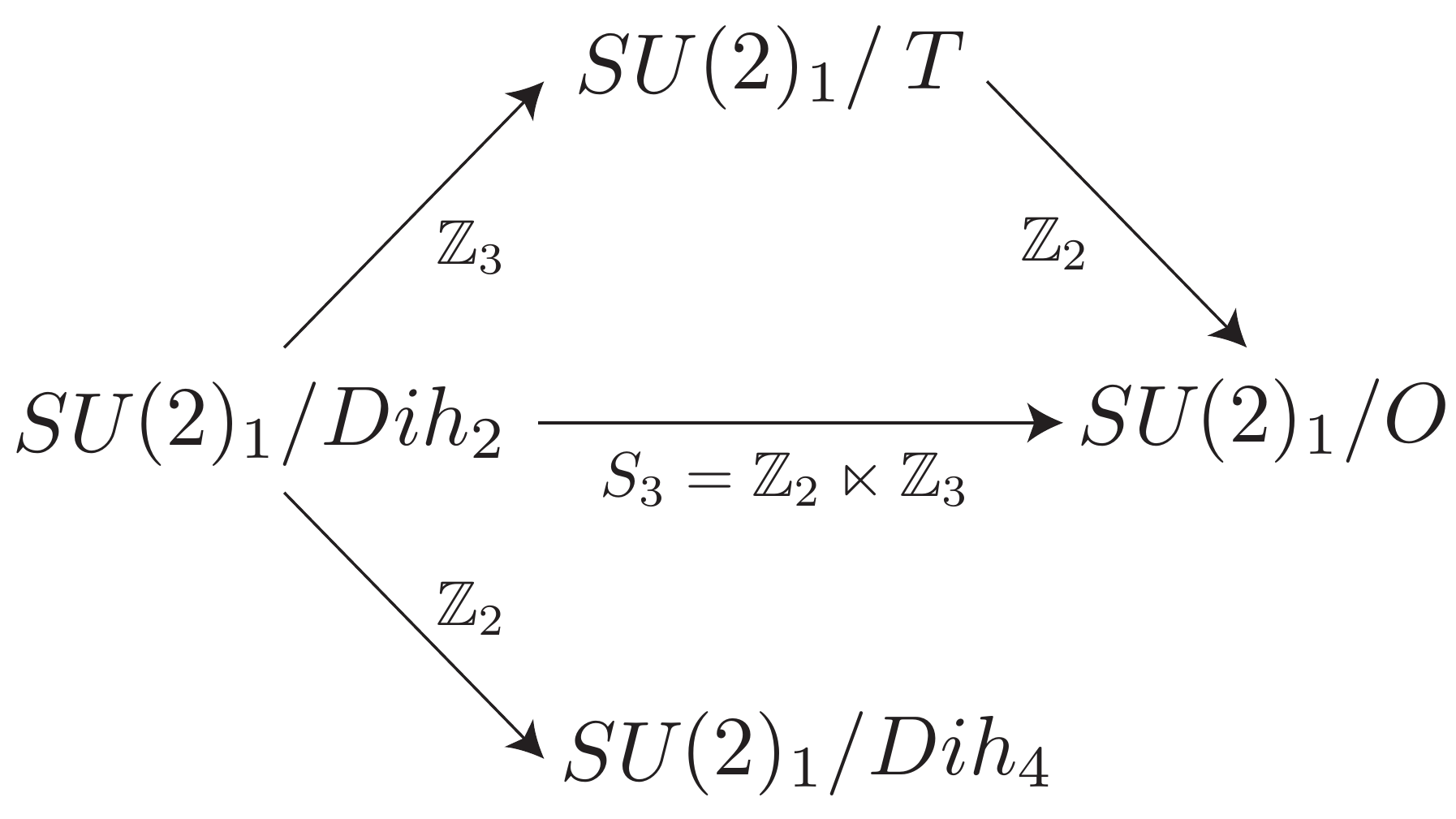}}}\label{eq:4statePottsgaugingdiagram}\end{align} In fact, we see that if we relate this back to the initial $SU(2)_1$ state then the tetrahedral group $T$ and octahedral group $O$ appear, just as we listed in the tower of symmetry subgroups \eqref{S4tower}. For instance, the threefold symmetry $\theta$ and a twofold symmetry $\alpha_a$ of $S_3$ shuffles the twist fields $\sigma_b$ of $Dih_2$ (see \eqref{S3operation4Potts}) according to how the $x,y,z$-axes are rotated by the threefold and twofold symmetries. The non-commutative product of group elements $\alpha_a\theta\alpha_a^{-1}=\theta^{-1}$ between twofold and threefold symmetries matches how a threefold rotation about the $(111)$-axis is reversed by a twofold rotation about the $(110)$-axis. Finally, similar to $SO(8)_1$, the threefold symmetry is broken if any one of the three twofold symmetries $\alpha_a$ is gauged. If we gauge one of the twofold symmetries we break out of the tower pattern and find instead a twist liquid which is related to the $SU(2)_1$ state through the gauging of the symmetry group $Dih_4.$ 
This dihedral group contains a special fourfold rotation, say about the $z$-axis, which is a combination of the twofold rotations about $(001)$ and $(110)$. The twofold $(001)$ axis corresponds to one of the non-Abelian anyon $\sigma_3$ in the ``4-state Potts" model, while the $(110)$-axis corresponds to one of the twofold symmetry $\alpha_3$ in \eqref{S3operation4Potts}.

Before continuing on to gauging the $S_3$ symmetry, let us make some comments about this appealing symmetry-tower structure. The $SU(2)_1/\Gamma$ series serves as an example that unifies a series of topological phases $\mathcal{B}/\Gamma_i$ related by gauging. In general, a \emph{solvable} tower of a group $\Gamma,$ is a series of subgroups \begin{align}\begin{diagram}1=\Gamma_0&\rInto^{G_1}&\Gamma_1&\rInto^{G_2}&\ldots&\rInto^{G_n}&\Gamma_n=\Gamma\end{diagram}\label{solvabletower}\end{align} where each $\Gamma_i$ is a {\em normal} subgroup of $\Gamma_{i+1},$ and the quotient $G_i=\Gamma_i/\Gamma_{i-1}$ is Abelian. The solvable tower then leads to a series of twist liquids \begin{align}\begin{diagram}\mathcal{B}&\rTo^{G_1}&\mathcal{B}/\Gamma_1&\rTo^{G_2}&\ldots&\rTo^{G_n}&\mathcal{B}/\Gamma_n\end{diagram}\label{solvabletowerTL}\end{align} starting with the globally $\Gamma$-symmetric parent state $\mathcal{B}$, where each intermediate step $\mathcal{B}/\Gamma_i$ is a $G_{i-1}$-twist liquid, but is also globally $G_i$-symmetric.


We can already learn much about this set of twist liquids from previous work in a different context. For example, Dijkgraaf {\em et.al.}~in Ref.~[\onlinecite{DijkgraafVafaVerlindeVerlinde99}] present the primary fields, as well as the fusion and scaling dimensions, of the orbifold CFT's $SU(2)_1/Dih_2$ and $SU(2)_1/T$. Later Cappelli and D'Appollonio in Ref.[\onlinecite{CappelliAppollonio02}] derived the modular $S$ and $T$ transformations of the $T,O,I$ orbifold CFT's of $SU(2)_1$. 
The corresponding $(2+1)D$ bulk twist liquid phases can be shown to carry an anyon content that matches the primary fields of the known orbifold CFT's. As such, the detailed Drinfeld calculation can, and will, be omitted. For now we will focus on the $S_3$-defect fusion category of the $(2+1)D$ ``$SU(2)_1/Dih_2$" parent state before gauging the anyonic symmetry. This new defect theory provides the input for the Drinfeld construction, and hence the related string-net model that contains the $S_3$-twist liquid ``$SU(2)_1/O$" as its ground state.

\subsection{The \texorpdfstring{$S_3$}{S3}-defect fusion category}
The non-Abelian state ``$SU(2)_1/Dih_2$" has the anyonic symmetry group $S_3=\{1,\theta,\overline\theta,\alpha_1,\alpha_2,\alpha_3\}$ that relabels its 11 anyons according to \eqref{S3operation4Potts}. This forms the defect fusion category \begin{align}\mathcal{C}_{S_3}=\mathcal{C}_1\oplus\mathcal{C}_\theta\oplus\mathcal{C}_{\overline\theta}\oplus\mathcal{C}_{\alpha_1}\oplus\mathcal{C}_{\alpha_2}\oplus\mathcal{C}_{\alpha_3}\end{align} where $\mathcal{C}_1$ is generated by the 11 anyons in the parent state ``$SU(2)_1/Dih_2$" listed in Table~\ref{tab:4statePottsanyons}, and the other sectors are generated by threefold ($\theta,\bar{\theta},\omega,\bar{\omega}$) and twofold ($\alpha,\mu$) twist defects: \begin{align}\begin{array}{*{20}c}\mathcal{C}_\theta=\langle\theta,\omega\rangle,\quad\mathcal{C}_{\overline\theta}=\langle\overline\theta,\overline\omega\rangle\hfill\\\mathcal{C}_{\alpha_a}=\left\langle\alpha_a^0,\alpha_a^1,\alpha_a^2,\alpha_a^3,\boldsymbol\mu_a\right\rangle\end{array}.\end{align} 
These defects can be understood as the fluxes of the octahedral point group. The defects $\theta$ and $\omega$ correspond to threefold rotations about diagonal axes like $(111)$, $\boldsymbol\mu_a$ corresponds to twofold rotations about axes such as $(110)$, and $\alpha_a$ correspond to fourfold rotations about the $x,y,z$-axes. We have chosen a bold symbol for $\boldsymbol\mu_a$ because, as we will see below, it will carry a larger quantum dimension than the $\alpha_a.$

We provide a detailed derivation of the relevant features of the defect fusion category in Appendix \ref{app:4statefusion}. For now we will simply list the important results. 
We begin with the threefold defects in $\mathcal{C}_\theta$.
There are two species of threefold defects $\theta$ and $\omega$ and they differ from each other by the fusion with the semion $\phi$ from the bosonic Laughlin state $SU(2)_1$, i.e.~$\theta\times\phi=\omega$. 
The quantum dimensions of the threefold defects (derived in Appendix \ref{app:4statefusion}) are  \begin{align}d_\theta=d_\omega=4.\end{align} This implies that the total quantum dimension for the defect sector $\mathcal{C}_\theta=\langle\theta,\omega\rangle$ is  \begin{align}\mathcal{D}_{\mathcal{C}_\theta}=\sqrt{d_\theta^2+d_\omega^2}=4\sqrt{2}=\mathcal{D}_0,\end{align} which is the same as the dimension of the parent state.

Next we describe the twofold defects in $\mathcal{C}_{\alpha_a}$. For each $a=1,2,3$ there are four species of twofold defects $\alpha^s_a$ for $s\in\mathbb{Z}_4=\{0,1,2,3\},$ which are distinguished by an algebraic structure shown in Appendix \ref{app:4statefusion}.
There is also another twofold defect $\boldsymbol\mu_a$, which is related to the others by \begin{align}&\boldsymbol\mu_a=\alpha_a^s\times\sigma_{a\pm1}=\alpha_a^s\times\tau_{a\pm1}\end{align} independent of $s=0,1,2,3$. 
The defect quantum dimensions are derived in Appendix \ref{app:4statefusion} and are found to be \begin{align}d_{\alpha_a^s}=2,\quad d_{\boldsymbol\mu_a}=4.\end{align} The total quantum dimension for the defect sector $\mathcal{C}_{\alpha_a}=\langle\alpha^{0}_a,\alpha^{1}_a,\alpha^{2}_a,\alpha^{3}_a,\boldsymbol\mu_a\rangle$ is therefore \begin{align}\mathcal{D}_{\mathcal{C}_{\alpha_a}}=\sqrt{4d_{\alpha_a^{s}}^2+d_{\boldsymbol\mu_a}^2}=4\sqrt{2}=\mathcal{D}_0,\end{align} which again is the dimension of the parent state. 

Finally, we note that similar to the $SO(8)_1$ state, the non-Abelian $S_3$-symmetry in the ``4-Potts" state implies non-commutative fusion rules between twofold and threefold defects. This is shown explicitly, along with the other fusion rules in the defect fusion category in Appendix~\ref{app:4statefusion}.

\subsection{Gauging the \texorpdfstring{$\mathbb{Z}_3$}{Z3} symmetry}
Gauging the threefold symmetry of the ``$SU(2)_1/Dih_2$" phase (i.e.~the chiral ``4-state Potts" phase) leads to the twist liquid ``$SU(2)_1/T$". There are three inequivalent possibilities labeled by $m=0,1,2$. They can be interchanged between one another by the addition of $\mathbb{Z}_3$-SPTs, which are classified cohomologically by $H^3(\mathbb{Z}_3,U(1))=\mathbb{Z}_3$ (see Section~\ref{sec:gaugingSPT}). Here we will omit the detailed Drinfeld derivations and only highlight the results. The exchange statistics and quantum dimensions of the  21 quasiparticles in the resulting twist liquid are summarized in Table~\ref{tab:QPsu(2)T}. 
\begin{table}[htbp]
\centering
\begin{tabular}{lll}
Anyons $\chi$ & Dimensions $d_\chi$ & Spins $\theta_\chi$\\\hline
$1_n$&$1$&$1$\\
$\Phi_n$&$2$&$e^{2\pi i\frac{1}{4}}$\\
$j=j_1+j_2+j_3$&$3$&$1$\\
$\sigma=\sigma_1+\sigma_2+\sigma_3$&$6$&$e^{2\pi i\frac{1}{16}}$\\
$\tau=\tau_1+\tau_2+\tau_3$&$6$&$e^{2\pi i\frac{9}{16}}$\\
$\omega_n$, $\overline\omega_n$&$4$&$e^{2\pi i\left(\frac{m}{9}+\frac{n}{3}\right)}$\\
$\theta_n$, $\overline\theta_n$&$4$&$e^{2\pi i\left(\frac{1}{4}+\frac{m}{9}+\frac{n}{3}\right)}$\\
\end{tabular}
\caption{The exchange statistics $\theta_\chi=e^{2\pi ih_\chi}$ (which should be distinguished from the anyons $\theta_n$) and quantum dimensions $d_\chi$ of quasiparticles in the $\mathbb{Z}_3$-twist liquid ``$SU(2)_1/T$". $n=0,1,2$ labels the charge components. $m=0,1,2$ is fixed and corresponds to the three (cohomologically) inequivalent twist liquids.}\label{tab:QPsu(2)T}
\end{table}

The $\mathbb{Z}_3$ charges are labeled by $1_n$ for $n=0,1,2$ modulo 3, where $1=1_0$ is the vacuum and $1_2=\overline{1_1}$. $\Phi_n=\Phi\times1_n$ are the three composites between the semion and   $\mathbb{Z}_3$ charges. The bosons $j_a,$ and twist fields $\sigma_a$, $\tau_a,$ are identified by the threefold symmetry and form super-sectors $j=j_1+j_2+j_3$, $\sigma=\sigma_1+\sigma_2+\sigma_3,$ and $\tau=\tau_1+\tau_2+\tau_3$. They cannot carry $\mathbb{Z}_3$ charge as threefold symmetry permutes the individual components. These quasiparticle-charge composites satisfy the fusion rules \begin{gather}1_n\times1_{n'}=1_{n+n'},\quad\Phi_n\times1_{n'}=\Phi_{n+n'}\nonumber\\j\times 1_n=j,\quad\sigma\times1_n=\sigma,\quad\tau\times1_n=\tau\nonumber\\\Phi_n\times\Phi_{n'}=1_{n+n'}+j,\quad j\times j=1_0+1_1+1_2+2j\nonumber\\j\times\Phi_n=\Phi_0+\Phi_1+\Phi_2\\j\times\sigma=\sigma+2\tau,\quad j\times\tau=\tau+2\sigma\nonumber\\\Phi_n\times\sigma=\Phi_n\times\tau=\sigma+\tau\nonumber\\\sigma\times\sigma=\tau\times\tau=\sum_n1_n+\sum_n\Phi_n+j+2\sigma+2\tau\nonumber\\\sigma\times\tau=\sum_n\Phi_n+2j+2\sigma+2\tau,\nonumber\end{gather} and their spin/statistics are inherited from the globally symmetric parent state ``$SU(2)_1/Dih_2$".

The defect sectors $\mathcal{C}_\theta=\langle\theta,\omega\rangle$ and $\mathcal{C}_{\overline\theta}=\langle\overline\theta,\overline\omega\rangle$ are associated to the flux-charge composites $\theta_n=\theta\times1_n$, $\omega_n=\omega\times1_n$ and their conjugates $\overline\theta_n=\overline\theta\times1_{-n}$, $\overline\omega_n=\overline\omega\times1_{-n}$. The fusion rules between a flux and a quasiparticle-charge composite can be inferred from the defect theory. \begin{gather}1_n\times\theta_{n'}=\theta_{n+n'},\quad1_n\times\omega_{n'}=\omega_{n+n'}\nonumber\\j\times\theta_n=\theta_0+\theta_1+\theta_2,\quad j\times\omega_n=\omega_0+\omega_1+\omega_2\nonumber\\\Phi_n\times\theta_{n'}=\omega_{n+n'-1}+\omega_{n+n'+1}\\\Phi_n\times\omega_{n'}=\theta_{n+n'-1}+\theta_{n+n'+1}\nonumber\\\sigma\times\theta_n=\tau\times\theta_n=\theta_0+\theta_1+\theta_2+\omega_0+\omega_1+\omega_2.\nonumber\end{gather} The fusion rules between a pair of fluxes are \begin{gather}\theta_n\times\overline\theta_{n'}=\omega_n\times\overline\omega_{n'}=1_{n-n'}+j+\sigma+\tau\nonumber\\\theta_n\times\overline\omega_{n'}=\Phi_{n-n'-1}+\Phi_{n-n'+1}+\sigma+\tau\label{thetathetafusionAG}\\\theta_n\times\theta_{n'}=\omega_n\times\omega_{n'}=\overline\omega_{m-n-n'}+\sum_{k}\overline\theta_k\nonumber\\\theta_n\times\omega_{n'}=\overline\theta_{m-n-n'}+\sum_k\overline\omega_k.\nonumber\end{gather}

We explicitly see that the exchange statistics of the threefold fluxes $\theta_n$ and $\omega_n,$ as well as the last two fusion rules in \eqref{thetathetafusionAG}, depend on $m=0,1,2$, which labels the three inequivalent twist liquids after gauging the $\mathbb{Z}_3$ symmetry of the ``4-state Potts" phase. For instance, the twist liquid for $m=1$ matches the primary field content of the $T$-orbifold of the $su(2)_1$ CFT in $(1+1)$D\cite{DijkgraafVafaVerlindeVerlinde99, CappelliAppollonio02}, while the other two choices do not. 

Additionally, the braiding $S$-matrix can be derived from the fusion and spin/statistics data using \eqref{braidingS}. Up to a normalization factor of $\mathcal{D}^{-1}$, the unitary $S$-matrix is
\begin{widetext}
\begin{align}
\begin{array}{c|ccccccccc}
&1_l&\Phi_l&j&\sigma&\tau&\omega_l&\overline\omega_l&\theta_l&\overline\theta_l\\\hline
1_k&1&2&3&6&6&4w^k&4w^{-k}&4w^k&4w^{-k}\\
\Phi_k&2&-4&6&0&0&-4w^k&-4w^{-k}&4w^k&4w^{-k}\\
j&3&6&9&-6&-6&0&0&0&0\\
\sigma&6&0&-6&6\sqrt{2}&-6\sqrt{2}&0&0&0&0\\
\tau&6&0&-6&-6\sqrt{2}&6\sqrt{2}&0&0&0&0\\
\omega_k&4w^l&-4w^l&0&0&0&4e^{\frac{2\pi im}{9}}w^{k+l}&4e^{-\frac{2\pi im}{9}}w^{-k-l}&4e^{\frac{2\pi im}{9}}w^{k+l}&4e^{-\frac{2\pi im}{9}}w^{-k-l}\\
\overline\omega_k&4w^{-l}&-4w^{-l}&0&0&0&4e^{-\frac{2\pi im}{9}}w^{-k-l}&4e^{\frac{2\pi im}{9}}w^{k+l}&4e^{-\frac{2\pi im}{9}}w^{-k-l}&4e^{\frac{2\pi im}{9}}w^{k+l}\\
\theta_k&4w^l&4w^l&0&0&0&4e^{\frac{2\pi im}{9}}w^{k+l}&4e^{-\frac{2\pi im}{9}}w^{-k-l}&-4e^{\frac{2\pi im}{9}}w^{k+l}&-4e^{-\frac{2\pi im}{9}}w^{-k-l}\\
\overline\theta_k&4w^{-l}&4w^{-l}&0&0&0&4e^{-\frac{2\pi im}{9}}w^{-k-l}&4e^{\frac{2\pi im}{9}}w^{k+l}&-4e^{-\frac{2\pi im}{9}}w^{-k-l}&-4e^{\frac{2\pi im}{9}}w^{k+l}
\end{array}\label{Smatrixsu(2)T}
\end{align}
\end{widetext}
where $w=e^{2\pi i/3}$, $k,l=0,1,2$ labels the $\mathbb{Z}_3$-charge components, and $m=0,1,2$ labels the three inequivalent twist liquid theories. The total quantum dimension is $\mathcal{D}=12\sqrt{2}$, which differs from the dimension, $\mathcal{D}_0=4\sqrt{2}$, of the parent state by the group order $|\mathbb{Z}_3|=3$.

\subsection{Gauging the full \texorpdfstring{$S_3$}{S3} symmetry}
Gauging the full $S_3$ symmetry of the ``4-state Potts" phase leads to the twist liquid state ``$SU(2)_1/O$". This can be done by gauging the $\mathbb{Z}_2$ conjugation symmetry of the ``$SU(2)_1/T$" twist liquid state presented previously (see Table~\ref{tab:QPsu(2)T}). The conjugation symmetry interchanges $1_1\leftrightarrow 1_2$, $\Phi_1\leftrightarrow\Phi_2$, $\theta_n\leftrightarrow\overline\theta_n,$ and $\omega_n\leftrightarrow\overline\omega_n$. There are, in total, six inequivalent gauging possibilities labeled by $m=0,1,2$ and $t=0,1$. They are related to each other by the addition of $S_3$-SPT's which are classified cohomologically by $H^3(S_3,U(1))=\mathbb{Z}_6=\mathbb{Z}_3\times\mathbb{Z}_2$. Again, we will not present the detailed Drinfeld derivations, but instead only highlight the resulting anyon content after gauging. The exchange statistics and quantum dimensions of the 28 quasiparticles of the $S_3$-twist liquid are summarized in Table~\ref{tab:QPsu(2)O}.
\begin{table}[htbp]
\centering
\begin{tabular}{lll}
Anyons $\chi$ & Dimensions $d_\chi$ & Spins $\theta_\chi$\\\hline
$1=1_0$&1&1\\
$z$&1&1\\
$1_1+1_2$&$2$&1\\
$\Phi_0$, $z\Phi_0$&$2$&$e^{2\pi i\frac{1}{4}}$\\
$\Phi_1+\Phi_2$&$4$&$e^{2\pi i\frac{1}{4}}$\\
$j$, $zj$&3&1\\
$\sigma$, $z\sigma$&6&$e^{2\pi i\frac{1}{16}}$\\
$\tau$, $z\tau$&6&$e^{2\pi i\frac{9}{16}}$\\
$\Omega_n=\omega_n+\overline\omega_n$&8&$e^{2\pi i\left(\frac{m}{9}+\frac{n}{3}\right)}$\\
$\Theta_n=\theta_n+\overline\theta_n$&8&$e^{2\pi i\left(\frac{1}{4}+\frac{m}{9}+\frac{n}{3}\right)}$\\
$\alpha^s=\alpha^s_1+\alpha^s_2+\alpha^s_3$&6&$e^{2\pi i\left[\frac{1}{64}+\frac{s(2s-1)}{8}+\frac{t}{4}\right]}$\\
$\alpha^{s+4}=z\alpha^s$&6&$e^{2\pi i\left[\frac{33}{64}+\frac{s(2s-1)}{8}+\frac{t}{4}\right]}$\\
$\boldsymbol\mu=\boldsymbol\mu_1+\boldsymbol\mu_2+\boldsymbol\mu_3$&12&$e^{2\pi i\left(\frac{1}{16}+\frac{t}{4}\right)}$\\
$z\boldsymbol\mu$&12&$e^{2\pi i\left(\frac{9}{16}+\frac{t}{4}\right)}$
\end{tabular}
\caption{The exchange statistics $\theta_\chi=e^{2\pi ih_\chi}$ and quantum dimensions $d_\chi$ of quasiparticles in the $S_3$-twist liquid ``$SU(2)_1/O$". $n=0,1,2$ labels the $\mathbb{Z}_3$-charge component of $\theta,\omega$, and $s=0,1,2,3$ labels the the $Dih_2$-charge component of $\alpha$. $m=0,1,2$ and $t=0,1$ are fixed and correspond to the six (cohomologically) inequivalent twist liquids.}\label{tab:QPsu(2)O}
\end{table}

The pure $S_3$-charges $1$, $z,$ and $1_1+1_2$ are associated to the three irreducible representations of the symmetry group $S_3=\{1,\theta,\theta^{-1},\alpha_1,\alpha_2,\alpha_3\}$. Each of the threefold fluxes $\Omega_n=\omega_n+\overline\omega_n,$ and $\Theta_n=\theta_n+\overline\theta_n$ is now a combination of the two components corresponding to the two elements in the conjugacy class $[\theta]=\{\theta,\theta^{-1}\}$. The two-component super-sectors $1_1+1_2$, $\Phi_1+\Phi_2$, $\Omega_n,$ and $\Theta_n$ cannot carry a twofold charge as each of their components is not fixed by the conjugation symmetry. The fusion rules from the previous $\mathbb{Z}_3$ twist liquid ``$SU(2)_1/T$" imply \begin{gather}(1_1+1_2)\times(1_1+1_2)=1+z+(1_1+1_2)\nonumber\\\Phi_0\times(1_1+1_2)=\Phi_1+\Phi_2\nonumber\\(\Phi_1+\Phi_2)\times(1_1+1_2)=\Phi_0+z\Phi_0+(\Phi_1+\Phi_2)\nonumber\\j\times(1_1+1_2)=j+zj\nonumber\\\sigma\times(1_1+1_2)=\sigma+z\sigma,\quad\tau\times(1_1+1_2)=\tau+z\tau\nonumber\\\Omega_n\times(1_1+1_2)=\Omega_{n-1}+\Omega_{n+1}\nonumber\\\Theta_n\times(1_1+1_2)=\Theta_{n-1}+\Theta_{n+1}\nonumber\\\Phi_0\times\Phi_0=1+j,\quad j\times\Phi_0=z\Phi_0+(\Phi_1+\Phi_2)\nonumber\\\sigma\times\Phi_0=\tau\times\Phi_0=\sigma+\tau\nonumber\\\Omega_n\times\Phi_0=\Theta_{n-1}+\Theta_{n+1},\quad\Theta_n\times\Phi_0=\Theta_{n-1}+\Omega_{n+1}\nonumber\\j\times j=1+(1_1+1_2)+j+zj\nonumber\\\sigma\times j=\sigma+\tau+z\tau,\quad\tau\times j=\tau+\sigma+z\sigma\nonumber\\\Omega_n\times j=\sum_{n'}\Omega_{n'},\quad\Theta_n\times j=\sum_{n'}\Theta_{n'}\nonumber\\\sigma\times\sigma=\tau\times\tau=1+(1_1+1_2)+\Phi_0+(\Phi_1+\Phi_2)\nonumber\\\quad\quad+\sigma+z\sigma+\tau+z\tau\nonumber\\\sigma\times\tau=\Phi_0+(\Phi_0+\Phi_1)+j+zj+\sigma+z\sigma+\tau+z\tau\nonumber\\\sigma\times\Theta_n=\tau\times\Theta_n=\sum_{n'}\Omega_{n'}+\Theta_{n'}.\nonumber\end{gather} The fusion rules between a pair of threefold fluxes become \begin{align}\Theta_n\times\Theta_n&=(1+z)\times(1+j+\sigma+\tau)\nonumber\\&\quad\quad+\Omega_{m-2n}+\sum_k\Theta_k\nonumber\\\Theta_n\times\Theta_{n\pm1}&=(1_1+1_2)+(1+z)\times(j+\sigma+\tau)\nonumber\\&\quad\quad+\Omega_{m-2n\mp1}+\sum_k\Theta_k\nonumber\\\Theta_n\times\Omega_n&=2(\Phi_1+\Phi_2)+(1+z)\times(\sigma+\tau)\nonumber\\&\quad\quad+\Theta_{m-2n}+\sum_k\Omega_k\nonumber\\\Theta_n\times\Omega_{n\pm1}&=(\Phi_1+\Phi_2)+(1+z)\times(\Phi_0+\sigma+\tau)\nonumber\\&\quad\quad+\Theta_{m-2n\mp1}+\sum_k\Omega_k.\nonumber\end{align}

Twofold fluxes $\alpha^s=\alpha_1^s+\alpha_2^s+\alpha_3^s,$ and $\boldsymbol\mu=\boldsymbol\mu_1+\boldsymbol\mu_2+\boldsymbol\mu_3,$ contain three components as there are three elements in the conjugacy class $[\alpha_a]=\{\alpha_1,\alpha_2,\alpha_3\}$. It is convenient to include the $\mathbb{Z}_2$-charge into the species labels of $\alpha$ so that \begin{align}\alpha^{s+4}=z\times\alpha^s\end{align} where $s$ now runs from 0 to 7 modulo 8. From the defect fusion rules $\mathcal{C}_0\times\mathcal{C}_{\alpha_a}\to\mathcal{C}_{\alpha_a}$, \begin{gather}\alpha^s\times(1_1+1_2)=\alpha^s+z\alpha^s,\quad\boldsymbol\mu\times(1_1+1_2)=\boldsymbol\mu+z\boldsymbol\mu\nonumber\\\Phi_0\times\alpha^s=\alpha^{s-1}+\alpha^{s+1},\quad\Phi_0\times\boldsymbol\mu=\boldsymbol\mu+z\boldsymbol\mu\nonumber\\j\times\alpha^s=\alpha^s+\alpha^{s-2}+\alpha^{s+2},\quad j\times\boldsymbol\mu=\boldsymbol\mu+2z\boldsymbol\mu\nonumber\\\sigma\times\alpha^s=\alpha^{-s}+\alpha^{-s+1}+\boldsymbol\mu+z\boldsymbol\mu\nonumber\\\sigma\times\boldsymbol\mu=\sum_{s=0}^7\alpha^s+\boldsymbol\mu+z\boldsymbol\mu.\nonumber\end{gather} The fusion rules between threefold and twofold defects $\mathcal{C}_\theta\times\mathcal{C}_{\alpha_a}\to\mathcal{C}_{\alpha_{a-1}}$ give \begin{gather}\Theta_n\times\alpha^s=\sum_{u=0}^3\alpha^{s+2u}+\boldsymbol\mu+z\boldsymbol\mu\nonumber\\\Omega_n\times\alpha^s=\sum_{u=0}^3\alpha^{s+2u+1}+\boldsymbol\mu+z\boldsymbol\mu\nonumber\\\boldsymbol\mu\times\Theta_n=\boldsymbol\mu\times\Omega_n=\sum_{s=0}^7\alpha^s+2\boldsymbol\mu+2z\boldsymbol\mu.\nonumber\end{gather} 

Finally, the fusion rules of a pair of twofold fluxes can also be deduced from the defect theory shown in Appendix \ref{app:4statefusion}: 
\begin{widetext}
\begin{gather}\alpha^{s_1}\times\alpha^{s_2}=\left(\begin{smallmatrix}1+(1_1+1_2)+j&\Phi_0+(\Phi_1+\Phi_2)&j+zj&z\Phi_0+(\Phi_1+\Phi_2)\\\Phi_0+(\Phi_1+\Phi_2)&1+(1_1+1_2)+j&\Phi_0+(\Phi_1+\Phi_2)&j+zj\\j+zj&\Phi_0+(\Phi_1+\Phi_2)&1+(1_1+1_2)+j&\Phi_0+(\Phi_1+\Phi_2)\\z\Phi_0+(\Phi_1+\Phi_2)&j+zj&\Phi_0+(\Phi_1+\Phi2)&1+(1_1+1_2)+j\end{smallmatrix}\right)+\left(\begin{smallmatrix}\sigma&\sigma&\tau&z\tau\\\sigma&\tau&z\tau&z\sigma\\\tau&z\tau&z\sigma&z\sigma\\z\tau&z\sigma&z\sigma&z\tau\end{smallmatrix}\right)+\sum_{k=0}^2\left(\begin{smallmatrix}\Theta_k&\Omega_k&\Theta_k&\Omega_k\\\Omega_k&\Theta_k&\Omega_k&\Theta_k\\\Theta_k&\Omega_k&\Theta_k&\Omega_k\\\Omega_k&\Theta_k&\Omega_k&\Theta_k\\\end{smallmatrix}\right)\nonumber\\\alpha^s\times\boldsymbol\mu=\sigma+z\sigma+\tau+z\tau+\sum_{k=0}^2\Theta_k+\sum_{k=0}^2\Omega_k\nonumber\\\boldsymbol\mu\times\boldsymbol\mu=1+(1_1+1_2)+j+2zj+\Phi_0+z\Phi_0+2(\Phi_1+\Phi_2)+(1+z)\times(\sigma+\tau)+2\sum_{k=0}^2\Theta_k+2\sum_{k=0}^2\Omega_k\nonumber
\end{gather}\end{widetext} where the rows and columns in the first equation are arranged in the order of $s_1,s_2=0,1,2,3$.

Again we notice that the exchange statistics of $\Omega_n,\Theta_n,\alpha^s,\boldsymbol\mu$ depend on the cohomology labels $(m,t)\in\mathbb{Z}_3\times\mathbb{Z}_2=H^3(S_3,U(1))$ that are associated to the six inequivalent $S_3$-twist liquids of the ``4-Potts" phase. For instance, the phase with $(m,t)=(1,0)$ corresponds to the $O$-orbifold of the $su(2)_1$ CFT in $(1+1)$D.\cite{CappelliAppollonio02} The braiding $S$-matrix can be evaluated from the exchange statistics and fusion rules by \eqref{braidingS}. Up to a normalization factor of $\mathcal{D}^{-1}$, the unitary $S$-matrix is 
\begin{widetext}
\begin{align}
\begin{array}{c|ccccccccccc}
&1^\pm&\vec{1}&\Phi_0^\pm&\vec{\Phi}&j^\pm&\sigma^\pm&\tau^\pm&\Omega_l&\Theta_l&\alpha^{s_2}&\boldsymbol\mu^{\varsigma_2}\\\hline
1^\pm&1&2&2&4&3&6&6&8&8&\pm6&\pm12\\
\vec{1}&2&4&4&8&6&12&12&-8&-8&0&0\\
\Phi_0^\pm&2&4&-4&-8&6&0&0&-8&8&\pm6\sqrt{2}(-1)^{s_2}&0\\
\vec{\Phi}&4&8&-8&-16&12&0&0&8&-8&0&0\\
j^\pm&3&6&6&12&9&-6&-6&0&0&\pm6&\mp12\\
\sigma^\pm&6&12&0&0&-6&6\sqrt{2}&-6\sqrt{2}&0&0&\pm\chi(s_2)&0\\
\tau^\pm&6&12&0&0&-6&-6\sqrt{2}&6\sqrt{2}&0&0&\pm\psi(s_2)&0\\
\Omega_k&8&-8&-8&8&0&0&0&c_m(k+l)&c_m(k+l)&0&0\\
\Theta_k&8&-8&8&-8&0&0&0&c_m(k+l)&-c_m(k+l)&0&0\\
\alpha^{s_1}&\pm6&0&\pm6\sqrt{2}(-1)^{s_1}&0&\pm6&\pm\chi(s_1)&\pm\psi(s_1)&0&0&c_t(s_1,s_2)&0\\
\boldsymbol\mu^{\varsigma_1}&\pm12&0&0&0&\mp12&0&0&0&0&0&12\sqrt{2}(-1)^{t+\varsigma_1+\varsigma_2}
\end{array}
\end{align}
\end{widetext}
Here, to cut down on the size of the matrix we have had to be creative with the anyon labeling scheme. First, we have  anyons labeled by $x^+=x$ and $x^-=zx$ (for $x=1,\Phi_0,j,\sigma,\tau$) which differ by the fusion with the $\mathbb{Z}_2$ charge $z.$ Next, $\boldsymbol\mu^\varsigma=\boldsymbol\mu$, $z\boldsymbol\mu$ for $\varsigma=0$, $1$ respectively; and $\vec{1}=1_1+1_2$, $\vec\Phi=\Phi_1+\Phi_2$ are the two supsersectors. The integers $k,l=0,1,2$ label the $\mathbb{Z}_3$-charge components of $\Omega$ and $\Theta$, while $s_1,s_2=0,1,\ldots,7$ label the eight $\alpha$'s. The entries $\chi,\psi,c_m,c_t$ are abbreviations for \begin{gather}
\chi(s)=12\cos\left(\frac{\pi}{8}-\frac{s\pi}{2}\right),\quad\psi(s)=12\sin\left(\frac{\pi}{8}-\frac{s\pi}{2}\right)\nonumber\\c_m(n)=16\cos\left[2\pi\left(\frac{2m}{9}+\frac{n}{3}\right)\right]\nonumber\\c_t(s_1,s_2)=12(-1)^t\cos\left[\frac{\pi(4s_1-1)(4s_2-1)}{16}\right].\nonumber\end{gather} These odd factors arise from Wilson loop calculations. For example, the factors $\chi(s)/d_\alpha$ and $\psi(s)/d_\alpha$ are the eigenvalues of the $\sigma$- and $\tau$-loops around the $\alpha$-defects in Eq.~\eqref{alphaloopvalues}. To summarize, the total quantum dimension of the $S_3$-twist liquid is $\mathcal{D}=24\sqrt{2}$, which $|S_3|=6$ times bigger than the dimension $\mathcal{D}_0=4\sqrt{2}$ of the parent state. This is what we expect if we naively apply our arguments based on Abelian parent theories. It is interesting to see that it holds just as well in this more complicated case.

\subsection{Comments on the bilayer toric code}\label{sec:bilayertoriccode}
We end this section on some observations about the bilayer toric code \begin{align}D(\mathbb{Z}_2\times\mathbb{Z}_2)&=\left(\mbox{Toric code}\right)_\uparrow\otimes\left(\mbox{Toric code}\right)_\downarrow\nonumber\\&=SO(8)_1^R\otimes SO(8)_1^L,\end{align} which is identical to a $\mathbb{Z}_2\times\mathbb{Z}_2$ gauge theory in $(2+1)$D. We label the two layers by $s=\uparrow,\downarrow,$ and each layer contains the anyon content $\{1,e_s,m_s,\psi_s\}$. Interestingly, as indicated in the equation above, this model can also be identified as two $SO(8)_1$-states labeled by $R$ and $L$ that indicating opposite chiralities. For example, the six fermions in the system decompose into two groups \begin{align}\begin{array}{*{20}c}\psi^R_1=\psi_\uparrow,&\psi^R_2=e_\uparrow\psi_\downarrow,&\psi^R_3=m_\uparrow\psi_\downarrow\\\psi^L_1=\psi_\downarrow,&\psi^L_2=e_\downarrow\psi_\uparrow,&\psi^R_3=m_\downarrow\psi_\uparrow.\end{array}\end{align} Just like the $SO(8)_1$-state and the ``4-Potts" state considered before, the bilayer toric code contains an $S_3$ anyonic symmetry generated by the layer flip operation \begin{align}\sigma:(e_\uparrow,m_\uparrow)\leftrightarrow(e_\downarrow,m_\downarrow),\quad\psi_a^R\leftrightarrow\psi_a^L,\end{align} and the threefold rotation \begin{align}\rho:(\psi_a^R,\psi_a^L)\to(\psi_{a+1}^R,\psi_{a-1}^L).\end{align} We also note that, in some exactly-solvable spin models on a honeycomb lattice, this $S_3$ symmetry can be realized as point group operations\cite{BombinMartin06, Bombin11, TeoRoyXiao13long}.

It was proposed in Ref.~[\onlinecite{MesarosKimRan13}] that gauging the $\mathbb{Z}_2$ bilayer symmetry of this system would lead to a twist liquid which matches the $D_4$ (dihedral group) discrete gauge theory. This is natural because the $\sigma$ symmetry flips the fluxes $m_\uparrow\leftrightarrow m_\downarrow$ in the parent state, and the group generated by $\sigma,m_\uparrow,m_\downarrow$ is exactly the semi-direct product $D_4=(\mathbb{Z}_2\times\mathbb{Z}_2)\rtimes\mathbb{Z}_2$. One can verify the anyon structure after gauging the anyonic symmetry via the Drinfeld construction. Indeed, the result  matches that of a $D_4$ gauge theory, but only when the Frobenius-Schur indicator is $\varkappa_\sigma=+1.$ On the other hand, the resulting twist liquid will correspond to the $Q_8$ gauge theory (group of quaternions) when $\varkappa_\sigma=-1.$ These two twist liquids differ by a $\mathbb{Z}_2$-SPT as described in Section~\ref{sec:anyonicsymmetries} and \ref{sec:gaugingSPT}, and can alternatively be understood by a $\mathbb{Z}_2^3$ Chern-Simons theory.\cite{Propitius-1995}

Similar to the chiral ``4-Potts" phase, the bilayer toric code belongs to a series of related twist liquids (c.f.~Eq.~\eqref{eq:4statePottsgaugingdiagram}) \begin{align}\vcenter{\hbox{\includegraphics[width=0.25\textwidth]{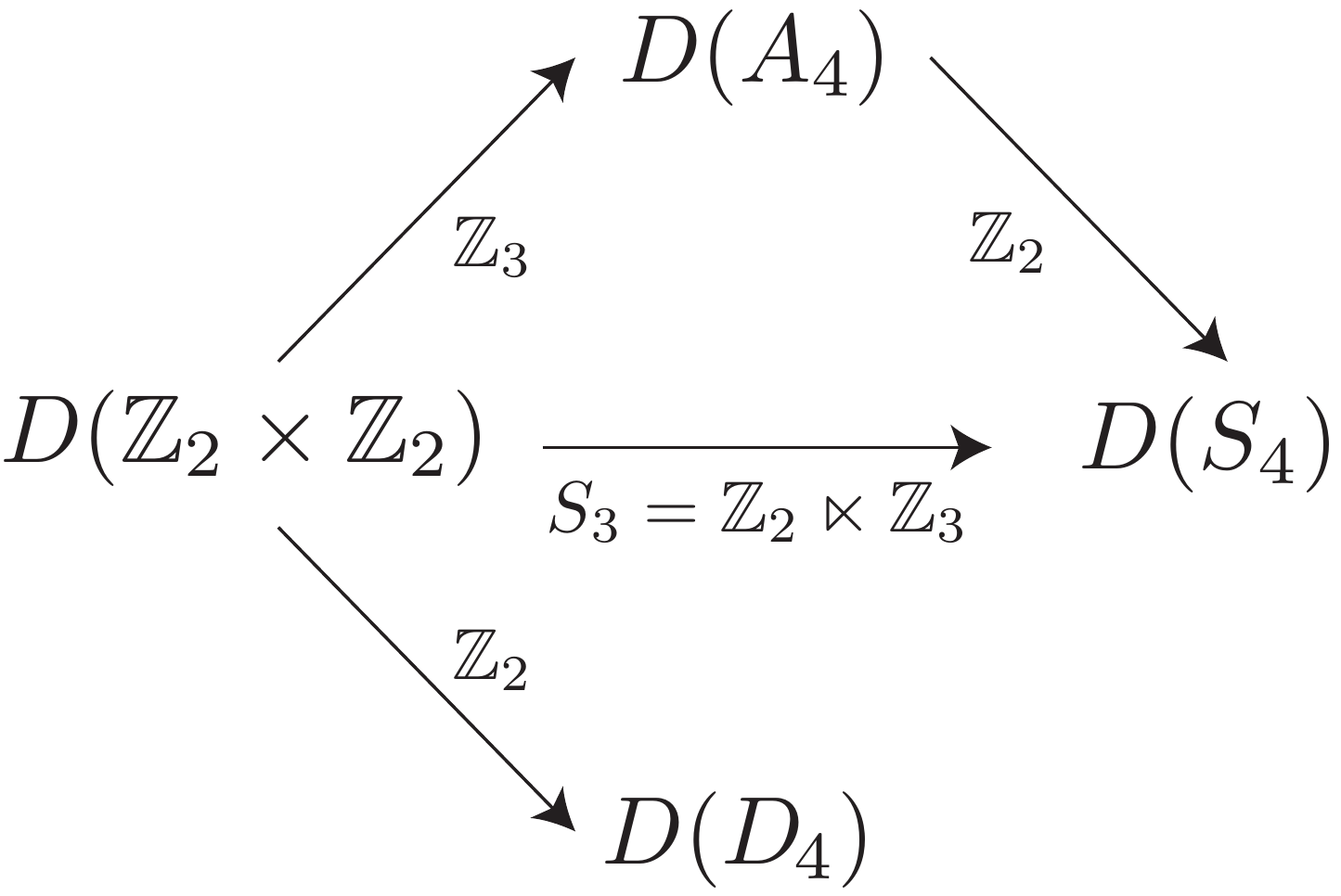}}}\end{align} associated to the solvable tower of symmetry groups (see Eqs.~\eqref{solvabletower} and \eqref{solvabletowerTL}): \begin{align}\begin{diagram}1&\rInto^{\mathbb{Z}_2^2}&\mathbb{Z}_2\times\mathbb{Z}_2&\rInto^{\mathbb{Z}_3}&A_4&\rInto^{\mathbb{Z}_2}&S_4\end{diagram}.\end{align} Here $S_4$ is identical to the cubic point group in $SO(3)$ and contains $A_4$, the tetrahedral point group. The $\mathbb{Z}_2\times\mathbb{Z}_2$ group represents twofold rotations about $x,y,z$. Each arrow \begin{align}D(H)\xrightarrow{G/H}D(G)\end{align} describes the gauging transition from the discrete gauge theory $D(H)$ with global $(G/H)$-anyonic symmetry to the twist liquid $D(G)$, which happens to also be a discrete gauge theory with gauge group $G$. For example, $A_4/(\mathbb{Z}_2\times\mathbb{Z}_2)=\mathbb{Z}_3$, $S_4/A_4=\mathbb{Z}_2$ and $S_4/(\mathbb{Z}_2\times\mathbb{Z}_2)=S_3$ show the connections between different global anyonic symmetries.

\section{Summary and Conclusion}\label{sec:conclusion}
For any article of this length it is difficult to summarize all of the important conclusions and results without sounding like a list. Instead we will try to recapitulate the main themes of our work in a slightly less formal tone, and touch upon the main technical results where appropriate. We start with the idea of anyonic symmetries. 
The importance of anyonic symmetries, and their key role in many newly understood physical phenomena, e.g. the theory of twist defects, inspired our work.
In fact, at its core, this article resulted from trying to answer a single question, ``what topological phase emerges when an anyonic symmetry is gauged?" 

Given a phase of matter with a discrete set of anyons, there are often subsets of anyons which have the same topological properties, e.g.~identical spin and braiding relations. Anyonic symmetry operations,\cite{Kitaev06,EtingofNikshychOstrik10,Bombin,YouWen,BarkeshliJianQi,TeoRoyXiao13long,khan2014,BarkeshliBondersonChengWang14} are essentially just rearrangements of the anyons within each subset. 
Physical manifestations of these anyonic symmetries come in the form of twist defects.\cite{Kitaev06, EtingofNikshychOstrik10, barkeshli2010, Bombin, Bombin11, KitaevKong12, kong2012A, YouWen, YouJianWen, PetrovaMelladoTchernyshyov14, BarkeshliQi, BarkeshliQi13, BarkeshliJianQi, MesarosKimRan13, TeoRoyXiao13long, teo2013braiding, khan2014, BarkeshliBondersonChengWang14} Given a topological phase, there is a corresponding set of twist defects, which are semi-classical point objects attached to a branch cut. A twist defect comes with two labels, the first is an anyonic symmetry operation. When an anyon passes through the branch cut of a twist defect it is acted upon by the corresponding symmetry operation, and can be converted (or not) to a different anyon.  The second label on a twist defect is its ``species." This label encodes which anyon has been attached to the defect. However, fusing the defect with different anyons does not always give rise to a different species. Instead, since certain anyons cannot be distinguished due to the non-trivial symmetry action of the defect, the species labels only run over a reduced set of anyon equivalence classes. In some cases the defect species is modified from fusion with an anyon, and colorfully, the defect is said to have ``mutated" its species.  

The theory of twist defects has a fusion structure as well.\cite{EtingofNikshychOstrik10, TeoRoyXiao13long, teo2013braiding, BarkeshliBondersonChengWang14} In systems with a global anyonic symmetry there are fusion rules for fusing the twist defects with the initial set of anyons. These could be called ``mutation" rules, as they can change the species of the defect. In addition to this, there are rules that are specified for the fusion of two defects. Twist defect fusion is an involved process, but we can break it down into pieces. First let us recall that the branch cuts associated to each defect fuse together to create a composite branch cut. This is mathematically represented by saying that $\widehat{M}\widehat{N}\to \widehat{MN}$ under fusion, where $\widehat{M}, \widehat{N}$ are in a quantum representation of the anyonic symmetry group. Unfortunately, there is already a complication at this stage because the outcome $\widehat{MN}$ is ambiguous, and could differ by a projective phase, i.e.~the symmetry operators could be in a projective representation. Hence, there are multiple possibilities for the defect-defect fusion structure, and the physically distinct possibilities are classified by the group cohomology $H^2(G,{\mathcal{B}}^{\times})$ where ${\mathcal{B}}^{\times}$ is the set of Abelian anyons in the parent state.  If a non-trivial choice in $H^2(G,{\mathcal{B}}^{\times})$ is used, then the anyonic symmetry becomes ``non-symmorphic" as discussed in Section \ref{sec:globalquantumsymmetries}. For Abelian parent states, these projective phases can be simply interpreted: they are an additional phase arising from an extra Abelian anyon that results from the branch-cut fusion. Hence, even if the branch cuts are from inverse symmetry operations, there can be a non-trivial Abelian anyon left over. Now remember that, attached to each twist defect, there is a species label which represents an equivalence class of anyons. Once all of the allowed possibilities for how the equivalence classes are fused are accounted for, one then fuses the result with the projective anyon at the branch-cut intersection. And the final result Eq. \ref{generaldefectfusion} is obtained. Note that we did not discuss the projective ambiguity at length for our major examples, because almost all of the anyonic symmetry groups we considered have $H^2(G,{\mathcal{B}}^{\times})=0.$

The final  data that can be derived purely from the anyon content of the parent state and the set of semiclassical twist defects, are the basis transformation $F$-symbols for the defects. Once these have been tabulated, the construction of the defect fusion category is complete, since we recall that there are no well-defined braiding operations for the semiclassical twist defects. Interestingly, given a fixed parent theory, there are also different physical choices for the $F$-symbols that can be derived. These choices are captured by the ambiguity in the Frobenius-Schur indicator, and they have a nice physical interpretation. The physical interpretation follows. First, the allowed set of $F$-symbols are classified by the group cohomology $H^3(G,U(1)).$ Perhaps seeming coincidental, this is the same classification for symmetry protected topological phases with global (non-anyonic) symmetry group $G$.~\cite{ChenGuLiuWen12, LuVishwanathE8, MesarosRan12, EssinHermele13, WangPotterSenthil13, BiRasmussenSlagleXu14, Kapustin14} These SPTs do not have an underlying topological order, and they can be stacked together with the parent state to produce a new state with the same symmetries, and the same set of anyons and twist defects. However, one notable change is that the $F$-symbols can be modified by a phase by this procedure if a non-trivial SPT is added. Hence, the classifications are equivalent. Now, when there is a twist defect it also passes through the extra SPT layer, which generates the extra phase in the $F$-symbols. As we recall, each physically different choice for the $F$-symbols gives rise to a different twist liquid state after the anyonic symmetry is gauged, just like gauging the different global $G$-symmetric SPT states will give rise to different topological orders. 

From this initial structure, we proceed by making an analogy between the set of twist defects in some parent topological phase, and the topological (flux) defects that can appear when a system maintains a conventional global symmetry. For example, a global $U(1)$ symmetry has magnetic flux defects, a global translation symmetry has dislocation defects, and a global rotation symmetry has disclination defects. All of these cases represent semi-classical (confined) defects. If a particle carries charge under these global symmetries, then its state will obtain a holonomic phase factor when it encircles a defect, e.g.~the conventional Aharonov-Bohm effect for a global $U(1)$ symmetry. When the global symmetry is gauged, the flux defects become excitations of the resulting discrete gauge theory. The difference between this, and our case of interest, is that the globally-symmetric state has some initial topological order with non-trivial anyon content, and the global (anyonic) symmetry acts non-trivially on the set of anyons.

To calculate the anyon content after gauging the anyonic symmetry, we proceed analogously to the gauging of a discrete non-anyonic global symmetry. For a discrete gauge theory,\cite{BaisDrielPropitius92,Bais-2007,Propitius-1995,PropitiusBais96,Preskilllecturenotes,Freedman-2004,Mochon04} the anyon content consists of the vacuum, pure fluxes, pure charges, and dyons (composites of flux and charge). We can generically denote these anyons with a pair $([M],\rho)$ where $[M]$ represents the flux, and $\rho$ represents the charge. For a twist liquid we can again have  the vacuum, pure fluxes, pure charges, and dyons, but there are additional anyons that come from the anyon content of the parent state. There are now pure quasiparticles, flux-quasiparticle composites, charge-quasiparticle composites, and flux-quasiparticle-charge composites (trions?). However, because the anyonic symmetry acts non-trivially on the parent anyons, the resulting theory is not just the product of the initial anyon content and a discrete $G$ gauge theory. When the anyonic symmetry acts non-trivially on an anyon it is no longer a well-defined object in the twist liquid. Instead we must consider the set of orbits, or super-sectors, of the action of $G$ on the set of anyons in the parent state. Each orbit is just a closed set of anyons that transform into each other under the symmetry. Ultimately, the anyons in the twist liquid are represented by the triplet $([M],\boldsymbol\lambda, \rho),$ where $[M]$ and $\rho$ still represent the flux and charge, while $\boldsymbol\lambda$ represents the quasiparticle super-sector/orbit. We showed in Section \ref{sec:QPstructuretwistliquid} that, given a choice for $[M],$ the allowed values for $\boldsymbol\lambda$ and $\rho$ are constrained, and given by a particular algorithm.

Given a generic Abelian parent state, we proved two important general properties of the resulting twist liquid. The first is that, given that the total quantum dimension of the parent state is ${\mathcal{D}}_0,$ the resulting total quantum dimension of the resulting twist liquid is ${\mathcal{D}}_0\vert G\vert$ for an anyonic symmetry group $G$ (c.f. Eq. \ref{TLdimension3}). This is analogous to the total quantum dimension of a discrete gauge theory, which increases to $\vert G \vert$ from its initial value of $1,$ when gauging a global symmetry $G.$ The second property is that the central charge of the twist liquid remains identical (possibly mod $8$) to that of the parent state (see Eq. \ref{chiralcinvariance}). One can heuristically understand this through the bulk-boundary correspondence since, as far as the edge is concerned, the edge theory is orbifolded by the anyonic symmetry gauging, which is a process that preserves the central charge. As an aside we note that the orbifold construction picks the edge of a particular twist liquid, and  does not seem to capture the cohomological possibilities mentioned above. Similarly, the unchanged central charge mirrors that of a discrete gauge theory where the central charge of the initial state is $0,$ and that of the gauged state is also $0.$ While we only explicitly proved these results for Abelian parent states, we showed in Section \ref{sec:4statePotts} that they also hold for an explicit non-Abelian parent state, the ``4-Potts" state. One should see Ref. \onlinecite{BarkeshliBondersonChengWang14}, for a more general discussion of non-Abelian systems. 

In addition to these general conclusions, we illustrated an explicit lattice model construction using the Levin-Wen string-net construction. This model is general enough to capture the conventional construction of a discrete gauge theory, as well as the twist liquid construction from the input of a defect fusion category. Also, while the string-net model generally results in a time-reversal invariant topological phase, we showed that one can systematically remove an inert part of the theory (via the relative Drinfeld construction) to reveal a chiral twist liquid of a chiral parent state.

All of these results are utilized in our lengthy examples section to illustrate some remarkable relationships between different topological phases related by gauging anyonic symmetries, and subsequently by the inverse process of anyon condensation. It is likely there are many more beautiful connections to be uncovered in future work, especially for non-Abelian parent theories.

\begin{acknowledgments}
We thank Maissam Barkeshli, Xiao Chen, Meng Cheng, Lukasz Fidkowski, Abhishek Roy, and Zhenghan Wang for useful discussions. 
This work was supported in part by the National Science Foundation, under grants  DMR-1408713 (EF) and DMR-1351895-CAR (TLH) at the University of Illinois, and by the Simons Foundation (JCYT). \end{acknowledgments}

\appendix
\section{Quick Review of Topological Field Theory}\label{app:reviewTQFT}

The anyon structure of a topological phase in $(2+1)$D is given by the fusion and braiding properties of its quasiparticles. An anyon (also known as a quasiparticle type or topological charge) is a class of gapped point-like excitations that have non-trivial statistical and braiding behavior. The set of anyons $\{1,{\bf a},{\bf b},\ldots\}$ form a collection of simple objects that generate a modular tensor category~\cite{Kitaev06, Turaevbook, BakalovKirillovlecturenotes, Wangbook, Walkernotes91, FreedmanLarsenWang00, Bondersonthesis} $\mathcal{B}$. A pair of objects in $\mathcal{B}$ can fuse, denoted by $\times$, or form a super-sector, denoted by $+$. There is a closed fusion algebraic structure \begin{align}{\bf a}\times{\bf b}=\sum_{\bf c}N_{\bf ab}^{\bf c}{\bf c}\label{appfusionrule}\end{align} where $N_{\bf ab}^{\bf c}$ is a non-negative integer that counts the fusion degeneracy -- the number of inequivalent ways for ${\bf a}$ and ${\bf b}$ to fuse into ${\bf c}$. Equivalently, $N_{\bf ab}^{\bf c}$ also counts the splitting degeneracy -- the number of ways for ${\bf c}$ to split into ${\bf a}$ and ${\bf b}$. Eq.~\eqref{appfusionrule} is associative so that ${\bf a}\times({\bf b}\times{\bf c})=({\bf a}\times{\bf b})\times{\bf c}$, i.e.~$\sum_{\bf d}N_{\bf ad}^{\bf e}N_{\bf bc}^{\bf d}=\sum_{\bf d}N_{\bf ab}^{\bf d}N_{\bf db}^{\bf e}$; and is commutative so that ${\bf a}\times{\bf b}={\bf b}\times{\bf a}$, i.e.~$N_{\bf ab}^{\bf c}=N_{\bf ba}^{\bf c}$. There is a unit object, referred to as the vacuum $1$, such that $1\times{\bf a}={\bf a}$ for any anyon ${\bf a}$. Moreover every anyon ${\bf a}$ must have a unique anti-particle $\bar{\bf a}$ so that $N_{{\bf a}\overline{\bf a}}^1=1$.

Let us fix a finite number of anyonic excitations $\{\lambda_1,\lambda_2,\ldots\}$ in a topological phase on a closed sphere, a quantum state can be specified by a {\em splitting tree} (or {\em fusion tree}) with known internal branches and vertices (see Fig.~\ref{fig:Fmoves} and Eq.~\eqref{splittingtreeapp} below). A splitting tree is a directed tree diagram, whose internal branches are labeled by anyons $x_v,$ and vertices -- which must be trivalent -- are labeled by {\em splitting states} $\mu_v$, to be explained below. The external branches are fixed by the known excitations $\lambda_i$. Each vertex describes the splitting of an anyon. It has one incoming branch $z,$ and two outgoing branches $x,y$ such that $N_{xy}^z>0$, i.e.~the fusion $x\times y\to z$ is admissible. The splitting degeneracy corresponds to a $N_{xy}^z$-dimensional space, whose orthonormal basis is labeled by $\mu_v=1,\ldots N_{xy}^z$. The set of admissible $|\{x_l\};\{\mu_v\}\rangle$ forms an orthonormal basis of the degenerate Hilbert space of quantum states with the fixed anyonic excitations $\{\lambda_1,\lambda_2,\ldots\}$.

Essentially, a splitting tree corresponds to a particular maximal set of mutually commuting statistical observables, i.e.~Wilson loops; and the internal branch and vertex labels correspond to simultaneous eigenvalues. For example, the fusion channel $x\times y\to z$ at a vertex determines the eigenvalue of the Wilson loop $\hat{\mathcal{W}}_w$ encirling $x$ and $y$ by the braiding between $w$ and $z$. Alternatively, a splitting tree describes a particular sequence of splittings that ultimately results in the creation of the excitations $\lambda_i$ on the external branches. Starting from the ground state $|GS\rangle$ with no excitations, anyons can be created by open Wilson string operators. For example, in the toric code (see Section~\ref{sec:Zkgaugetheory}) the vacuum $1$ can be split into a pair of plaquette excitations, say $m\times m$, by a string of $\sigma_{x,z}$ operators $\hat{L}^{mm}_1$ connecting them. One of the plaquette excitations can subsequently be divided into $m\to e\times\psi$ by a string operator $\hat{L}^{e\psi}_m$ that branches from $m$ to $e$ and $\psi$. The quantum state with excitations $e,\psi,m$ can then be created by $\hat{L}^{e\psi}_m\hat{L}^{mm}_1|GS\rangle$. Notice that there are multiple ways an $m$-string can branch into $e$ and $\psi$. For instance, the $e$ and $\psi$ strings can braid and give a minus sign. The particular splitting operator $\hat{L}^{e\psi}_m$ specifies the phase of the quantum state. On the other hand, the string operators can be deformed by plaquette operators and still act identically on the ground state. These form equivalence classes of splitting string operators, such as $[\hat{L}^{e\psi}_m]$, with fixed open ends.

In more exotic cases, splitting can be degenerate and different branching operators can give orthogonal states. For example the $SU(3)_3$-state discussed in Section~\ref{sec:gaugingZ3} supports an anyon associated to the eight dimensional adjoint representation which obeys the degenerate fusion rule ${\bf 8}\times{\bf 8}=1+{\bf 10}+\overline{\bf 10}+{\bf 8}+{\bf 8},$ such that $N_{\bf 8,8}^{\bf 8}=2$. There are two linearly independent local operators that split ${\bf 8}$ into ${\bf 8}\times{\bf 8}$. In general, the splitting $z\to x\times y$ at a vertex with degeneracy $N_{xy}^z$ is associated to a {\em splitting space} $V^{xy}_z$. It contains equivalence classes of local operators $[(\hat{L}^{xy}_z)_\mu]$ -- referred to as {\em splitting states} -- that connect an incoming $z$ to the outgoing $x$ and $y$. The collection of linearly independent splitting operators $\{(\hat{L}^{xy}_z)_\mu:\mu=1,\ldots,N_{xy}^z\}$ spans the splitting space $V^{xy}_z$, which forms an irreducible representation of the algebra of Wilson operators around $x$ and $y$: \begin{align}\hat{\mathcal{W}}(\hat{L}^{xy}_z)_\mu\hat{\mathcal{W}}^{-1}=\sum_{\nu}\mathcal{W}^\nu_\mu(\hat{L}^{xy}_z)_\nu.\end{align} A quantum state with known excitations is in general constructed by piecing the splitting string operators together with matching boundary conditions: \begin{align}&\left|\vcenter{\hbox{\includegraphics[width=0.1\textwidth]{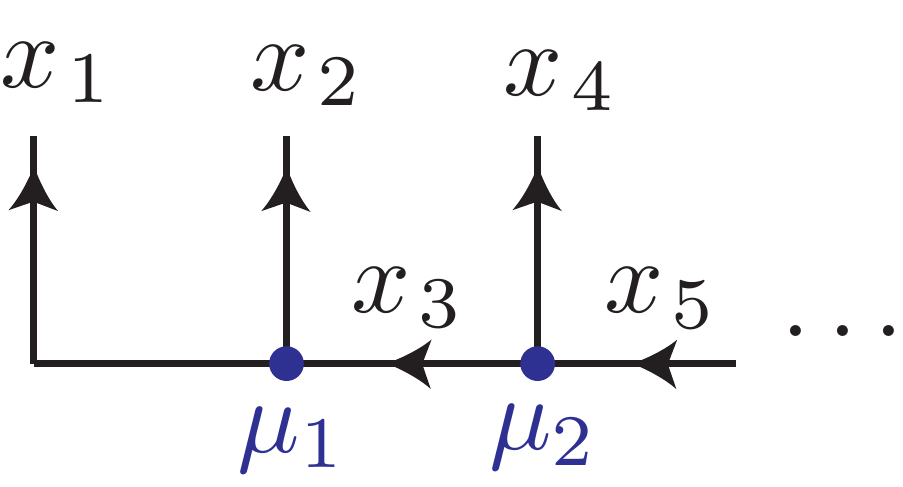}}}\right\rangle\nonumber\\&\propto\mathop{\sum_{boundary}}_{conditions}\left[(L^{x_1x_2}_{x_3})_{\mu_1}\right]\otimes\left[(L^{x_3x_4}_{x_5})_{\mu_2}\right]\otimes\ldots|GS\rangle.\label{splittingtreeapp}\end{align}

In an Abelian theory, where fusion is single-channel ($N_{\bf ab}^{\bf c}=0$ or $1$), the quantum state \eqref{splittingtreeapp} is completely determined (up to a $U(1)$-phase) by its external branch labels, as all internal branch labels are fixed by the fusion rules. This means that the quantum state is completely determined by the anyon types of  the excitations. This is not true for non-Abelian theories where fusion rules can be multi-channeled. For example, a pair of Ising anyons can fuse into an even or odd fermion parity channel, $\sigma\times\sigma=1+\psi$;\cite{MooreRead} Fibonacci anyons obey the fusion rule $\tau\times\tau=1+\tau$.\cite{ReadRezayi} These non-Abelian anyons give rise to degeneracies of quantum states since the internal branches and vertices of \eqref{splittingtreeapp} can carry different labels.

The quantum dimension $d_x$ of an anyon $x$ is defined to be the positive number that counts the quantum state degeneracy, which is proportional to $d_x^N$, of a closed system with $N$ type $x$ quasiparticle excitations in the thermodynamic limit when $N\to\infty$. Using the Perron-Frobenius theorem,\cite{Kitaev06} the quantum dimension $d_x$ is given by the largest (absolute) eigenvalue of the fusion matrix $N_x=\left(N_{xy}^z\right)$. For example, an anyon is Abelian if and only if it has unit quantum dimension. The quantum dimension of an Ising anyon is $d_\sigma=\sqrt{2},$ and that of a Fibonacci anyon is $d_\tau=(1+\sqrt{5})/2$. In general, the fusion rules \eqref{appfusionrule} require \begin{align}d_xd_y=\sum_{z}N_{xy}^zd_z.\end{align} The total quantum dimension of the anyon theory is defined by $\mathcal{D}=\sqrt{\sum_xd_x^2}$.

Fusion associativity ${\bf a}\times({\bf b}\times{\bf c})=({\bf a}\times{\bf b})\times{\bf c}$ in the quantum state level is realized as basis transformations between different splitting trees (see Fig.~\ref{fig:Fmoves}). Primitive basis transformations are known as $F$-symbols. They relate  \begin{align}\left|\vcenter{\hbox{\includegraphics[width=0.5in]{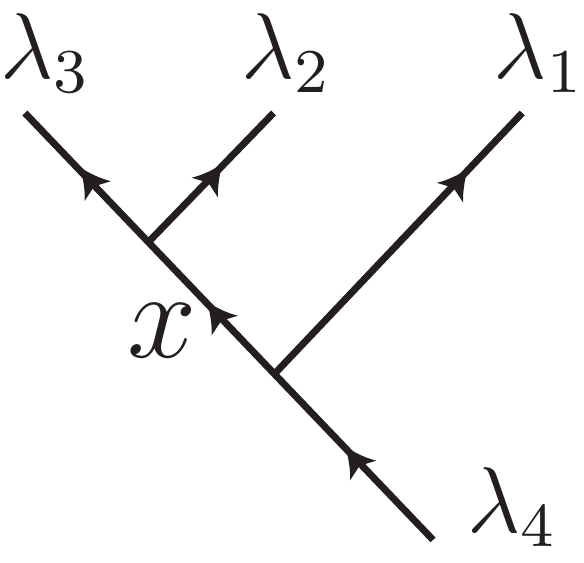}}}\right\rangle=\sum_{y}\left[F^{\lambda_3\lambda_2\lambda_1}_{\lambda_4}\right]_{x}^{y}\left|\vcenter{\hbox{\includegraphics[width=0.5in]{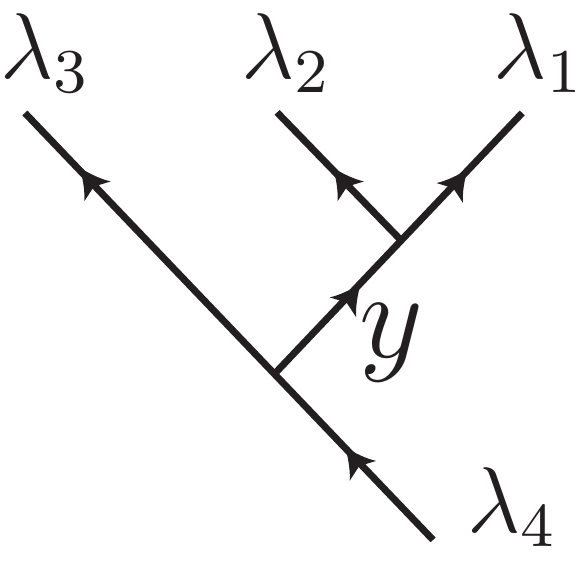}}}\right\rangle.\label{Fsymboldef}\end{align} where $F$-matrix entries are given by the inner product \begin{align}\left[F^{\lambda_3\lambda_2\lambda_1}_{\lambda_4}\right]_{x}^{y}=\left\langle\vcenter{\hbox{\includegraphics[width=0.5in]{F2}}}\right.\left|\vcenter{\hbox{\includegraphics[width=0.5in]{F1}}}\right\rangle .\label{Fdefinition}\end{align} Here the symbols $\mu_j$ for vertex degeneracies are suppressed. There are consistency relations, referred to as the pentagon equations, that the $F$-symbols have to obey (see Fig.~\ref{fig:Fpentagon}).\cite{Kitaev06} They are required to ensure that any sequence of $F$-moves that connects two fixed initial and final splitting trees will give the identical overall basis transformation.\cite{MacLanebook}

There are particle-antiparticle duality relationships in a fusion theory. For instance $x$ and $\bar{x}$ must have identical quantum dimensions. At the quantum state level, reversing the worldline of an anyon $x$ and replacing it by its conjugate $\bar{x}$ may result in an overall phase. This is determined by the bending diagram \eqref{bendingdef}, or precisely the Frobenius-Schur (FS) indicator\cite{FredenhagenRehrenSchroer92,Kitaev06} \begin{align}\varkappa_x=\frac{[F^{x\overline{x}x}_x]_1^1}{|[F^{x\overline{x}x}_x]_1^1|}\in U(1).\end{align}

As a simple example, we consider four Ising anyons on a closed sphere, each is associated to a Majorana zero mode $\gamma_j$, for $j=1,2,3,4$. The twofold ground state degeneracy can be labeled by the fermion parity of a pair of Ising anyons, say $(-1)^{F_{12}}=i\gamma_1\gamma_2$. The total parity in a closed system is fixed, and is taken to be even $(-1)^{F_{12}+F_{34}}=(-1)^{F_{14}+F_{23}}=-\gamma_1\gamma_2\gamma_3\gamma_4=+1$. One can also pick another fermion parity, say $(-1)^{F_{23}}=i\gamma_2\gamma_3$, to label the states. As $(-1)^{F_{12}}$ and $(-1)^{F_{23}}$ anticommute, they do not share simultaneous eigenvalues, and the even and odd parity states with respect to these two operators are related by the  non-diagonal transformation: \begin{align}\left|\vcenter{\hbox{\includegraphics[width=0.08\textwidth]{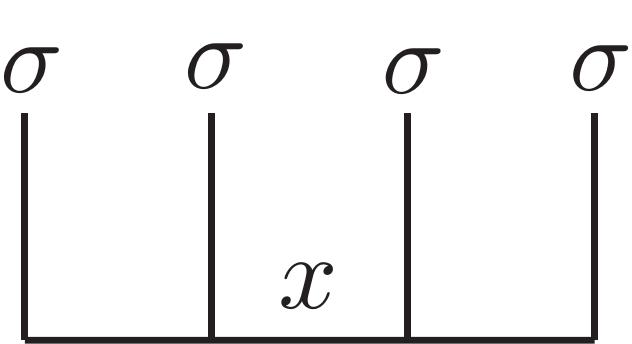}}}\right\rangle=\sum_{y=1,\psi}\left[F^{\sigma\sigma\sigma}_\sigma\right]_x^y\left|\vcenter{\hbox{\includegraphics[width=0.08\textwidth]{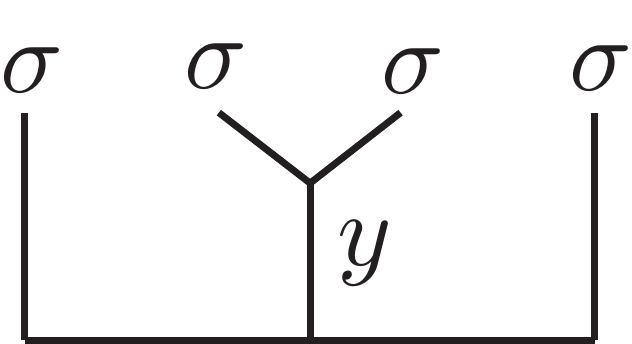}}}\right\rangle\end{align} where the $F$-matrix is  \begin{align}F^{\sigma\sigma\sigma}_\sigma=\frac{1}{\sqrt{2}}\left(\begin{array}{*{20}c}1&1\\1&-1\end{array}\right)\end{align} with its rows and columns arranged according to $x,y=1,\psi$ which label the parities $(-1)^{F_{12}}$ and $(-1)^{F_{23}}$ respectively. 

A fusion category is a theory that encodes associative fusion rules \eqref{appfusionrule} and a consistent collection of basis transformations \eqref{Fsymboldef}. In addition to this structure, an anyon theory also contains information about exchange and braiding. Fusion commutativity $x\times y=y\times x$ at the quantum state level translates to another kind of basis transformation. They are generated by the $R$-symbols \begin{gather}\vcenter{\hbox{\includegraphics[width=0.05\textwidth]{R1}}}=R^{xy}_z\vcenter{\hbox{\includegraphics[width=0.05\textwidth]{R2}}}\label{Rsymboldefapp}\\\vcenter{\hbox{\includegraphics[width=0.05\textwidth]{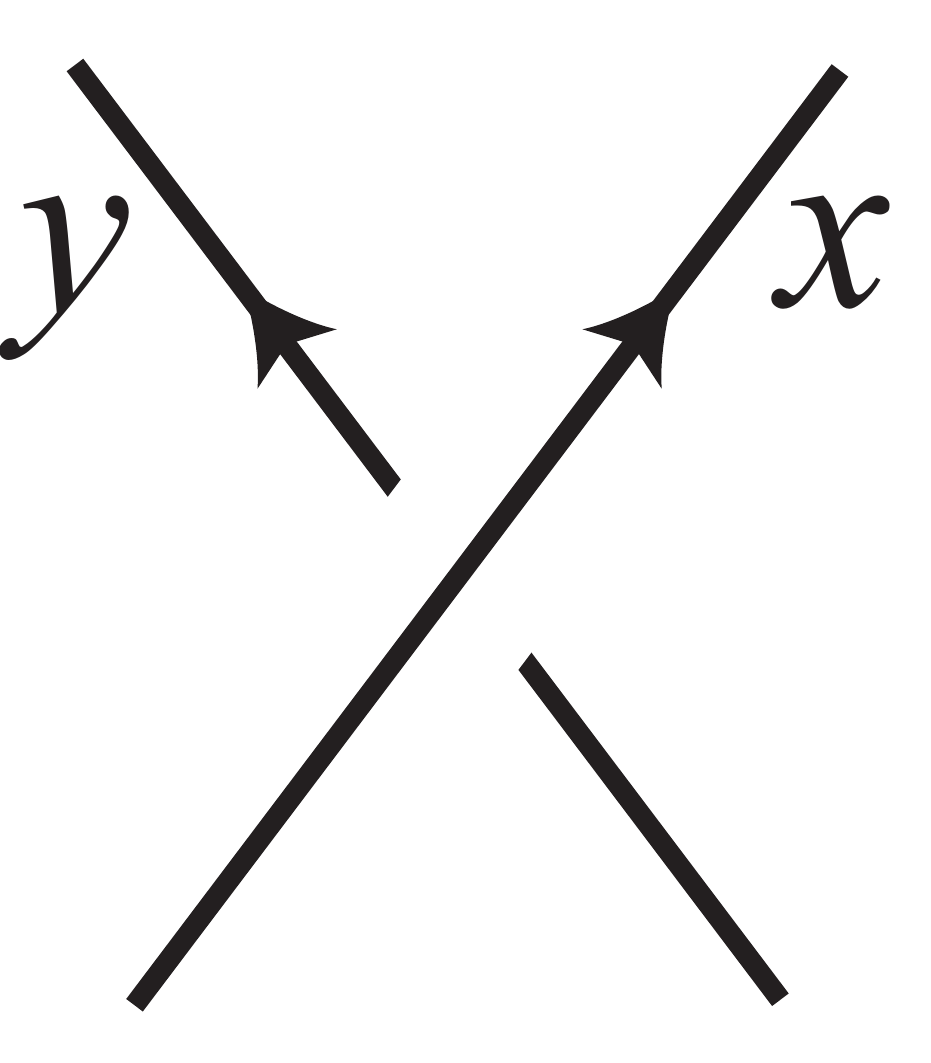}}}=\sum_{z}\sqrt{\frac{d_z}{d_xd_y}}R^{xy}_z\vcenter{\hbox{\includegraphics[width=0.05\textwidth]{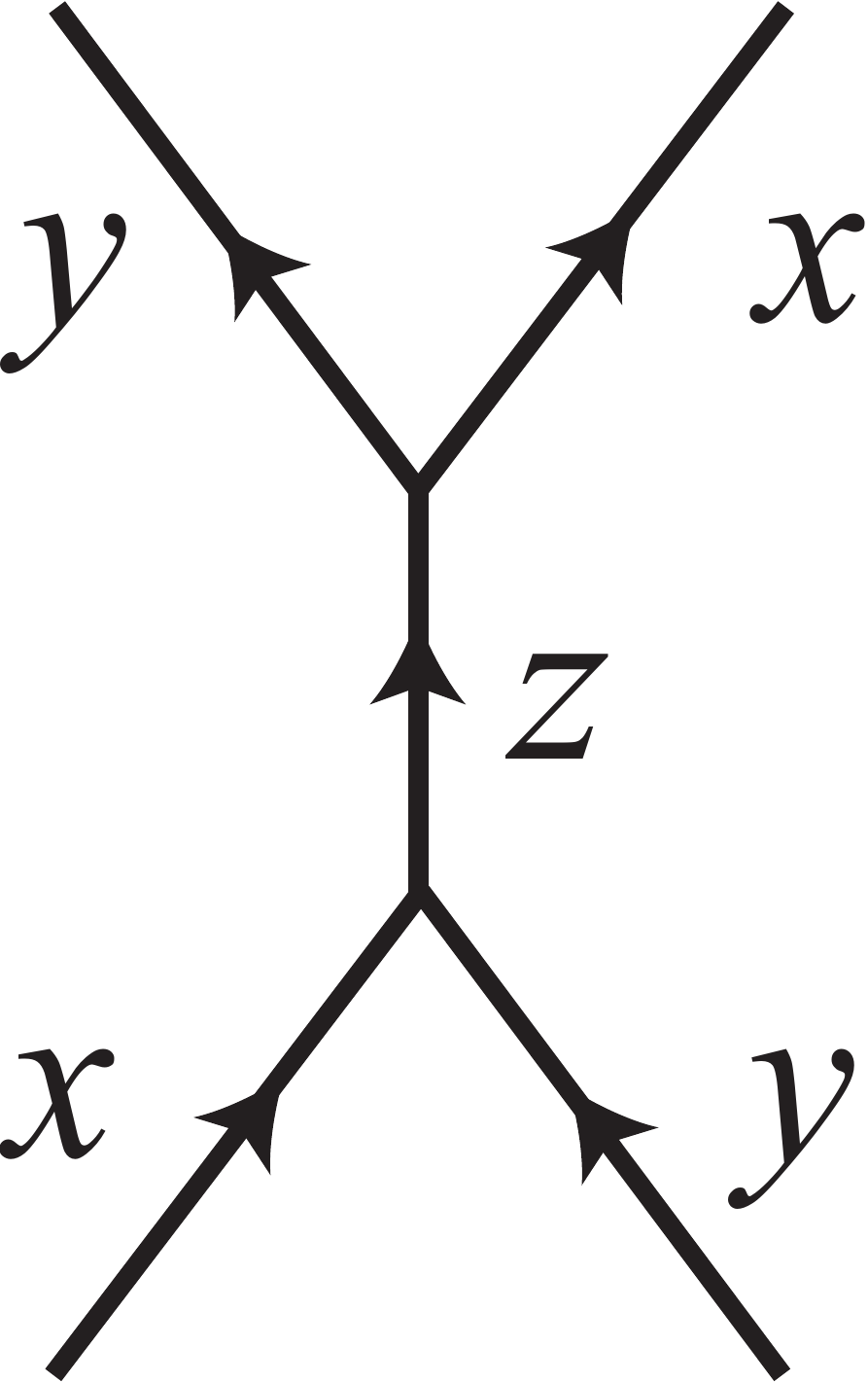}}}\label{Rsymboldefapp2}\end{gather} when the splitting $z\to x\times y$ is admissible. When splitting is degenerate, vertices in \eqref{Rsymboldefapp} and \eqref{Rsymboldefapp2} should be labeled, and $R^{xy}_z$ would be a $(N_{xy}^z)\times(N_{xy}^z)$ unitary matrix. 

These exchange $R$-symbols follow the consistency relations called the {\em hexagon identity}\cite{Kitaev06} (see Eq.~\eqref{hexagoneq} and Fig.~\ref{fig:hexagon1}). Essentially they guarantee fusion and exchange are compatible by requiring that the successive exchanges between $x,y$ and between $x,z$ are overall equivalent to the exchange between $x$ and $w=y\times z$. An anyon theory is therefore a {\em braided} fusion category that is equipped with a consistent exchange structure.

The spin statistics theorem equates the $\pi$-exchange phase \begin{align}\theta_x=\frac{1}{d_x}\sum_{y}d_y\mbox{Tr}\left(R^{xx}_y\right)\end{align} with the $2\pi$ twist $\theta_x=e^{2\pi ih_x}=\varkappa_x(R^{x\overline{x}}_1)^\ast$, where $h_x$ is referred to as the spin. Particle $x$ has identical spin to its conjugate $\bar{x}$. Exhange phases are related to $2\pi$ braiding phases by the {\em ribbon identity} \begin{align}R^{xy}_zR^{yx}_z=\frac{\theta_z}{\theta_x\theta_y}\openone\label{ribbonapp}\end{align} (see also Eq.~\eqref{ribbon}) where $\openone$ is the $(N_{xy}^z)\times(N_{xy}^z)$ identity matrix. The braiding between anyons can be summarized by the average \begin{align}S_{xy}=\frac{1}{\mathcal{D}}\sum_{z}d_{z}\mbox{Tr}\left(R^{xy}_{z}R^{yx}_{z}\right)=\frac{1}{\mathcal{D}}\sum_{z}d_{z}N^{z}_{xy}\frac{\theta_{z}}{\theta_{x}\theta_{y}}\label{braidingSapp}\end{align} which are the matrix elements of the modular $S$-matrix. The $2\pi$ twists give the modular $T$-matrix $T_{xy}=\theta_{x}\delta_{xy}$. They projectively represent the modular group $SL(2;\mathbb{Z})$, i.e. the group of automorphisms of the torus, and obey the group relation \begin{align}(ST)^3=e^{2\pi i\frac{c_-}{8}}S^2\label{ST3=S2}\end{align} where $C=S^2$ is the conjugation matrix that relates $x\leftrightarrow\bar{x}$, squares to identity, and commutes with both $S$ and $T$. The projective phase is associated to the chiral central charge $c_-=c_R-c_L$ (mod 8) of the CFT along the $(1+1)$D boundary of the topological phase. Eq.~\eqref{ST3=S2} is equivalent to the Gauss-Milgram formula \eqref{GaussMilgram}.


\section{Defect Fusion Category of the Toric Code}\label{app:toricdefect}

Here we explain the fusion rules and the basis transformation ($F$-symbols) in the defect theory of the  toric code. They were used in Section~\ref{sec:Zkgaugetheory}, and were previously evaluated in Refs.~\onlinecite{TeoRoyXiao13long, teo2013braiding}. A quick derivation will be provided in this Appendix. We first begin with the fusion rules between a pair of twofold defects: \begin{align}\sigma_0\times\sigma_0=\sigma_1\times\sigma_1=1+\psi,\quad\sigma_0\times\sigma_1=e+m\label{toriccodedefectfusion2a}.\end{align} For example the first equation in \eqref{toriccodedefectfusion2a} states that two defects with identical species can fuse into either the vacuum $1$ or fermion $\psi$ channel. This matches that of a pair of Ising anyons. The different fusion channels on the right side can be distinguished by Wilson operators $\mathcal{W}_{\bf b}$. The Wilson loops represent quasiparticle strings ${\bf b}=1,e,m,\psi=(0,0),(1,0),(0,1),(1,1)$ that encircle the defect pair, and their eigenvalues $\mathcal{W}_{\bf b}=e^{2\pi i{\bf b}^TK^{-1}{\bf a}}$ read-off the fusion channel ${\bf a}=1,e,m,\psi$ in \eqref{toriccodedefectfusion2a}, where the $K$-matrix is given by $K=2\sigma_x$ in the Chern-Simons description \eqref{toriccodeCSaction}. \begin{align}\vcenter{\hbox{\includegraphics[width=0.18\textwidth]{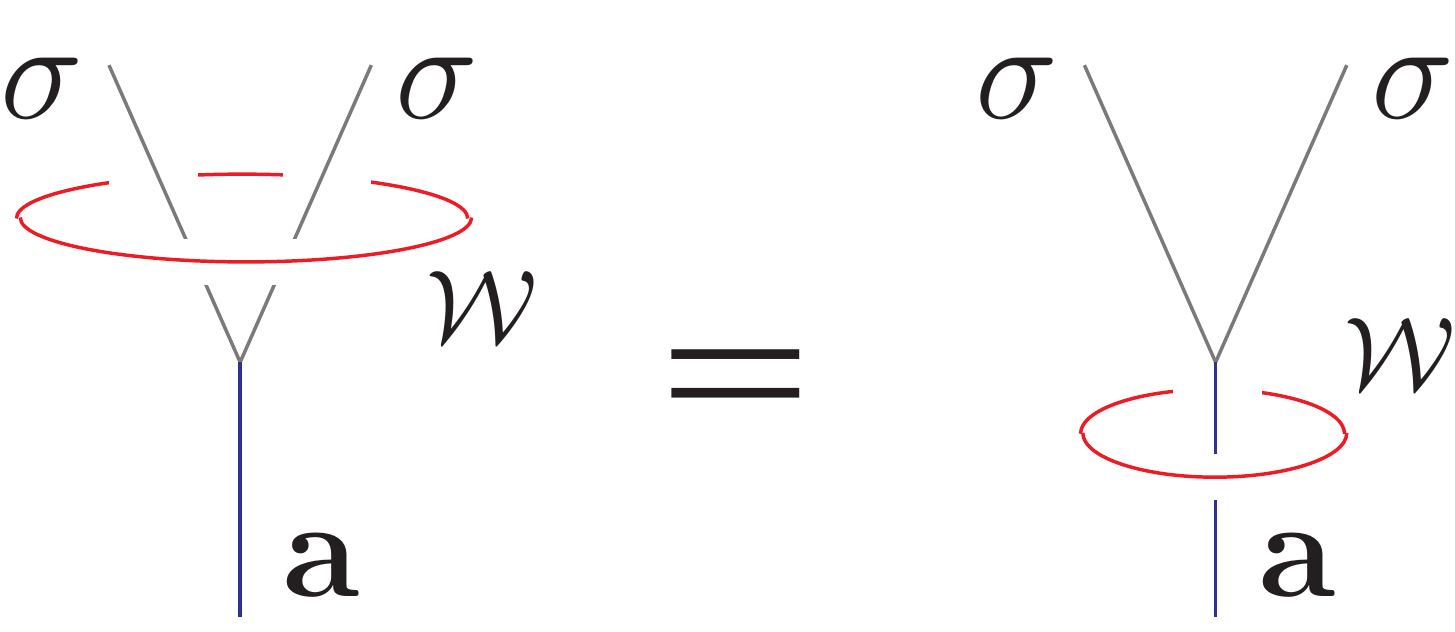}}}\end{align}

The admissible fusion channels are, however, restricted by the species of the defect pair. For instance, certain Wilson loops can be generated by double loop operators around defects, e.g. \begin{align}\vcenter{\hbox{\includegraphics[width=0.4\textwidth]{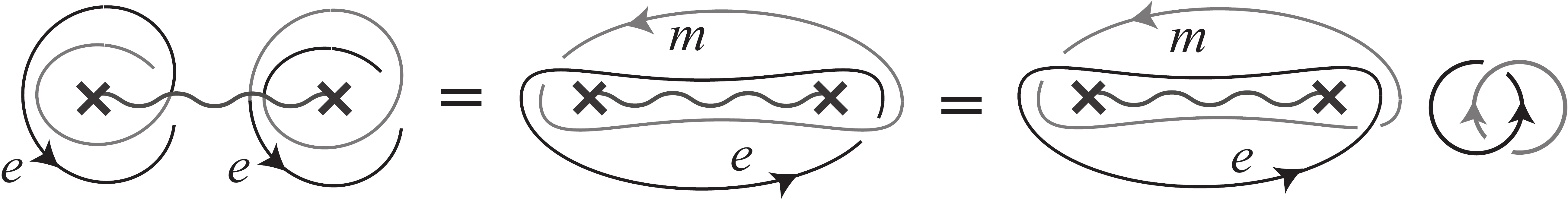}}}\label{doubleloopeq}\end{align} where the double loop operator $\Theta$ around each defect (see also Fig.~\ref{fig:defect1}) already has a fixed eigenvalue that depends on the species label $\lambda$ in $\sigma_\lambda$. By multiplying usual plaquette stabilizers, the $\Theta$ operator can be deformed and shrunk, into $\pm iP_{\pentagon}$, where $P_{\pentagon}$ is the plaquette operator at the dislocation (see Fig.~\ref{fig:toriccode}). The species label $\lambda$, or equivalently the eigenvalue of $\Theta=\pm iP_{\pentagon}=(-1)^\lambda i$, is therefore fixed by Hamiltonian of the lattice model. Eq.~\eqref{doubleloopeq} then restricts the fusion channels of a defect pair to \eqref{toriccodedefectfusion2a}.

\begin{figure}[htbp]
\includegraphics[width=0.5\textwidth]{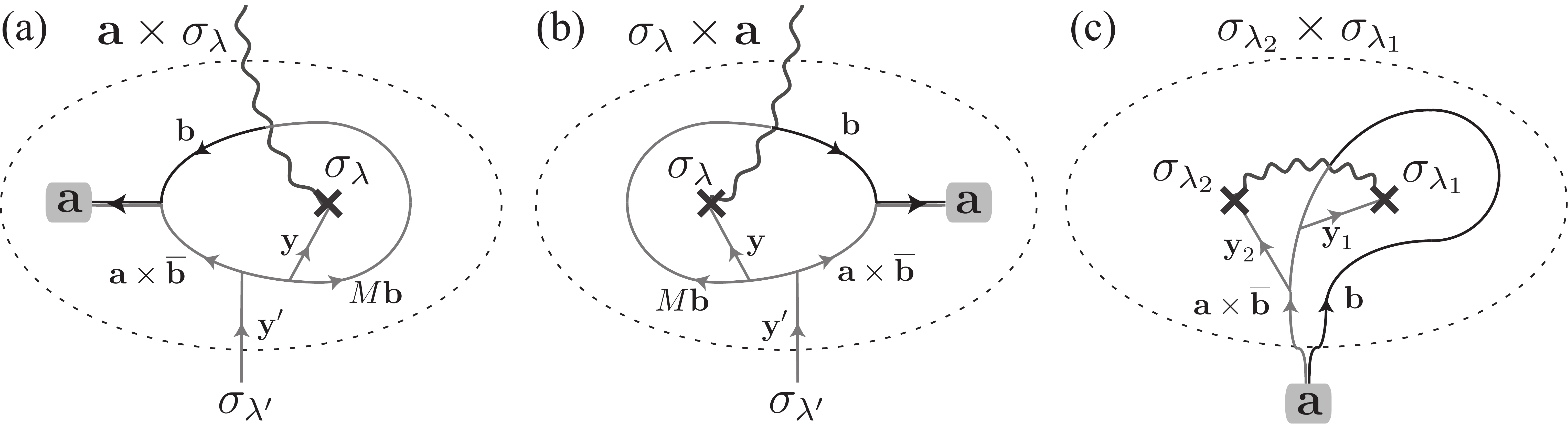}
\caption{Quasiparticle string configurations for splitting states $L^{x_1x_2}_{x_3}$. Branch cuts (wavy lines) switch labels ${\bf b}\to M{\bf b}$ of passing quasiparticles according to the e-m symmetry $M:e\leftrightarrow m$. ${\bf b}=1$ when ${\bf a}=1,e$, or ${\bf b}=e$ when ${\bf a}=m,\psi$. String ${\bf y}$ determines species of the defect where it ends. ${\bf y}=1$ for $\sigma_0$ and ${\bf y}=e$ for $\sigma_1$.}\label{fig:splittingstatestoriccode}
\end{figure}

The idea of {\em splitting states} can be generalized to defect theories. We demonstrate the splitting of a fermion into a pair of bare defects, $\psi\to\sigma_0\times\sigma_0$. $\psi$ can be created from the ground state by an open (semi-infinite) Wilson string of $\sigma_{x,z}$ operators that ends at $\psi$ (see left side of \eqref{splittingdefinition} below). Splitting is the procedure of cutting out a disc containing $\psi,$ and then replacing the interior by a new lattice with two dislocation defects $\sigma$. \begin{align}\vcenter{\hbox{\includegraphics[width=0.3\textwidth]{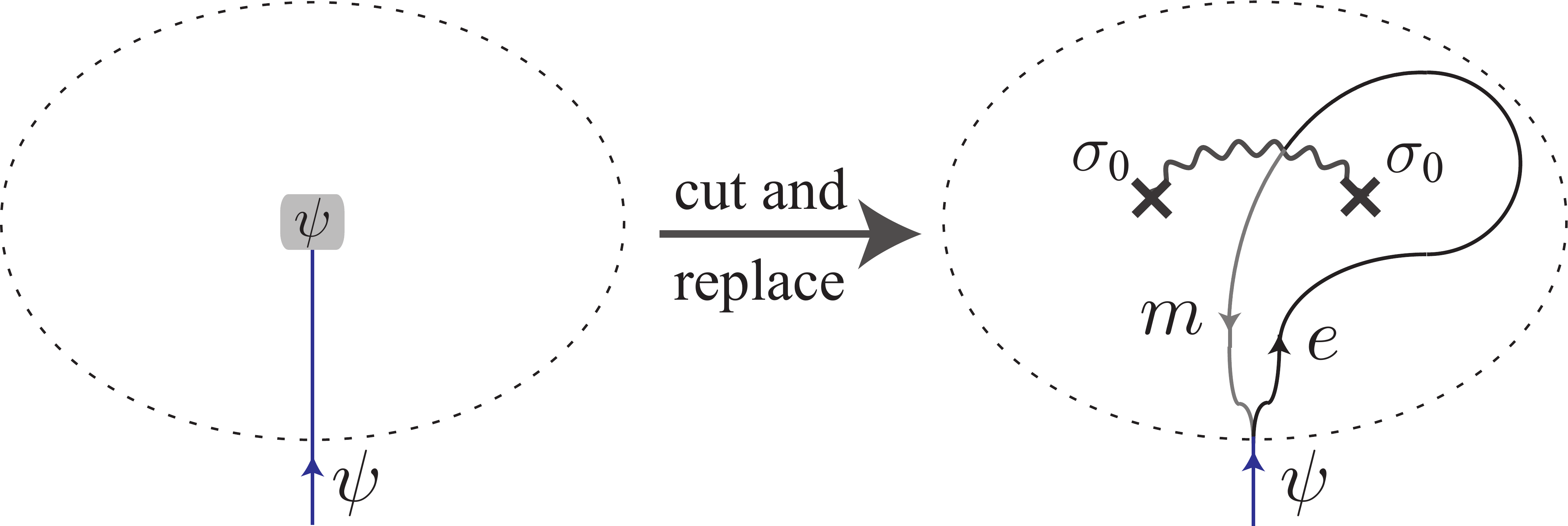}}}.\label{splittingdefinition}\end{align} The replacement must match the boundary condition (along the dashed line) so that the quasiparticle string continues into the domain. However, the string cannot terminate, otherwise there would be another fermion in the new domain, and splitting would be $\psi\to\sigma\times\sigma\times\psi$ instead of $\psi\times\sigma\times\sigma$. This means the Wilson string must wind non-trivially around the defects. One particular possibility is presented in \eqref{splittingdefinition}. A splitting state $[L^{\sigma_0,\sigma_0}_\psi]$ is the equivalence class of such string configurations. For example, the string could deform inside the region and would correspond to the same splitting state. However, the mirror image of \eqref{splittingdefinition} is an inequivalent string pattern, and differs from the original by an $e$-loop around the defect pair. The $e$-loop cuts the $\psi$-string and gives an additional minus sign from crossing. Fig.~\ref{fig:splittingstatestoriccode} picks the string patterns of particular splitting states for $\sigma_{\lambda'}\to{\bf a}\times\sigma_{\lambda}$, $\sigma_{\lambda'}\to\sigma_\lambda\times{\bf a}$ and ${\bf a}\to\sigma_{\lambda_1}\times\sigma_{\lambda_2}$. 

Now let us describe the basis transformations that determine the set of $F$-symbols. Given a particular configuration of defects and quasiparticles, an eigenstate of a maximal set of commuting Wilson loop operators can be specified by the intermediate channels of a fusion tree. For example, the eigenvalues of the following three Wilson loops encodes the same information as the three Abelian anyons ${\bf a}_1,{\bf a}_2,{\bf a}_3$ in the fusion tree:
\begin{align}
\vcenter{\hbox{\includegraphics[width=0.23\textwidth]{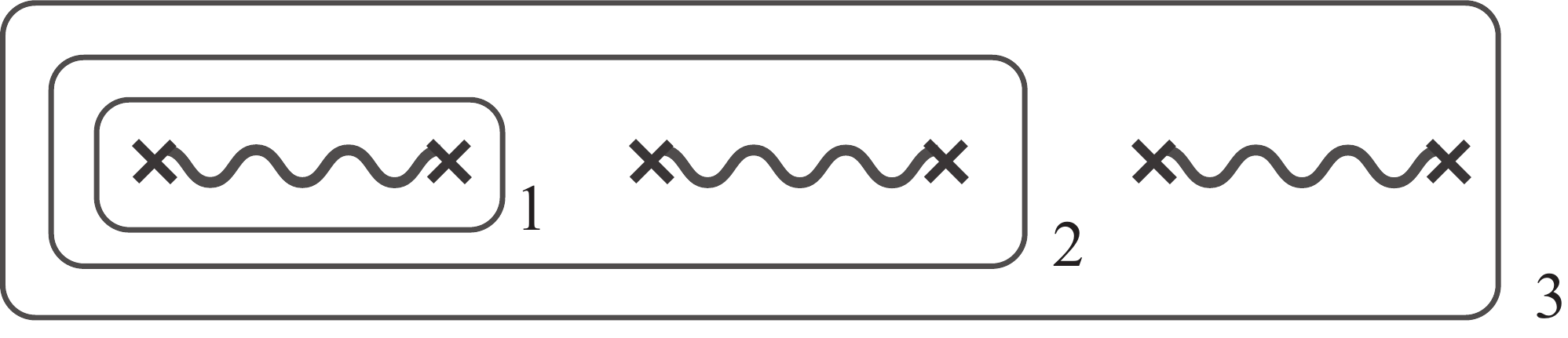}}}\leftrightarrow\left|\vcenter{\hbox{\includegraphics[width=0.16\textwidth]{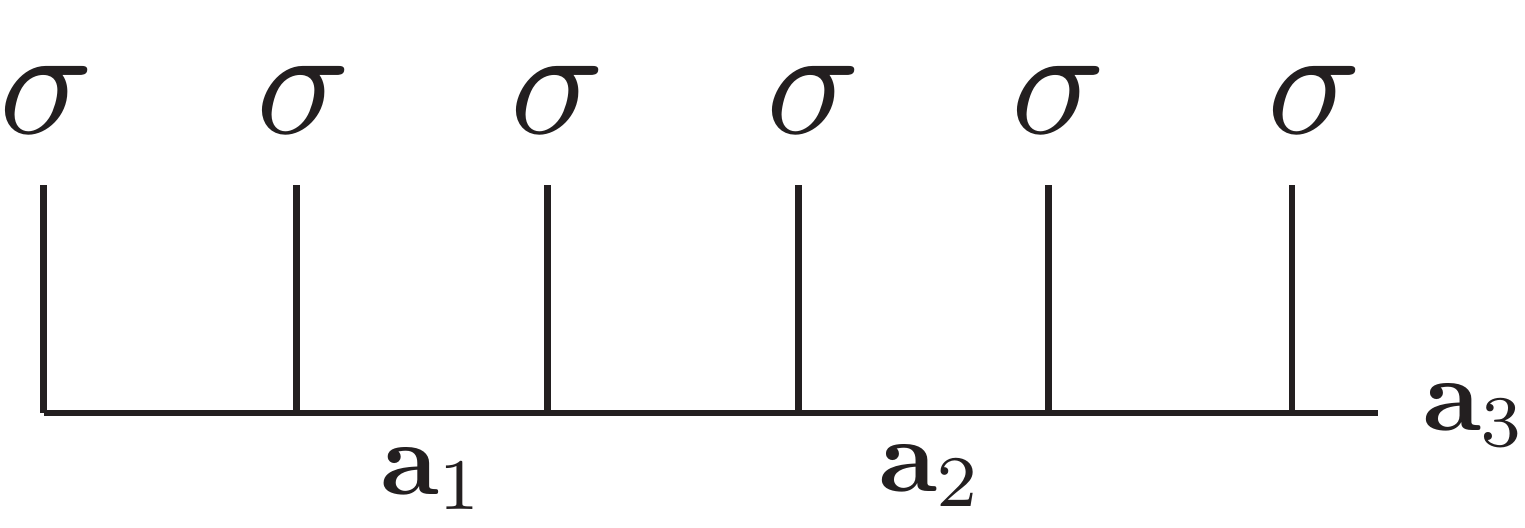}}}\right\rangle
\end{align}
 Each leg of a fusion tree is labeled by a quasiparticle or defect. All vertices must be trivalent and admissible according to the fusion rules. Each vertex associated to a splitting state is diagrammatically represented by specific quasiparticle string patterns as shown in Fig.~\ref{fig:splittingstatestoriccode}. Piecing together the local string patterns of each vertex gives the global string structure. The strings must be attached continuously by matching boundary conditions between splitting states (c.f.~\eqref{splittingtreeapp}).
For example, by patching the splitting state string patterns of $\psi\to\sigma_0\times\sigma_0$ and $\sigma_0\to\psi\times\sigma_0$ from Fig.~\ref{fig:splittingstatestoriccode}(a) and (c), \begin{align}\left|\vcenter{\hbox{\includegraphics[width=0.07\textwidth]{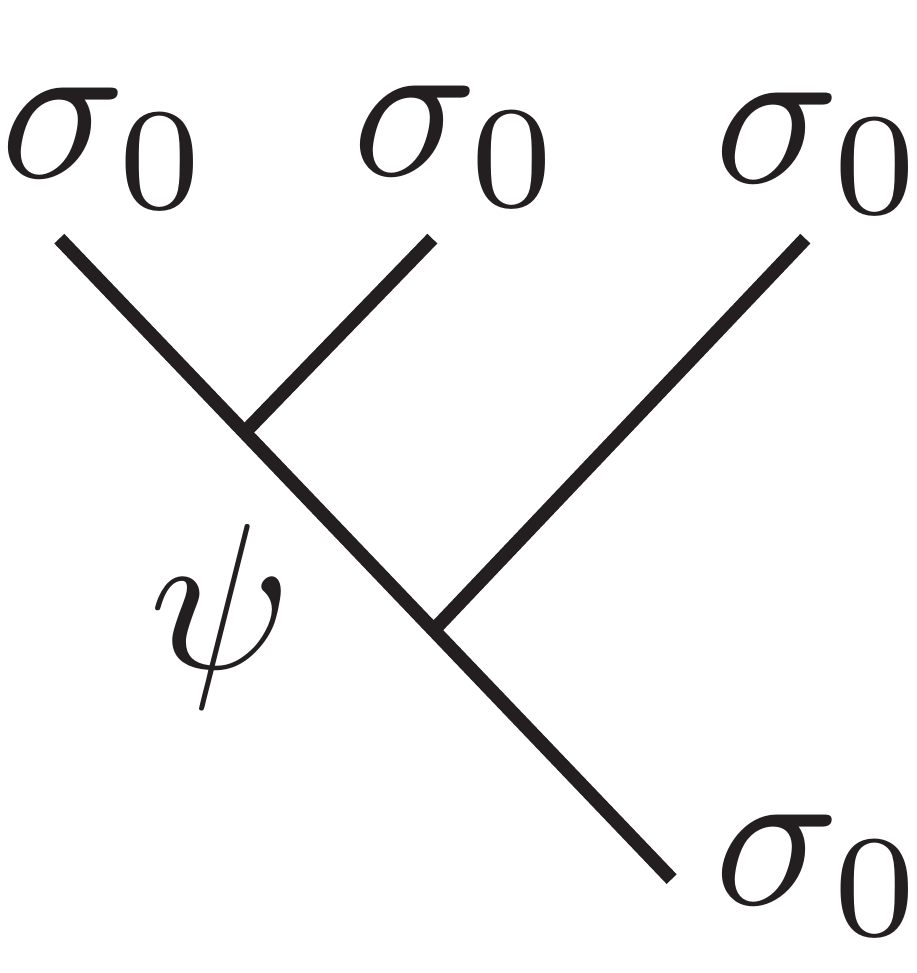}}}\right\rangle=\vcenter{\hbox{\includegraphics[width=0.25\textwidth]{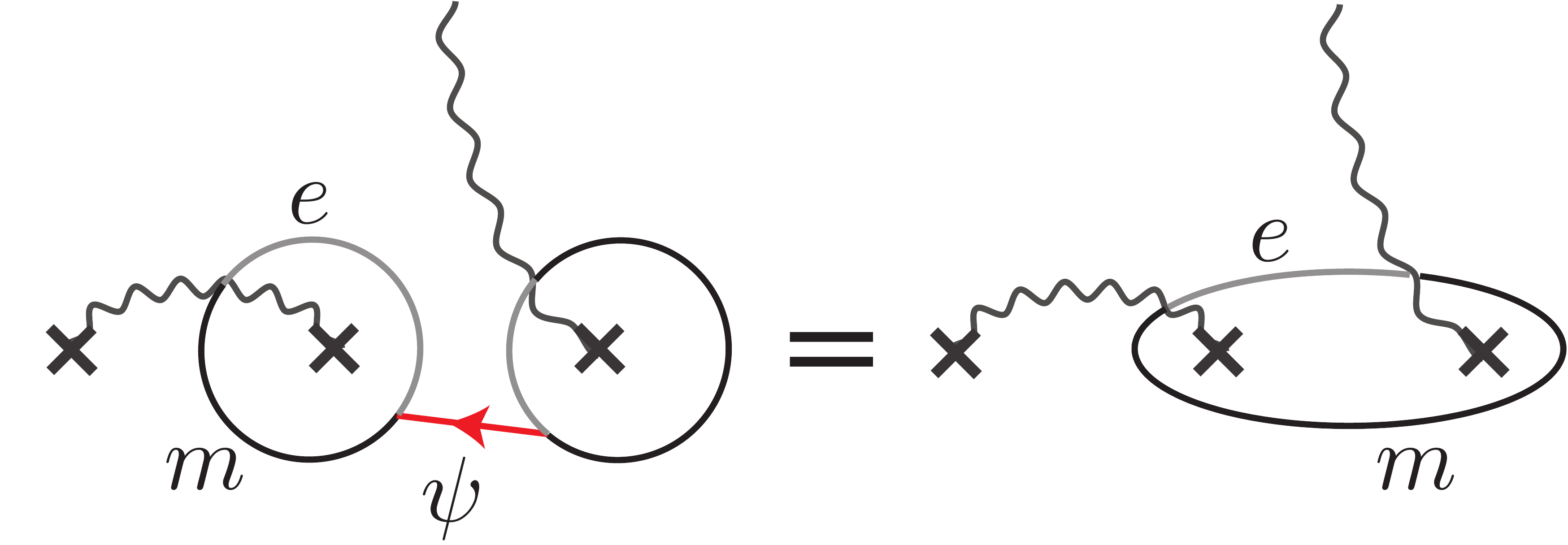}}}.\label{Fsss1toriccode}\end{align} 

The vacuum splitting $1\to\sigma_0\times\sigma_0$ is diagramatically represented by a pair of defects joined together by a branch cut with no quasiparticle strings. The original vacuum requires all Wilson loops $\mathcal{W}$ around the pair to condense to the ground state with a trivial phase, i.e. act trivially on the ground state. In the case of three defects, different configurations of branch cuts correspond to different ground states. For instance, \begin{align}\left|\vcenter{\hbox{\includegraphics[width=0.07\textwidth]{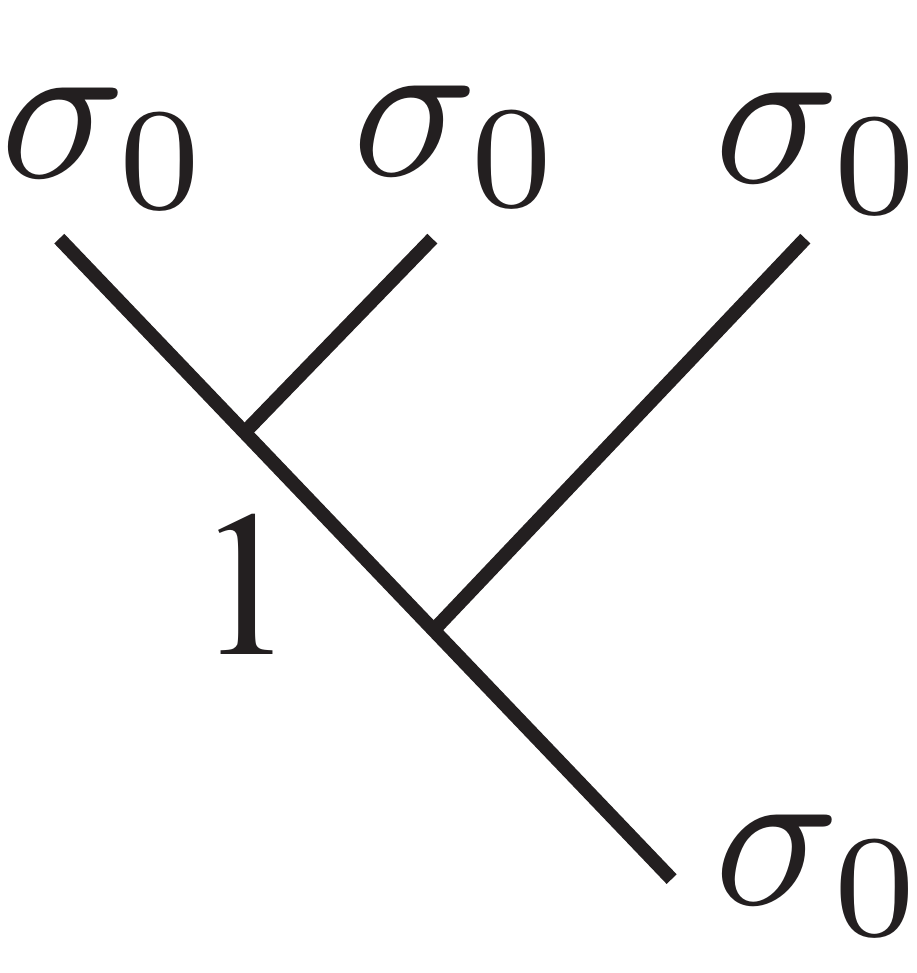}}}\right\rangle&=\vcenter{\hbox{\includegraphics[width=0.1\textwidth]{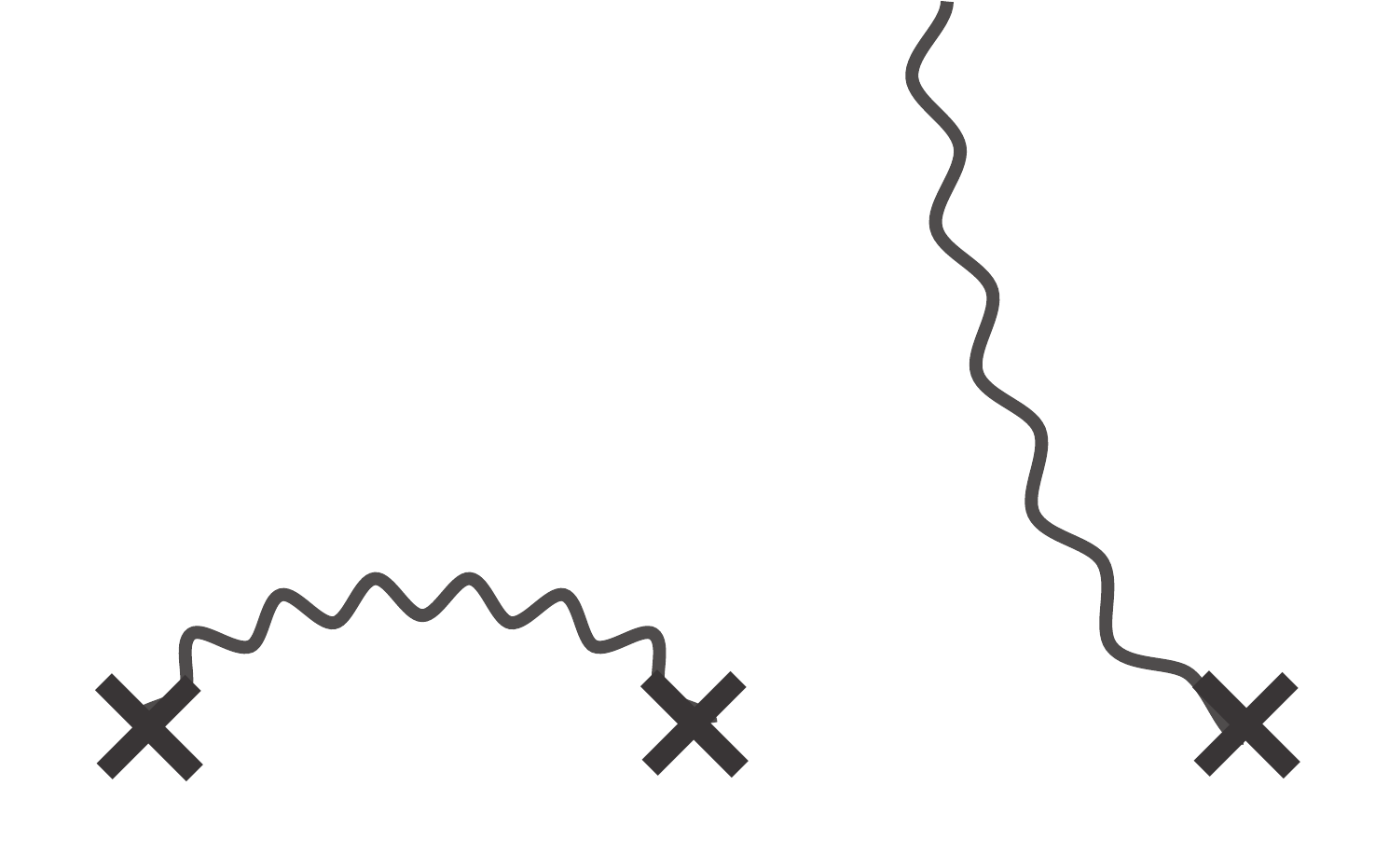}}}=\frac{(-1)^s}{\sqrt{2}}\sum_{{\bf a}=1,m}\vcenter{\hbox{\includegraphics[width=0.1\textwidth]{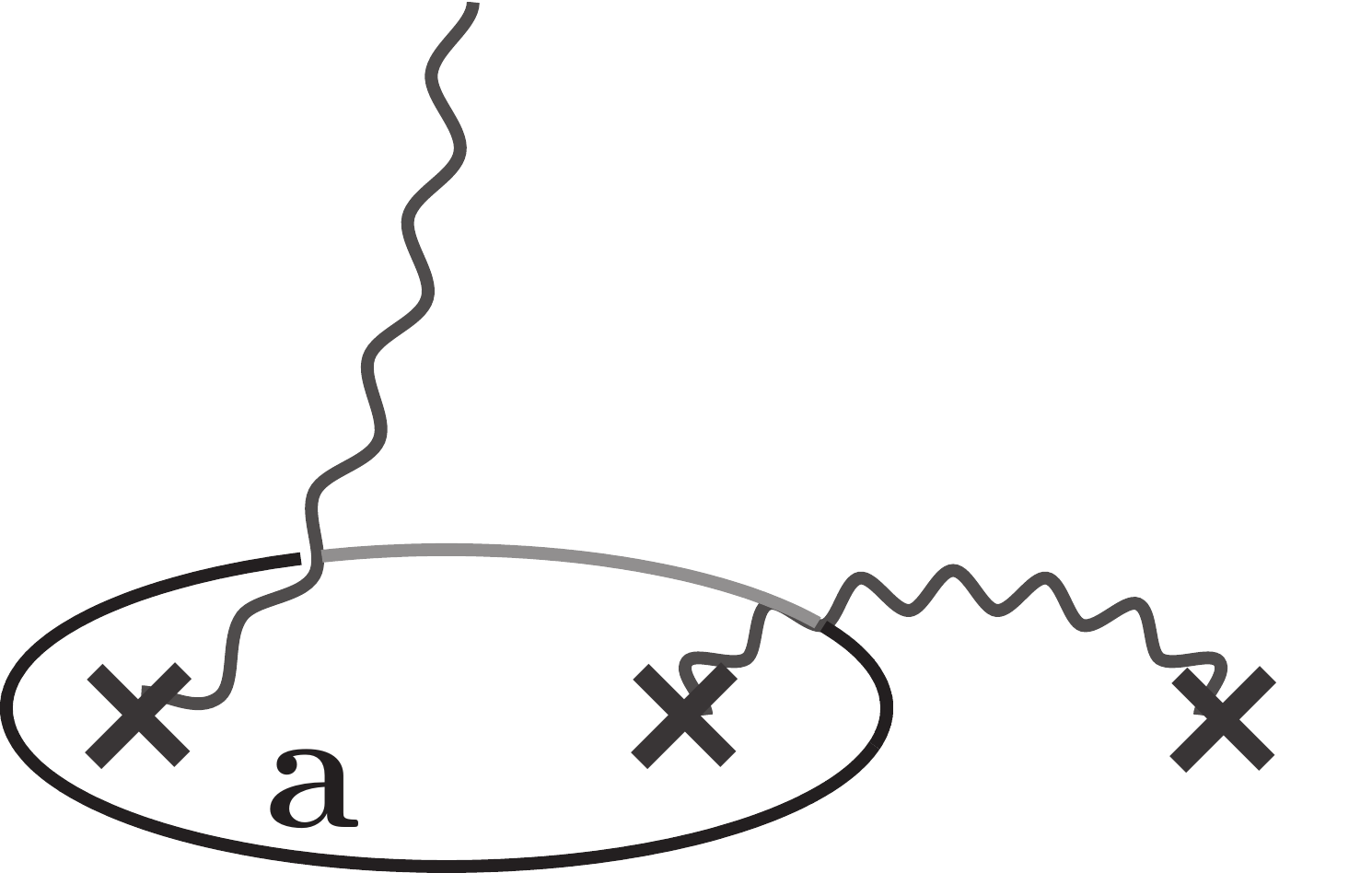}}}\nonumber\\&=\frac{(-1)^s}{\sqrt{2}}\left[\vcenter{\hbox{\includegraphics[width=0.1\textwidth]{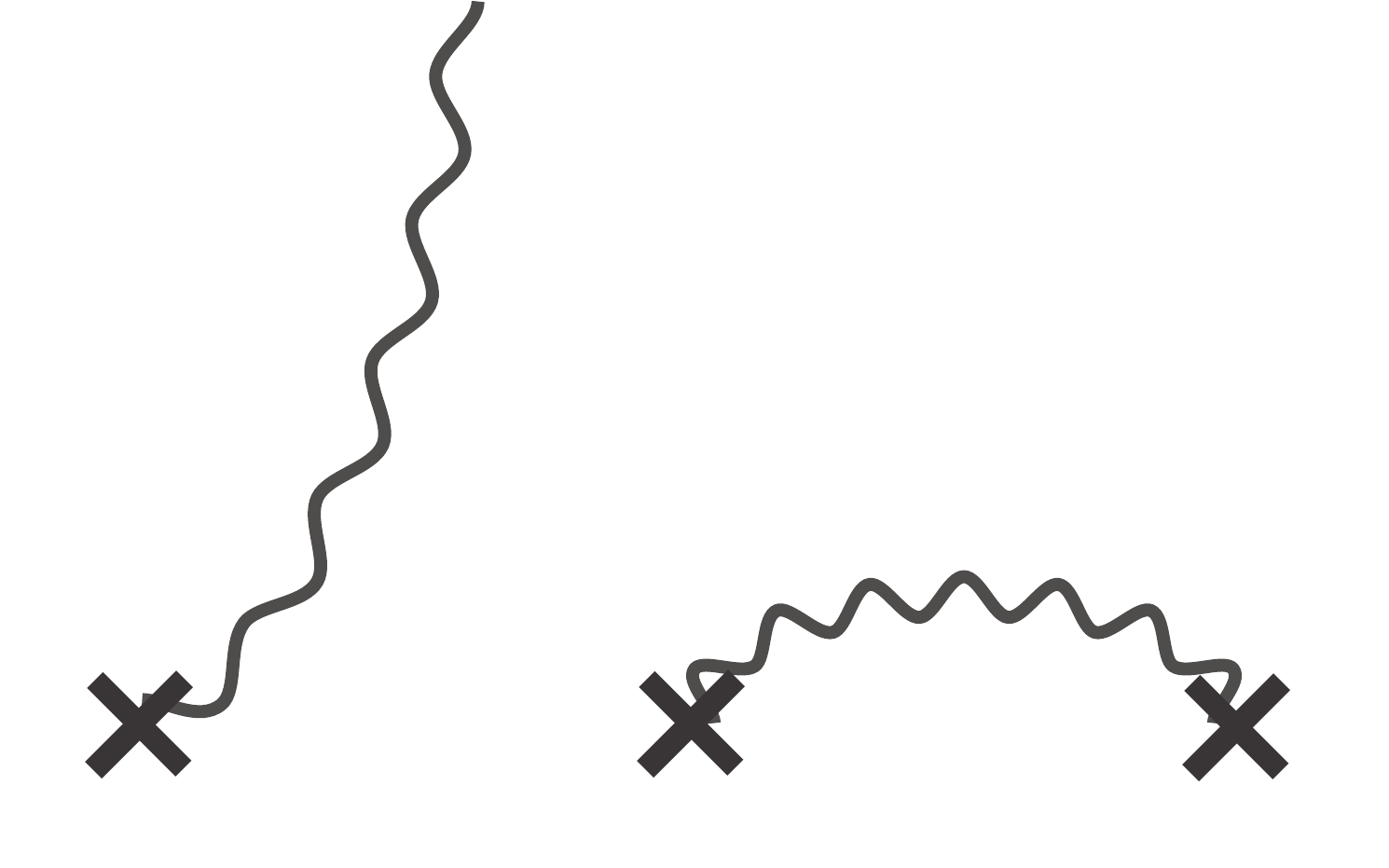}}}\right]+\frac{(-1)^s}{\sqrt{2}}\left[\vcenter{\hbox{\includegraphics[width=0.1\textwidth]{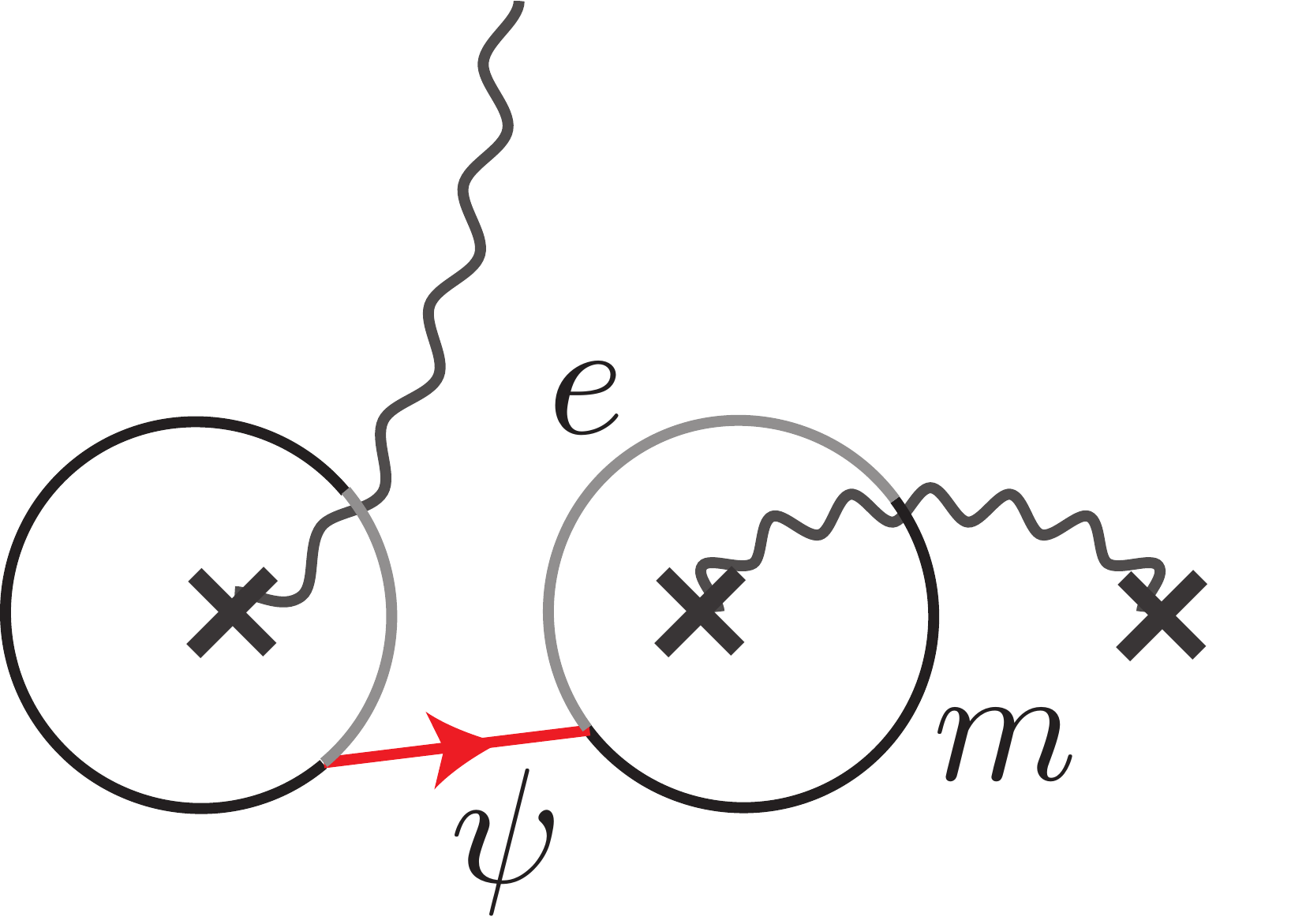}}}\right]\nonumber\\&=\frac{(-1)^s}{\sqrt{2}}\left|\vcenter{\hbox{\includegraphics[width=0.07\textwidth]{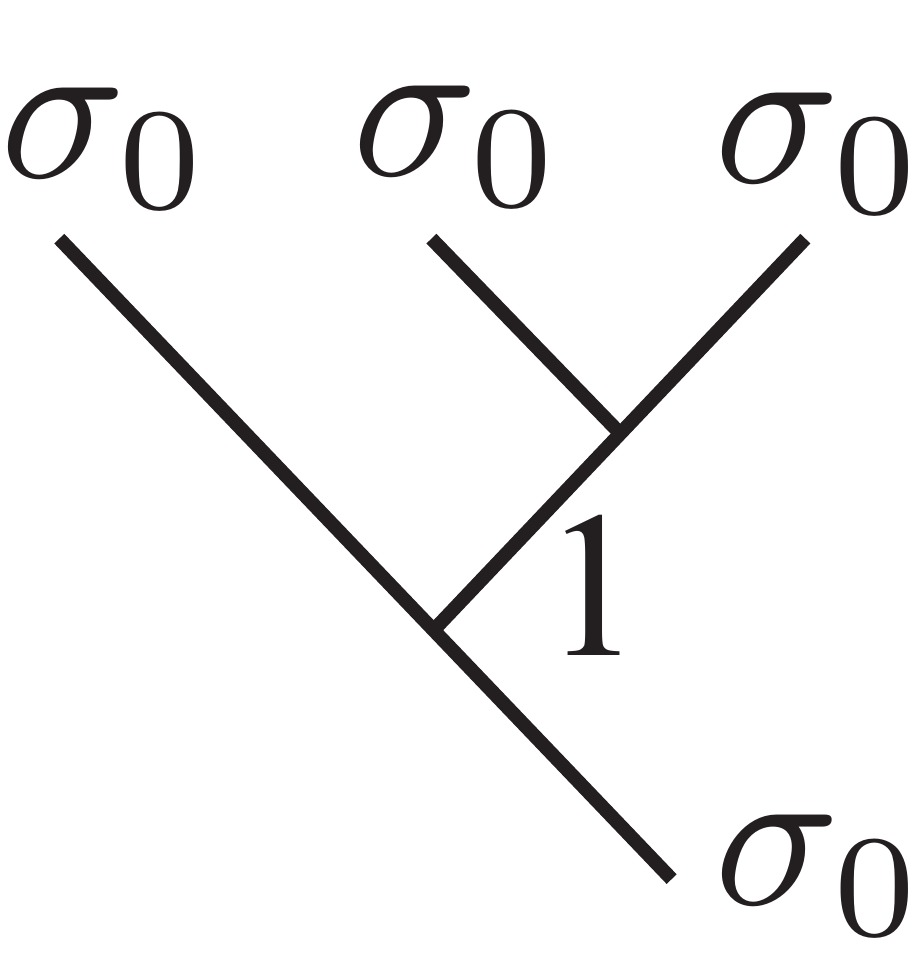}}}\right\rangle+\frac{(-1)^s}{\sqrt{2}}\left|\vcenter{\hbox{\includegraphics[width=0.07\textwidth]{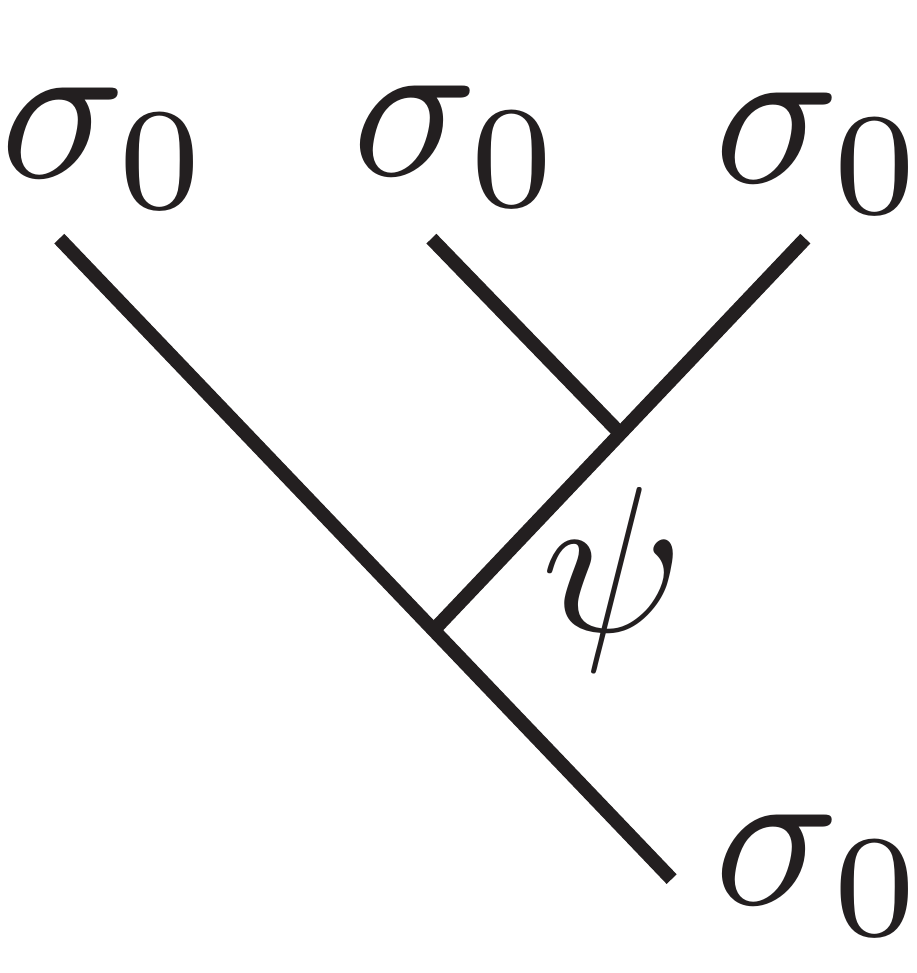}}}\right\rangle\label{Fsss2toriccode}\end{align} where the sum of the Wilson operators ${\bf a}=1,m$ around the first two defects forms a projection operator and forces the vacuum fusion channel. For the toric code, the sign is positive ($s=0$). However when the lattice model is modified by a $\mathbb{Z}_2$-SPT, the sign is flipped ($s=1$). This sign determines the Frobenius-Schur indicator $\varkappa_\sigma=(-1)^s$ (see Eq.~\eqref{IsingFSindicator}), which is classified by $H^3(\mathbb{Z}_2,U(1))=\mathbb{Z}_2$ (see Sections~\ref{sec:defectclassificationobstruction} and \ref{sec:gaugingSPT}).

Now combining \eqref{Fsss1toriccode} and \eqref{Fsss2toriccode}, and keeping track of the crossing phase of the Wilson loops, we have \begin{align}\left|\vcenter{\hbox{\includegraphics[width=0.07\textwidth]{Fsss0toriccode}}}\right\rangle&=\frac{(-1)^s}{\sqrt{2}}\sum_{{\bf a}=1,m}\vcenter{\hbox{\includegraphics[width=0.1\textwidth]{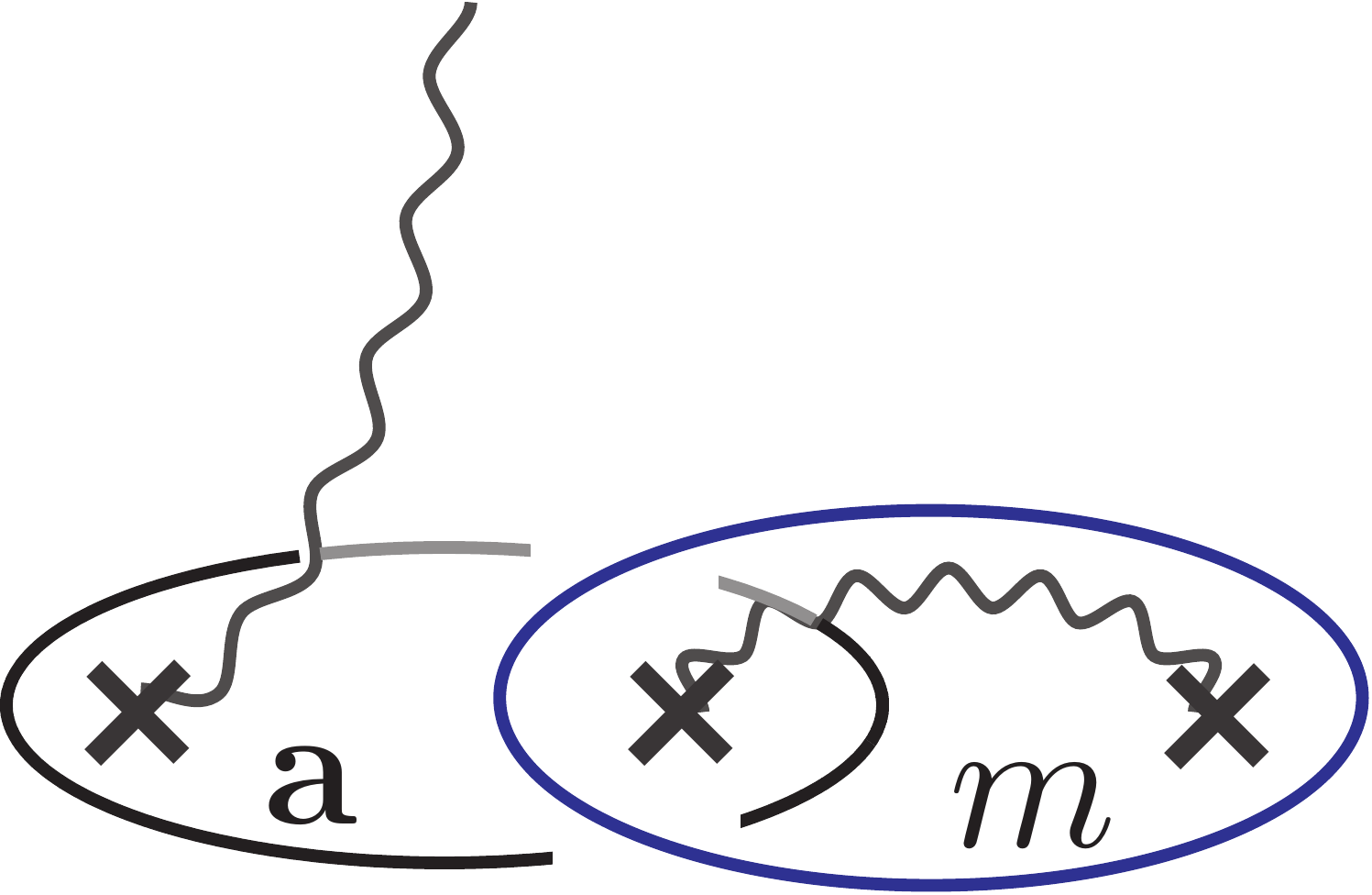}}}\nonumber\\&=\frac{(-1)^s}{\sqrt{2}}\left[\vcenter{\hbox{\includegraphics[width=0.1\textwidth]{Fsss8toriccode}}}\right]-\frac{(-1)^s}{\sqrt{2}}\left[\vcenter{\hbox{\includegraphics[width=0.1\textwidth]{Fsss5toriccode}}}\right]\nonumber\\&=\frac{(-1)^s}{\sqrt{2}}\left|\vcenter{\hbox{\includegraphics[width=0.07\textwidth]{Fsss6toriccode}}}\right\rangle-\frac{(-1)^s}{\sqrt{2}}\left|\vcenter{\hbox{\includegraphics[width=0.07\textwidth]{Fsss7toriccode}}}\right\rangle.\label{Fsss3toriccode}\end{align}
Eqs.~\eqref{Fsss2toriccode} and~\eqref{Fsss3toriccode} describe the basis transformations between different fusion trees. They transform between eigenstates of different sets of Wilson operators. In general, basis transformations are generated by $F$-symbols defined in \eqref{Fsymboldef}. 

Following a similar procedure to that of Eqs.~\eqref{Fsss1toriccode}, \eqref{Fsss2toriccode}, and \eqref{Fsss3toriccode}, we can obtain a consistent set of $F$-symbols with arbitrary admissible labels $\lambda_i=1,e,m,\psi,\sigma_0,\sigma_1$ along the external legs. The fusion trees \begin{align}\left|\vcenter{\hbox{\includegraphics[width=0.5in]{F1}}}\right\rangle&=\mathop{\sum_{boundary}}_{condition}\left[L^{\lambda_3\lambda_2}_{x}\right]\otimes\left[L^{x\lambda_1}_{\lambda_4}\right]|GS\rangle\nonumber\\\left|\vcenter{\hbox{\includegraphics[width=0.5in]{F2}}}\right\rangle&=\mathop{\sum_{boundary}}_{condition}\left[L^{\lambda_3y}_{\lambda_4}\right]\otimes\left[L^{\lambda_2\lambda_1}_{y}\right]|GS\rangle\end{align} can be diagrammatically represented by patching the splitting states defined in Fig.~\ref{fig:splittingstatestoriccode} together with matching boundary conditions. Their overlap can be derived by keeping track of how quasiparticle strings are deformed and intersect. The results for the F-symbols of the defect fusion category based on the toric code anyonic symmetry are summarized in Table~\ref{tab:Fsymbolstoriccode}.

\section{\texorpdfstring{$F$}{F}-Symbols for the \texorpdfstring{$SO(8)_1$}{SO(8)} Defect Fusion Category}\label{app:so8fsymbols}

In this Appendix, we explain the defect $F$-symbols relevant in gauging the threefold symmetry of $SO(8)_1$ in Section~\ref{sec:so(8)symmetry}. 
To express the $F$-symbol basis transformations in a simple notation, it is convenient to represent the Abelian quasiparticles by two dimensional $\mathbb{Z}_2$-valued vectors ${\bf a}=(0,0)=1$, $(1,0)=\psi_1$, $(0,1)=\psi_2,$ and $(1,1)=\psi_3$. In this notation, the threefold symmetry, for example, is represented by the $\mathbb{Z}_2$-valued matrix \begin{align}\Lambda_3=\left(\begin{array}{*{20}c}0&-1\\1&-1\end{array}\right)\equiv\left(\begin{array}{*{20}c}0&1\\1&1\end{array}\right).\end{align} The exchange $R$-symbols \eqref{Rsymboldefapp} between Abelian anyons in $SO(8)_1$ can be chosen to be \begin{align}R^{{\bf a}{\bf b}}=(-1)^{{\bf a}^T\sigma_x\Lambda_3^2{\bf b}},\label{Rso8abelianapp}\end{align} so that the braiding phase, $\mathcal{D}S_{{\bf a}{\bf b}}=R^{{\bf a}{\bf b}}R^{{\bf b}{\bf a}}=(-1)^{{\bf a}^T\sigma_x{\bf b}},$ agrees with that from the conventional $K$-matrix description \eqref{so(8)Kmatrix}. 

The $\mathbb{Z}_3$-defect theory contains the quasiparticles $1,\psi_1,\psi_2,\psi_3$ of the parent state, and threefold defects $\rho,\overline\rho$. Fusion rules can be found in Section~\ref{sec:so(8)defectcategory}. It is worth noting that the $SO(8)_1$-state can be embedded as one of the chiral sectors of the bilayer toric code $D(\mathbb{Z}_2)\otimes D(\mathbb{Z}_2)\approx SO(8)_1^L\otimes SO(8)_1^R$, which can be realized as a lattice model.\cite{BombinMartin06, TeoRoyXiao13long} In this model the twist defects manifest as lattice disclinations and dislocations, and the defect $F$-symbols were previously computed in Ref.[\onlinecite{TeoRoyXiao13long}]. Here we sketch the results in one of the chiral $SO(8)_1$ sector.

First, the $R$-symbol  in Eq.~\ref{Rso8abelianapp} is bilinear in ${\bf a}$ and ${\bf b},$ and therefore the $F$-symbols involving only Abelian quasiparticles are trivial: $F^{{\bf a}{\bf b}{\bf c}}_{{\bf a}+{\bf b}+{\bf c}}=1$.
Next, the $F$-symbols for threefold defects are determined by fixing the splitting states. The splitting of $\rho\to{\bf a}\times\rho$ and ${\bf a}\to\rho\times\overline\rho$ are defined by the Wilson string structures \begin{align}\left[\vcenter{\hbox{\includegraphics[width=0.3in]{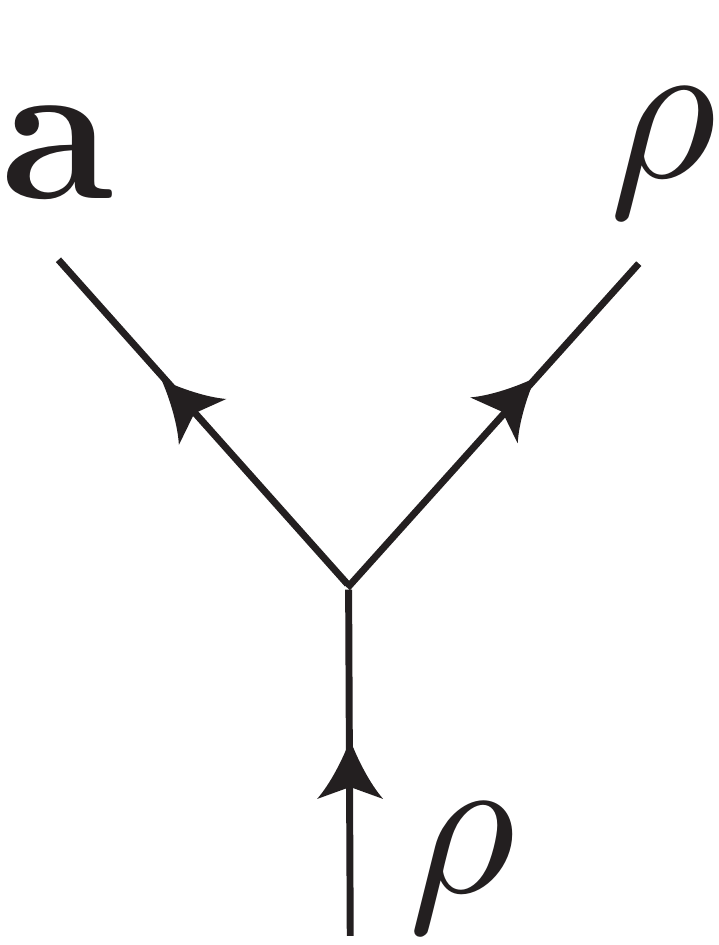}}}\right]=\vcenter{\hbox{\includegraphics[width=0.7in]{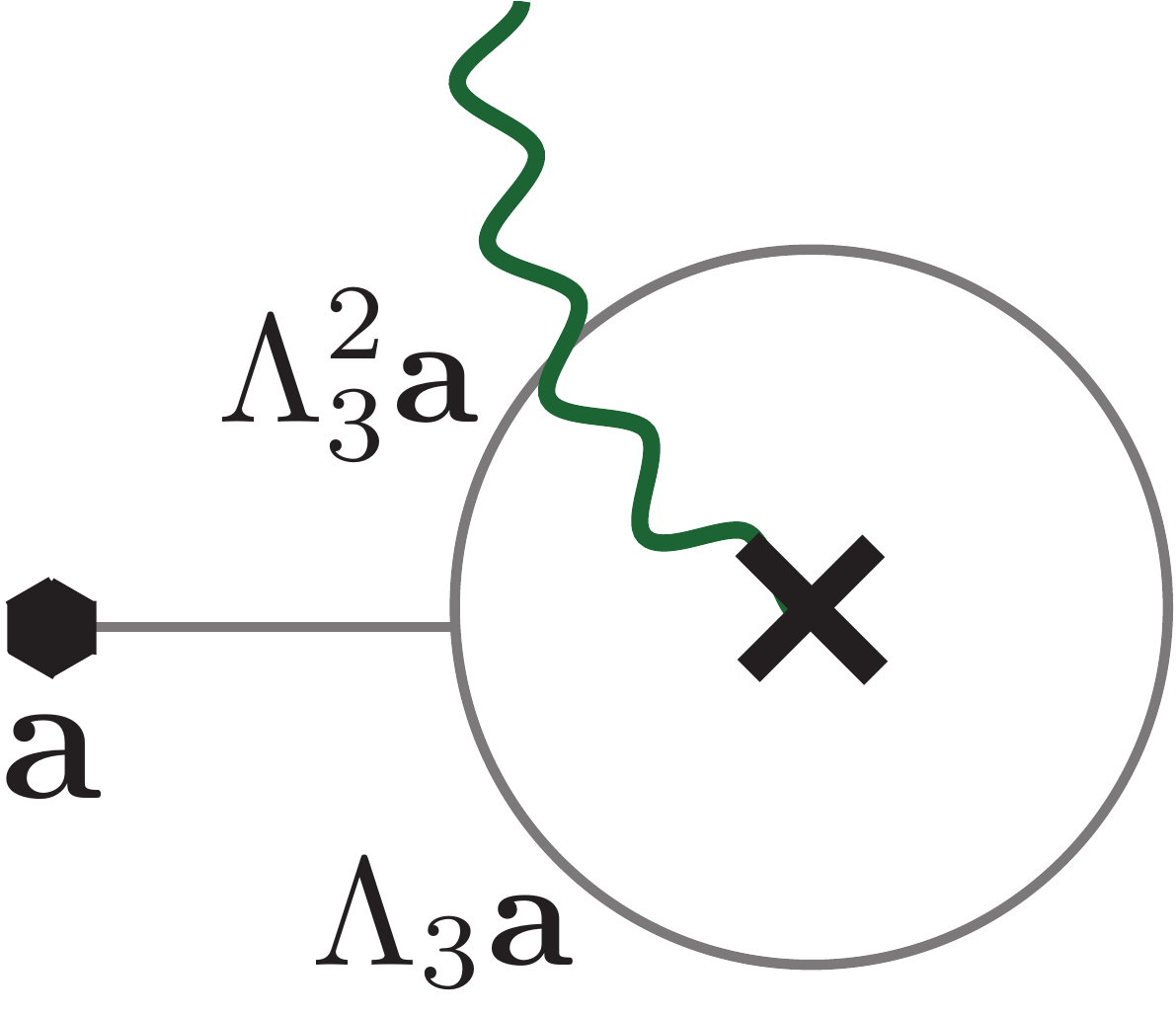}}},\quad\left[\vcenter{\hbox{\includegraphics[width=0.3in]{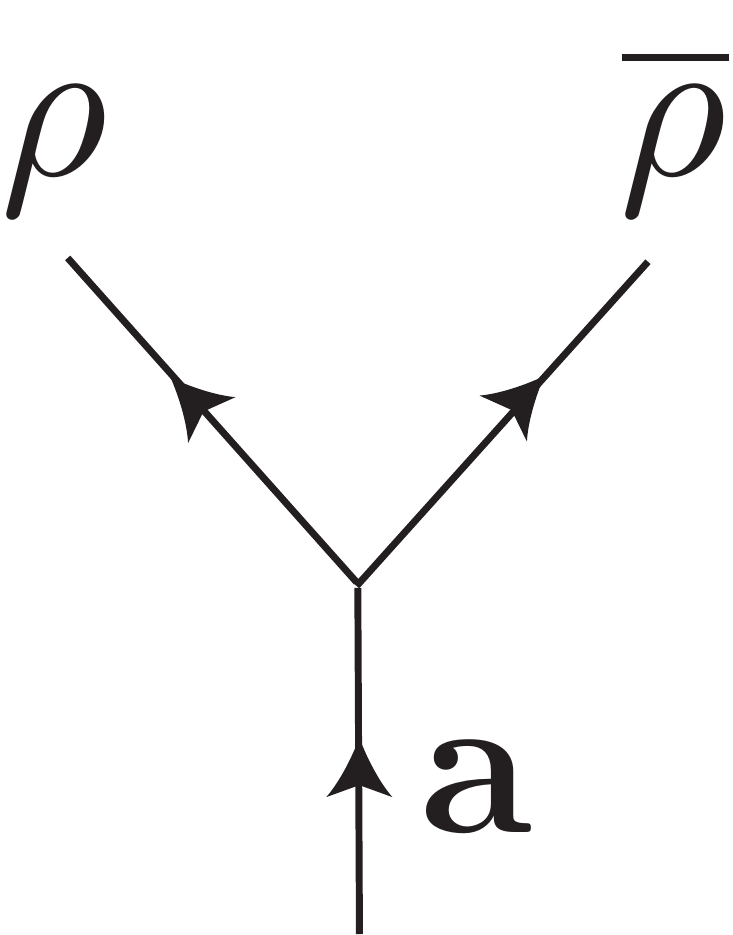}}}\right]=\vcenter{\hbox{\includegraphics[width=0.7in]{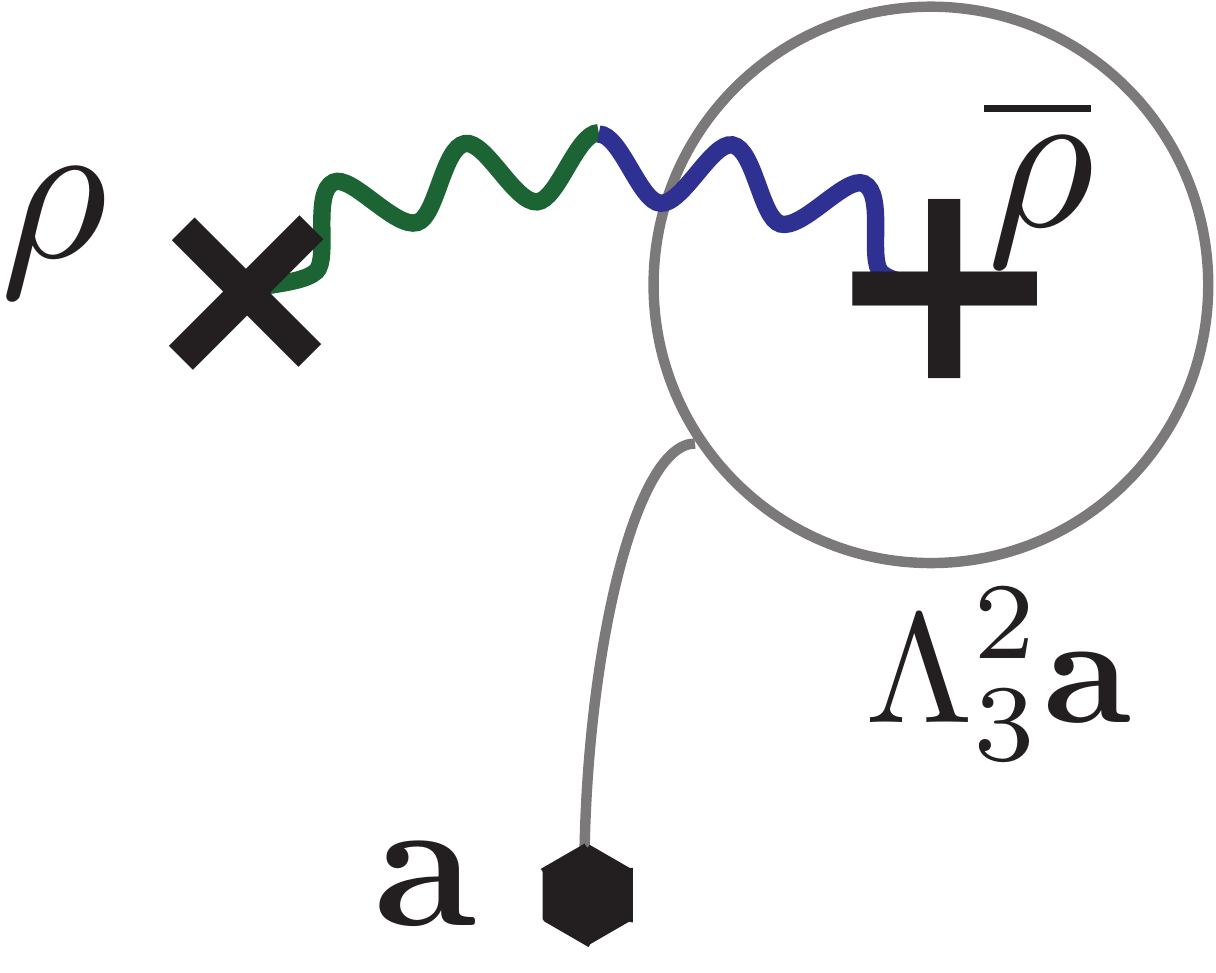}}}.\label{so(8)splittings}\end{align} Here a quasiparticle string changes type ${\bf a}\to\Lambda_3{\bf a}$ (or ${\bf a}\to\Lambda_3^{-1}{\bf a}$) as it goes from left to right across a $\rho$-branch cut represented by a green curvy line (resp.~a $\overline\rho$-branch cut represented by a blue curvy line).

Let $\mathcal{W}_{\bf b}$ be the ${\bf b}$-Wilson loop around the defect pair $\rho\times\overline\rho$. It commutes with the splitting state $1\to\rho\times\overline\rho$ in \eqref{so(8)splittings} because there are no Wilson strings intersecting $\mathcal{W}_{\bf b}$ and the overall vacuum state has unit eigenvalue with respect to $\mathcal{W}_{\bf b}$. On the ground state, this means \begin{align}\left|\vcenter{\hbox{\includegraphics[width=0.7in]{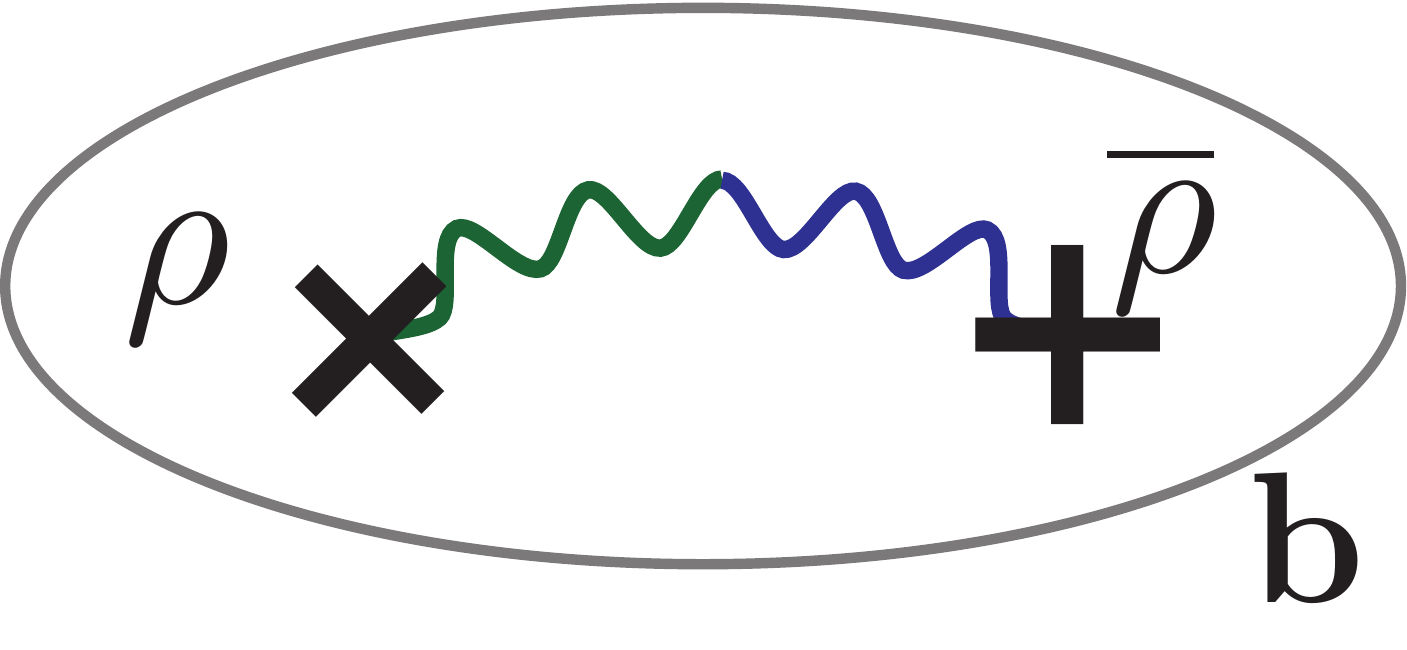}}}\right\rangle=\left|\vcenter{\hbox{\includegraphics[width=0.6in]{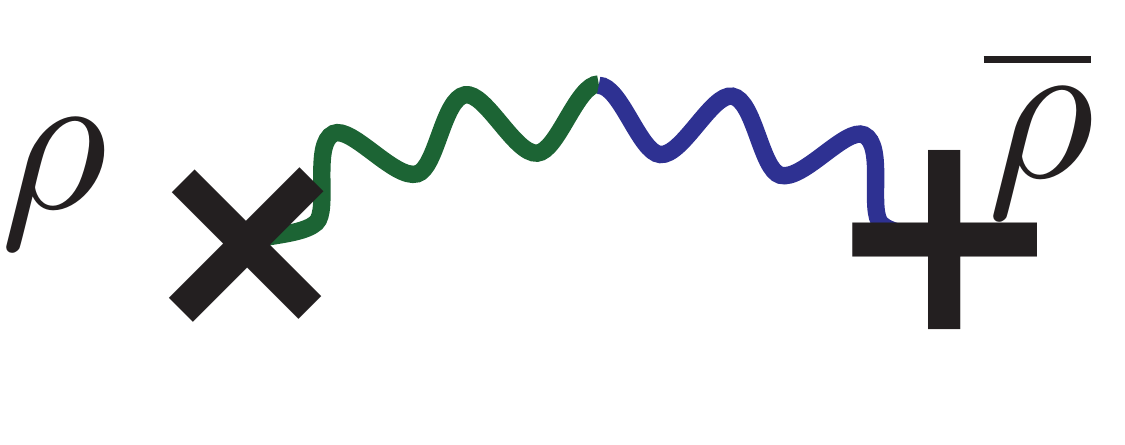}}}\right\rangle.\end{align} 

The two dimensional fusion degeneracy $N^{\overline\rho}_{\rho\rho}=2$ arises from an irreducible representation of the non-commuting algebra of Wilson operators \begin{align}\widehat{\mathcal{A}}_{\bf a}\left|\vcenter{\hbox{\includegraphics[width=0.3in]{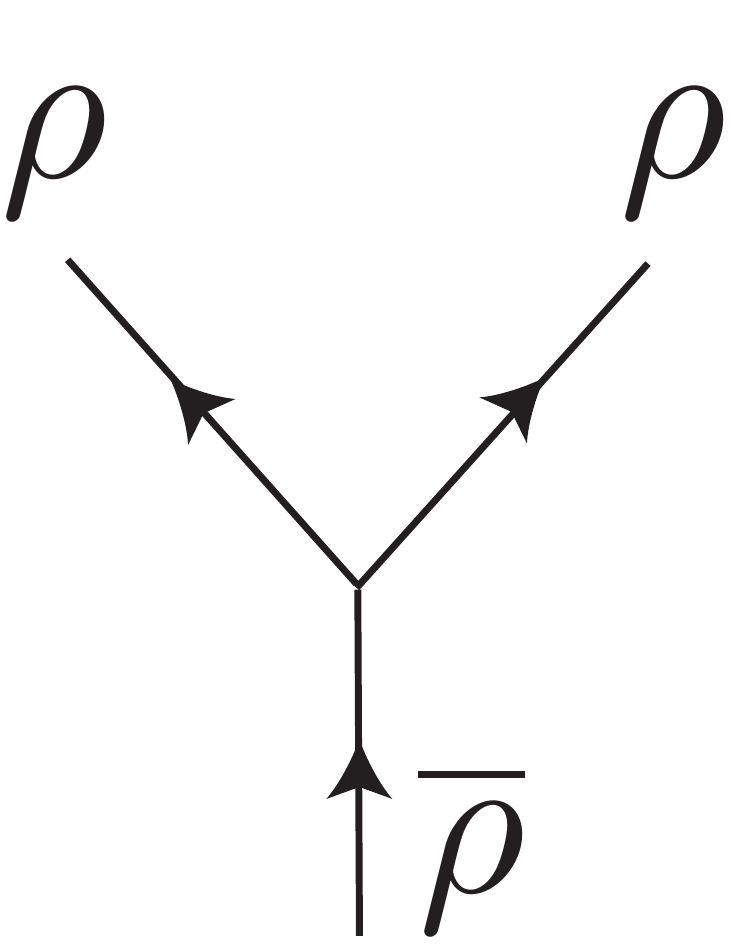}}}\right\rangle=\vcenter{\hbox{\includegraphics[width=0.8in]{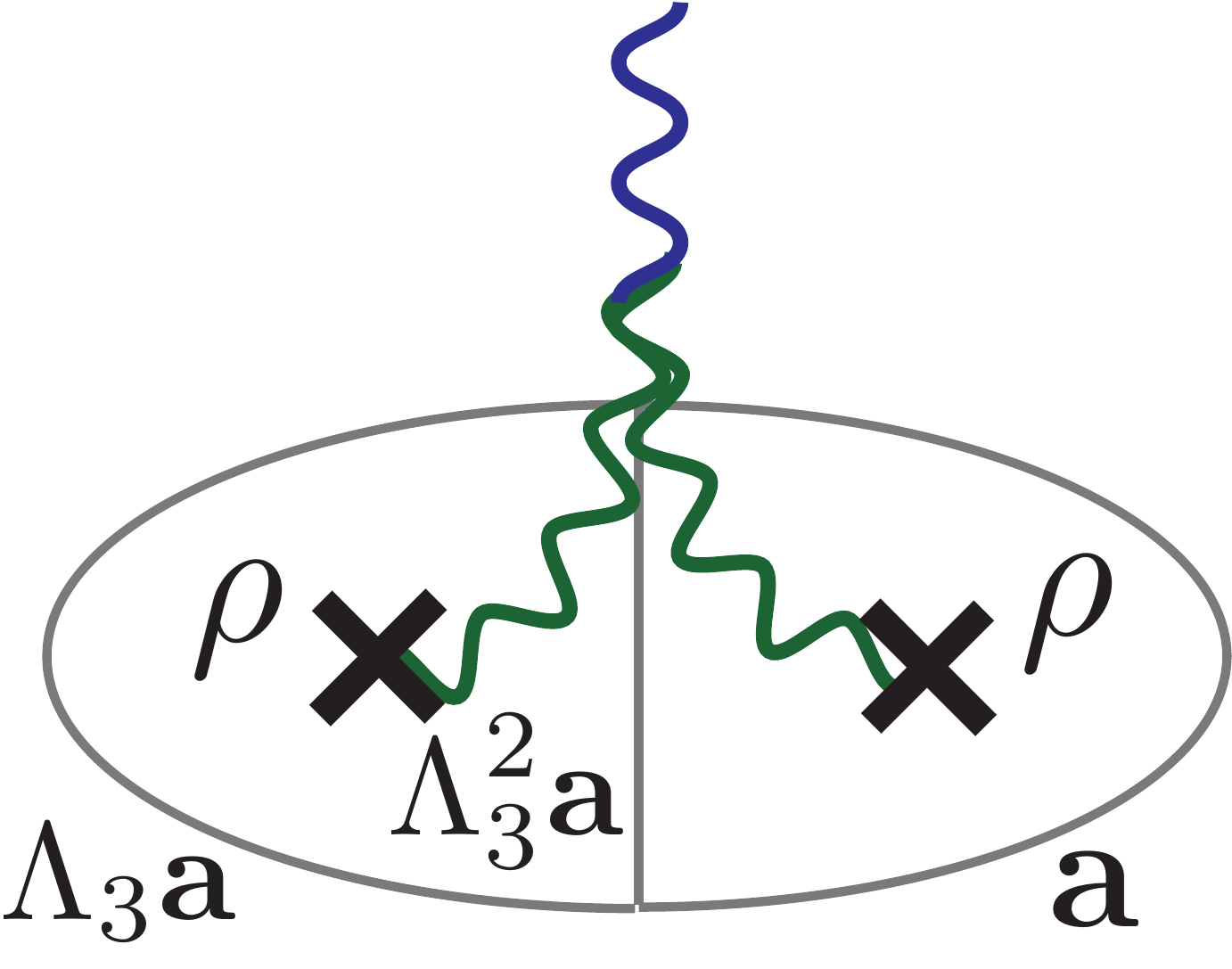}}}.\label{so(8)splittingrhorho}\end{align} By keeping track of the crossing phases of the Wilson strings, we find that the operators obey the algebraic relation \begin{align}\widehat{\mathcal{A}}_{\bf a}\widehat{\mathcal{A}}_{\bf b}=R^{{\bf a}{\bf b}}\widehat{\mathcal{A}}_{{\bf a}+{\bf b}}\label{so(8)AWilsonalgebra}\end{align} for $\widehat{\mathcal{A}}_a=1$. In particular, they satisfy the Clifford relation $\{\widehat{\mathcal{A}}_{\psi_i},\widehat{\mathcal{A}}_{\psi_j}\}=-2\delta_{ij},$ and exhibit the group structure of unit quaternions, $Q_8=\{\pm1,\pm\widehat{\mathcal{A}}_{\psi_1},\pm\widehat{\mathcal{A}}_{\psi_2},\pm\widehat{\mathcal{A}}_{\psi_3}\}$.  As such, they can be represented by $2\times2$ Pauli matrices \begin{align}\mathcal{A}_{\psi_1}=i\sigma_x,\quad\mathcal{A}_{\psi_2}=-i\sigma_y,\quad\mathcal{A}_{\psi_3}=i\sigma_z.\label{so(8)Arepapp}\end{align} In certain scenarios, it may be more appropriate to use a basis $|\pm\rangle$ which is symmetric under the threefold symmetry so that $\langle\pm|\widehat{\mathcal{A}}_{\psi_j}|\pm\rangle=\pm i/\sqrt{3}$ and $\langle+|\widehat{\mathcal{A}}_{\psi_j}|-\rangle=\sqrt{2/3}e^{2\pi ji/3}$. In addition to this algebraic structure, the splitting of $\rho\to\overline\rho\times\overline\rho$ has the same two-fold degeneracy. We thus have additional Wilson operators $\overline{\mathcal{A}}_{\bf a},$ which are defined similarly to \eqref{so(8)splittingrhorho}, except the threefold symmetry $\Lambda_3$ should be replaced by its inverse $\Lambda_3^{-1}=\Lambda_3^2$. The algebraic structure for $\overline{\mathcal{A}}_{\bf a}$ is identical to that of $\mathcal{A}_{\bf a}$ in \eqref{so(8)AWilsonalgebra}. Hence, we will not make the distinction between $\mathcal{A}$ and $\overline{\mathcal{A}}$ unless necessary. 


The $F$-symbols for threefold defects in $SO(8)_1$ are listed in Table~\ref{tab:so(8)Fsymbols}. They are evaluated by matching the Wilson strings of the splitting states $x\times(y\times z)$ and $(x\times y)\times z,$ as highlighted in Section~\ref{sec:twistdefects}, as well as the previous Appendix. The $F$-transformations can be understood as a representation of the double cover of $A_4,$ which is the group of even permutations of 4 elements, or equivalently, the rotation group $T$ of a tetrahedron. The three $\mathcal{A}_{\psi_i}$ represent $\pi$-rotations about the $x$, $y,$ or $z$ axes and \begin{align}F^{\rho\rho\rho}_1=\exp\left[\frac{\pi}{3}\left(\frac{\mathcal{A}_{\psi_1}+\mathcal{A}_{\psi_2}+\mathcal{A}_{\psi_3}}{\sqrt{3}}\right)\right]\label{Frrrapp}\end{align} for instance, represents a $2\pi/3$ rotation about the $(111)$-axis. 

We focus on the derivations of the $F$-symbols in table~\ref{tab:so(8)Fsymbols} that are relevant in solving the \hexeq's in Section~\ref{sec:so(8)symmetry}, namely $F^{{\bf ab}\rho}_\rho$, $F^{{\bf a}\rho{\bf b}}_\rho$, $F^{\rho{\bf ab}}_\rho$ and $F^{\rho\rho\rho}_{\bf a}$. Applying the first splitting state in \eqref{so(8)splittings}, the successive splittings $\rho\to{\bf a}\times\rho\to{\bf a}\times({\bf b}\times\rho)$ associate the Wilson string structure \begin{align}\left|\vcenter{\hbox{\includegraphics[width=0.07\textwidth]{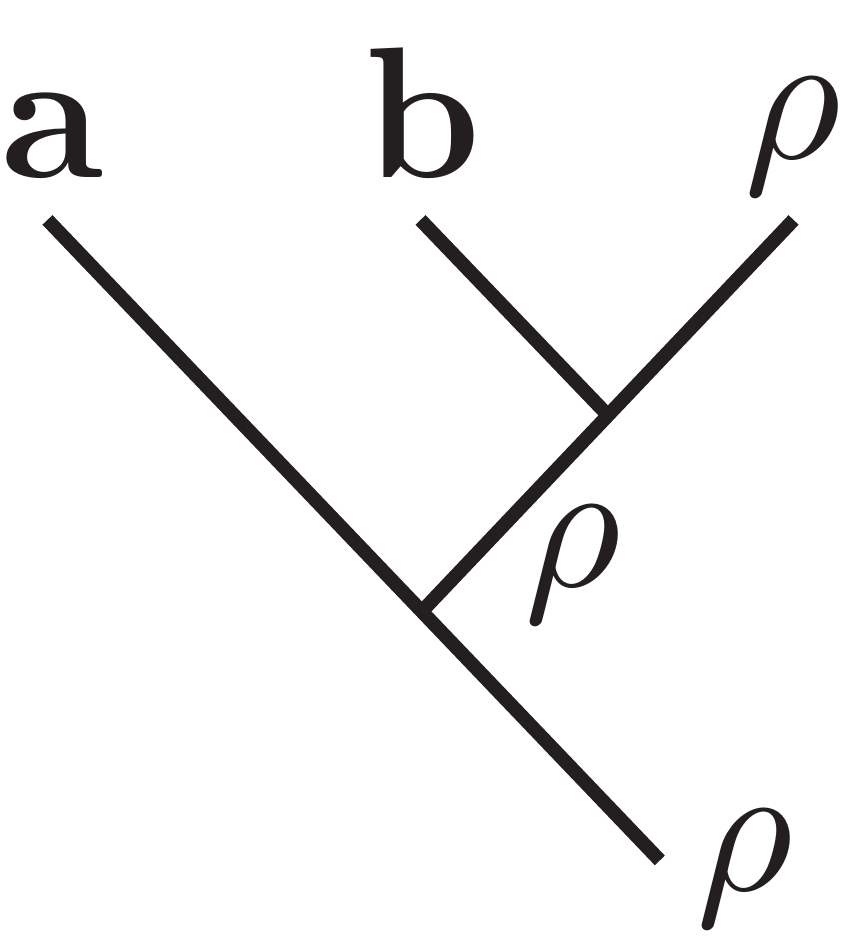}}}\right\rangle=\left[\vcenter{\hbox{\includegraphics[width=0.1\textwidth]{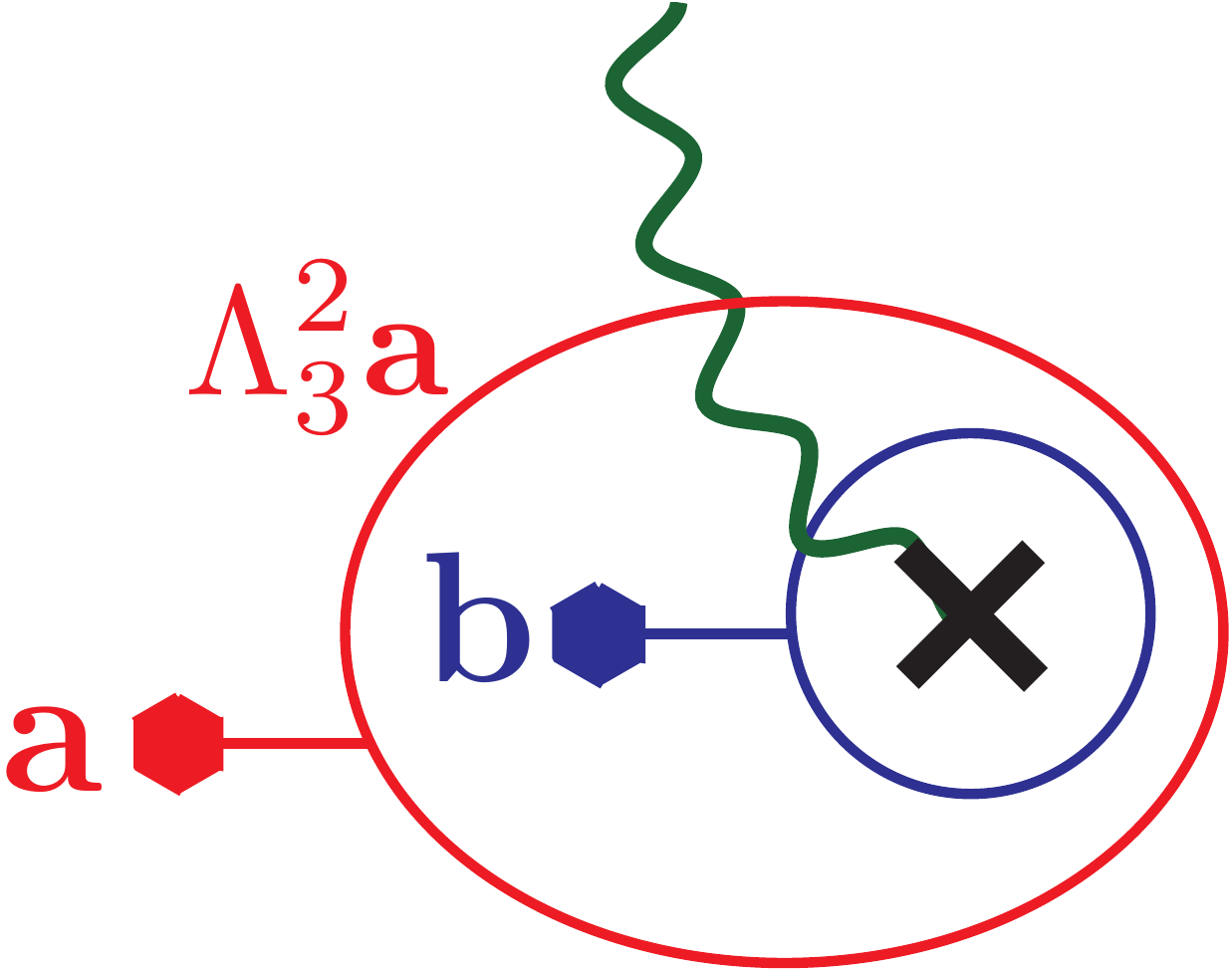}}}\right]|GS\rangle\end{align} where the $\Lambda_3^2{\bf a}$ string can be deformed by the Wilson loop condensate $|GS\rangle$ at the background to go underneath the ${\bf b}$ string. The two set of strings can then be put on the same level by a $R$-move \begin{align}\left[\vcenter{\hbox{\includegraphics[width=0.1\textwidth]{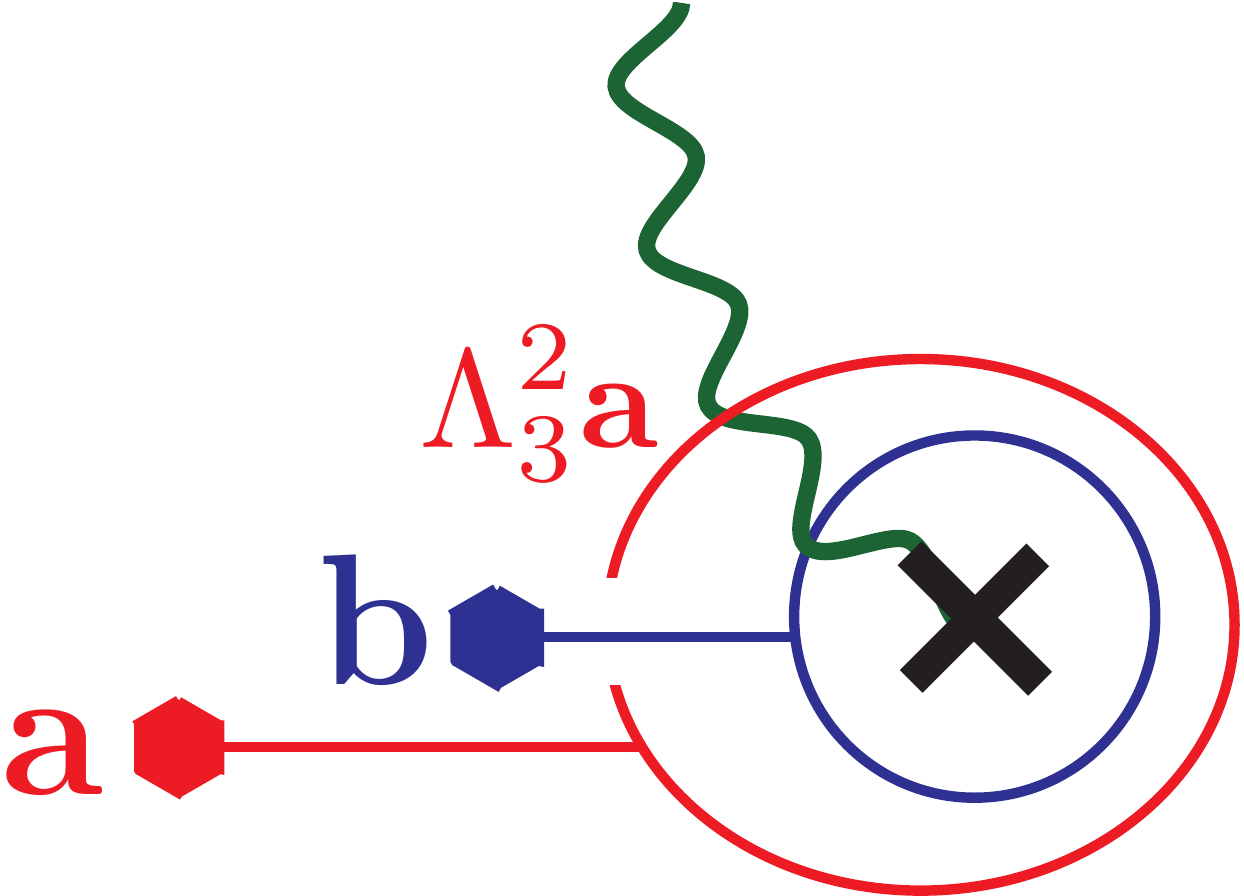}}}\right]&=R^{{\bf b}(\Lambda_3^2{\bf a})}\left[\vcenter{\hbox{\includegraphics[width=0.1\textwidth]{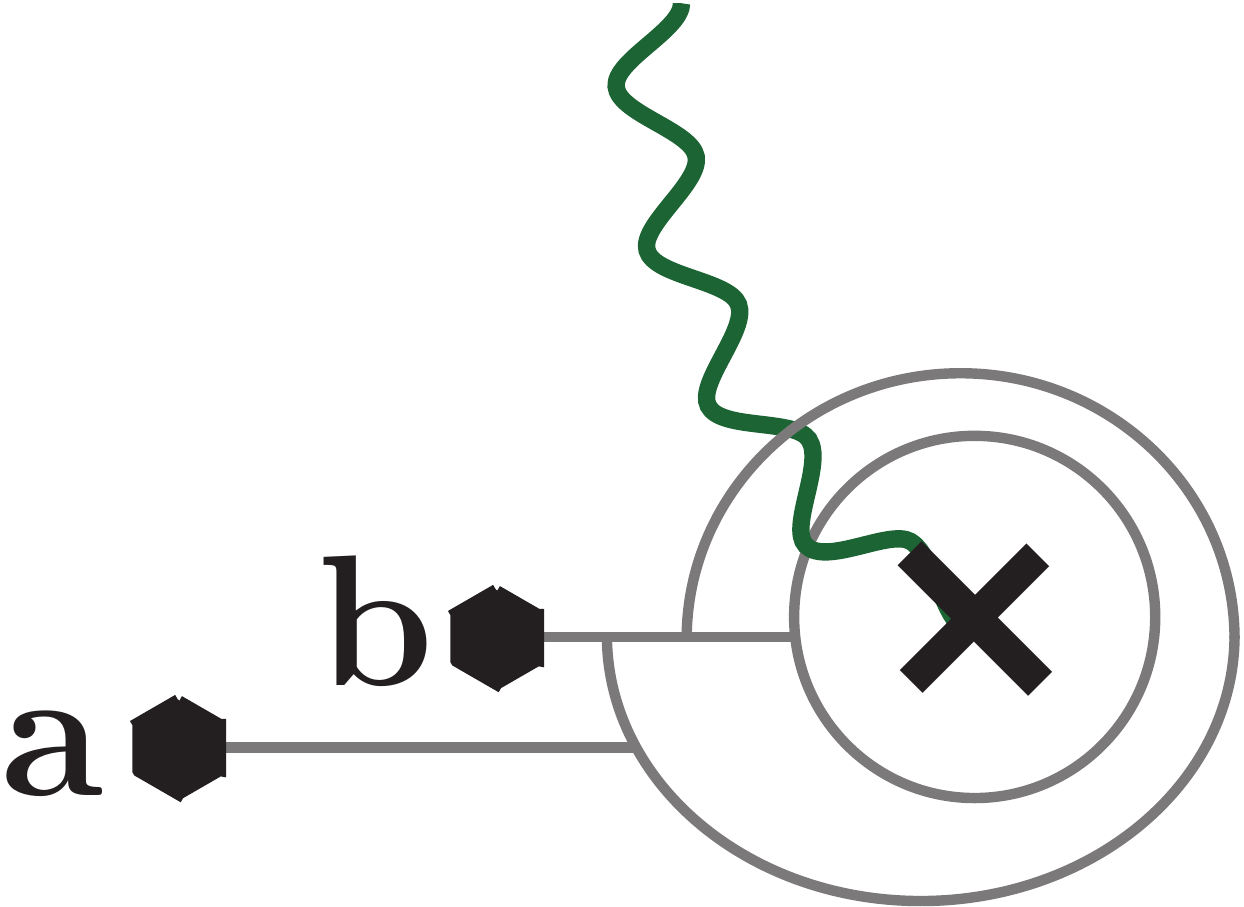}}}\right]\nonumber\\&=R^{{\bf b}(\Lambda_3^2{\bf a})}\left[\vcenter{\hbox{\includegraphics[width=0.08\textwidth]{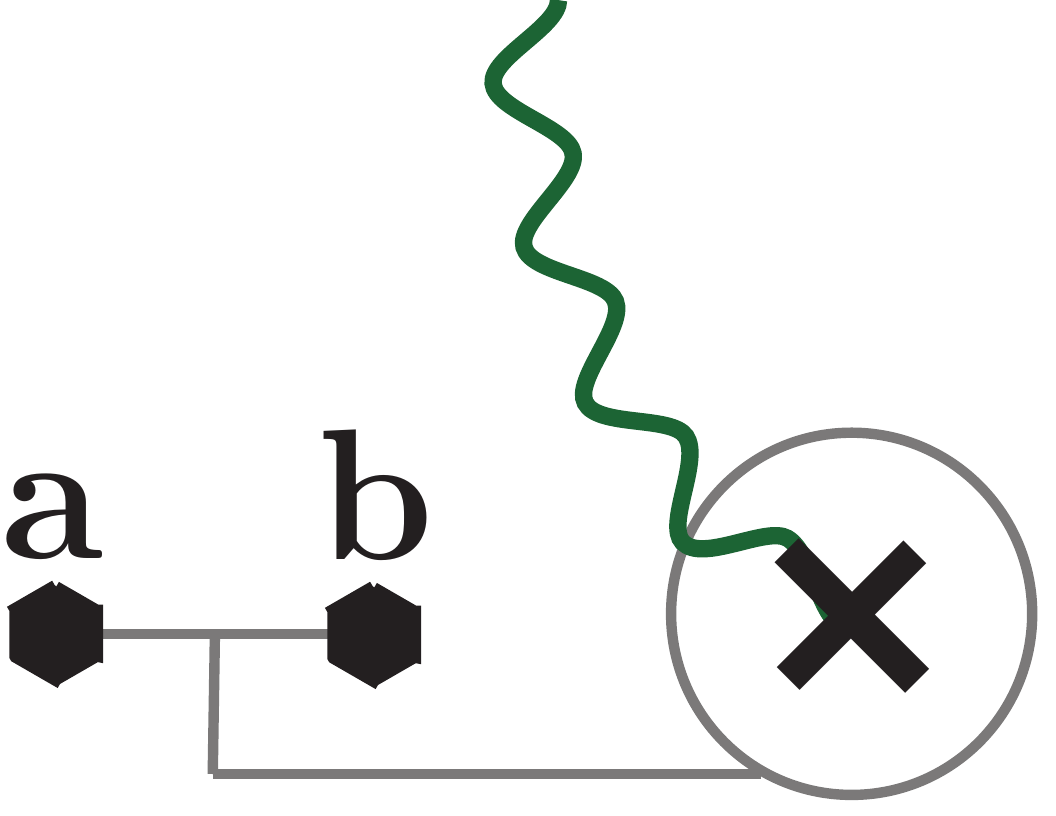}}}\right]\end{align} where the last step is done by a series of $F$-moves, which give only trivial phases $F^{\bf abc}_{\bf d}=1$. The final product is exactly the Wilson string structure of the splitting $\rho\to({\bf a}\times{\bf b})\times\rho$. Thus we arrive at the $F$-symbol $F^{{\bf ab}\rho}_\rho=R^{{\bf b}(\Lambda_3^2{\bf a})}$. $F^{{\bf a}\rho{\bf b}}_\rho=\mathcal{D}S_{(\Lambda_3{\bf a}){\bf b}}$ and $F^{\rho{\bf ab}}_\rho=R^{{\bf a}(\Lambda_3{\bf b})}$ can be derived in a similar manner.

Next we consider the splitting of a quasiparticle ${\bf b}$ into the triplet $\rho\times\rho\times\rho$. Recall there is a 2-fold degenerate vertex space assoiates to each splitting $\overline\rho\to\rho\times\rho$. The $F$-symbol $F^{\rho\rho\rho}_{\bf b}$ is a $2\times2$ basis transformation between the vertex space $V^{12}$ of the first pair of $\rho$'s and that of the last pair $V^{23}$. There are three types of closed Wilson loops \begin{gather}\widehat{\mathcal{A}}_{\bf a}^{12}=\vcenter{\hbox{\includegraphics[width=0.1\textwidth]{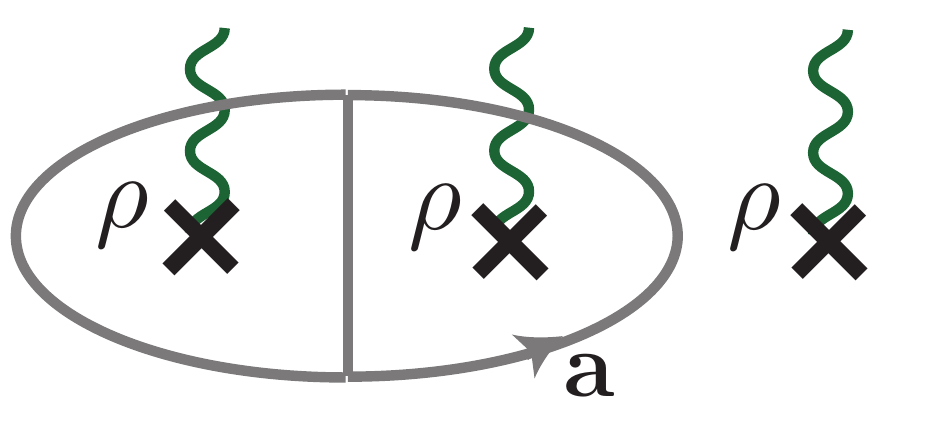}}},\quad\widehat{\mathcal{A}}_{\bf a}^{23}=\vcenter{\hbox{\includegraphics[width=0.1\textwidth]{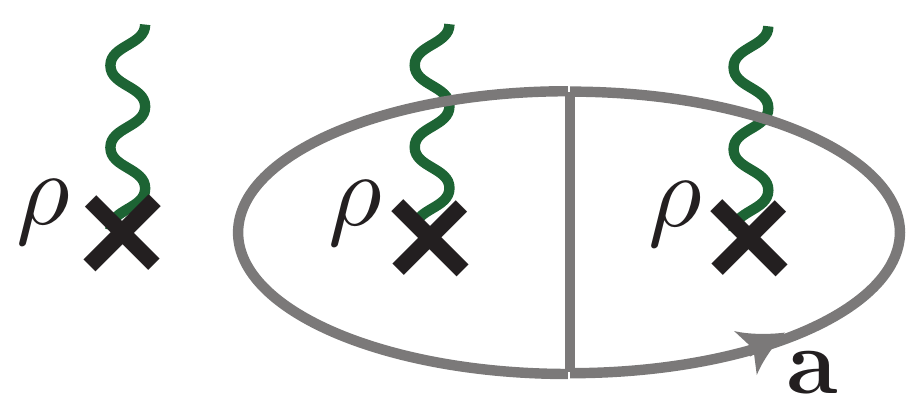}}},\nonumber\\\widehat{\mathcal{W}}_{\bf a}=\vcenter{\hbox{\includegraphics[width=0.12\textwidth]{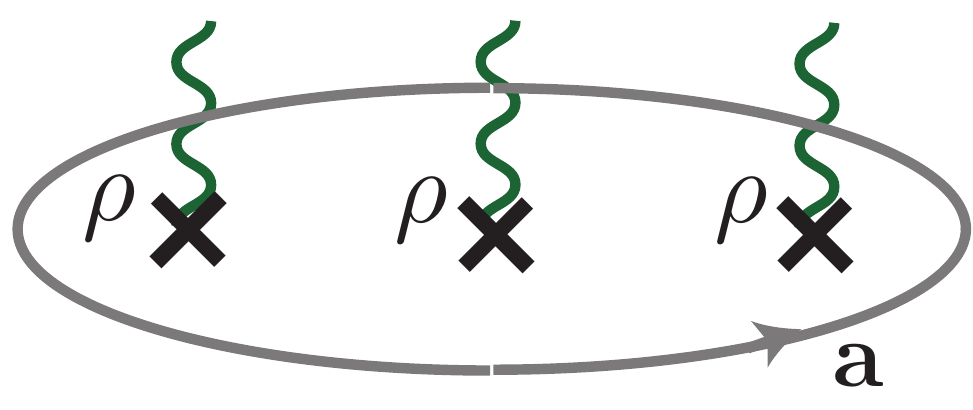}}}\end{gather} where $\mathcal{W}_{\bf a}$ has the fixed the eigenvalue $\mathcal{D}S_{\bf ab}=(-1)^{{\bf a}^T\sigma_x{\bf b}}$ because of the overall channel ${\bf b}$. Applying \eqref{Rso8abelianapp} and $F^{\bf abc}_{\bf d}=1$, the Wilson operators overlap and obey \begin{align}\widehat{\mathcal{A}}_{\Lambda_3^{-1}{\bf a}}^{12}\widehat{\mathcal{A}}_{\bf a}^{23}=\widehat{\mathcal{A}}_{\bf a}^{23}\widehat{\mathcal{A}}_{\Lambda_3^{-1}{\bf a}}^{12}=-\mathcal{W}_{\bf a}=-(-1)^{{\bf a}^T\sigma_x{\bf b}}\label{AAWrelation}\end{align} For example the trivial overall channel ${\bf b}=0$ requires the identification $\mathcal{A}^{12}_{\bf a}=\mathcal{A}^{23}_{\Lambda_3{\bf a}}$. This threefold cyclic rotation of ${\bf a}\to\Lambda_3{\bf a}$ can be generated in the quantum state level by noticing that \begin{align}\mathcal{A}_{\Lambda_3{\bf a}}=F^{\rho\rho\rho}_1\mathcal{A}_{\bf a}{F^{\rho\rho\rho}_1}^\dagger\end{align} where the $(111)$-rotation $F^{\rho\rho\rho}_1$ is represented by \eqref{Frrrapp}. $F^{\rho\rho\rho}_1$ therefore provides the basis transformation that connects the vertex spaces $V^{12}$ and $V^{23}$. For general overall channel ${\bf b}$, $F^{\rho\rho\rho}_{\bf b}=\mathcal{A}_{\bf b}F^{\rho\rho\rho}_1$ represents threefold rotation about the $(\bar{1}11)$, $(1\bar{1}1)$ or $(11\bar{1})$ axis because of the the extra sign from $\mathcal{W}_{\bf a}$.

Finally we elaborate on the $F$-symbol \begin{align}\left[F^{\overline\rho\rho\rho}_\rho\right]_{\bf c}^{\overline\mu\mu}=\left\langle\vcenter{\hbox{\includegraphics[width=0.05\textwidth]{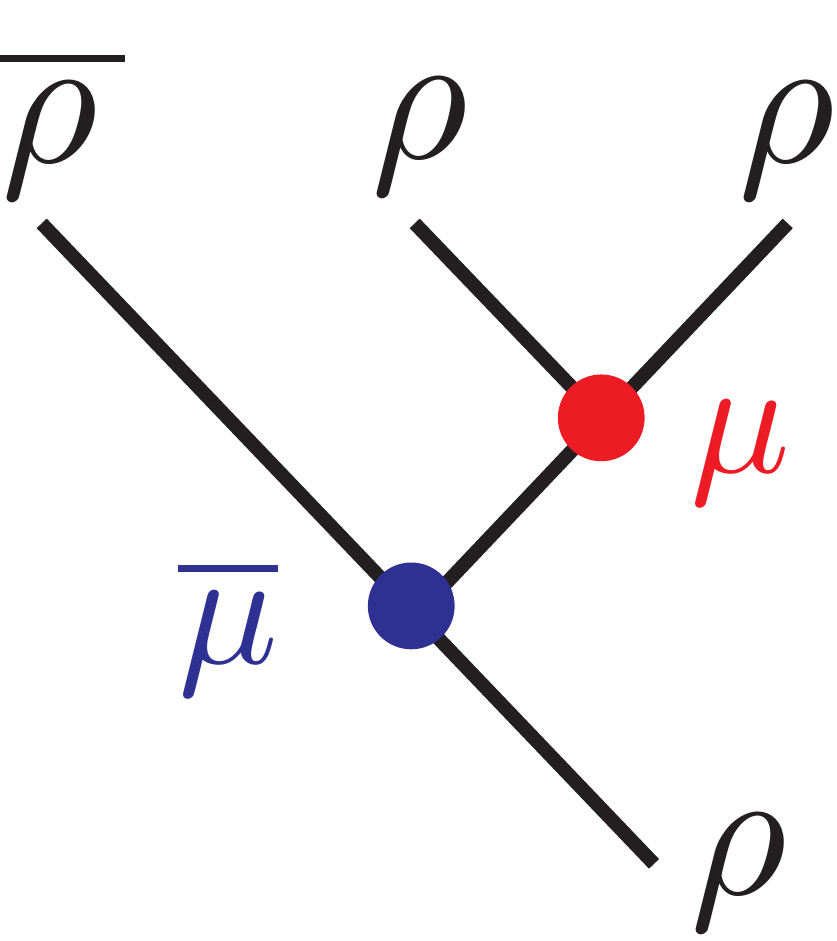}}}\right|\left.\vcenter{\hbox{\includegraphics[width=0.05\textwidth]{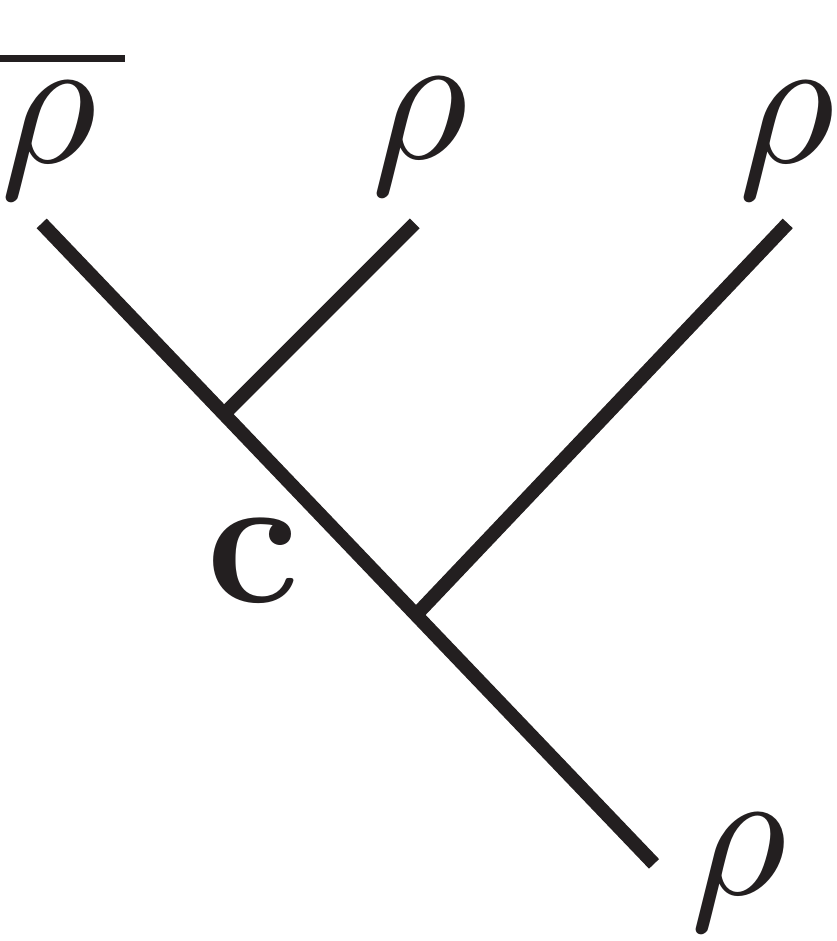}}}\right\rangle\end{align} although it was not used in the \hexeq's considered in Section~\ref{sec:so(8)symmetry}. The successive splittings $\rho\to\overline\rho\times\overline\rho\to\overline\rho\times(\rho\times\rho)$ give the 2-fold degenerate splitting spaces $V^{\bar\rho\bar\rho}$ and $V^{\rho\rho}$ spanned by $|\overline\mu\rangle=|\bar{0}\rangle,|\bar{1}\rangle$ and $|\mu\rangle=|0\rangle,|1\rangle$ respectively. There are three closed Wilson operators \begin{gather}\mathcal{A}_{\bf a}=\vcenter{\hbox{\includegraphics[width=0.1\textwidth]{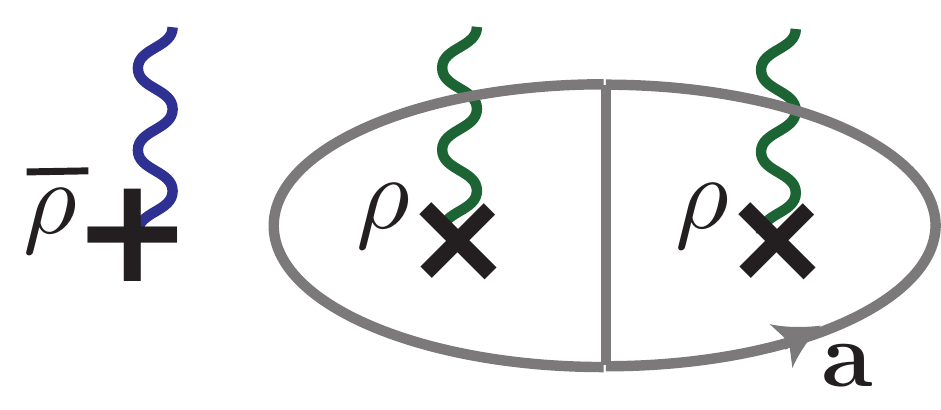}}},\quad\overline{\mathcal{A}}_{\bf a}=\vcenter{\hbox{\includegraphics[width=0.11\textwidth]{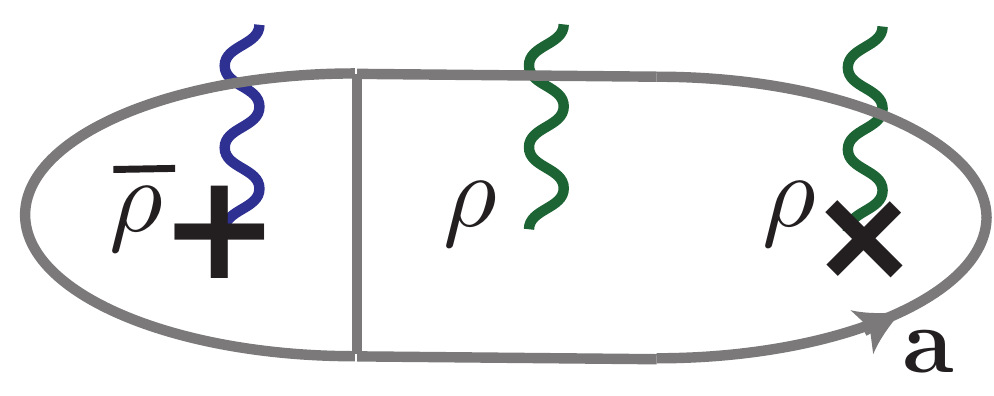}}},\nonumber\\\mathcal{W}_{\bf a}=\vcenter{\hbox{\includegraphics[width=0.1\textwidth]{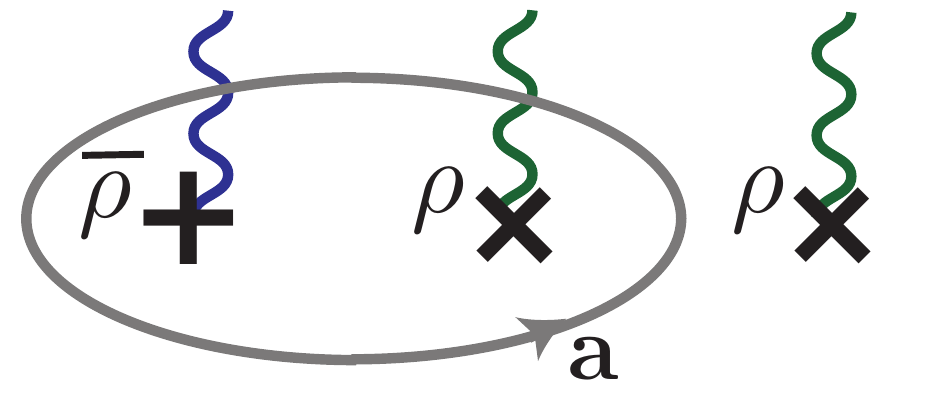}}}\end{gather} where $\mathcal{A}_{\bf a}$ and $\overline{\mathcal{A}}_{\bf b}$ commute and satisfy \begin{align}\mathcal{W}_{\bf a}=\mathcal{A}_{\Lambda_3{\bf a}}\overline{\mathcal{A}}_{\Lambda_3{\bf a}}.\label{WAbAapp}\end{align} For instance, for vacuum intermediate splitting channel ${\bf c}=0$, we have trivial braiding $\mathcal{W}_{\bf a}=\mathcal{D}S_{\bf ac}=1$. Eq.\eqref{WAbAapp} becomes the identification $\mathcal{A}_{\bf a}=-\overline{\mathcal{A}}_{\bf a}$ and restricts $V^{\bar\rho\bar\rho}\otimes V^{\rho\rho}$ onto a ray generated by the singlet $(|\bar{1}0\rangle-|\bar{0}1\rangle)/\sqrt{2}$. The equivalence $V^{\bar\rho\bar\rho}\otimes V^{\rho\rho}\cong\mbox{Hom}(V^{\rho\rho},V^{\bar\rho\bar\rho})$ translates the singlet into the $2\times2$ matrix \begin{align}\left[F_\rho^{\overline\rho\rho\rho}\right]_1=\frac{1}{\sqrt{2}}\left(\begin{array}{*{20}c}0&1\\-1&0\end{array}\right)=\frac{-1}{\sqrt{2}}\mathcal{A}_{\psi_2}\end{align} where the rows and columns are arranged in the order of $|\bar{1}\rangle,|\bar{0}\rangle$ and $|1\rangle,|0\rangle$ respectively. In general, $\left[F_\rho^{\overline\rho\rho\rho}\right]_{\bf c}$ with an arbitrary intermediate Abelian channel ${\bf c}$ can be represented by a $2\times2$ matrix presented in table~\ref{tab:so(8)Fsymbols}. $\left[F_\rho^{\rho\rho\overline\rho}\right]^{\bf c}$ can be derived by a similar manner.

\section{Fusion Rules  and Properties of the Defect Fusion Category of the ``4-Potts" State}\label{app:4statefusion}
In this appendix we will give the detailed derivation of the fusion rules of the defect fusion category for the ``4-Potts" state. We recall from earlier that  the defect fusion category is \begin{align}\mathcal{C}_{S_3}=\mathcal{C}_1\oplus\mathcal{C}_\theta\oplus\mathcal{C}_{\overline\theta}\oplus\mathcal{C}_{\alpha_1}\oplus\mathcal{C}_{\alpha_2}\oplus\mathcal{C}_{\alpha_3}\end{align} where $\mathcal{C}_1$ is generated by the 11 anyons in the parent state ``$SU(2)_1/Dih_2$" listed in Table~\ref{tab:4statePottsanyons}, and the other sectors are generated by threefold and twofold twist defects: \begin{align}\begin{array}{*{20}c}\mathcal{C}_\theta=\langle\theta,\omega\rangle,\quad\mathcal{C}_{\overline\theta}=\langle\overline\theta,\overline\omega\rangle\hfill\\\mathcal{C}_{\alpha_a}=\left\langle\alpha_a^0,\alpha_a^1,\alpha_a^2,\alpha_a^3,\boldsymbol\mu_a\right\rangle\end{array}.\end{align} 

We first explain the properties of the threefold defects in $\mathcal{C}_\theta$. The only non-trivial anyon in ``$SU(2)_1/Dih_2$" unaltered by a threefold symmetry is the semion super-sector $\Phi$. With this anyon we can  define a \emph{closed} Wilson loop \begin{align}\Theta_\Phi=\vcenter{\hbox{\includegraphics[width=0.05\textwidth]{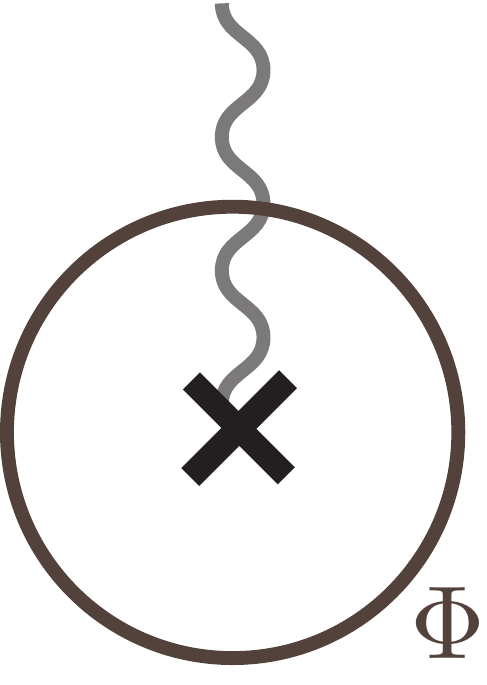}}}\end{align} around a threefold defect. From the $F$-symbols $F^{\Phi\Phi\Phi}_\Phi$ in \eqref{4PottsFsymbols}, it squares to unity. \begin{align}\Theta_\Phi^2&=\vcenter{\hbox{\includegraphics[width=0.05\textwidth]{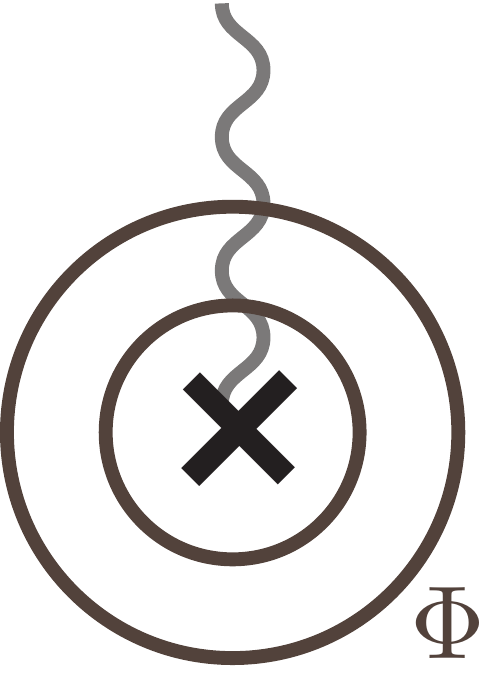}}}=\frac{1}{2}\left[\vcenter{\hbox{\includegraphics[width=0.06\textwidth]{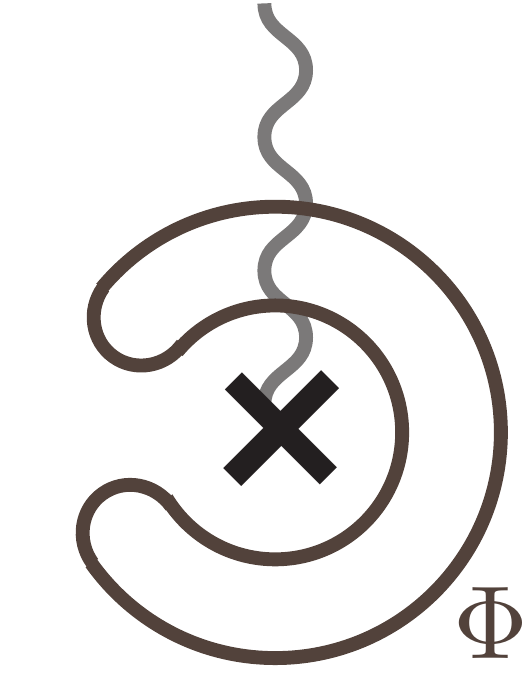}}}+\sum_{a=1}^3\vcenter{\hbox{\includegraphics[width=0.065\textwidth]{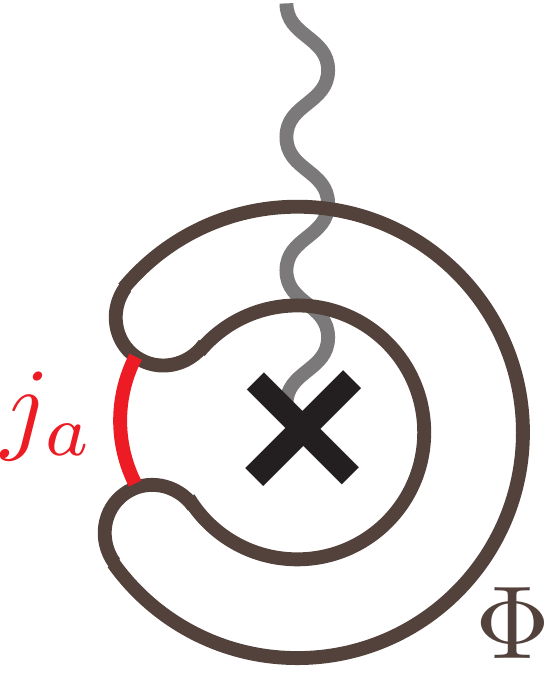}}}\right]\nonumber\\&=\frac{1}{2}\vcenter{\hbox{\includegraphics[width=0.05\textwidth]{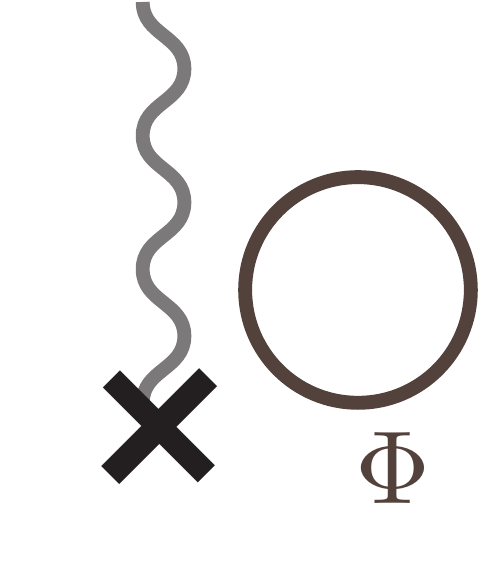}}}=\frac{d_\Phi}{2}\vcenter{\hbox{\includegraphics[width=0.033\textwidth]{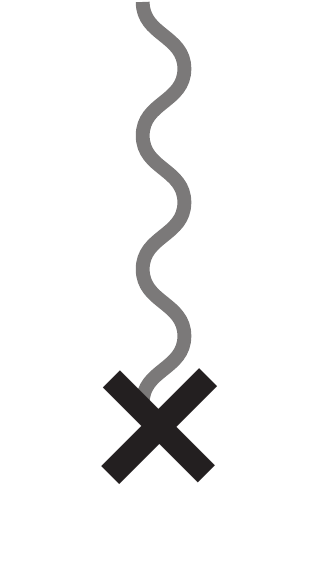}}}=1\label{ThetaPhisquare}\end{align} where each summand vanishes because the internal channel $j_a$ can be moved through the branch cut and changed so that the diagram vanishes: \begin{align}\vcenter{\hbox{\includegraphics[width=0.065\textwidth]{ThetaPhi21}}}=\vcenter{\hbox{\includegraphics[width=0.09\textwidth]{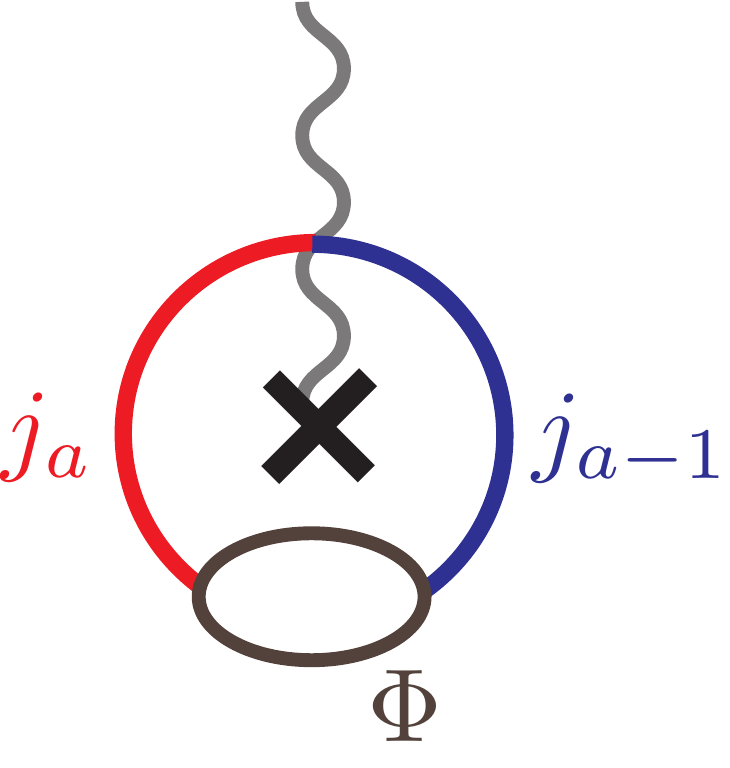}}}=0.\end{align} Eq.~\eqref{ThetaPhisquare} shows that the $\Phi$-loop has two possible eigenvalues $\Theta_\Phi=1$ and $-1$, which distinguish the two species of threefold defects $\theta$ and $\omega$ respectively: \begin{align}\vcenter{\hbox{\includegraphics[width=0.05\textwidth]{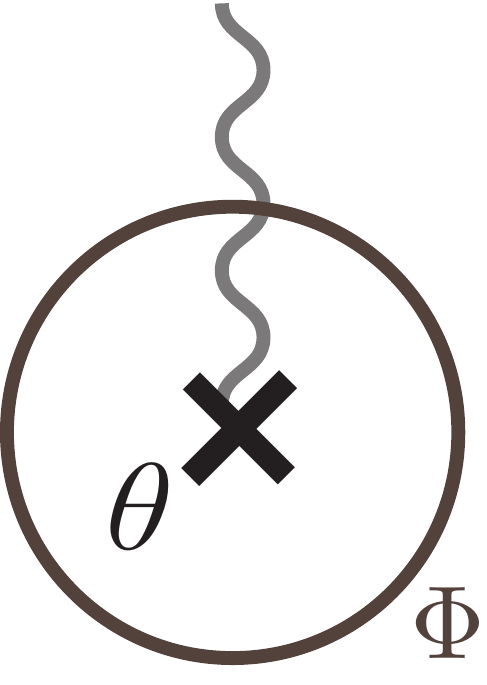}}}=+1,\quad\vcenter{\hbox{\includegraphics[width=0.05\textwidth]{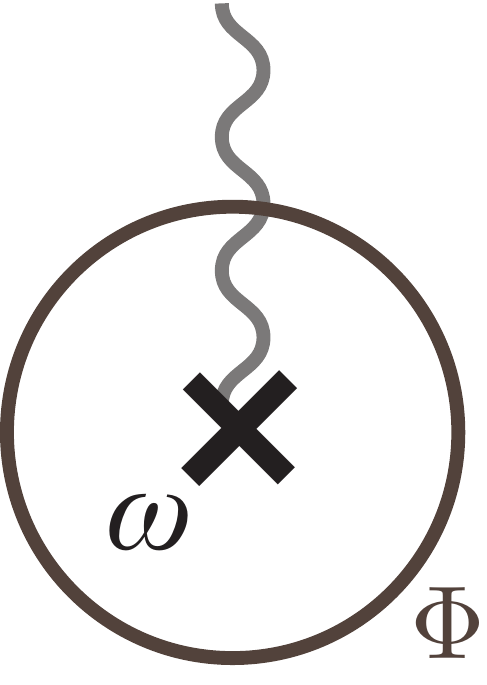}}}=-1.\label{Philoopapp}\end{align} The two species of threefold defects differ from each other by the fusion with the semion $\phi$ from the bosonic Laughlin state $SU(2)_1$, i.e.~$\theta\times\phi=\omega$. In the chiral ``4-Potts" state, the semion super-sector $\Phi$ is two dimensional, and gives rise to a fusion degeneracy \begin{align}\theta\times\Phi=2\omega,\quad\omega\times\Phi=2\theta.\end{align} This means the two threefold defects $\theta$ and $\omega$ differ from each other by a semion string $\phi$. The additional $\phi$-string crosses the $\Phi$-loop around the defect in \eqref{Philoopapp}, and gives the extra minus sign for $\Theta_\Phi$. 

Fusing a threefold defect with the bosons $j_a$ will not alter $\Theta_\Phi$ as the braiding phase between $\Phi$ and $j_a$ is trivial, and therefore \begin{align}\theta\times j_a=\theta,\quad\omega\times j_a=\omega.\end{align} Fusion associativity -- $x\times(y\times z)=(x\times y)\times z$ -- requires \begin{align}\theta\times\sigma_a=\omega\times\sigma_a=\theta\times\tau_a=\omega\times\tau_a=\theta+\omega.\end{align}


From the $S_3$-group product relation $\theta^2=\overline\theta$, a pair of threefold defects fuses to its anti-partner, i.e.~$\theta\times\theta\to\overline\theta$ (or $\overline\omega$). The splitting states of $\theta\times\theta$ are labeled by the good quantum numbers of the closed Wilson operators \begin{align}\hat{\mathcal{A}}_{j_a}=\vcenter{\hbox{\includegraphics[width=0.1\textwidth]{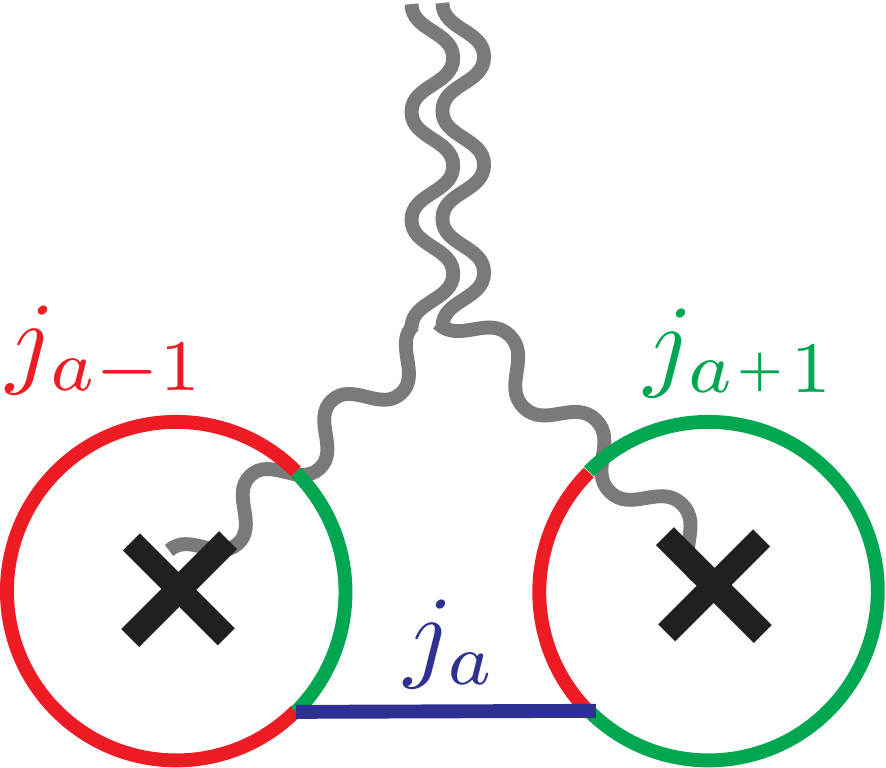}}}.\end{align} They mutually commute and satisfy the relations \begin{align}\hat{\mathcal{A}}_{j_1}\hat{\mathcal{A}}_{j_2}=\hat{\mathcal{A}}_{j_3},\quad\hat{\mathcal{A}}_{j_a}^2=1.\end{align} The four splitting states are specified by the simultaneous eigenvalues $\hat{\mathcal{A}}_{j_a}=(-1)^{s_a}$ for $(-1)^{s_1+s_2}=(-1)^{s_3}$. The fusion channels of $\theta\times\theta$ can be distinguished by a $\Phi$-loop, which, from the $F$-symbols in \eqref{4PottsFsymbols}, decomposes into \begin{gather}\vcenter{\hbox{\includegraphics[width=0.1\textwidth]{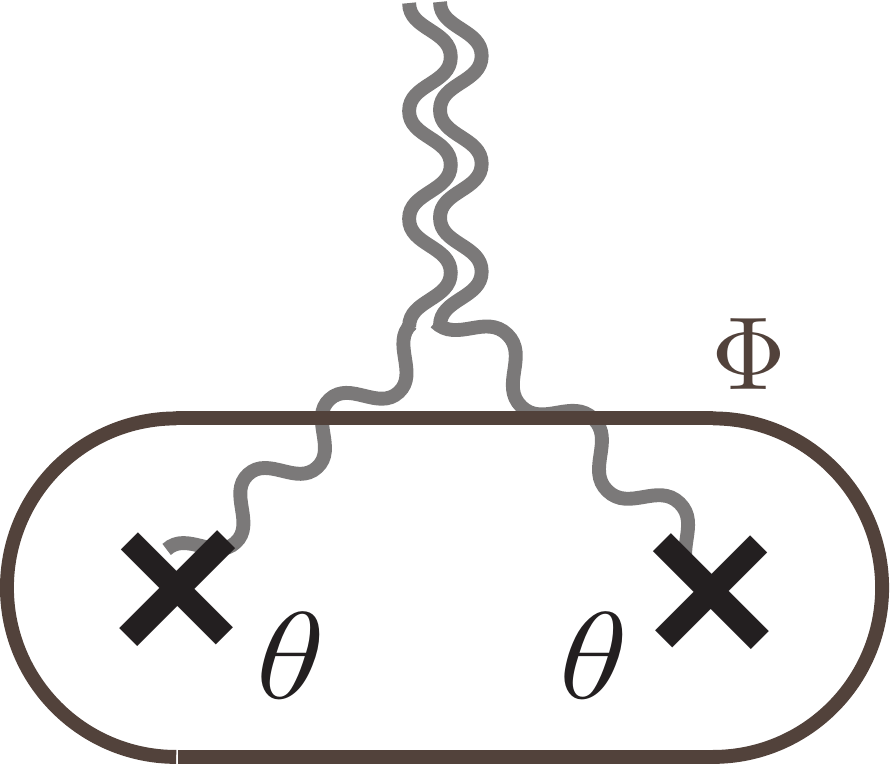}}}=\frac{1}{2}\left[\vcenter{\hbox{\includegraphics[width=0.1\textwidth]{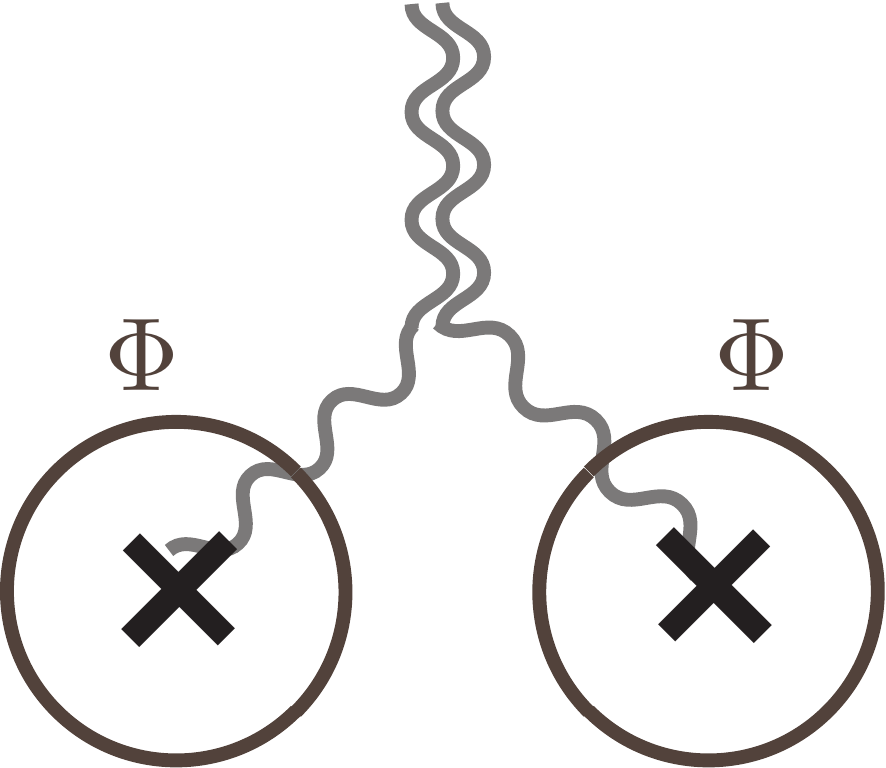}}}+\sum_{a=1}^3\vcenter{\hbox{\includegraphics[width=0.1\textwidth]{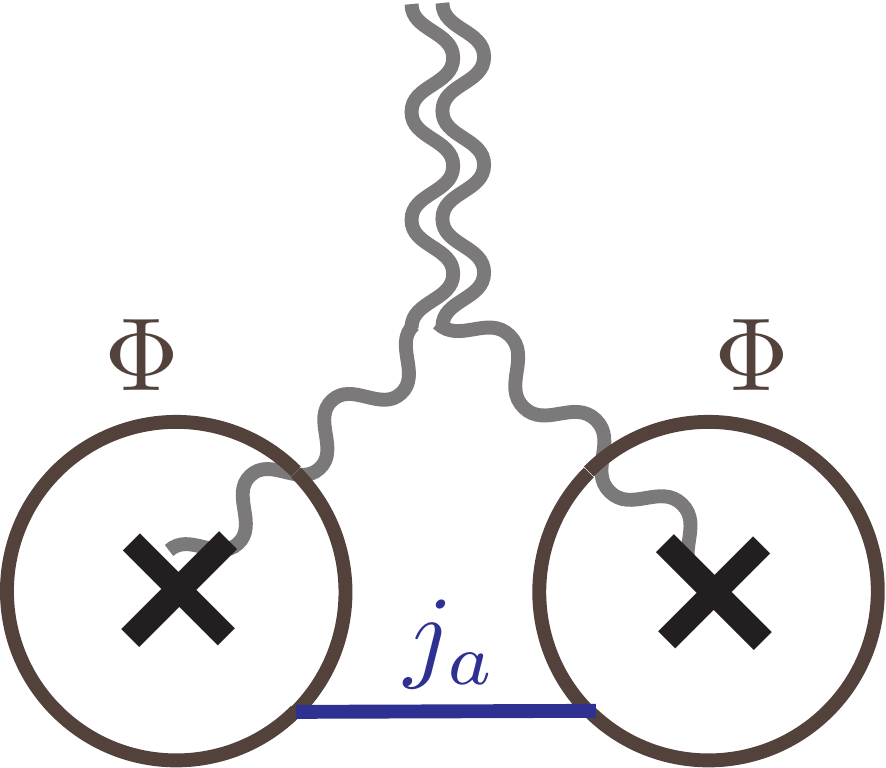}}}\right]\nonumber\\=\frac{1}{2}\left[\vcenter{\hbox{\includegraphics[width=0.1\textwidth]{APhi1}}}-\sum_{a=1}^3\vcenter{\hbox{\includegraphics[width=0.11\textwidth]{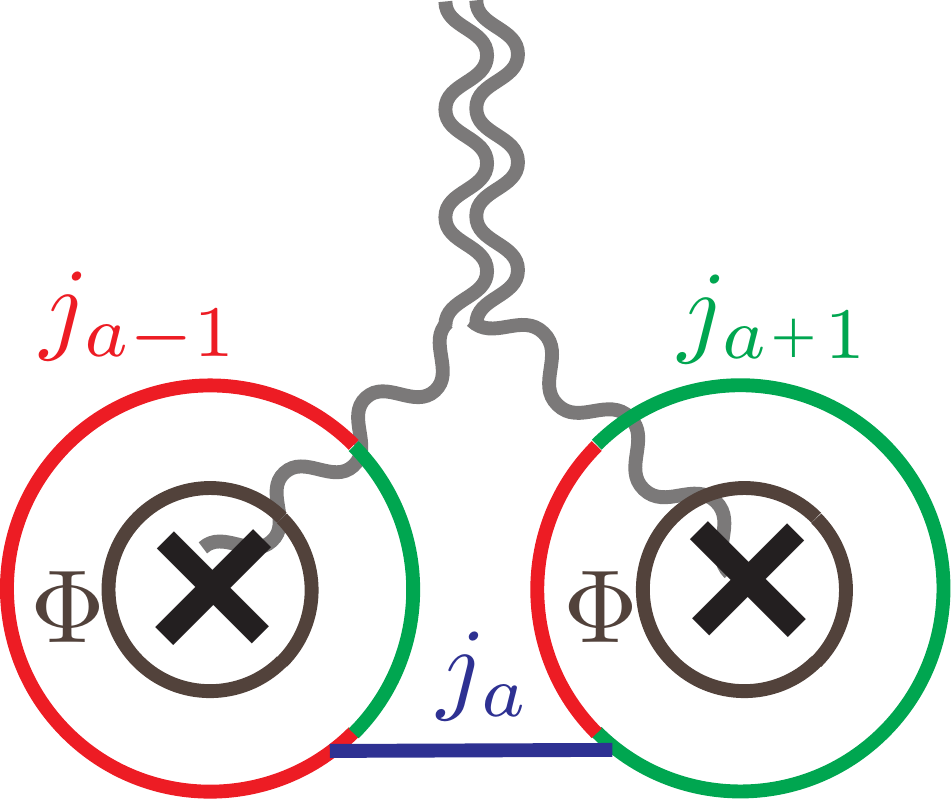}}}\right],\end{gather} where the small $\Phi$-loops around the $\theta$ defects are absorbed by the ground state. Hence \begin{align}\vcenter{\hbox{\includegraphics[width=0.1\textwidth]{APhi}}}&=\frac{1}{2}\left[\vcenter{\hbox{\includegraphics[width=0.1\textwidth]{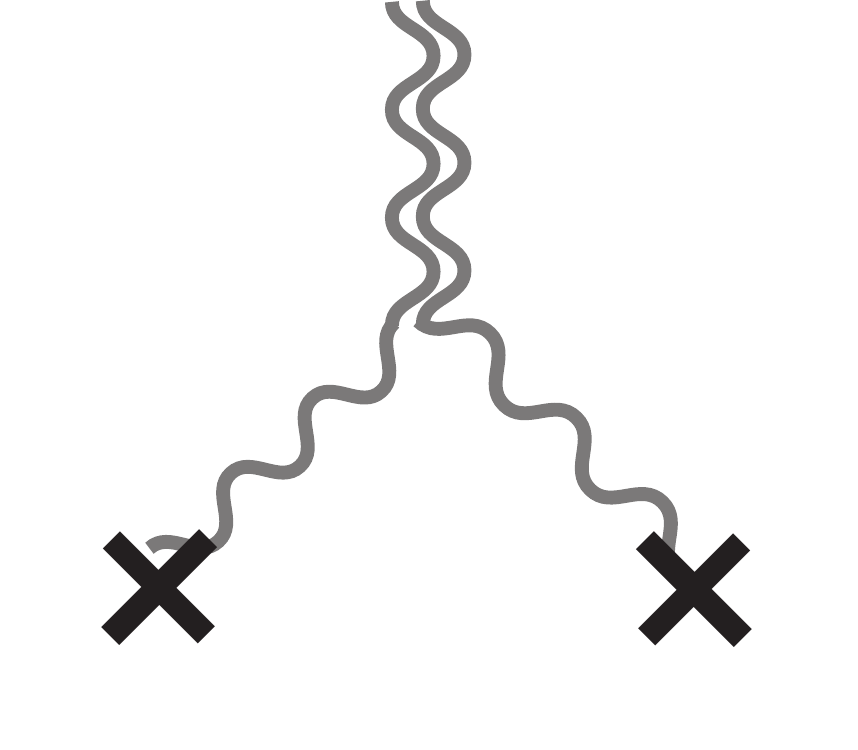}}}-\sum_{a=1}^3\vcenter{\hbox{\includegraphics[width=0.1\textwidth]{Aj}}}\right]\nonumber\\&=\frac{1}{2}\left(1-\sum_{a=1}^3\hat{\mathcal{A}}_{j_a}\right),\end{align} which is $-1$ if $(-1)^{s_1}=(-1)^{s_2}=1$, or $+1$ if otherwise. This shows the fusion rules \begin{align}\theta\times\theta=\omega\times\omega=\overline\omega+3\overline\theta,\quad\theta\times\omega=\overline\theta+3\overline\omega.\label{thetaxtheta}\end{align} The threefold fusion degeneracy is protected by the  closed Wilson loop algebra \begin{align}\hat{\mathcal{A}}_{\sigma_a}=\vcenter{\hbox{\includegraphics[width=0.1\textwidth]{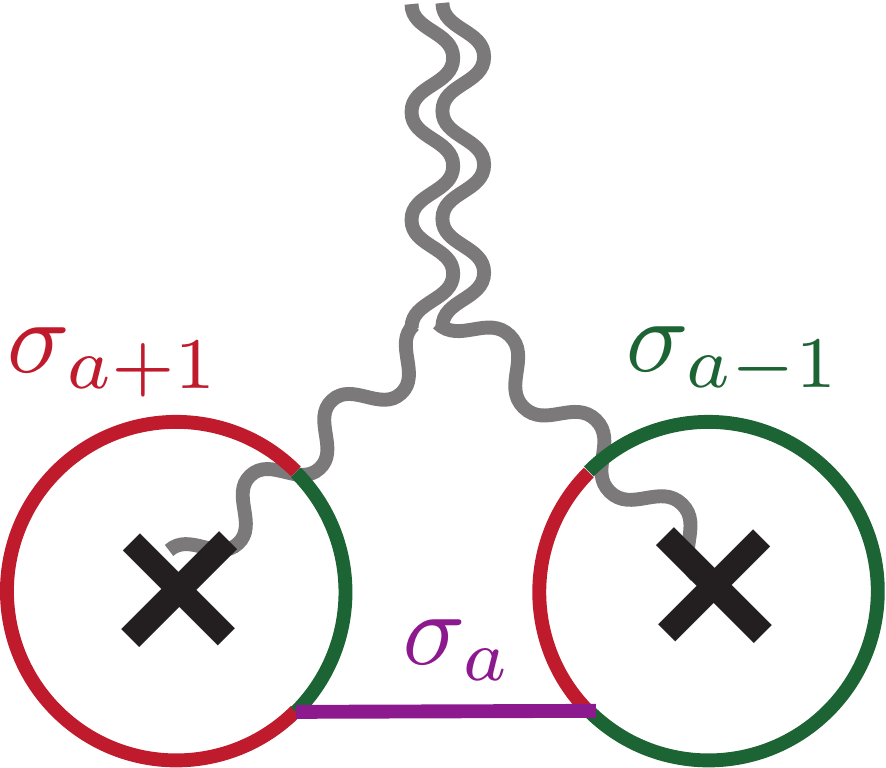}}},\end{align} which do not commute with the $\hat{\mathcal{A}}_{j_b}$'s. We also have \begin{align} \hat{\mathcal{A}}_{\sigma_a}\hat{\mathcal{A}}_{j_b}=-(-1)^{\delta_{ab}}\hat{\mathcal{A}}_{j_b}\hat{\mathcal{A}}_{\sigma_a}\end{align} due to the braiding phase between $\sigma_a$ and $j_b$.

Eq.~\eqref{thetaxtheta} enforces that the quantum dimension of the threefold defects \begin{align}d_\theta=d_\omega=4.\end{align} This is consistent with the fusion rules between conjugate pairs \begin{align}\theta\times\overline\theta&=\omega\times\overline\omega=1+\sum_{a=1}^3j_a+\sum_{a=1}^3\sigma_a+\sum_{a=1}^3\tau_a\\\theta\times\overline\omega&=\omega\times\overline\theta=2\Phi+\sum_{a=1}^3\sigma_a+\sum_{a=1}^3\tau_a\end{align} where the non-trivial fusion channels of $\theta\times\overline\theta\to\chi_a$ can be generated by the Wilson structure \begin{align}\vcenter{\hbox{\includegraphics[width=0.13\textwidth]{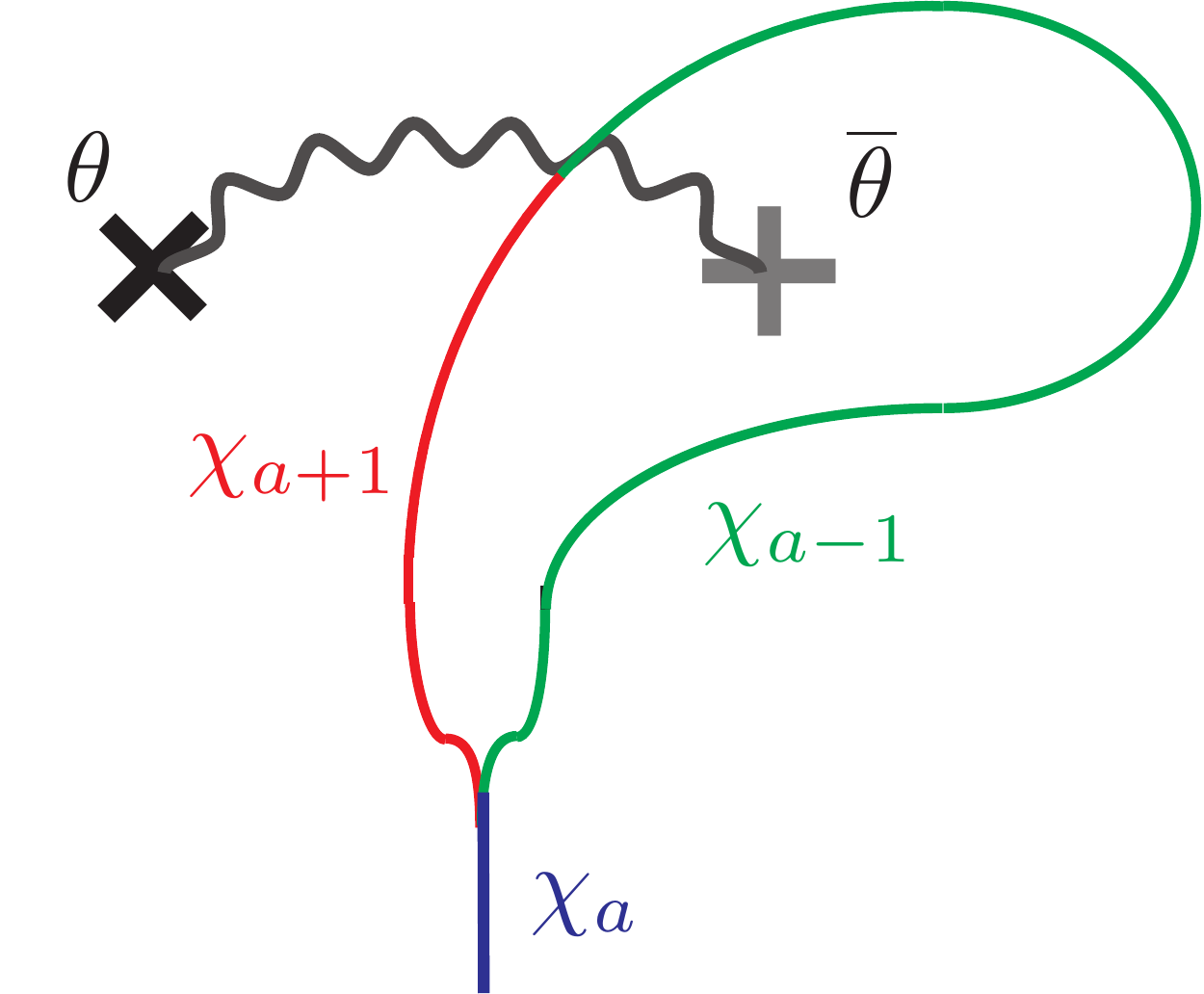}}}\end{align} for $\chi_a=j_a,\sigma_a,\tau_a$. The total quantum dimension for the defect sector $\mathcal{C}_\theta=\langle\theta,\omega\rangle$ is therefore \begin{align}\mathcal{D}_{\mathcal{C}_\theta}=\sqrt{d_\theta^2+d_\omega^2}=4\sqrt{2}=\mathcal{D}_0\end{align} which is also the total quantum dimension of the parent state.

Next, we explain the properties of the twofold defects in $\mathcal{C}_{\alpha_a}$. The non-trivial anyons in ``$SU(2)_1/Dih_2$" unaltered by the twofold symmetry $\alpha_a$ are $j_a$, $\Phi$, $\sigma_a,$ and $\tau_a$. These form the set of closed Wilson loops \begin{align}\Theta_{j_a}=\vcenter{\hbox{\includegraphics[width=0.05\textwidth]{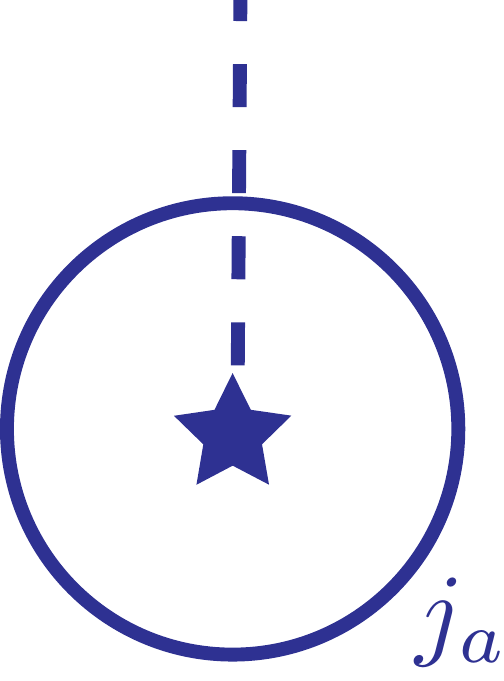}}},\quad\Theta_\Phi=\vcenter{\hbox{\includegraphics[width=0.05\textwidth]{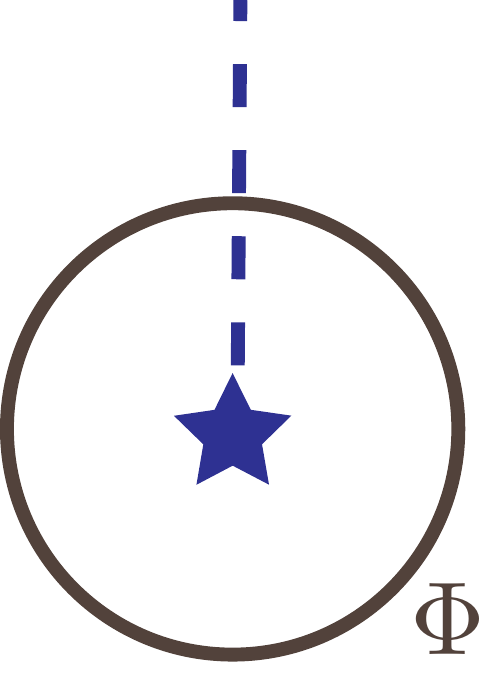}}},\nonumber\\\Theta_{\sigma_a}=\vcenter{\hbox{\includegraphics[width=0.05\textwidth]{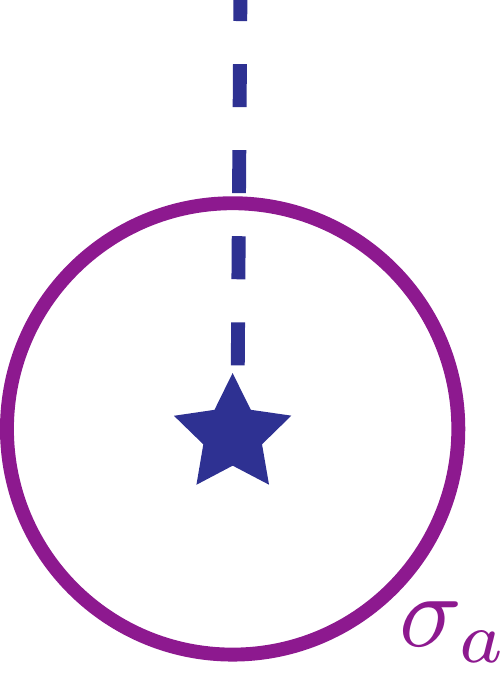}}},\quad\Theta_{\tau_a}=\vcenter{\hbox{\includegraphics[width=0.05\textwidth]{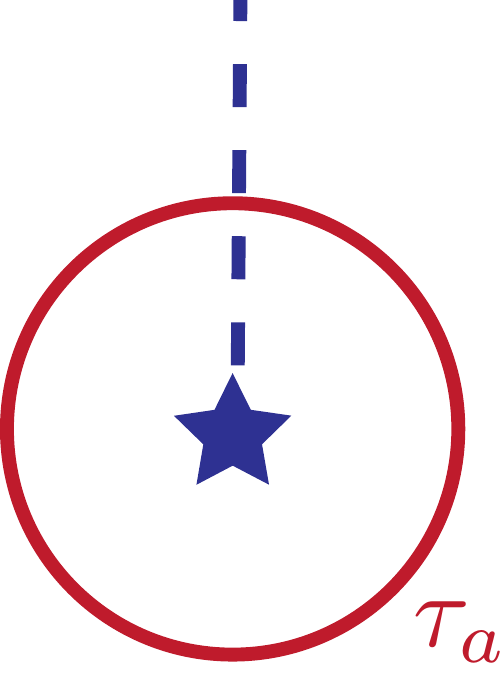}}}\end{align} around the twofold defect. They mutually commute as they can pass through one another without intersecting, but they are not independent. From the fusion rules \eqref{4statePottsfusion}, and the $F$-symbols in \eqref{4PottsFsymbols}, they satisfy \begin{gather}\Theta_{j_a}\Theta_\Phi=\Theta_\Phi,\quad\Theta_{j_a}\Theta_{\sigma_a}=\Theta_{\sigma_a},\quad\Theta_{j_a}\Theta_{\tau_a}=\Theta_{\tau_a}\nonumber\\\Theta_{j_a}^2=1,\quad\Theta_\Phi^2=1+\Theta_{j_a}\nonumber\\\Theta_{\sigma_a}\Theta_{\tau_a}=\Theta_\Phi\label{alphaloopalgebra}\\\Theta_{\sigma_a}^2=1+\Theta_{j_a}+\Theta_\Phi,\quad\Theta_{\tau_a}^2=1+\Theta_{j_a}-\Theta_\Phi\nonumber\\\Theta_\Phi\Theta_{\sigma_a}=\Theta_{\sigma_a}+\Theta_{\tau_a},\quad\Theta_\Phi\Theta_{\tau_a}=\Theta_{\sigma_a}-\Theta_{\tau_a}.\nonumber\end{gather} For example, $\Theta_\Phi^2$ can be evaluated by $F^{\Phi\Phi\Phi}_\Phi$ \begin{align}\vcenter{\hbox{\includegraphics[width=0.05\textwidth]{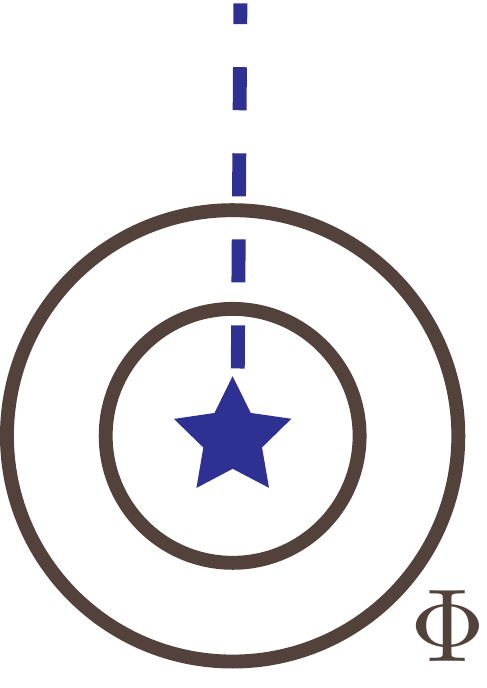}}}&=\frac{1}{2}\left[\vcenter{\hbox{\includegraphics[width=0.06\textwidth]{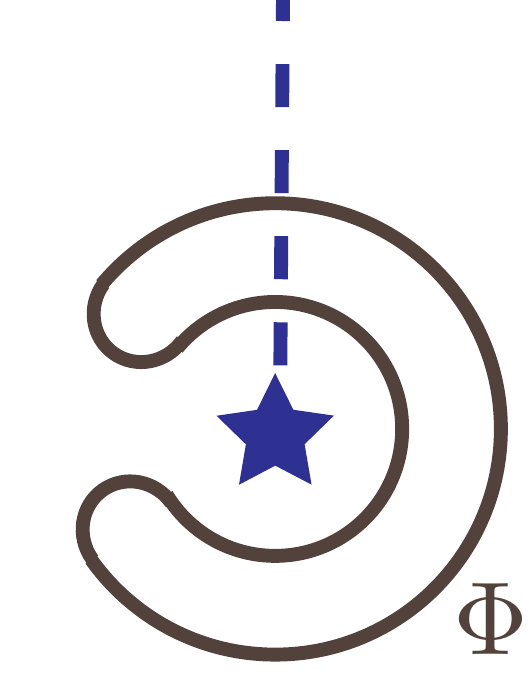}}}+\sum_{b=1}^3\vcenter{\hbox{\includegraphics[width=0.065\textwidth]{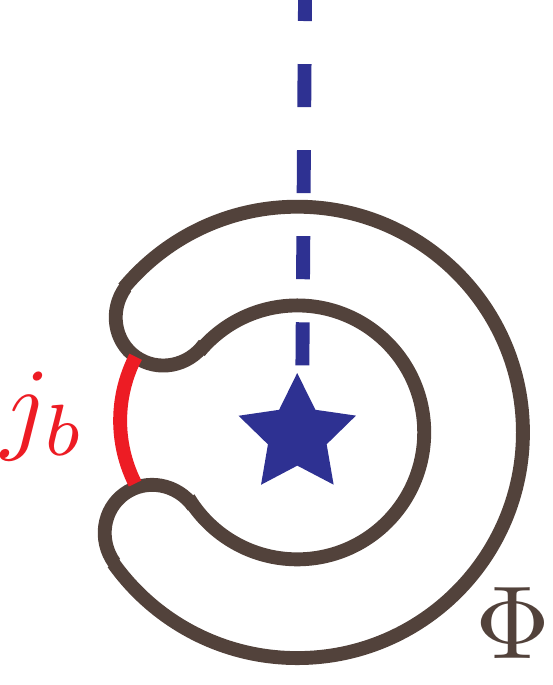}}}\right]\nonumber\\&=\frac{1}{2}\left[\vcenter{\hbox{\includegraphics[width=0.05\textwidth]{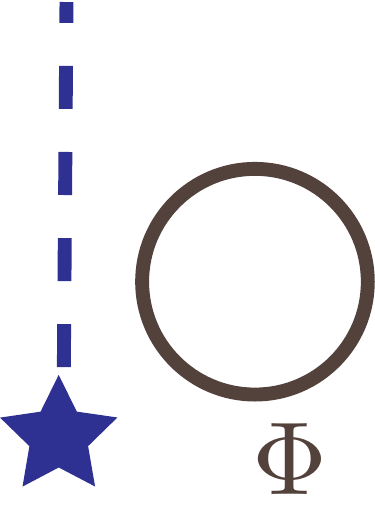}}}+\vcenter{\hbox{\includegraphics[width=0.065\textwidth]{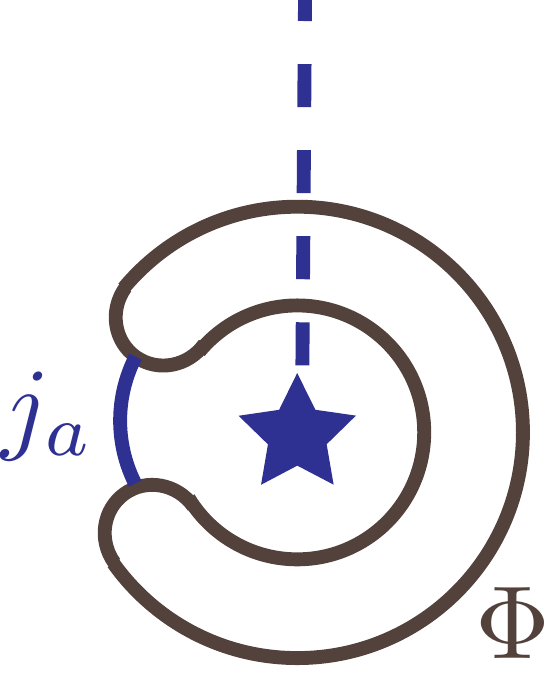}}}\right]=\vcenter{\hbox{\includegraphics[height=0.05\textwidth]{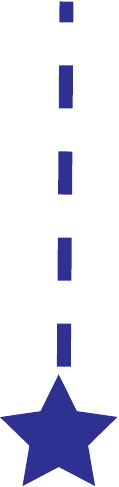}}}+\vcenter{\hbox{\includegraphics[width=0.05\textwidth]{alphaja}}}\end{align} where the other two summands involving $j_{a\pm1}$ vanish because their internal channels can be dragged past the branch cut and changed to lead to a vanishing diagram: \begin{align}\vcenter{\hbox{\includegraphics[width=0.065\textwidth]{alphaPhi21}}}=\vcenter{\hbox{\includegraphics[width=0.09\textwidth]{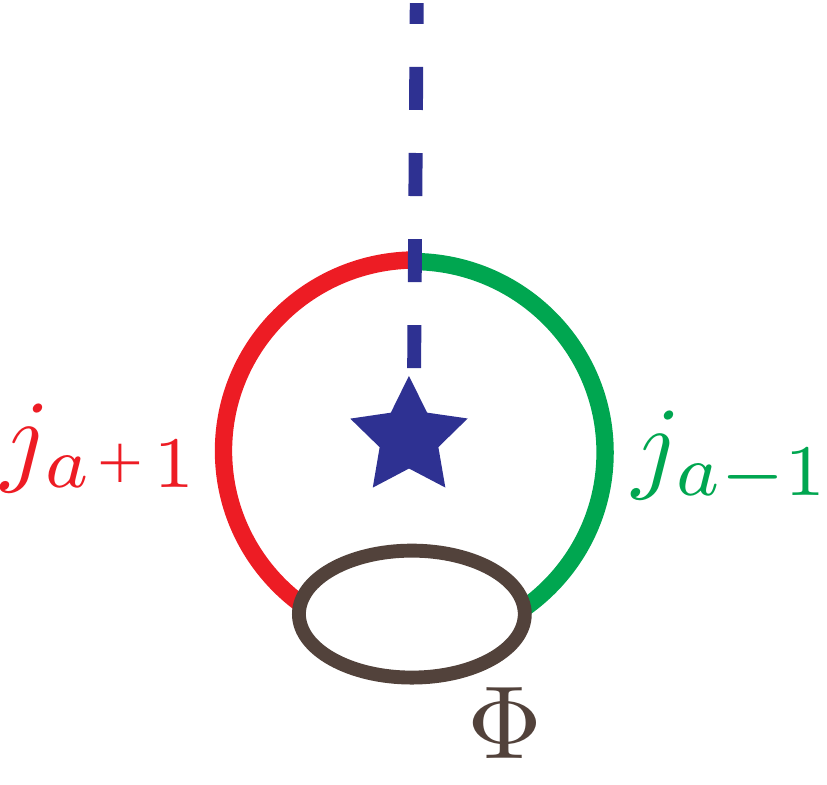}}}=0.\end{align} The Wilson loop $\Theta_{\sigma_a}^2$ can be evaluated by $F^{\sigma_a\sigma_a\sigma_a}_{\sigma_a}$ \begin{align}\vcenter{\hbox{\includegraphics[width=0.05\textwidth]{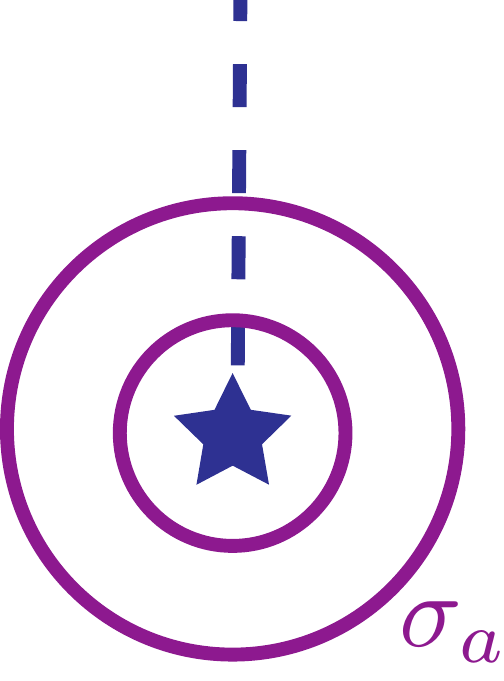}}}&=\frac{1}{2}\left[\vcenter{\hbox{\includegraphics[width=0.06\textwidth]{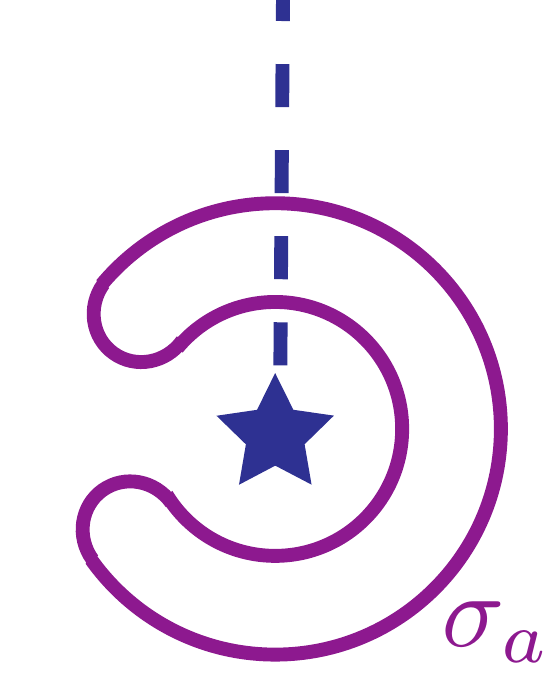}}}+\vcenter{\hbox{\includegraphics[width=0.065\textwidth]{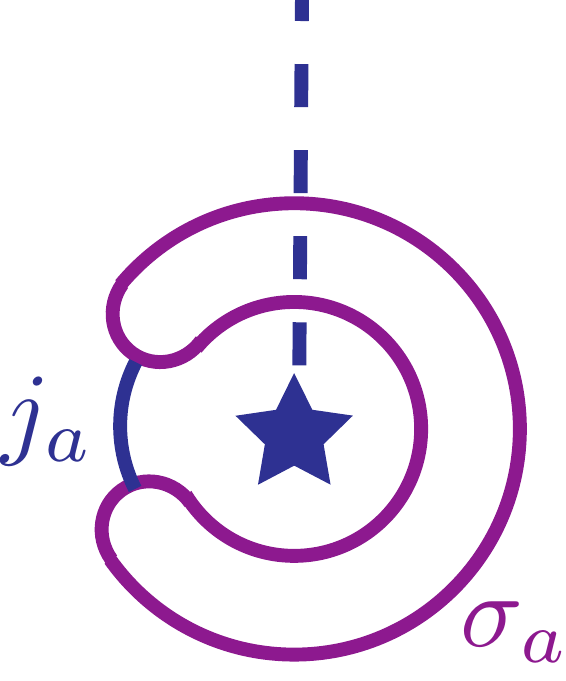}}}+\sqrt{2}\vcenter{\hbox{\includegraphics[width=0.065\textwidth]{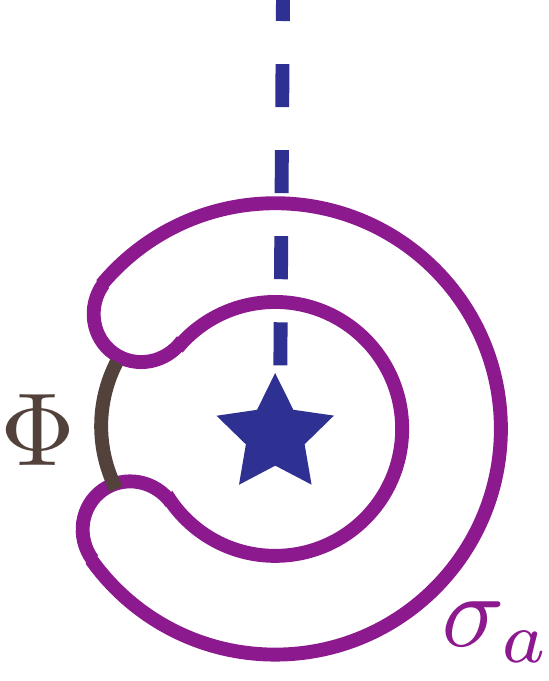}}}\right]\nonumber\\&=\vcenter{\hbox{\includegraphics[height=0.05\textwidth]{alphaPhi32}}}+\vcenter{\hbox{\includegraphics[width=0.05\textwidth]{alphaja}}}+\vcenter{\hbox{\includegraphics[width=0.05\textwidth]{alphaPhi}}}.\end{align} Other products of \eqref{alphaloopalgebra} are determined by associativity, i.e.~$\Theta_x(\Theta_y\Theta_z)=(\Theta_x\Theta_y)\Theta_z,$ and by a redefinition of signs in front of the Wilson operators $\Theta_x\to-\Theta_x$ if necessary. 

The Wilson loops $\Theta_{j_a},\Theta_{\Phi},\Theta_{\sigma_a},\Theta_{\tau_a}$ form an associative and commutative ring with identity. The species labels of twofold defects are representations of this ring. Let us see what this implies. First, $\Theta_{j_a}$ has eigenvalues $\pm1$ since it squares to unity. The product relations \eqref{alphaloopalgebra} then force the eigenvalues of the other Wilson operators to satisfy \begin{align}&\left\{\begin{array}{*{20}c}\Theta_\Phi=\sqrt{2}(-1)^s\hfill\\\Theta_{\sigma_a}=2\cos\left(\frac{\pi}{8}-\frac{s\pi}{2}\right)\\\Theta_{\tau_a}=2\sin\left(\frac{\pi}{8}-\frac{s\pi}{2}\right)\end{array}\right.,\quad\mbox{for }\Theta_{j_a}=+1\label{alphaloopvalues}\\&\Theta_\Phi=\Theta_{\sigma_a}=\Theta_{\tau_a}=0,\quad\mbox{for }\Theta_{j_a}=-1\label{muloopvalues}\end{align} for $2\cos(\pi/8)=\sqrt{2+\sqrt{2}}$, $2\sin(\pi/8)=\sqrt{2-\sqrt{2}}$.

The first possibility, which is listed in Eq.~\eqref{alphaloopvalues}, gives four twofold defects $\alpha^s_a,$ one for each value of $s\in\mathbb{Z}_4=\{0,1,2,3\}$. The $\Theta_\Phi$ and $\Theta_{\sigma_a}$ loops change signs upon fusing the defect with anyons that intersect non-trivially with $\Phi$ and $\sigma_a$. Fusion associativity requires \begin{gather}\alpha^s_a\times j_a=\alpha^s_a,\quad\alpha^s_a\times j_{a\pm1}=\alpha^{s+2}_a\nonumber\\\alpha^s_a\times\Phi=\alpha^{s+1}_a+\alpha^{s-1}_a\\\alpha^s_a\times\sigma_a=\alpha^{-s}_a+\alpha^{-s+1}_a.\nonumber\end{gather} The second possibility, which is listed in Eq.~\eqref{muloopvalues}, describes another twofold defect $\boldsymbol\mu_a$, which is related to the others by \begin{align}&\boldsymbol\mu_a=\alpha_a^s\times\sigma_{a\pm1}=\alpha_a^s\times\tau_{a\pm1}\label{musigmarelation}\\&\boldsymbol\mu_a\times\sigma_{a\pm1}=\boldsymbol\mu_a\times\tau_{a\pm1}=\sum_{s=0}^3\alpha_a^s.\end{align} 

The fusion channels of the a defect pair $\alpha^{s_1}_a\times\alpha^{s_2}_a$ are distinguished by the mutually commuting Wilson loops \begin{align}\mathcal{W}_x=\vcenter{\hbox{\includegraphics[width=0.08\textwidth]{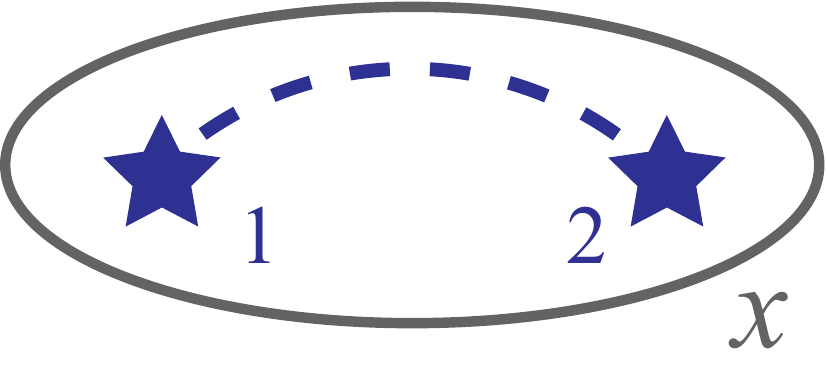}}}\end{align} where $x$ are anyons in the ``4-Potts" state. This Wilson algebra depends only on $\mathcal{W}_{j_{a-1}}$ and $\mathcal{W}_{\sigma_{a-1}}$ as the others are generated by the product relations: \begin{gather}\mathcal{W}_{j_a}=\Theta^1_{j_a}\Theta^2_{j_a}=1,\quad\mathcal{W}_{j_{a+1}}=\Theta^1_{j_a}\Theta^2_{j_a}\mathcal{W}_{j_{a-1}}=\mathcal{W}_{j_{a-1}}\nonumber\\\mathcal{W}_\Phi=\frac{1}{2}\Theta^1_\Phi\Theta^2_\Phi\left(1+\mathcal{W}_{j_{a-1}}\right)\nonumber\\\mathcal{W}_{\sigma_a}=\frac{1}{2}\left(\Theta^1_{\sigma_a}\Theta^2_{\sigma_a}+\Theta^1_{\tau_a}\Theta^2_{\tau_a}\mathcal{W}_{j_{a-1}}\right)\\\mathcal{W}_{\sigma_{a+1}}\mathcal{W}_{\sigma_{a-1}}=\frac{1}{\sqrt{2}}\left(\mathcal{W}_{\sigma_a}+\mathcal{W}_{\tau_a}\right)\nonumber\\\mathcal{W}_{\tau_a}=\mathcal{W}_{\sigma_a}\mathcal{W}_{j_{a-1}},\quad\mathcal{W}_{\tau_{a\pm1}}=\mathcal{W}_{\sigma_{a\pm1}}\mathcal{W}_{j_a}=\mathcal{W}_{\sigma_{a\pm1}}\nonumber\end{gather} where $\Theta^1$ (or $\Theta^2$) is the Wilson loop around only the first (resp.~second) defect. These product relations can be evaluated by using the $F$-symbols in \eqref{4PottsFsymbols}. For example, the $\Phi$-loop can be resolved into \begin{align}\vcenter{\hbox{\includegraphics[width=0.08\textwidth]{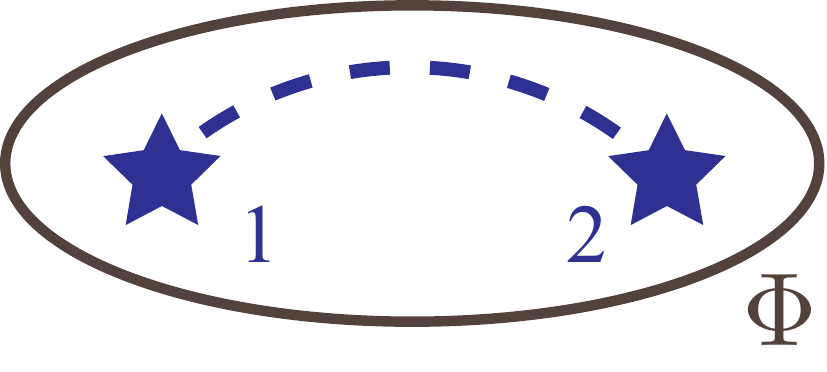}}}&=\frac{1}{2}\left[\vcenter{\hbox{\includegraphics[width=0.08\textwidth]{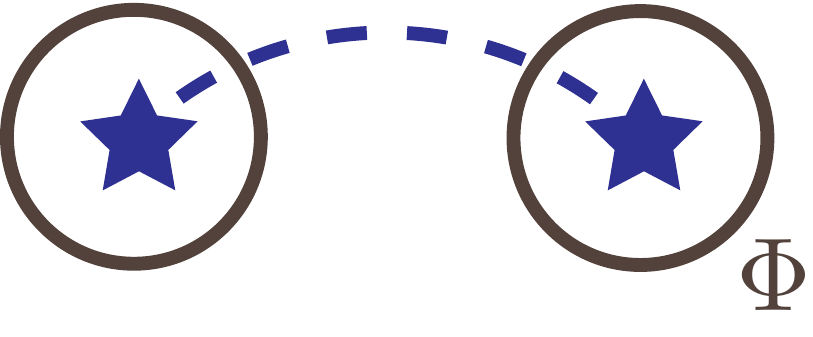}}}+\sum_{b=1}^3\vcenter{\hbox{\includegraphics[width=0.08\textwidth]{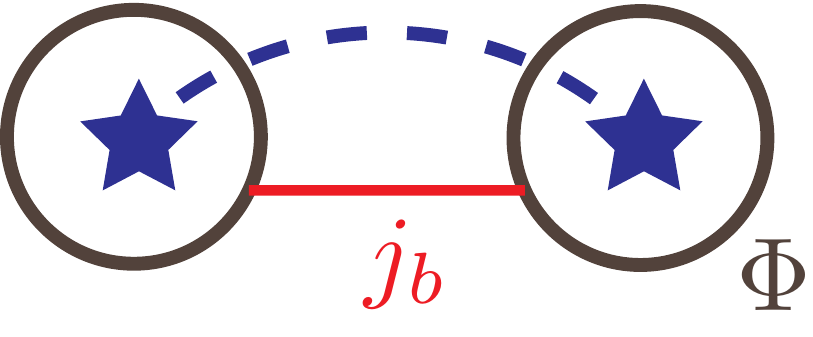}}}\right]\nonumber\\&=\frac{1}{2}\left[\vcenter{\hbox{\includegraphics[width=0.08\textwidth]{WPhi1}}}+\vcenter{\hbox{\includegraphics[width=0.09\textwidth]{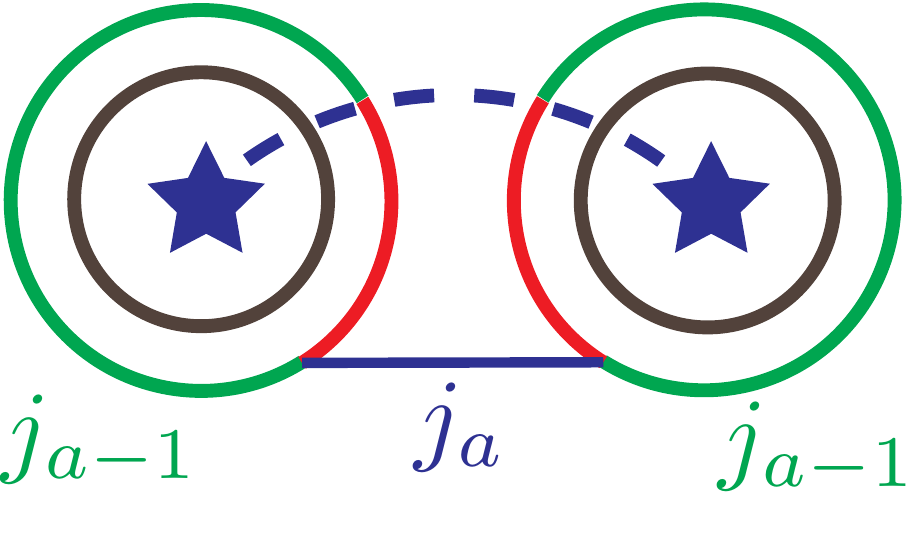}}}\right]\nonumber\\&=\frac{1}{2}\left[\vcenter{\hbox{\includegraphics[width=0.08\textwidth]{WPhi1}}}+\vcenter{\hbox{\includegraphics[width=0.09\textwidth]{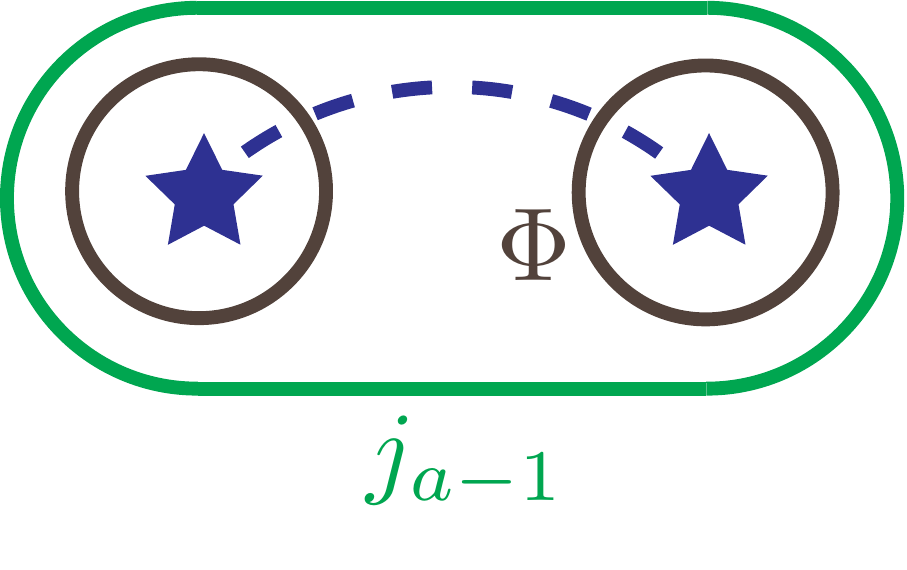}}}\right]\end{align} where the $j_b=j_{a\pm1}$ terms in the summand vanish. And the $\sigma_a$-loop is \begin{align}\vcenter{\hbox{\includegraphics[width=0.08\textwidth]{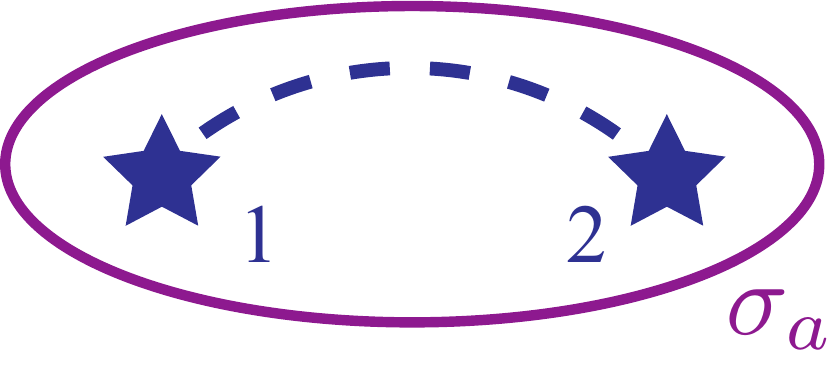}}}&=\frac{1}{2}\left[\vcenter{\hbox{\includegraphics[width=0.08\textwidth]{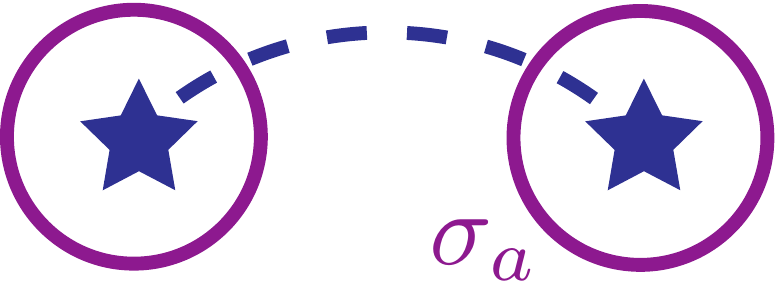}}}+\vcenter{\hbox{\includegraphics[width=0.08\textwidth]{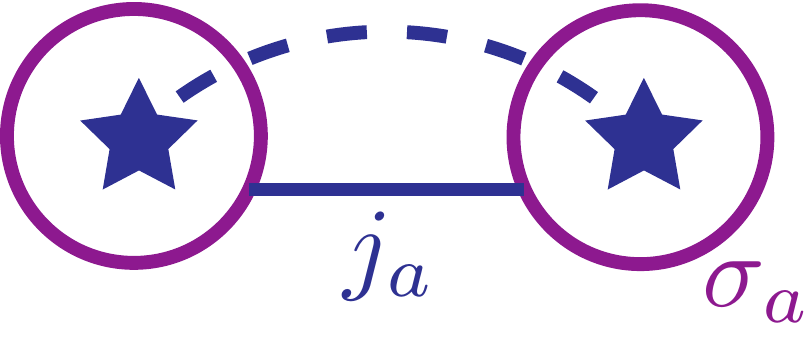}}}+\sqrt{2}\vcenter{\hbox{\includegraphics[width=0.08\textwidth]{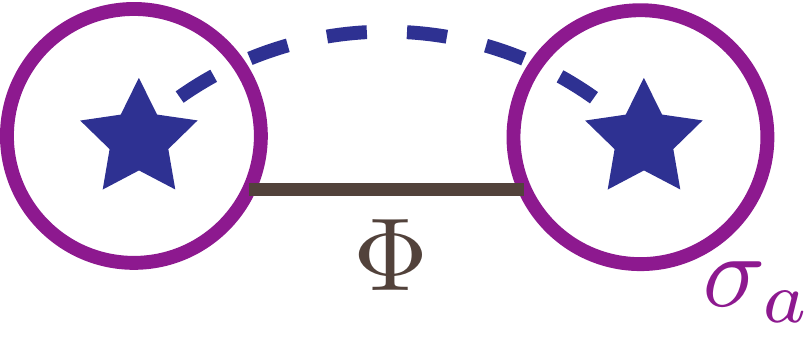}}}\right]\nonumber\\&=\frac{1}{2}\left[\vcenter{\hbox{\includegraphics[width=0.08\textwidth]{Wsigma1}}}+\vcenter{\hbox{\includegraphics[width=0.09\textwidth]{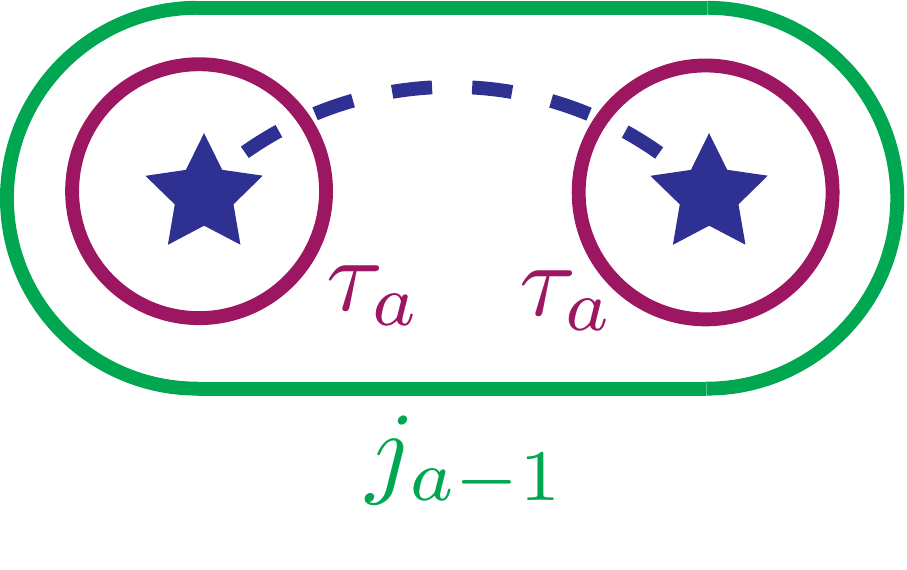}}}\right]\end{align} where the last term involving the intermediate channel $\Phi$ vanishes.

The fusion channels of $\alpha^{s_1}_a\times\alpha^{s_2}_a\to y$ are determined by the eigenvalues of the Wilson loop \begin{align}\mathcal{W}_x=\frac{1}{d_y}\mathcal{D}_0S_{xy}\end{align} where the $S$-matrix is given in \eqref{4statePottsSmatrix}, and $d_y$ is the quantum dimension of the fusion channel $y$. The fusion rules are summarized by \begin{align}\alpha_a^s\times\alpha_a^s&=\left\{\begin{array}{*{20}c}1+j_a+\sigma_a,&\mbox{for $s$ even}\\1+j_a+\tau_a,&\mbox{for $s$ odd}\end{array}\right.\nonumber\\\alpha_a^s\times\alpha_a^{s+2}&=\left\{\begin{array}{*{20}c}j_{a-1}+j_{a+1}+\tau_a,&\mbox{for $s$ even}\\j_{a-1}+j_{a+1}+\sigma_a,&\mbox{for $s$ odd}\end{array}\right.\label{alphaalpha}\\\alpha_a^s\times\alpha_a^{s+1}&=\left\{\begin{array}{*{20}c}\Phi+\sigma_a,&\mbox{for $s$ even}\\\Phi+\tau_a,&\mbox{for $s$ odd}\end{array}\right. .\nonumber\end{align} 

The fusion rule for a pair of $\boldsymbol\mu_a$'s can be deduced using \eqref{musigmarelation} \begin{align}\boldsymbol\mu_a\times\boldsymbol\mu_a=1+\sum_{b=1}^3j_b+2\Phi+2\sigma_a+2\tau_a.\label{mumu}\end{align} 

Eqs.~\eqref{alphaalpha} and \eqref{mumu} imply the defect quantum dimensions are \begin{align}d_{\alpha_a^s}=2,\quad d_{\boldsymbol\mu_a}=4.\end{align} The total quantum dimension for the defect sector $\mathcal{C}_{\alpha_a}=\langle\alpha^{0}_a,\alpha^{1}_a,\alpha^{2}_a,\alpha^{3}_a,\boldsymbol\mu_a\rangle$ is therefore \begin{align}\mathcal{D}_{\mathcal{C}_{\alpha_a}}=\sqrt{4d_{\alpha_a^{s}}^2+d_{\boldsymbol\mu_a}^2}=4\sqrt{2}=\mathcal{D}_0,\end{align} which again is the total quantum dimension of the parent state. 

Finally, we describe the fusion rules between twofold and threefold defects. Similar to the $SO(8)_1$ state, the non-abelian $S_3$-symmetry in the ``4-Potts" state implies non-commutative fusion rules: \begin{gather}\alpha_a^{s_1}\times\alpha_{a+1}^{s_2}=\left\{\begin{array}{*{20}c}\theta,&\mbox{for $s_1+s_2$ even}\\\omega,&\mbox{for $s_1+s_2$ odd}\end{array}\right.\nonumber\\\alpha_a^{s_1}\times\alpha_{a-1}^{s_2}=\left\{\begin{array}{*{20}c}\overline\theta,&\mbox{for $s_1+s_2$ even}\\\overline\omega,&\mbox{for $s_1+s_2$ odd}\end{array}\right.\nonumber\\\alpha_a^s\times\theta=\overline\theta\times\alpha_a^s=\alpha_{a+1}^s+\alpha_{a+1}^{s+2}+\boldsymbol\mu_{a+1}\nonumber\\\alpha_a^s\times\omega=\overline\omega\times\alpha_a^s=\alpha_{a+1}^{s-1}+\alpha_{a+1}^{s+1}+\boldsymbol\mu_{a+1}\nonumber\\\boldsymbol\mu_a\times\theta=\boldsymbol\mu_a\times\omega=2\boldsymbol\mu_{a+1}+\sum_{s=0}^3\alpha_{a+1}^s\nonumber\\\theta\times\boldsymbol\mu_a=\omega\times\boldsymbol\mu_a\times=2\boldsymbol\mu_{a-1}+\sum_{s=0}^3\alpha_{a-1}^s.\nonumber\end{gather} The first and second identities come from the the product relation $\alpha_a\alpha_{a+1}=\theta$ and $\alpha_a\alpha_{a-1}=\theta^{-1}$ of the symmetry group $S_3$, where the $\theta$ or $\omega$ channel is distinguished by the $\Phi$-loop \begin{align}\vcenter{\hbox{\includegraphics[width=0.09\textwidth]{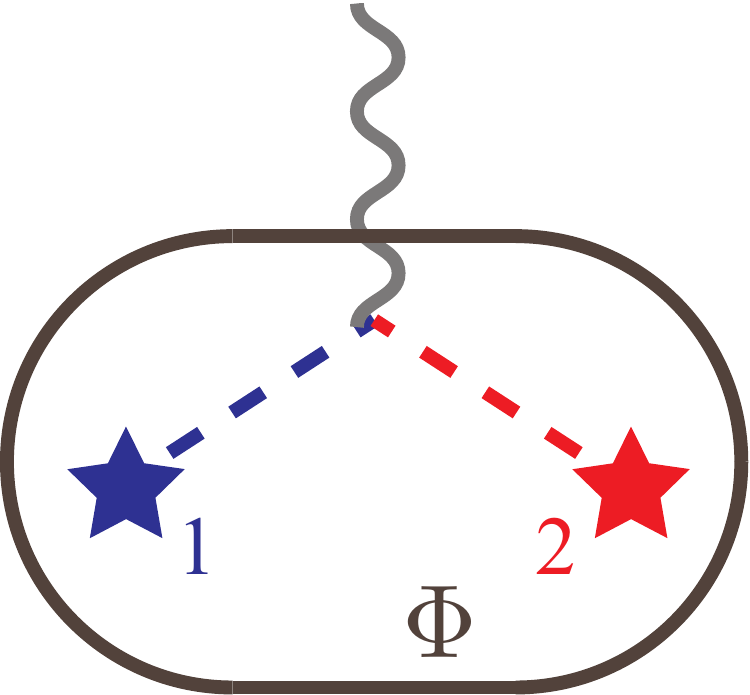}}}&=\frac{1}{2}\left[\vcenter{\hbox{\includegraphics[width=0.09\textwidth]{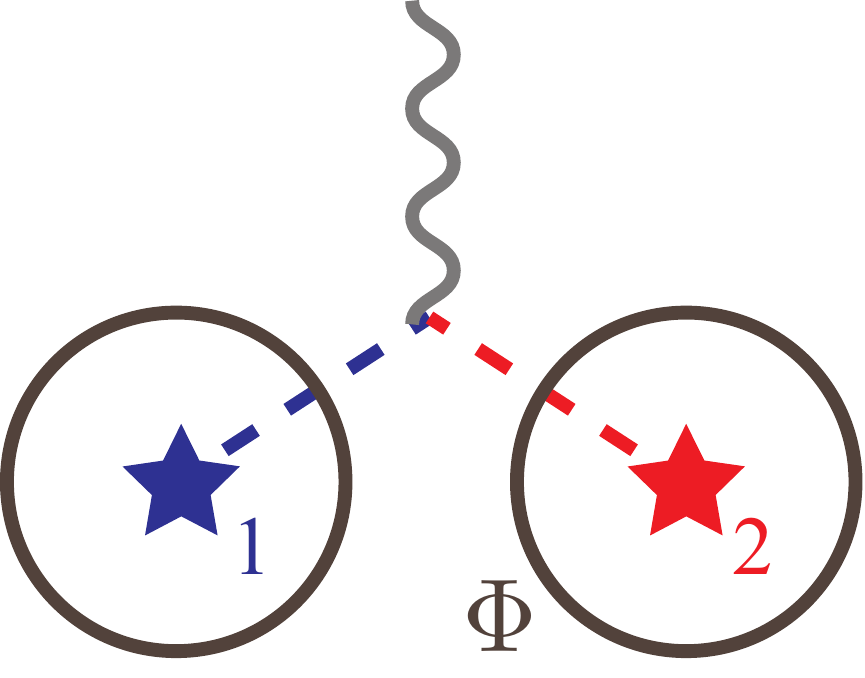}}}+\sum_{b=1}^3\vcenter{\hbox{\includegraphics[width=0.09\textwidth]{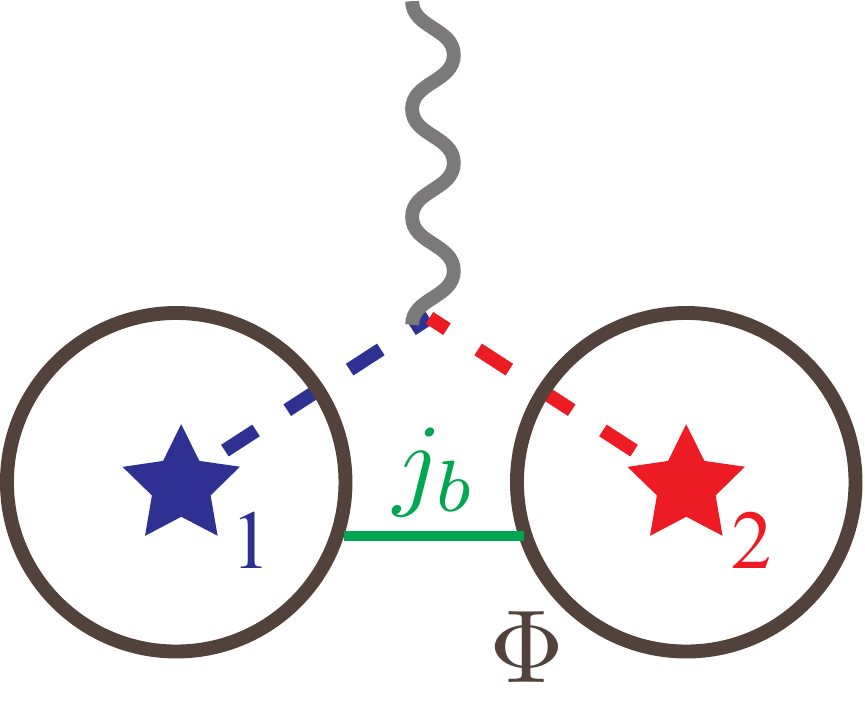}}}\right]\nonumber\\&=\frac{1}{2}\Theta^1_\Phi\Theta^2_\Phi=(-1)^{s_1+s_2}.\end{align} The other identities are consequences of fusion associativity.


\begin{thebibliography}{157}
\bibitem [{foo({\natexlab{a}})}]{footnotequantumop}%
  \BibitemOpen
  \bibinfo {note} {A non-trivial anyon
  cannot appear on its own in a closed system. A global quantum operator or
  state is a linear combination of tensor products of local ones over matching
  boundary conditions.}\BibitemShut {Stop}%
\end{thebibliography}
\end{document}